\documentclass[11pt,a4paper]{article}

\pdfoutput=1
\usepackage{multirow,fancyvrb}
\usepackage{empheq,enumerate,amsbsy}
\usepackage{upgreek,bbm,mathtools,subfigure}

\usepackage{jheppub}

\usepackage{%
  MnSymbol}
\usepackage{natbib}

\makeatletter
\newcommand{\m}[1]{%
  \begingroup
  \def\FV@Space{ }%
  \mathcode`\_="8000 %
  \do@@us %
  $#1$%
  \endgroup
}
\newcommand{\do@@us}{%
  \begingroup\lccode`~=`\_ \lowercase{\endgroup\let~}\sb
}
\makeatother

\newcommand{\lL}[1][a]{\ensuremath{\hat\lambda_{#1}}}
\newcommand{\lM}[1][i]{\ensuremath{\hat\nu_{#1}}}
\newcommand{\uL}[1][a]{\ensuremath{\hat\lambda^\star_{#1}}}
\newcommand{\uM}[1][i]{\ensuremath{\hat\nu^\star_{#1}}}
\newcommand{\x}{\ensuremath{\mathsf{x}}}
\newcommand{\y}{\ensuremath{\mathsf{y}}}
\newcommand{\z}{\ensuremath{\mathsf{z}}}
\newcommand{\be}{\begin{eqnarray}}
\newcommand{\ee}{\end{eqnarray}}
\newcommand{\Pexp}[2]{\ensuremath{\mathrm{P}\,e^{\int_{#1}^{#2}A(v)\mathrm{d}v}}}

\newcommand{\bi}{\ensuremath{\mathbbm i}}

\newcommand{\nI}{\ensuremath{|I|}}
\newcommand{\nA}{\ensuremath{|A|}}
\newcommand{\dbless}[1]{{\leftarrow#1}}
\newcommand{\ket}[1]{\ensuremath{\left|#1\right\rangle}}

\newcommand{\Bset}{\ensuremath{\mathcal{B}}}
\newcommand{\Fset}{\ensuremath{\mathcal{F}}}
\newcommand{\Sset}{\ensuremath{\mathcal{S}}}

\newcommand{\Kone}{\ensuremath{K_1}}
\newcommand{\Ktwo}{\ensuremath{K_2}}
\newcommand{\Mone}{\ensuremath{M_1}}
\newcommand{\Mtwo}{\ensuremath{M_2}}

\newcommand{\so}{{\mathsf{so}}}%
\newcommand{\su}{{\mathsf{su}}}%
\newcommand{\psu}{{\mathsf{psu}}}%
\newcommand{\GL}{{\mathsf{GL}}}%
\newcommand{\gl}{{\mathsf{gl}}}%
\renewcommand{\sl}{{\mathsf{sl}}}%
\newcommand{\SL}{{\mathsf{SL}}}%
\newcommand{\SU}{{\mathsf{SU}}}%
\newcommand{\bzdQ}{q}%

\newcommand{\secref}[1]{section~\ref{#1}}
\newcommand{\appref}[1]{appendix~\ref{#1}}

\newcommand{\HQ}[1]{Q^{#1}}%
\newcommand{\HP}[1]{P^{#1}}%
\newcommand{\HQs}[2]{(\HQ{#1})^{#2}}%
\newcommand{\HPs}[2]{(\HP{#1})^{#2}}%
\newcommand{\hHQ}[1]{{\bzdQ}^{#1}}%
\newcommand{\hHQs}[2]{(\hHQ{#1})^{#2}}%
\newcommand{\nQ}{\mathsf{q}}
\def\tm{{\mathfrak{m}}}
   \def\bT{{\bf T}}
    \def\bQ{{\bf Q}}
    \def\bP{{\bf P}}

      \def\bp{{\bf p}}
        \def\e{{\epsilon}}
\newcommand{\fQ}{{\mathcal
    Q}}
\newcommand{\p}{\partial}

  \def\es{{\emptyset}}
   \def\bT{{\bf T}}
    \def\bQ{{\bf Q}}
    \def\bP{{\bf P}}
    \def\bp{{\bf p}}

    \def\tx{{\mathsf{x}}}
  \def\ty{{\mathsf{y}}}
    \def\tG{{G}}
  
    \def\tm{{\mathfrak{m}}}
    \def\PR{{\mathsf{\Pi}}}

\newcommand{\CO}{{\mathcal O}}

\newcommand{\CP}{{\mathcal{P}}}

\DeclareMathOperator{\Pf}{Pf}

\title{T-system  on T-hook: Grassmannian Solution and Twisted
  Quantum Spectral Curve}

\author[a]{\hfill Vladimir Kazakov}
\author[b]{\hfill S\'ebastien Leurent}
\author[c,d]{\hfill Dmytro Volin}
\author{\hfill\rule{0pt}{0pt}\\[.5cm]
\hfill    \includegraphics{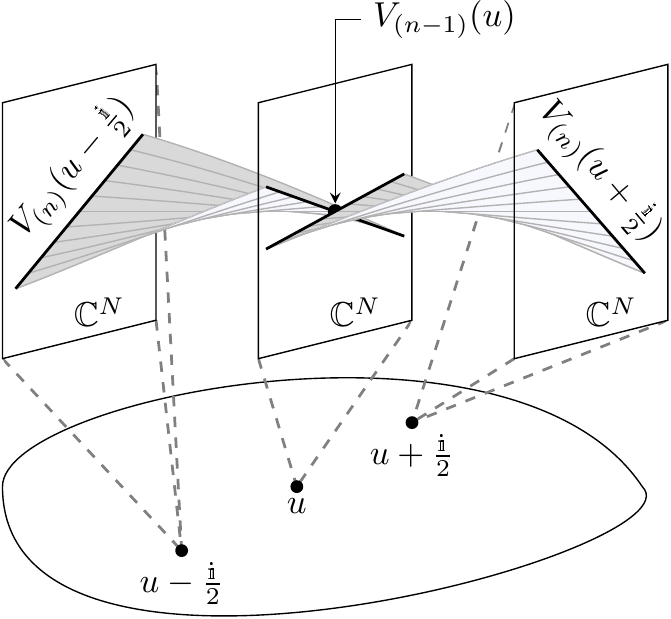}\hfill}

\affiliation[a]{LPT, \'Ecole Normale Superieure, 24, rue Lhomond 75005 Paris, France\\ \& Universit\'e Paris-VI, Place Jussieu, 75005 Paris, France}
\affiliation[b]{Institut de Math\'ematiques de Bourgogne, UMR 5584 du CNRS,\\ Univ. Bourgogne Franche-Comt\'e, 9 avenue Alain Savary, 21000 DIJON, France.}
\affiliation[c]{School of Mathematics, Trinity College Dublin, College
  Green, Dublin 2, Ireland.}
\affiliation[d]{Nordita
KTH Royal Institute of Technology and Stockholm University,\\
Roslagstullsbacken 23, SE-106 91 Stockholm, Sweden
}

\abstract{ We propose an efficient grassmannian formalism for solution of bi-linear finite-difference Hirota equation (T-system) on T-shaped lattices related to the space of highest weight representations of \(\gl(K_1,K_2|M)\) superalgebra. The formalism is inspired by the quantum fusion procedure known from the integrable spin chains and is based on exterior forms of Baxter-like Q-functions. We find a few new interesting relations among the exterior forms of Q-functions and reproduce, using our new formalism, the Wronskian determinant solutions of Hirota equations known in the literature. Then we generalize this construction to the twisted Q-functions and demonstrate the subtleties of untwisting procedure on the examples of rational quantum spin chains with twisted boundary conditions. Using these observations, we generalize the recently discovered, in our paper with N.~Gromov, AdS/CFT Quantum Spectral Curve for exact planar spectrum of AdS/CFT duality to the case of arbitrary Cartan twisting of AdS\(_5\times\)S\(^5\) string sigma model. Finally, we successfully probe this formalism by reproducing the energy of gamma-twisted BMN vacuum at single-wrapping orders of weak coupling expansion. }

\newcommand{\xx}{{\bf x}}

\newcommand{\yy}{{\bf y}}

\DeclareMathOperator*{\Det}{det}
\DeclareMathOperator*{\sdet}{sdet}

\newcommand{\bC}{\mathbb{C}}

\newcommand{\bg}{\zeta} %

\newcommand{\chs}[1]{{\chi_{{\rule{0pt}{1.7ex}#1}}}}

\newcommand{\commented}[1]{{}}

\newcommand{\THook}{\texorpdfstring{\ensuremath{\mathbb T}}{T}-hook}
\newcommand{\LHook}{\texorpdfstring{\ensuremath{\mathbb L}}{L}-hook}
\newcommand{\fHook}{fat hook}
\newcommand{\FHook}{Fat hook}

\usepackage{amssymb}
\usepackage{makeidx}
\makeindex
\usepackage{amsmath}

\newcommand{\gp}{g_{\vphantom{1}(+)}}
\newcommand{\gm}{g_{\vphantom{1}(-)}}
\newcommand{\gpm}{g_{\vphantom{1}(\pm)}}
\usepackage{varioref}

\begin{document}

\VerbatimFootnotes

 \maketitle

\section{Introduction}
In 1931, Hans Bethe analysed the very first example of a quantum integrable model -- Heisenberg SU(2) XXX spin chain -- and showed that it can be reduced to algebraic equations which now bear his name \cite{Bethe:1931hc}. The roots of these equations,  called Bethe roots, enter the observable quantities only through their symmetric combinations. This is one of many reasons to work with the Baxter Q-polynomial -- a polynomial with zeros at Bethe roots, \(Q(u)=\prod_{k=1}^L(u-u_k)\). Later, several different techniques have been developed to determine \(Q(u)\). For instance, instead of the Bethe equations one can use the Baxter equation
\be\label{Baxter}
\phi(u+\frac {\bi}2) Q(u+{\bi})+\phi(u-\frac {\bi}2) Q(u-{\bi})=T(u)\,Q(u)\,,\ \ \ \phi(u)=u^L\,,
\ee
and search for such solutions that \(Q(u)\) and \(T(u)\) are both polynomials.

Another reformulation of the same problem is to demand the Wronskian identity\be\label{Wronskian}
W\equiv\frac 1{\phi(u)}
\left|
\begin{matrix}
Q_1(u+\frac {\bi}2) & Q_1(u-\frac {\bi}2)
\\
Q_2(u+\frac {\bi}2) & Q_2(u-\frac {\bi}2)
\end{matrix}
\right|=1
\ee
to be satisfied. Indeed, it is easy to show that for any two solutions \(Q_1(u)\), \(Q_2(u)\) of the Baxter equation the Wronskian combination \(W\) is an \(\bi\)-periodic function. We can further normalize the solutions so as to put \(W=1\),  resulting in \eqref{Wronskian}. Then it is enough to demand that both \(Q_1\) and \(Q_2\) solving \eqref{Wronskian} are polynomials to get solutions equivalent to the polynomial solutions of \eqref{Baxter}. On this example we see that there are actually two  Q-functions appearing. \\[0em]

The Wronskian condition \eqref{Wronskian} can be interpreted in a natural geometric way. Consider \(\mathbb{C}^2\) and denote by \(\zeta_1\) and \(\zeta_2\) two basis vectors in it. Then we can introduce a one-form
\be
Q_{(1)}\equiv Q_1\zeta_1+Q_2\zeta_2\,.
\ee
Multiplication by \(Q_{(1)}\) defines an embedding of the complex line \(\mathbb C\) into \(\mathbb{C}^2\) with image \(V_{(1)}\equiv\{\lambda\,Q_{(1)}|\lambda\in\mathbb C\}\subset \mathbb C^2\). This image can be characterized as the set of points \({\bf x}\) satisfying\footnote{We denote by the wedge symbol an arbitrary bilinear antisymmetric product such that \(\zeta_1\wedge \zeta_2\neq 0\). Consequently, one has \({\bf x}\wedge{\bf y}=\det({\bf x},{\bf y})\, \zeta_1\wedge \zeta_2\).
} \(Q_{(1)}\wedge {\bf x}=0\). In this context, \(Q_i\) play the role of Pl\"ucker coordinates.

The Wronskian condition can be written as
\be\label{WronskainD}
Q_{(1)}\left(u+\frac {\bi}2\right)\wedge Q_{(1)}\left(u-\frac {\bi}2\right)=\phi(u)\,\zeta_1\wedge \zeta_2\,.
\ee
First, it implies that the lines \(V_{(1)}(u+\frac {\bi}2)\) and \(V_{(1)}(u-\frac {\bi}2)\) are not collinear. Second, we demand that the embedding is polynomial (i.e. realised by Pl\"ucker coordinates being polynomial functions of \(u\)) and,  as a consequence, \(\phi(u) \)  in \eqref{WronskainD} is a polynomial which we denote as  \(\phi(u)=\prod_{k=1}^L(u-\theta_k).\)
\begin{figure}[t]
\begin{centering}
\includegraphics[width=0.4\textwidth]{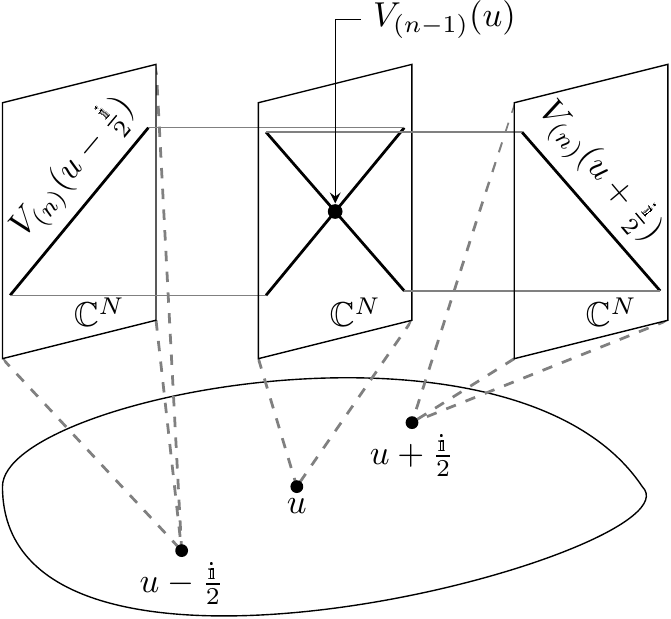}
\caption{\label{fig:intersection} Q-functions define a fibration of grassmannians over the Riemann surface of the spectral parameter \(u\). Relation between grassmannians of different rank is restricted by \eqref{eq:81D}.}
\end{centering}
\end{figure}

We are ready to establish the following map: to each polynomial embedding \(V_1(u)\), such that \(V_1(u+\frac {\bi}2)\cap V_1(u-\frac {\bi}2)=\{0\}\)    should
correspond an eigenstate of the SU(2) XXX spin chain of length \(L\) in the fundamental representation with inhomogeneities \(\theta_1,\theta_2,\ldots,\theta_L\). The correspondence is established after factoring out elementary symmetry transformations, as it will be described in the text.

In this way, we reformulated the solution of XXX spin chain in a geometric fashion. This point of view can be generalised to integrable systems with a higher rank symmetry algebras of \(gl\) type as follows. Denote by \(V_{(n)}\) an \(n\)-dimensional linear subspace of \(\mathbb C^N\), i.e. a point in the Grassmannian \(\mathbf{G}^n_N\). \(V_{(n)}(u)\) is a function of the spectral parameter \(u\). Consider a collection \(V_{(0)}(u)\,,V_{(1)}(u)\,,\ldots,V_{(N)}(u)\) of all possible subspaces and demand the property
\begin{align} \label{eq:81D}
    V_{(n)}(u+{\bi}/2)\cap V_{(n)}(u-{\bi}/2)=V_{(n-1)}(u)\,,\quad  \forall n\in\{1,2,\dots,N-1\}
\end{align}
to hold for any \(u\) save a discrete number of points, see Fig.~\ref{fig:intersection}.

We will advocate in this article that solving equation \eqref{eq:81D} supplemented with appropriate analytic constraints is equivalent to finding the spectrum of certain integrable models. For the case of compact rational spin chains equation \eqref{eq:81D} is an analog of fusion procedure and the analytic constraints are reduced to the demand that Q-functions, which are defined as Pl\"ucker coordinates for \(V_{(n)}\), are polynomials in \(u\). However,  this example is not unique. Equation \eqref{eq:81D} appears to be generic and applies to many quantum integrable systems, including (1+1)-dimensional QFT's, with \(\gl(N)\) symmetry or \(\gl(k|N-k)\) super-symmetry, or even for non-compact (super)algebras \(\su(K_1,K_2|M)  \).    It is closely related  to the fact that the transfer-matrices and their eigenvalues, such as the T-function of eq.\eqref{Baxter}, satisfy the so-called  Hirota bi-linear finite-difference equation \eqref{eq:1} which, as we will see later, can be solved in terms of Wronskian expressions through a finite number of Q-functions. The Q-functions are not obliged to be polynomials, as it is the case in integrable non-compact spin chains and (1+1)-dimensional QFT's. Moreover, there are situations when an approach similar to the coordinate or algebraic Bethe ansatz is not known, and  yet the equation \eqref{eq:81D} holds.

Moreover, the equation  \eqref{eq:81D} is also central to the spectral problem of integrable two-dimensional quantum field theories, and in particular sigma-models. It even allows for a concise and efficient  description  for exact spectrum of energies (anomalous dimensions)  of \(AdS_5/CFT_4\) duality. It is  because the quantum spectral curve (QSC) of the model, describing the dynamics of quantum conservation laws,  is most adequately formulated in terms of the Q-system based on equation   \eqref{eq:81D} and  related to \(\mathsf{psu}(2,2|4)\) superconformal symmetry algebra \cite{Gromov2014a,Gromov:2014caa}. 

Since \eqref{eq:81D} is such a generic equation expected to appear  in virtually all quantum integrable models its properties deserve to be studied in detail, which is one of the main goals of this paper.

One should always bear in mind that Q-functions is a way to introduce a {\it coordinate} system, hence they are not defined uniquely. For instance, we can  replace \(Q_2\to Q_2+{\rm const}\,Q_1\) without any consequence for the Wronskian condition \eqref{Wronskian}, and the possible linear transformations are not exhausted by this example. In addition, the overall rescaling of all Q-functions by any function of \(u\) does not affect the embeddings \(V_{(n)}\). In \secref{sec:HiroQSys} we will construct the T-functions as determinants of Q-functions; \(T(u)\) in the Baxter equation \eqref{Baxter} is one of them: \(T(u)\zeta_1\wedge\zeta_2=Q_{(1)}(u+{\bi})\wedge Q_{(1)}(u-{\bi})\). T-functions should be thought as certain volume elements in \(\mathbb{C}^N\), i.e. they are represented by a full form. They are invariant under rotation of the basis but still transform under rescalings. The fully invariant objects are Y-functions which are certain ratios of T-functions. Although the description in terms of Y's is a more invariant way to parameterise the system, the description in terms of \(V_n(u)\)  has an important advantage since usually the analytic properties of  Q-functions, directly related to T- and Y-functions by Wronskian solutions, are significantly simpler than the ones of T's or Y's.

In this article we discuss the following applications of the proposed approach. In \secref{sec:Hirota-equat-strip} we show how the Hirota equation (T-system) for integrable systems with \(gl(N)\) type of symmetry is solved in terms of Q-functions and also discuss how the Wronskian-type formulation \eqref{eq:81D} is related to higher-rank Baxter equations. This is a quite well established topic in the literature, in particular its geometric interpretation can be easily spotted from discussion in \cite{Krichever:1996qd}.  We include it into the paper as a simple example which  contains the guiding lines useful  for the further generalizations to supergroups and noncompact representations.

Then, in \secref{sec:Hirota-equation-an}, we generalise the \(\gl(N)\)  solution and show how to get from our formalism  the generic Wronskian solution of  Hirota equation with the boundary conditions of the  ``T-hook'' type,  describing the  weight space of highest weight non-compact representations   appearing in integrable models with \(\su(K_1,K_2|M)\) symmetry. Note that the T-hook itself was  first proposed as  a formulation of  AdS\(_5\)/CFT\(_4\) Y-system \cite{Gromov:2009tv} with superconformal \(\mathsf{psu}(2,2|4) \) symmetry.  The generic symmetry algebra \(\su(K_1,K_2|M)\)  also includes  two interesting particular cases: the compact supersymmetric algebra \(\su(K|M)\) and the non-compact one \(\su(K_1,K_2)\), the latter should be relevant for Toda-like systems. We emphasize here a remarkable fact that the supersymmetric generalization still relies on the same equation \eqref{eq:81D}, with \(N=K_1+K_2+M\). However, a convenient way to properly treat it  is to choose a subspace  \(\mathbb{C}^M\) in \(\mathbb{C^N}\) and work with Q-functions in  specially re-labeled Grassmannian coordinates obtained by a  Hodge-duality transformation in \(\mathbb{C}^M\). 

The Wronskian solution of  Hirota equations on ``T-hook'' was given for the most interesting case of \(\psu(2,2|4)\) symmetry by Gromov,  two of the authors, and Tsuboi in \cite{Gromov:2010km}, and then it was presented for the generic case in the work of Tsuboi \cite{Tsuboi:2011iz}. We believe that the formalism of exterior forms  developed here presents these results in a much more concise and geometrically transparent way. We also establish several interesting new relations among the Q-functions, especially elegantly written in terms of the exterior forms. Some of them have been extensively used in the study of the Q-system emerging in AdS/CFT integrability case \cite{Gromov2014a,Gromov:2014caa}.

\begin{figure}[t]
\begin{centering}
\includegraphics[width=0.4\textwidth]{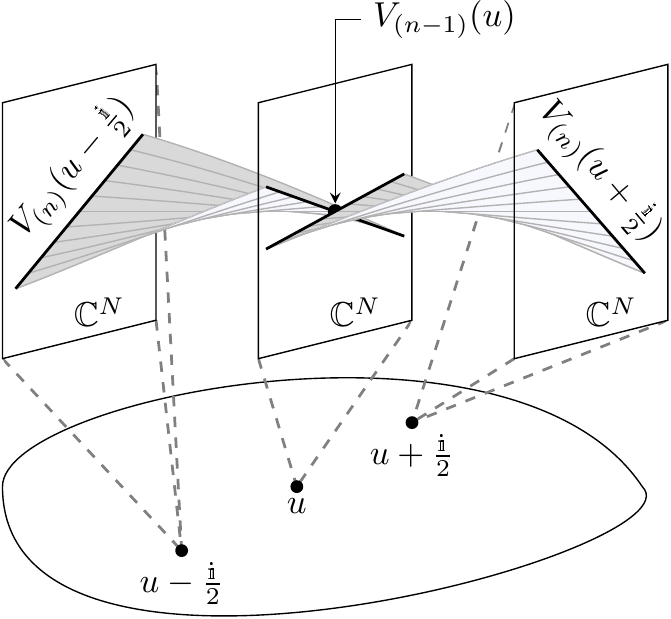}
\caption{\label{fig:intersectiong} Deformation of the fibration by introducing a connection. This connection ``rotates'' the spaces \(V_{(n)}\) via the parallel transport from point \(u\pm\frac\bi 2\) to point \(u\) where the equation (\ref{eq:81D}) can be used.}
\end{centering}
\end{figure}

In \secref{sec:twisted-case} we discuss how the construction can be amended to include the case of integrable spin chains with twisted boundary conditions. It happens in a very natural way: One should gauge the global rotational \(\GL(N)\) symmetry w.r.t. the space of spectral parameter,  making it local  and hence  introducing a new object: a holomorphic connection \(A\). The non-local relation \eqref{eq:81D} is modified  by inserting a parallel transport \(P \exp\left[\int_{u-\frac {\bi}2}^{u+\frac {\bi}2} A(u')du'\right]\) of the plane \(V_{(n)}(u-{\bi}/2)\), so that the intersection in \eqref{eq:81D} naturally happens at the same point (see fig.~\ref{fig:intersectiong}). This parallel transport precisely realizes the twisting.

The new properties emerging in the twisted case are thoroughly studied, mainly on the examples of rational spin chains. Especial attention is paid to the untwisting limit which is singular and quite non-trivial. In particular, we give a detailed description how relation between the asymptotics at infinity and the  representation theory depends on the presence or absence of particular twists.

The Wronskian solution of Hirota equation \eqref{eq:1} in the case of
super-conformal algebra \(\su(2,2|4)\) and the  grassmannian structure of
the underlying Q-system have played an important role in the discovery
of the most advanced version of equations for the exact spectrum of
anomalous dimensions in planar \(\mathcal{N}=4\) SYM theory -- the quantum spectral
curve (QSC) \cite{Gromov2014a,Gromov:2014caa}. In fact, many of the
relations discussed and re-derived in the present paper in terms of
the very efficient formalism of exterior forms have been already
present in \cite{Gromov:2014caa} in the  coordinate form.  As an
interesting generalization of the QSC, we will present in \secref{twistedQSC} its twisted
version, in the presence of  all (3+3) angles describing the gamma
deformation and a non-commutativity deformation of the original \(\mathcal{N}=4\)
SYM theory \cite{Lunin:2005jy,Frolov:2005dj,Beisert:2005if}. The corresponding \(\bP-\mu\) and \(\bQ-\omega\) equations of  \cite{Gromov2014a,Gromov:2014caa}, as well as all Pl\"ucker QQ-relations, 
will be essentially unchanged and the whole difference with the
untwisted case will reside in the large \(u\) asymptotics of
Q-functions with respect to the spectral parameter \(u\), which are modified due to the presence of twists by certain exponential factors. This is the only change in the analytic properties of QSC due to the twisting.  The algebraic part of the twisted QSC formulation will be simply a particular \((2|4|2)\) case of the twisted version of the general \((K_1|M|K_2)\)
Q-system presented in this paper.

Finally, in \secref{sec:BMNvac} we probe our conjectures for twisted QSC on an interesting case  of \(\gamma\)-deformed BMN vacuum of this AdS/CFT duality.
 For a particular case, \(\beta\)-deformation, the Y-system and
 T-system for the twisted case were formulated and tested in
 \cite{Gromov2011,Arutyunov2011} (see also \cite{Ahn:2010yv,Ahn2011} at the level of the
 S-matrix).  We reproduce by our method the one-wrapping terms in the  energy of this state,  known by direct solution of TBA equations  \cite{deLeeuw:2011rw,Ahn:2011xq}, which was also known by the direct perturbation theory computation\cite{Fokken2014a}. 
 A potential  advantage of our method is the possibility to find the next corrections to this state on a regular basis, by the methods similar to \cite{Marboe:2014gma,Gromov:2015vua} as well as application of the efficient numerical procedure of \cite{Gromov:2015wca}, but this is beyond the scope of the current paper. 
\section{Algebraic properties of Q-system and solution of Hirota equations}
\label{sec:HiroQSys}In this section we show how the Q-system is used to solve  Hirota equations on \((K_1|M|K_2)\) \(\mathbb{T}\)-hooks. We also establish notations and algebraic properties of the Q-system. Although this solution was already presented in the literature \cite{Tsuboi:2011iz}, we take a look on it from a different, more geometric point of view, and we believe it will be a useful contribution to the subject as the technicality of the involved formulae is significantly reduced and the solution itself is made more transparent.

\subsection{Hirota equation in historical perspective}
\label{sec:introduction}

 The bi-linear discrete Hirota equation, sometimes also called the Hirota-Miwa equation  \cite{Hirota,Jimbo:1983if,Kuniba2011}
\begin{equation}
  \label{eq:1}
  T_{a,s}(u+\tfrac {\bi} 2)  T_{a,s}(u-\tfrac \bi 2) = T_{a+1,s}(u) T_{a-1,s}(u)+T_{a,s+1}(u) T_{a,s-1}(u)\,
\end{equation}
appears in numerous quantum and classical integrable systems.
In these notations, typically used in the context of quantum integrable spin chains and sigma-models,  
\(T_{a,s}(u)\) are complex-valued functions of two integer indices \(a\) and \(s\) parameterising a \(\mathbb{Z}^2\) lattice, and of a  parameter \(u\in\bC\) usually called spectral parameter.  Although the  parameter \(u\) enters the equation only with discrete shifts and hence can be treated as another discrete variable, the analytic dependence of \(T_{a,s}\) on \(u\) is an important piece of information used  to specify the physical model. We will exploit this analytic dependence starting from  \secref{sec:twisted-case}.

\begin{figure}[t]
 \centering
\subfigure[\(\gl(N)\) Strip]{\label{fig:glnstrip}
\hspace{.4cm}\includegraphics{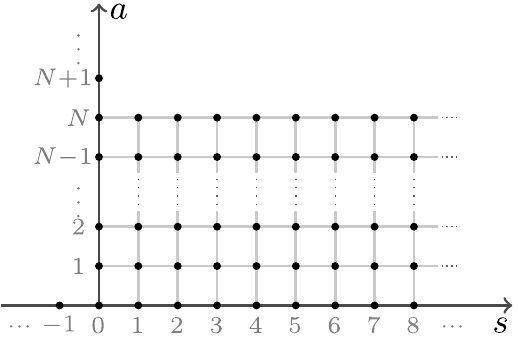}\hspace{.4cm}}\hspace{1.3cm}
\subfigure[Identification of a node of the strip to a rectangular
Young diagram]{\label{fig:identification}
\hspace{.4cm}\includegraphics{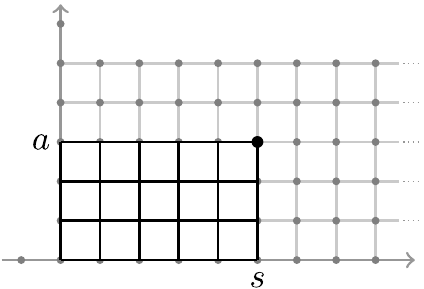}\hspace{.4cm}}

\caption{The Young diagrams of compact representations of       \(\gl(N) \) group are confined to a half-strip, depicted on fig.(a), of width \(N\) on infinite representational \((a,s)\)-lattice. The vertices within this strip are in one-to-one correspondence with rectangular Young tableux of size \(a\times s\), as depicted on fig.(b), as well as  with corresponding   characters or T-functions. 
}
 \label{fig:Strip}
\end{figure}

In integrable quantum spin chains with \(\gl(N)\) symmetry, \(T_{a,s}\) appears to be the transfer matrix in the representation \(s^a\), with the \(a\times s\) rectangular Young diagram, as shown in Fig.~\ref{fig:identification}, while \(u\)  plays the role of the spectral parameter. Equation \eqref{eq:1} describes the fusion procedure among these transfer-matrices \cite{kulish1982gl_3,reshetikhin1983functional,Klumper:1992vt,Kuniba:1993cn,Krichever:1996qd}. The statement generalises to supersymmetric case \cite{Tsuboi:1997iq,Tsuboi:1998ne,Kazakov:2007na,Tsuboi:2009ud} and, with a particular modification of \eqref{eq:1}, to other semi-simple Lie algebras, see \cite{Klumper:1992vt,Kuniba:1993cn,Krichever:1996qd,Kuniba2011} and the references therein.  In integrable 2d CFT's at finite size or finite temperature, and in particular in 2d sigma-models, this Hirota equation first appeared in relation to quantum KdV \cite{Bazhanov:1994ft}  and more recently it was successfully used for the finite size analysis, including excited states, for the \(\SU(N)\times \SU(N)\) principal chiral field (PCF)  and some related models  \cite{Gromov:2008gj,Kazakov:2010kf}. It was also proposed as a  version of the AdS/CFT  Y-system \cite{Gromov:2009bc}   appearing in the spectral problem of the planar \(\mathcal{N}=4\) SYM theory and  it was successfully exploited there for extracting many non-trivial results at arbitrary strength of the 't~Hooft coupling and in various physically interesting limits \cite{Beisert:2010jr}. The finite-difference Hirota equation \eqref{eq:1} is also related  in different way to the classical integrability, besides the standard classical limit \(\hbar\to 0\) of the original quantum system.  It can be obtained from  the canonical Hirota equation for  \(\tau\)-function of classical integrable hierarchies of PDE's   by introduction of discrete Miwa variables \cite{Jimbo:1983if}. And in particular, a generating series of transfer-matrices of \(\gl(N)\) quantum Heisenberg spin chains  can be interpreted as a \(\tau\)-function of the mKP hierarchy \cite{2011arXiv1112.3310A}.

As was shown in the past, Hirota equation admits general and  exact solutions for specific boundary conditions on the \(\mathbb{Z}^2\) lattice. In particular, if one demands \(T_{a<0,s}=0\) then all T-functions can be expressed explicitly in terms of \(T_{0,s}\) and \(T_{1,s}\) by
\begin{eqnarray}T_{a,s}(u)=\frac{\underset{1\le i,j\le
       a}{\det}T_{1,s+i-j}\left(u+\bi\frac{1+a-i-j}2\right)}{\prod\limits_{k=1}^{a-1}T_{0,s}\left(u+̱i\frac{a+1-2k}2\right)}%
   \label{GJTfunction}\,,\end{eqnarray}
which is a particular case of  the Cherednik-Bazhanov-Reshetikhin (CBR) determinant
\cite{Bazhanov:1989yk,Cherednik1987,Kazakov:2010kf,Kazakov:2007na} formulae. This determinant relation is a generic solution of the Hirota equation in the sense it can be proven recursively in \(a\) assuming \(T_{a\geq 0,s}\neq 0\); if \(T_{a,s}=0\) for some positive \(a\) then \eqref{GJTfunction} may be violated, however in practice this affects only  T's which do not have an explicit physical interpretation, and we choose to define these T's such that \eqref{GJTfunction} holds.

If we impose a more severe restriction on T's and demand them to be non-zero only in the black nodes of Fig.~\ref{fig:glnstrip} (i.e. for \(s=0\) or \(a\geq 0\) or \(s>0, N\geq a\geq 0\)) then we get the Hirota equation appearing in integrable models with \(\gl(N)\) symmetry and related to the compact representations of the latter. For such boundaries, we can recognise in CBR determinants a quantum generalisation of  standard Gambelli-Jacobi-Trudi formulae for characters of \(\gl(N)\) irreps. The analog of  \eqref{GJTfunction} looks especially simple
\begin{eqnarray}
  \chs{a,s}({G})=
\mathrm{det}(\chs{1,s+j-k}(G))_{1\leq j,k\leq a}\,,
  \label{GJTcharacter}\end{eqnarray}
where \(G=\{\x_1,\dots,\x_N\}\) is a   Cartan subgroup element. This character satisfies the simplified Hirota equation\footnote{It is sometimes called the Q-system in the mathematical literature. We will avoid this in order ot to confuse it with the Baxter's Q-functions \(Q_I(u)\) which we use all over the paper. We rather call the collection of these Q-functions as the Q-system.}
\be\label{eq:SHE}
\chs{a,s}(G) \chs{a,s}(G)=\chs{a+s,s}(G) \chs{a-1,s}(G)+\chs{a,s+1}(G) \chs{a,s-1}(G)\,;
\ee
it can be derived directly from \eqref{WeylCharacter} due to the Jacobi relation for  determinants (see e.g. the appendix of \cite{Kazakov:2007na}). 

\begin{table}
  \centering
\renewcommand{\arraystretch}{1.5}
  \begin{tabular}{|c|c|}
\hline
    Characters of the \(GL(N)\) group&T-functions on the \(GL(N)\) strip
    \\ \hline \hline
\(2^{\rm{nd}}\) Weyl formula&Cherednik-Bazhanov-Reshetikhin formula\\
\(\chs{\lambda}({G})=\underset{1\leq j,k\leq|\lambda|}{\det}
\chs{(\lambda_k+j-k)}
(G)%
\)%
&
  \(T_\lambda(u)=%
\underset{1\le j,k\le|\lambda|}{\det}
      T_{(\lambda_{k}
      +j-k)}\left(u+\bi \tfrac
      {\lambda_k+1+|\lambda|-|\lambda'|-j-k} 2\right)%
\)%
\\[.2cm]
\hline
\(1^{\rm{st}}\)  Weyl formula&Wronskian expression\\
  \(\chs{\lambda}(G)=\frac{\displaystyle
\underset{1\le j,k\le N}{\det}
 ~ \x_k^{\lambda_j+N-j}%
}{\displaystyle\underset{1\le j,k\le N}{\det}%
 ~   \x_k^{
N-j}%
}\)%
&%
  \(T_\lambda(u)=%
\underset{1\le j,k\le N}{\det}
  Q_k(u+\bi \tfrac
 {2\lambda_j-2j+1+|\lambda|-|\lambda'|}2)%
\)%
\\\hline
  \end{tabular}
  \caption{Expression of the \(GL(N)\) characters and their
    generalization to T-functions.
Representations are labeled by Young diagrams
\(\lambda=(\lambda_1,\lambda_2,\dots,\lambda_{|\lambda|})\), and
\(|\lambda'|\) denotes \(\lambda_1\). Characters \(\chi_\lambda(g)\) are
written in terms of the eigenvalues \((\x_1,\x_2,\dots,\x_N)\) of a group element \(g\). The CBR formula and Wronskian expression of
T-functions are written in this table under specific gauge
constraint. In other gauges they hold up to division by \(T_{(0)}\) or
\(Q_\emptyset\), cf.  \eqref{GJTfunction} and (\ref{eq:34a}), the normalisation is clarified in \secref{sec:gauge-symm-hirota}.
}
  \label{tab:compare}
\end{table}

In the case of characters, we know that there exists a more explicit, Weyl formula expressing the character as  a  determinant involving the Cartan elements:
\begin{align}
  \chs{a,s}(G)&=\frac{\det\limits_{1\le
  j,k\le N}\x_k^{N-j+s\,\Theta_{a,j}}}{\det\limits_{1\le j,k\le N}\x_k^{
N-j}},&\textrm{ where }\Theta_{i,j}\equiv
\begin{cases}
  1 &\textrm{if } i\ge j\\
  0 &\textrm{if } i < j
\end{cases}\,.
   \label{WeylCharacter}\end{align}

It is clear  that it should be possible to generalize the Weyl formula from characters to T-functions.  Such a quantum generalization was known since quite a while~\cite{Krichever:1996qd} in terms of the Wronskian-type determinant:
\begin{equation} 
  T_{a,s}(u)=
\det\limits_{1\le
  j,k\le N} Q_k\left(u+\bi \tfrac
 {a+1+s(2\Theta_{a,j}-1)}2-\bi\,j\right)\,.
 \label{WeylFunction}
 \end{equation}
It gives, up to rescaling of T-functions, the general
 solution of Hirota equation for a half-strip boundary conditions of fig.~\ref{fig:Strip} in terms of \(N\) independent
 Q-functions \(Q_1(u),\dots,Q_{N}(u)\). More precisely, it applies for
 the semi-infinite rectangular domain \(s\geq 0, N\geq a\geq 0\); the
 rest of non-zero T-functions, corresponding to the black nodes of
 fig.~\ref{fig:Strip},  \(a=0, s<0\) and  \(s=0,   a> N\)  are %
easily restored\footnote{Indeed, the Hirota equation gives
  \(T_{0,s-1}=T_{0,s}^+T_{0,s}^-/T_{0,s+1}\), and
  \(T_{a+1,0}=T_{a,0}^+T_{a,0}^-/T_{a-1,0}\) for \(a>0\), allowing to
  iteratively restore the boundary T-functions.
}.

The parallels between character formulae and T-functions (or, when meaningful, transfer matrices) extend beyond the rectangular representations \(s^a\), the equivalent formulae  for arbitrary finite-dimensional representations of \(\gl(N)\) algebra are summarised in table~\ref{tab:compare}.

\begin{figure}[t]
  \centering{  \renewcommand{\k}{3}
  \renewcommand{\m}{4}
\subfigure[\(\gl(K|M)\) ``{\fHook}'' (with \(K=\k\), \(M=\m\))]{\label{fig:LHook}
\includegraphics{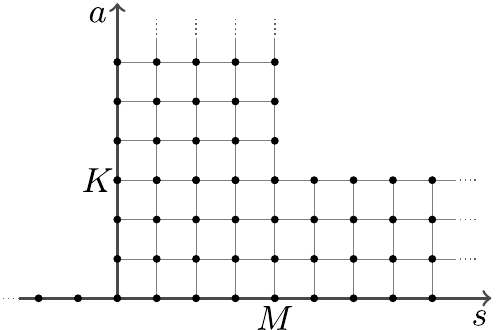}}\hspace{.7cm}
  \newcommand{\ko}{3}
  \newcommand{\kt}{2}
\newcommand{\mo}{4}
  \newcommand{\mt}{1}
\subfigure[{\THook} of size \((K_1|M_1+M_2|K_2)\) (where
{\Kone}=\ko, {\Ktwo}=\kt, {\Mone}=\mo{} and {\Mtwo}=\mt)]{
\label{fig:THook}\includegraphics{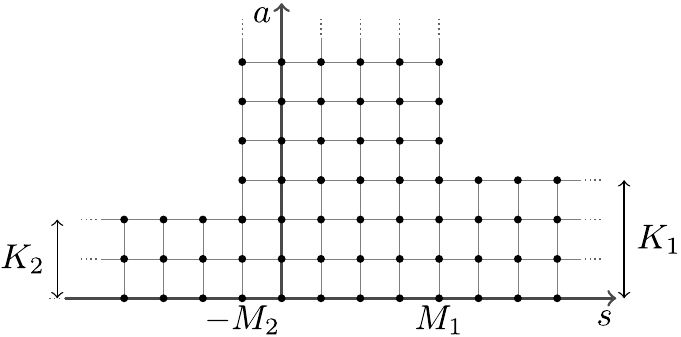}
}
 \caption{  \label{fig:LTHook}\emph{{\FHook}} and \emph{{\THook}}, for supersymmetric symmetry groups.
 }}

\end{figure}

The Gambelli-Jacobi-Trudi-type formulae \eqref{GJTcharacter} and their quantum counterpart \eqref{GJTfunction} remain unchanged if one generalises the symmetry to the case of superalgebras of \(\gl\) (or rather \(\sl\)) type, including  the non-compact cases. They are used, however, under different boundary conditions outlined in figure~\ref{fig:LTHook}.

The super-analogues of Weyl-type formulae are not obtained by a straightforward generalisation, yet they are also known. For the compact case \(\su(K|M)\) the determinant expressions for characters were established in \cite{Moens}. In the non-compact case   \(\su(K_1,K_2|M)\)  certain expression for characters were given in \cite{Kwon06} and their determinant version for the case of rectangular representations\footnote{Determinant character formulae for non-rectangular highest-weight representation were not published explicitly to our knowledge.} was elaborated in \cite{Gromov:2010vb}. The generalization to the quantum case was first presented for finite-dimensional irreps of \(\su(K|M)\) in \cite{Tsuboi:2009ud}, then for
\(\su(2,2|4)\) in \cite{Gromov:2010km} (this is the most interesting case for physics as it is realised in the context of AdS/CFT integrability, see a review \cite{Beisert:2010jr} for introduction into the subject) and finally generalized to any
\(\su(K_1,K_2|M)\) in  \cite{Tsuboi:2011iz}.  In the case of  rectangular
\(s^a\) irrep, the formulae of \cite{Tsuboi:2011iz} give the generic
(up to a gauge transformation, as explained below) Wronskian solution
of Hirota equation \eqref{eq:1} within the \((K_1|M|K_2)\)-hook
presented in fig.~\ref{fig:THook}  (which was also called {\THook} due
to its shape). The   so-called {\fHook} of the  fig.~\ref{fig:LHook},
which we also call {\LHook}, is a particular  case \(K_2=0\) of
\(\su(K_1,K_2|M)\)    corresponding to the compact representations of
\(\su(K|M)\).

In \cite{Tsuboi:2011iz}, the Weyl-type solution of Hirota equation is
presented in terms of an explicit finite determinant and it summarises
the whole progress achieved in this field. However, the corresponding
expressions are extremely bulky which somewhat obscures their nice
geometric and algebraic properties.  The main aim of this section is
to present a more concise and more intuitive formalism, based on the
exterior forms of Baxter-type Q-functions. It will clarify the
Grassmannian nature of Wronskian solutions for T-functions on
supergroups and allow simple and general proofs for these formulae. We
will re-derive several relations already proven in
\cite{Tsuboi:2011iz} in this new language and present some new useful
relations.

\subsection{Notations}
\label{sec:notations}

The Wronskian solution of Hirota equation \eqref{eq:1}  with boundary conditions shown in figures~\ref{fig:glnstrip} and \ref{fig:LTHook} will be written in subsequent sections in terms of a set of Q-functions \(Q_{b_1b_2\dots}\)  which are labeled by several indices \(b_k\) and which are antisymmetric under permutations of these indices. There exist relations between the Q-functions, and there are two equivalent ways to formulate them: either as an algebraic statement -- the ``QQ-relations'' -- or as a geometric statement -- in terms of the intersection property \eqref{eq:81D}.

Algebraically, the QQ-relations read (in the non-super-symmetric case of \secref{sec:Hirota-equat-strip}) \cite{Pronko:1999gh,Bazhanov:2001xm,Dorey:2006an,Belitsky:2006cp,Gromov:2007ky,Bazhanov:2008yc}
\begin{align}
  \label{eq:2}
  Q_{A}Q_{Abc}
&=Q_{Ab}^{+}
Q_{Ac}^{-}-
Q_{Ab}^{-}
Q_{Ac}^{+}\,.
\end{align}
All other  QQ-relations derived below ultimately follow from \eqref{eq:2}, hence we will pause for a while to accurately introduce the notational conventions related to \eqref{eq:2} and to Q-system in general.

The Q-functions are functions of the spectral parameter \(u\). This dependence is typically assumed implicitly, and the shifts of \(u\) are denoted following the convention
\begin{align}
  \label{eq:13}
  f^{[\pm n]}&=f(u\pm n \tfrac \bi 2)\,,&   f^{\pm }&=f(u\pm \tfrac
  \bi 2)\,,
  & \bi\equiv\sqrt{-1}\,.
\end{align}

The indices \(b\), \(c\) in \eqref{eq:2} take value in
the ``bosonic'' set \({\Bset}=\{1,2,\dots,N\}\). The multi-index \(A\) of the bosonic set can for instance
contain one single index \(a\in {\Bset}\), or no index
at all (it is then denoted as \(A=\emptyset\)), or all indices (which is
denoted as \(A={\Bset}=\bar \emptyset\)), etc.
 The multi-index \(2,1\) is
different from the multi-index \(1,2\) (one has \(Q_{2,1}=-Q_{1,2}\)), and
we will say that the multi-index \(A=a_1a_2\dots a_n\) is sorted if
\(\forall k<n,~a_k< a_{k+1}\). The sum over all sorted multi-indices of length \(n\) is denoted by \(\sum_{|A|=n}\).

For a multi-index \(A\),  \(\{\!A\}\) means the associated set
(for instance \(\{2,1\}=\{1,2\}\)), and we  denote by
\(\bar A\) the sorted multi-index obeying \(\{\!\bar A\}={\Bset}\setminus\{\!A\}\) %
(for instance \(\overline {1,3}\equiv
2,4,5,\dots,N\)).

\begin{figure}
  \centering
\includegraphics{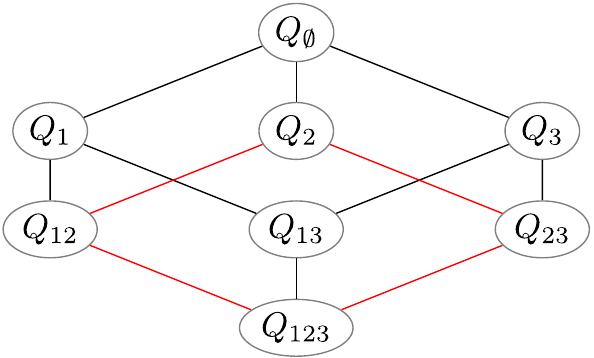}  \caption{Hasse diagram for \(\gl(3)\) Q-functions.}
  \label{fig:HasseQ}
\end{figure}

There are \(2^N\) different Q-functions corresponding to the different subsets of {\Bset}. They can  be arranged as a Hasse diagram forming an N-dimensional  hypercube, see figure~\ref{fig:HasseQ}.  Each facet of the Hasse diagram is associated with a QQ-relation: for instance the bottom facet in figure~\ref{fig:HasseQ} is associated to the relation
\begin{equation}
  \label{eq:227}
  Q_2Q_{123}=Q_{12}^+Q_{23}^--Q_{12}^-Q_{23}^+\,,
\end{equation}
which is the case\footnote{More precisely, (\ref{eq:2}) gives the relation \(Q_2Q_{213}=Q_{21}^+Q_{23}^--Q_{21}^-Q_{23}^+\), which is equivalent due to the antisymmetry.} \(A=2\), \(b=1\), \(c=3\) in (\ref{eq:2}).

Given a basis of  \(N\) independent elements
\(\bg_1,\bg_2,\dots,\bg_N\) and an associative antisymmetric bilinear product ``\(\wedge\)'', we also introduce the \(n\)-form
\begin{align}
  \label{eq:7}
  Q_{(n)}&=\sum_{|A|=n} Q_A ~ \bg_A,&\textrm{where
  }\bg_{b_1b_2\dots b_n}\equiv& \bg_{b_1}\wedge
  \bg_{b_2}\wedge\dots\wedge \bg_{b_n},&\textrm{(and }\bg_{\emptyset}&=1\textrm{).}
\end{align}
With explicit indices, \eqref{eq:7} reads: \(Q_{(n)}=\sum\limits_{1\leq b_1 < b_2<\dots< b_n\leq N}Q_{b_1b_2\dots b_n}\, \bg_{b_1}\wedge \bg_{b_2}\wedge\dots \wedge \bg_{b_n}\).

We also introduce the Hodge dual \(\star \omega\) of an arbitrary
\(n\)-form \(\omega\) as the linear transformation such that
\begin{align}
\label{eq:52}
\star \bg_A &=%
\epsilon_{A \bar A}\, \bg_{\bar A}\,,
\end{align}
where \(\epsilon_{b_1b_2\dots b_N}\) is the completely antisymmetric
tensor with the sign choice \(\epsilon_{12\dots N}=1\). %
For instance this definition gives \(\star \bg_{13}=-\bg_{2,4,5,\cdots,N}\).

The Hodge-dual Q-functions are denoted using the super-script labelling:  
\begin{align} \label{eq:75}
\HQ A &\equiv %
\epsilon^{\bar A A}\, Q_{\bar A}\,,& \textrm{so that}\ \ \ \ 
\star Q_{(n)}&=\sum_{|A|=|{\Bset}|-n} \HQ A ~ \bg_A.
\end{align}
The sign convention for the completely antisymmetric tensor \(\epsilon^{b_1b_2\dots b_N}\)  is also \(\epsilon^{12\ldots N}=1\). We will interchangeably use upper- and lower-indexed \(\epsilon\) to emphasise the covariance in relations.
 
Note that the inverse Hodge-dual operation given by
\begin{align} \label{eq:75D}
Q_{A\vphantom{\bar A}} &= %
\epsilon_{A\bar A}\, \HQ {\bar A}\,
\end{align}
has certain difference in signs compared to \eqref{eq:75}.

\paragraph{Pl\"ucker identity.}

Throughout this text, we will frequently use \emph{Pl\"ucker
identities}. The simplest one is 
\begin{multline}
\label{Plucker_Hodge}  \star(\bg_{b_1}\wedge
\bg_{b_2}\wedge\dots
\wedge\bg_{b_{N}} )\,\, \star (\bg_{c_1}\wedge \bg_{c_2}\wedge\dots\wedge\bg_{c_{N}} )=\\
=\sum_{a=1}^{n}\star( \bg_{b_1}\wedge\dots\wedge\bg_{b_{{N}-1}}
\wedge\bg_{c_a})\,\, \star (\bg_{c_1}\wedge\dots
\wedge\bg_{c_{a-1}}\wedge\bg_{b_{N}}\wedge\bg_{c_{a+1}}\wedge\dots\wedge\bg_{c_{N}}
)\,,\end{multline}
where the Hodge operation ``\(\star\)'' simply transforms each product \(\bg_{b_1}\wedge
\bg_{b_2}\wedge\dots\wedge\bg_{b_{N}}\) into the number
\(\epsilon_{b_1b_2\dots b_N}\). 

More generally, one has
\begin{multline}\label{Plucker}
\star(\xx_{1}\wedge \xx_{2}\wedge\dots\wedge\xx_{{N}} )\,\,\star(\yy_{1}\wedge \yy_{2}\wedge\dots\wedge\yy_{{N}} )=\\
=\sum_{a=1}^{N}\star( \xx_{1}\wedge\dots\wedge\xx_{{N}-1} \wedge \yy_{a})\,\,\star(\yy_{1}\wedge\dots \wedge\yy_{a-1}\wedge \xx_{{N}}\wedge\yy_{a+1}\wedge\dots\wedge\yy_{{N}} )\,,
\end{multline}
where \(\xx_i=\sum_{j=1}^{N}x_{i,j}\bg_j\) and
\(\yy_i=\sum_{j=1}^{N}y_{i,j}
\bg_j\) are arbitrary sets of
vectors.

\paragraph{Asymptotics.}
\label{sec:asymptotics}

The asymptotic behavior of functions at large \(u\) will have some importance later on in this article. We will then use the notation \(f\simeq g\) to say that \(\lim\limits_{|u|\to\infty}\frac f g=1\) and \(f\sim g\) to say that there exists \(\alpha\in\mathbb C^{\times}\) such that \(\lim\limits_{|u|\to\infty}\frac f g=\alpha\,.\)

\subsection{QQ-relations and flags of \texorpdfstring{$\mathbb C^N$}{C\^{}N}}
\label{sec:solut-qq-relat}

The geometric counterpart of the algebraic relation \eqref{eq:2} is the intersection condition \eqref{eq:81D}. Our nearest goal is to justify this statement.

The functions \(Q_A\) with \(|A|=n\) should be thought as Pl\"ucker  coordinates of the hyperplane \(V_{(n)}\); they define \(V_{(n)}\) as the collection of points \({\bf x}\) that satisfy \(Q_{(n)}\wedge{\bf x}=0\). Note that for a generic \(n\)-form \(\omega_n\) the condition \(\omega \wedge{\bf x}=0\) does not define an \(n\)-dimensional hyperplane (for instance if \(\omega=\zeta_1\wedge\zeta_2+\zeta_3\wedge\zeta_4\), the condition is satisfied only by \({\bf x}=0\)). However, as it will become clear in this subsection, the relation \eqref{eq:2} insures that the \(Q_{A}\) are indeed the Pl\"ucker coordinates of \(n\)-dimensional hyperplanes.

To derive \eqref{eq:2} from the intersection condition \eqref{eq:81D} we note that the latter can be equivalently reformulated as the following union property
\begin{align}\label{eq:80}
  \forall n\in\{1,2,\dots,N-1\},\qquad\qquad V_{(n)}^++ V_{(n)}^-=V_{(n+1)}\,,
\end{align}
which implies, in particular, that the sequence   \(\{0\}\equiv V_{(0)}\subset V_{(1)}^+\subset \dots \subset V_{(N)}^{[+N]}\equiv\mathbb C^N\) is a maximal flag
  of \(\mathbb C^N\). The union property should hold for almost all values of the spectral parameter save a discrete set of points.

Since \(V_{(1)}\) is a line there exists a
one-form
\begin{align}\label{eq:83}
  Q_{(1)}&=\sum_{a=1}^N Q_a \bg_a\,&\textrm{such that
  }\ \ \ V_{(1)}=\left\{\xx\in\mathbb C^N ~;~ Q_{(1)}\wedge {\bf x}=0\right\}\,.
\end{align}
This defines \(Q_{(1)}\) up to a normalisation,  i.e. up to the
transformation \(Q_{(1)}(u)\mapsto f(u)Q_{(1)}(u)\,,\) where \(f\) is a
\(\mathbb{C}\)-valued function of \(u\).  Next, one can
immediately see from (\ref{eq:80}) that
\begin{align}\label{eq:85}
V_{(n)}&=V_{(1)}^{[n-1]}+V_{(1)}^{[n-3]}+\cdots +V_{(1)}^{[-n+1]}\\
&=\left\{\xx\in\mathbb C^N ~;~
Q_{(1)}^{[n-1]}\wedge Q_{(1)}^{[n-3]}\wedge \cdots \wedge
Q_{(1)}^{[-n+1]} \wedge {\bf x}=0\right\}\,.\label{eq:3}
\end{align}
We can therefore define the forms \(Q_{(n)}\) by the relation
\begin{align}
 \label{eq:82} Q_{(n)}&=f_n~Q_{(1)}^{[n-1]}\wedge Q_{(1)}^{[n-3]}\wedge \cdots \wedge
Q_{(1)}^{[-n+1]}\,&\textrm{if }n>1\,,
\end{align}
where \(f_n(u)\) is a normalisation freedom that we will have to
fix.

The definition (\ref{eq:82}) enforces the
coordinates \(Q_A\) to obey the relation
\begin{align}
\label{eq:86}    Q_{A}Q_{Abc}\frac{f_{\nA+1}^+f_{\nA+1}^-}{f_{\nA} f_{\nA+2}}
&=
Q_{Ab}^{+} Q_{Ac}^{-}-
Q_{Ab}^{-}
Q_{Ac}^{+}\,,
\end{align}
a proof is given in \appref{sec:proof-qq-relation}, and it is based on a simple application of the Pl\"ucker identity (\ref{Plucker}).

The equation (\ref{eq:2}) corresponds to a particular choice of
normalisation such that \({f_{\nA+1}^+f_{\nA+1}^-}={f_{\nA}
  f_{\nA+2}}\), i.e. \(f_n=\frac{g^{[+n-1]}}{g^{[1-n]}}\) for some function
\(g\). Note that \eqref{eq:2} can be modified if one decides to use a different prescription for \(f_n\); equation \eqref{eq:86} is an invariant version of \eqref{eq:2}. Still, we stick to the normalisation choice of \eqref{eq:2} in this paper, this is also a common choice in the literature.

Plugging the expression \(f_n=\frac{g^{[+n-1]}}{g^{[1-n]}}\) into
(\ref{eq:82}) and using \(g^-/g^+=f_0\equiv Q_\emptyset\), we finally get
\begin{equation}
  \label{eq:5}
  Q_{(n)}%
=\frac{Q_{(1)}^{[n-1]}\wedge Q_{(1)}^{[n-3]}\wedge\dots \wedge Q_{(1)}^{[1-n]}}{
\prod\limits_{1\leq k \leq n-1}{Q_{\emptyset}^{[n-2 k]}}
}%
\end{equation}
or equivalently, when written in terms of coordinates,
\begin{equation}
  \label{eq:4}
  Q_{b_1b_2\dots b_n}=\frac{{\displaystyle \det_{1\leq j,k\leq
      n}Q_{b_j}^{[n+1-2 k]}}}{
\prod\limits_{1\leq k \leq n-1}{Q_{\emptyset}^{[n-2 k]}}
}\,.
\end{equation}
It is easy to see that the above expression is the general solution to 
QQ-relation (\ref{eq:2})\footnote{
The statement is true if there is no \(A\) such that \(Q_A=0\). For instance, if \(N=4\), and
\(Q_{\emptyset}=1\), \(Q_{(1)}=\sum_{i=1}^4 \bg_i\), \(Q_{(2)}=0=Q_{(3)}\),
\(Q_{(4)}=\bg_1 \wedge \bg_2 \wedge \bg_3 \wedge \bg_4\), then  the QQ-relations (\ref{eq:2})
hold, whereas (\ref{eq:5}) do not hold. A singular situation with \(Q_A=0\) may appear in practical applications, we observed it in cases related to short representations of supersymmetric algebra, see \secref{sec:su21example}. In the situations we encountered, \eqref{eq:5} holds even if \(Q_A=0\) for some \(A\).
}, which proves that the geometric statement
(\ref{eq:81D}) is equivalent to  QQ-relation (\ref{eq:2}).

In what precedes, we defined the Q-system by a very simple 3-terms bilinear relation (\ref{eq:2}). It  implies many other, in general multilinear, equations relating Q-functions. Equation  (\ref{eq:4}) is one example of such a relation and a few other relations are given throughout the text and in \appref{sec:index-splitting}.

\subsection{Hodge duality map}
Whereas the form \(Q_{(n)}\) defines a plane \(V_{(n)}\) of dimension \(n\) in \(\mathbb{C}^N\), it can be also used to define a plane of dimension \(N-n\) in the dual space. It is easy to see that the intersection condition \eqref{eq:81D} and the union condition \eqref{eq:80} exchange their roles in the dual space and, hence, we can devise a Q-system for the dual geometric construction which, quite naturally, is simply given by  Hodge-dual Q-functions \eqref{eq:75}. In practice, this means that Q-functions with upper indices obey exactly the same algebraic relations as the Q-functions with lower indices. For instance, one can derive
\begin{equation}
\label{eq:76}
  \HQ{b_1b_2\dots b_n}=\frac{{\displaystyle \Det_{1\leq j,k\leq
      n}\HQs{b_j}{[n+1-2 k]}}}{
\prod\limits_{1\leq k \leq n-1}{Q_{\bar \emptyset}^{[n-2 k]}}
}\,,
\end{equation}
etc.

Note that, technically speaking, Hodge duality is not a symmetry of a given Q-system, in the sense that it relates Q-functions with different set of indices. We can think about it as a map, a natural way to construct another collection of Q-functions obeying~(\ref{eq:2}) -- i.e. another Q-system -- differing from the original one by a  relabelling of the Q-functions.

\subsection{Symmetry transformations on Q-systems}
In this section we discuss other symmetries of the equation (\ref{eq:2}). Like the Hodge transformation, they map a given set of Q-functions (Q-system) to another Q-system. By contrast with the Hodge transformation, which maps the spaces \(V_{(n)}\) to the dual space, the transformations we will consider essentially leave the spaces \(V_{(n)}\) invariant.

We have seen that the QQ-relations is a way to rewrite the geometric intersection property in a coordinate form. But any coordinatisation is sensible to a choice of basis, hence there exist transformations which change a basis but do not affect the relation \eqref{eq:2} itself. These basis-changing transformations of Q-system are of two types: rescalings and rotations. 

\subsubsection{Rescalings (gauge transformations)}
Pl\"ucker coordinates are projective: rescaling them does not change the point in Grasmannian that they define. Hence the transformation \(Q_A\to g_{|A|} Q_A\) is a symmetry of the QQ-relation (\ref{eq:2}). As we saw in the last section, this rescaling, defined by arbitrary \(N+1\) functions \(g_{0}(u),g_1(u),\ldots,g_{N}(u)\), modifies \(f_i\) in \eqref{eq:86}. As we agreed to work in the normalisation compatible with \eqref{eq:2}, only \(2\) out of \(N+1\) functions remain independent. We can summarize the admissible rescalings that preserve \eqref{eq:2} in a compact form as
\begin{align}\label{eq:79}
  Q_A &\mapsto \gp^{[\,\nA\,]}\gm^{[\,-\nA\,]}\,Q_A\,,
\end{align}
where \(\gpm\) are certain combinations of \(g_i\).

These rescaling transformations are also known as {\it gauge} symmetries of the Q-system. Indeed, they are local transformations because \(\gpm\) depend on \(u\).

\subsubsection{Rotations}
\label{sec:rotat-q-funct}

One can also rotate\footnote{In this article we allowed a freedom of speech to call any linear non-degenerate transformation as rotation. There is no metric to preserve, hence it does not lead to confusion.} the basis frame, that is to choose different basis vectors \(\zeta_1,\ldots,\zeta_N\). However we cannot rotate the frames independently at different values of the spectral parameter as the QQ-relations are non-local. Therefore, the following transformation
\begin{subequations}\label{eq:QRotate}
\begin{align}
  \tilde  Q_{b}&\mapsto\sum_{c \in {\Bset}}
  h_{bc}\,Q_{c}\,  
\end{align}
of single-indexed Q-functions together with the transformation
\begin{align}
  \tilde Q_{b_1b_2\dots b_n}&=\sum_{c_1,c_2,\dots,c_n \in {\Bset}}
  h_{b_1c_1}^{[n-1]}h_{b_2c_2}^{[n-1]}\dots h_{b_nc_n}^{[n-1]}\,
  Q_{c_1c_2\dots c_n}
\end{align}
\end{subequations}
of multi-indexed Q-functions is a symmetry of the QQ-relation (\ref{eq:2}) if \(h_{bc}\) are \(\bi\)-periodic functions of \(u\):
\begin{align}\label{eq:per}
h_{bc}^+=h_{bc}^-\,.
\end{align}
The transformations \eqref{eq:QRotate} will be called
H-transformations \cite{Gromov2011a} or simply \emph{rotations}.

Note that the case \(h_{bc}=h\,\delta_{bc}\) can be viewed as a particular case of the rescaling symmetry with \(\gp\gm=1\) and \(\gp^+\gm^-=h\). Hence one may restrict to the case of unimodular H-transformations:
\begin{align}
\label{eq:74}
  \det\limits_{1\le b,c\le N}h_{bc}&=1.
\end{align}

In contrast to two local rescaling symmetries, rotations should be thought as a global symmetry. Indeed, periodic functions, e.g. \eqref{eq:per}, in the case of finite-difference equations  play the same role as constants in the case of differential equations. Eventually, we will gauge the rotational symmetry, in order to formulate a twisted Q-system in \secref{sec:twisted-case}. But until then, this symmetry will remain global.

\subsection{Solution of Hirota equation on a strip}
\label{sec:Hirota-equat-strip}

This section is devoted to solving the Hirota equation \eqref{eq:1} on a strip. One case of our interest is the semi-infinite strip of figure \ref{fig:Strip} which corresponds to compact representations of \(\gl(N)\). We remind that in this figure T-functions are identically zero outside the nodes denoted by black dots. The solution for these boundary conditions had been already written in \cite{Krichever:1996qd} and then analysed in a handful of follow-up works. We revise this case as a warm-up for our subsequent studies of T-systems related to non-compact supergroups. 

\begin{figure}[t]
  \centering
\includegraphics{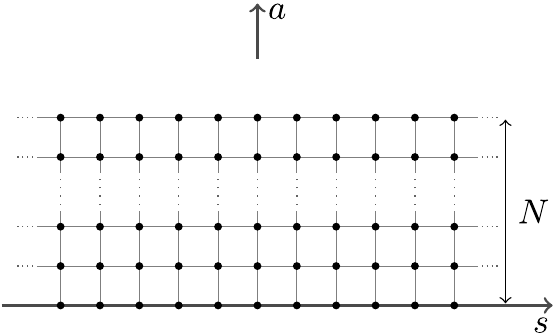}
 \caption{Infinite horizontal strip
 }
  \label{fig:InfStrip}
\end{figure}

The semi-infinite strip should be thought as a special reduction\footnote{Up to minor adjustments, namely the question is about the vertical line \(s=0\) in figure~ \ref{fig:Strip}. We can  replace this line, for the same solution of Hirota equation,  by the horizontal line \(a=N\) and demand that T-functions are non-zero on this horizontal line instead.} of an infinite horizontal strip shown in figure~\ref{fig:InfStrip}, i.e. related to the solution such that \(T_{a,s}\) is
identically zero outside the band  \(0\le a\le N\). We write down the generic solution for this case as well. It was already successfully used in \cite{Gromov:2008gj,Kazakov:2010kf,Caetano2010} for the study of TBA and physical Y-system for the spectrum of principal chiral field (PCF) model at finite space circle.

\ \\
On the infinite horizontal strip of figure \ref{fig:InfStrip}, the
generic solution to the Hirota equation is given by\footnote{The only purpose of Hodge \(\star\)-operation is to convert \((N)\)-forms to \((0)\)-forms,
as \(\star(\zeta_1\wedge
 \zeta_2\wedge\dots\wedge\zeta_N)=1\).}
\begin{align}\label{eq:87}
  &\framebox{\ensuremath{\displaystyle T_{a,s}=\star\left(Q_{(a)}^{[+s]}\wedge
    P_{(N-a)}^{[-s]}\right)}}&\textrm{when }&0\le a\le N&\textrm{and
  }T_{a,s}&=0\textrm{ otherwise.}
\end{align}
By letters \(P\) and \(Q\) we denote two independent sets of Q-functions, each of them expressed through
(\ref{eq:5})\footnote{This means in particular that
\(P_{(n)}
=\frac{P_{(1)}^{[n-1]}\wedge P_{(1)}^{[n-3]}\wedge\dots \wedge P_{(1)}^{[1-n]}}{
\prod_{1\leq k \leq n-1}{P_{\emptyset}^{[n-2 k]}}
}\).
}.

\ \\
On the semi-infinite strip %
of figure \ref{fig:glnstrip}%
, a
solution to the Hirota equation is given by:
\begin{align}
\label{eq:34}
  &\framebox{\ensuremath{\displaystyle T_{a,s}=\star\left(Q_{(a)}^{[+s]}\wedge Q_{(N-a)}^{[-s-N]}\right)}}&\textrm{when }&s\ge 0
  \textrm{ and } 0\le a\le N
\end{align}
In components, the last relation becomes
\begin{align}\label{eq:92}
  T_{a,s}&=(-1)^{a(N-a)}\sum_{|A|=a} Q_A^{[+s]}\HQs{A}{[-s-N]}\,.
\end{align}
The solution \eqref{eq:34}  has to be supplemented with \(T_{0,s}=T_{0,s+1}^+T_{0,s+1}^-/T_{0,s+2}=\star Q_\emptyset ^{[+s]} Q_{(N)}^{[-s-N]}\) for \(s<0\)
and \(T_{a,0}=T_{a-1,0}^+ T_{a-1,0}^- / T_{a-2,0}=\star Q_\emptyset ^{[-a]} Q_{(N)}^{[a-N]}\) for \(a>N\).%

We will now discuss what are the symmetry transformations of Hirota equation and of formulae  \eqref{eq:87} and \eqref{eq:34}, then we will give  a proof that \eqref{eq:87} and \eqref{eq:34}  are indeed the generic solution of the Hirota equation on the corresponding strips.

\subsubsection{Gauge symmetry of the Hirota equation}\label{sec:gauge-symm-hirota}
Hirota equation, for any ``shape'' of non-zero T-functions, is invariant under the transformation
\begin{align}
\label{eq:73}  T_{a,s}(u)&\mapsto g_{(++)}^{[+a+s]} g_{(+-)}^{[+a-s]} g_{(-+)}^{[-a+s]}
  g_{(--)}^{[-a-s]} T_{a,s}(u)\,,
\end{align}
where \(g_{(++)}\), \(g_{(+-)}\), \(g_{(-+)}\) and \(g_{(--)}\) are four arbitrary
functions of the spectral parameter \(u\).
This transformation is usually called the gauge
transformation. 

One can reformulate the Hirota equation \eqref{eq:1} as a Y-system:   \begin{equation}
  \frac{Y_{a,s}(u+\tfrac {\bi} 2)  Y_{a,s}(u-\tfrac \bi 2)}{Y_{a+1,s}(u) Y_{a-1,s}(u)} = \frac{(1+Y_{a,s+1}(u) )(1+Y_{a,s-1}(u))}{(1+Y_{a+1,s}(u))(1+ Y_{a-1,s}(u))}\,,
\end{equation}
using the Y-functions defined as
\(Y_{a,s}=\frac{T_{a,s}^+ T_{a,s}^-}{T_{a+1,s}T_{a-1,s}}\). The Y-functions are obviously
invariant under the gauge transformation \eqref{eq:73}. Typically, physically relevant quantities can be expressed  only through the gauge-invariant functions.

\ \\
If the gauge functions \(g_{(\pm\pm)}\) are \(\bi\)-periodic, i.e. if they
obey \(g^+=g^-\), then the gauge transformation is the multiplication of
\(T_{a,s}(u)\) with a single \(\bi\)-periodic function. Such transformation will
be called a \emph{normalization}. For instance\footnote{For instance \((-1)^{a(N-a)}\) can be written as \(\left((-1)^{(N-1)\bi u}\right)^{[+a]}/\left((-1)^{(N-1)\bi u}\right)^{[-a]}\).
One should note that in this example of normalization, the functions \(g_{(\pm\pm)}\) are not all periodic, but their product is.}, the prefactor \((-1)^{a(N-a)}\) in \eqref{eq:92} can be removed by an appropriate normalisation.

\ \\ As T-functions are determinants of Q-functions, unimodular rotations of the Q-basis have no effect on T-functions. By contrast, the rescaling gauge transformation of Q-system precisely generates gauge transformations of the T-functions. Indeed, one can spot from  \eqref{eq:87} that the rescaling\begin{subequations}
  \label{eq:142}\begin{gather}
    Q_A \mapsto g_ 1^{[\nA]}g_2^{[-\nA]}Q_A\,,\hspace{2cm} P_A \mapsto
    g_3^{[\nA]}g_4^{[-\nA]}P_A\,
    \end{gather}
induces  the following gauge transformation\footnote{This
  transformation clearly matches (\ref{eq:73}) up to relabeling the
  functions \(g\) and their shifts.} 
  \begin{gather}
    T_{a,s}\mapsto g_1^{[a+s]}g_3^{[N-a-s]} g_2^{[-a+s]}g_4^{[a-s-N]}
    T_{a,s}\,.
  \end{gather}
\end{subequations}
In a more restrictive case of \eqref{eq:34}, the rescaling of Q-functions generates only two gauge transformations:
\begin{align}\label{eq:g1}
  Q_A \mapsto& g_
  1^{[\nA]}g_2^{[-\nA]}Q_A\,,&T_{a,s}\mapsto&g_1^{[a+s]}g_1^{[-a-s]}
  g_2^{[-a+s]}g_2^{[+a-s-2N]} T_{a,s}\,.
\end{align}
In fact, the solution \eqref{eq:34} is written in a specific so-called Wronskian gauge in which \(T_{1,0}=T_{0,-1}\) and \(T_{N+1,0}=T_{N,1}\). In arbitrary gauge, the the semi-infinite strip solution should be written as
\begin{align}
\label{eq:34a}
  T_{a,s}&=\frac{f_1^{[a+s]}f_2^{[a-s-N]}}{f_1^{[-a-s]}f_2^{[-a+s-N]}}%
  \star\left(Q_{(a)}^{[+s]}\wedge Q_{(N-a)}^{[-s-N]}\right)&\textrm{when }&s\ge 0
  \textrm{ and } 0\le a\le N\,,
\end{align}
where \(f_1\) and \(f_2\) are two additional arbitrary functions of the
spectral parameter \(u\). Hence, obviously, we speak about \eqref{eq:34} as a general solution modulo gauge symmetry.

One can use \eqref{eq:g1} to set, for instance,
\(Q_{\emptyset}=Q_{\bar \emptyset}=1\). We note that if
\(Q_{\emptyset}=1\) then the expression (\ref{eq:34}) becomes a
determinant
\(T_{a,s}(u)=\det\limits_{1\leq j,k \leq N}Q_k(u+\bi \tfrac
 {2s
\Theta_{a,j}
-2j+1+a-s}2)\)
which coincides (for rectangular Young diagrams) with the determinant
expression written in table \ref{tab:compare}.

\subsubsection{Proof A: existence of solutions to Hirota equation %
}
\label{sec:proof-Hirota-equat}
Let us first prove that (\ref{eq:87}) provides a solution to the
Hirota equation (\ref{eq:1}). Since the Hirota equation is invariant
under the gauge transformations \eqref{eq:142}, it is sufficient to
prove that it is satisfied when \(P_\emptyset=Q_\emptyset=1\).

 We can start by writing
\begin{align}
  \label{eq:16}
  T_{a,s}^-T_{a,s}^+&= %
    {\star (\xx_{1}\wedge
  \xx_{2}\wedge\dots\wedge\xx_{{{N}}})\,\,\star
(\yy_{1}\wedge
  \yy_{2}\wedge\dots\wedge\yy_{{{N}}}) }\,,
\end{align}
where
\begin{align}
&&(\xx_1,\xx_2,\dots,\xx_a) &=(Q_{(1)}^{[+a+s-2]},Q_{(1)}^{[+a+s-4]},\dots,Q_{(1)}^{[-a+s]})\,,\\
&&
(\xx_{a+1},\xx_{a+2},\dots,\xx_N)&=(P_{(1)}^{[-s+N-a-2]},P_{(1)}^{[-s+N-a-4]},\dots,P_{(1)}^{[-s-N+a]})\,,\\
&&
(\yy_1,\yy_2,\dots,\yy_a)&=(Q_{(1)}^{[+a+s]},Q_{(1)}^{[+a+s-2]},\dots,Q_{(1)}^{[-a+s+2]})\,,
\\ &&
(\yy_{a+1},\yy_{a+2},\dots,\yy_N)&=(P_{(1)}^{[-s+N-a]},P_{(1)}^{[-s+N-a-2]},\dots,P_{(1)}^{[-s-N+a+2]})\,.
\end{align}

We can use (\ref{Plucker}), and notice that \(N-2\) terms of the sum in
the r.h.s. vanish because they contain a factor \(\xx_k\wedge\yy_{k+1}\)
(which is zero if \(k\ne a\)). This gives
 \begin{multline}
     T_{a,s}^-T_{a,s}^+=
\star
(\xx_{1}\wedge \dots \wedge
  \xx_{N-1}\wedge\yy_{{{1}}})\,\,\star
(\xx_N\wedge\yy_{2}\wedge\dots\wedge\yy_{{{N}}}) \qquad \qquad\\\qquad\qquad \quad  + \star
(\xx_{1}\wedge \dots \wedge
  \xx_{N-1}\wedge\yy_{{{a+1}}})\,\,\star
(\yy_1\wedge\dots\wedge\yy_{a}\wedge
\xx_N\wedge \yy_{a+2}\wedge
\dots\wedge\yy_{{{N}}})
\\[-.6cm]
\end{multline}
\begin{align}
 & \textrm{i.e.} &
  T_{a,s}^-T_{a,s}^+&=T_{a+1,s}T_{a-1,s} + T_{a,s-1}T_{a,s+1}\,,
\end{align}
which proves that the Hirota equation (\ref{eq:1}) is then satisfied
for \(0<a<N\). Also, the Hirota equation reduces to
\(T_{a,s}^+T_{a,s}^-=T_{a,s+1}T_{a,s-1}\) (resp \(0=0\)) if \(a=0\) or \(a=N\)
(resp \(a<0\) or \(a>N\)), so that it
is clearly satisfied at the boundaries of the strip as well.

It is also clear that the T-functions given by (\ref{eq:34}) obey the
Hirota equation for \(s>0\), since they are a particular case of
(\ref{eq:87}).  At the line \(s=0\), the Hirota equation reduces (if \(a>0\)) to
\(T_{a,0}^+T_{a,0}^-=T_{a+1,0}T_{a-1,0}\), and it indeed holds because
\(T_{a,0}=Q_{\bar\emptyset}^{[a-N]}\). Similarly it holds on the line \(a=0\) (arbitrary \(s\)), explicit formulae are given after \eqref{eq:92}. Now it is immediate to see that it holds
if we put \(T_{a,s}=0\) outside the black dots of figure~\ref{fig:Strip}.

\subsubsection{Proof B: uniqueness of the solution to Hirota equation%
}
\label{sec:ident-q-funct}

We showed above that if T-functions are expressed
 by the Wronskian ansatz \eqref{eq:87} (resp \eqref{eq:34}) then they
 obey the Hirota equation. We now focus on the opposite question:
 given a solution of the Hirota equation, does there exist Q-functions
 such that \eqref{eq:87} (resp \eqref{eq:34}) holds?

The answer is generically yes, as one can convince oneself by a simple
counting argument: If the functions \(T_{a,s}\) are non-zero within the
infinite strip of figure~\ref{fig:InfStrip} then
a solution of the Hirota equation  is characterized by the \(2N+2\)
independent functions \(T_{a,0}\) and \(T_{a,1}\) (where \(0\le a\le N\)), whereas the T-functions
written in \eqref{eq:87} are characterized by the \(2N+2\) independent
function \(Q_{\emptyset}\), \(Q_1\), \(Q_2\), \(\dots
\), \(Q_N\), \(P_\emptyset\), \(P_1\), \(P_2\), \(\dots\), \(P_N\). Similarly in
the case of the semi-infinite strip of figure \ref{fig:glnstrip}, the
solution of the Hirota equation is characterized by the \(N+3\)
independent functions \(T_{0,0}\), \(T_{1,0}\) and \(T_{a,1}\), whereas the T-functions
written in \eqref{eq:34a} are characterized by the \(N+3\) independent
function \(f_1\), \(f_2\), \(Q_{\emptyset}\), \(Q_1\), \(Q_2\), \(\dots
\), \(Q_N\).

In this subsection, we however provide a constructive proof that
Q-functions exist for a generic solution of Hirota equation. We will
focus on the case of the infinite strip, whereas the generalization to
the semi-infinite strip is done in \appref{sec:q-functions-semi}.

Furthermore, there exist degenerate solutions of
the Hirota equation, for which some T-functions are identically zero,
which cannot be expressed in terms of Q-functions by the Wronskian
expression \eqref{eq:87}. An example of this is given in \appref{sec:example-non-wronsk}.

\paragraph{Construction of the Q-functions}

Let us first notice that if \(T_{a,s}\) is given by (\ref{eq:87}), then
the single-indexed Q-functions are solutions of the
following finite-difference ``Baxter equation'' \cite{Krichever:1996qd} (see explainations below):
\begin{align}
\label{eq:37h}
\left(\sum_{r=0}^N Q_{(1)}^{[+2r]} \psi_r\right) \wedge
\left(\sum_{r=0}^N
T_{1,s_0+r}^{[-s_0+r]}\psi_r\right)
\wedge \left(\sum_{r=0}^N T_{1,r+s_0+1}^{[-s_0-1+r]} \psi_r\right) \wedge
\dots \wedge \left(\sum_{r=0}^N T_{1,r+s_0+N-1}^{[-s_0-N+1+r]} \psi_r\right)&=0
\end{align}
for any \(s_0\in\mathbb Z\),
where \(\psi_0\), \(\psi_1\), \(\ldots\), \(\psi_{N}\) are a set of
variables such that the antisymmetric product
\(\psi_0\wedge\psi_1\wedge\dots\wedge\psi_N\) %
does not vanish. For instance, if
\(N=2\), this equation takes the form
\begin{align} &&
\forall s_0\in& \mathbb{N},&
  \forall i \in& \{1,2\},& &&
  \begin{vmatrix}
    Q_i & Q_i^{[+2]}& Q_i^{[+4]}\\
T_{1,s_0}^{[-s_0]}&T_{1,s_0+1}^{[-s_0-1]}&T_{1,s_0+2}^{[-s_0-2]}\\
T_{1,s_0+1}^{[-s_0-1]}&T_{1,s_0+2}^{[-s_0-2]}&T_{1,s_0+3}^{[-s_0-3]}
  \end{vmatrix}
&=0\,. &&
\end{align}
The equation (\ref{eq:37h}) is a consequence of (\ref{eq:87}): indeed,
(\ref{eq:87}) implies that \(\forall k \in
\{0,1,\ldots,N-1\},\,\sum_{r=0}^N T_{1,r+s_0+k}^{[-s_0-k+r]}
\psi_r=\sum_{a=1}^N \alpha_{a,k} \left(\sum_{r=0}^N Q_a^{[+2r]}
  \psi_r\right)\), where \(\alpha_{a,k}=\HPs{a}{[-2s_0-2k]}\).This
implies that all
the \(N+1\) vectors in the wedge product (\ref{eq:37h}) are linear
combinations of  \(N\) vectors \(\sum_{r=0}^N Q_i^{[+2r]}
  \psi_r\), hence the vanishing of the l.h.s. of (\ref{eq:37h}).

Let us now show that this Baxter equation (\ref{eq:37h}) can be used
to define the Q-functions for a generic solution of Hirota equation,
and express the T-functions by the relation (\ref{eq:87}).
To this end, we assume that for a given value of \(s_0\)  the Baxter
equation (\ref{eq:37h}) has \(N\) independent solutions %
\(Q_1\),
\(Q_2\), \dots \(Q_N\). We also assume that for this value of \(s_0\), the
vectors \(\vec T_k\equiv\sum_{r=0}^N
T_{1,s_0+r+k}^{[r-s_0-k]}\psi_r\) (where \(k=0,1,\dots,N-1\)) are
independent.\footnote{
While we use the \emph{forms} notation \eqref{eq:7} for combinations
of the basis elements \(\bg_A\), we use the arrow for combinations
of the variables \(\psi_k\).
}
Then the equation (\ref{eq:37h}) states that  \(N\) independent
vectors \(\vec Q_a\equiv\sum_{r=0}^N Q_{a}^{[+2r]} \psi_r\) belong to the
\(N\)-dimensionnal space spanned by the vectors \(\vec T_k\), which implies
that the \(\vec T_k\) are linear combinations of them, i.e. there exists
functions
\(\alpha_{i,k}(u)\) such that \(\vec T_k
=\sum_{i=1}^N \alpha_{a,k} \,\vec Q_a\) (for \(k=0,1,\dots,N-1\)), i.e. such that
\begin{align} \label{eq:88}
  T_{1,r+s_0+k}^{[-s_0-k+r]}
&=\sum_{a=1}^N \alpha_{a,k} Q_a^{[+2r]}\,,&k&=0,1,\ldots,N-1\,,&r&=0,\dots,N\,.
\end{align}

One can see that the coefficients \(\alpha_{a,k}\) are not independent:
for any \(k\ge 1\) (and any \(r=0,\dots,N-1\)), if we plug the condition \(T_{1,r+s_0+k}^{[-s_0-k+r]}
= \left(T_{1,r+1+s_0+k-1}^{[-s_0-(k-1)+r+1]}\right)^{[-2]}\) into
(\ref{eq:88}), we obtain \(\sum_{a=1}^N \left(\alpha_{a,k}-\alpha_{a,k-1}^{[-2]}\right)
Q_a^{[+2r]}=0\) . Hence the independence of \(Q_1\), \(Q_2\), \(\ldots\), \(Q_N\)
implies that \(\alpha_{i,k}=\alpha_{i,k-1}^{[-2]}\), i.e.
\begin{align}
  \alpha_{i,k}=\alpha_{i,0}^{[-2k]}\,.
\end{align}

We therefore define \(Q_{\emptyset}\), \(\HP\emptyset\) and the
functions \(\HP a\) (where \(1\le a\le N\)) as follows\footnote{The
  existence of two functions \(Q_{\emptyset}\) and \(\HP\emptyset\) such
  that \(T_{0,s}=Q_{\emptyset}^{[+s]}\HPs\emptyset{[+s]}\) is a
  consequence of the Hirota equation at \(a=0\), which reads
  \(T_{0,s}^+T_{0,s}^-=T_{0,s+1}T_{0,s-1}\).
}:
\begin{align}
  T_{0,s}&=Q_{\emptyset}^{[+s]}\HPs\emptyset{[-s]}\,&\HP a&=\alpha_{a,0}^{[+2s_0]}\,.
\end{align}
One defines the Q- and P-functions for arbitrary multi-indices by
(\ref{eq:4}) and by applying (\ref{eq:76}) for the functions \(P\).

Then the functions \(\tilde T_{a,s}=\star\left(Q_{(a)}^{[+s]}\wedge
    P_{(N-a)}^{[-s]}\right)\) provide a solution to the Hirota
  equation, which coincides with \(T_{a,s}\) when \(a=0\) and when \(a=1\)
  and \(s=s_0,s_0+1,\dots,s_0+2N-1\). It is then easy to see that one
  can iteratively show that \(\tilde T_{a,s}=T_{a,s}\) using the Hirota
  equation (assuming that \(T_{a,s}\) is generic, i.e. \(T_{a,s}\neq 0\) for all \(a\),
  \(s\) inside the infinite strip). This
  concludes the proof that, with the functions \(P\) and \(Q\) defined above,
  \(T_{a,s}\) is given by the relation (\ref{eq:87}).

\subsection{On finite-difference (Baxter) equation and B\"acklund transforms}
\label{sec:backl-transf-qq}

In the previous sections, we reproduced the previously-known generic solution 
\cite{Krichever:1996qd} 
of Hirota equation, using  fact that this solution is of a Wronskian type.%

There exists also  an interpretation of the Q-functions from a B\"acklund
flow
\cite{Krichever:1996qd,Zabrodin:1996vm,Kazakov:2007fy,Zabrodin:2007rq}. Here
we remind the main points of this construction, as it gives
an intersting point of view on the Wronskian solution. In particular, we will relate it to the known method of ``variation of constants'', a standard trick used in the resolution of differential or difference equation.

In the proof for wronskian relation \(T\to Q\)  above, the existence of  finite-difference equation \eqref{eq:37h} (Baxter equation) played the decisive role. Let aside for a while  the goal of solving Hirota equation and discuss some generic finite-difference equation of the \(N\)-th order:
\be\label{eq:genN}
\sum_{n=0}^{N} c_n Q^{[2n]}=0\,.
\ee
If \(Q_1,Q_2,\ldots,Q_N\) are \(N\) independent solutions, the equation can be also rewritten as
\begin{align} 
&&
  \begin{vmatrix}\label{eq:detBax}
    Q & Q^{[+2]}&  \ldots & Q^{[2N]}\\
    Q_1 & Q_1^{[+2]}&  \ldots & Q_1^{[2N]}\\
    \ldots & \ldots & \ldots & \ldots \\
    Q_N & Q_N^{[+2]}&  \ldots & Q_N^{[2N]}\\
  \end{vmatrix}
&= 0\,. &&
\end{align}

Suppose we know one solution of \eqref{eq:genN}, say \(Q_1\). What simplification in the search for other solutions could we made? The standard trick (known as ``variation of the constant'') is to write the ansatz \(Q=\Psi\,Q_1\) and to derive the equation on \(\Psi\). After simple manipulations, this new equation can be written as an equation of degree \(N-1\) for the function 
\be\label{eq:linBax}
W=Q_1Q_1^{[2]}(\Psi-\Psi^{[2]})\,,
\ee 
where we introduced the prefactor \(Q_1Q_1^{[2]}\) for further convenience. The message is clear: we reduced the problem of solving a degree-\(N\) equation to solving the equation of degree \(N-1\) plus solving the linear equation \eqref{eq:linBax}. The linear equation can be always solved, at least in terms of a semi-infinite sum\footnote{For instance, solving equations like \eqref{eq:linBax} is routinely performed in the perturbative computation of the AdS/CFT spectrum \cite{Marboe:2014gma}.}  -- the analog of integration in the case of differential equations.

Notably for us, the determinant representation for the equation on \(W\) is
\begin{align} 
&&
  \begin{vmatrix}\label{eq:detBaxN}
    W & W^{[+2]}&  \ldots & W^{[2N-2]}\\
    Q_{12}^{\vphantom{[}+} & Q_{12}^{[3]}&  \ldots & Q_{12}^{[2N-1]}\\
    \ldots & \ldots & \ldots & \ldots \\
    Q_{1N}^{\vphantom{[}+} & Q_{1N}^{[3]}&  \ldots & Q_{1N}^{[2N-1]}\\
  \end{vmatrix}
&= 0\,, &&
\end{align}
which implies that \(N-1\) solutions for \(W\) are \(Q_{12}^+,Q_{13}^+,\ldots, Q_{1N}^+\), where \(Q_{ab}\) are precisely the Q-functions that we discuss in this paper (in the gauge \(Q_{\emptyset}=1\), i.e. \(Q_{ab}=Q_a^+Q_b^--Q_a^-Q_b^+\)).

Obviously, the argument is repeated recursively. If we happened to find one solution for \(W\), say  \(Q_{12}^+\), we can further reduce the degree of equation by one and get the equation which is solved by \(Q_{123}^{[2]},Q_{124}^{[2]},\ldots, Q_{12N}^{[2]}\) etc. 

Hence, resolution of {\it any} degree-\(N\) finite-difference equation is inherently linked to the construction of a Q-system. A part of this construction is to determine \(N\) Q-functions in the set \(Q_{b_1}\,,\ Q_{b_1b_2}\,,\ \ldots\,,\ Q_{b_1\ldots b_N}\). The B\"acklund flow precisely realises this goal, but now for Q-functions of the specific finite-difference equation \eqref{eq:37h}.

A B\"acklund transform (one step in the B\"acklund flow) is introduced as follows. For any Wronskian solution \(T_{a,s}\)  \eqref{eq:34} of the Hirota equation on the
semi-infinite \(\gl(N)\)-strip of figure \ref{fig:glnstrip},
one can notice that for any \(b\in\Bset\), the function
\begin{align}
  F_{a,s}=%
  &\star\left(Q_{(a)}^{[+s]}\wedge
  Q_{(N-1-a)}^{[-s-N+1]}\wedge \bg_b\right)&\textrm{when }&s\ge 0 \textrm{ and }
  0\le a\le N-1\label{eq:71}
\end{align}
is a solution of the Hirota equation on the \(GL(N-1)\)-strip of figure
\ref{fig:glnstrip}, which obeys the Lax pair condition \cite{Krichever:1996qd}
\begin{subequations}\label{eq:126}
  \begin{empheq}[left=\empheqlbrace]{align}
    \label{eq:125}
    T_{a+1,s}F_{a,s}^+-T_{a,s}^+F_{a+1,s}&=T_{a+1,s-1}^+F_{a,s+1}\\
\label{eq:130}
    T_{a,s+1}F_{a,s}^--T_{a,s}^-F_{a,s+1}&=T_{a+1,s}F_{a-1,s+1}^-\,.
  \end{empheq}
\end{subequations}
If \(T\) obeys the Hirota equation and \(F\) obeys \eqref{eq:126}, then \(F\) is
called the B\"acklund transform of \(T\), and it automatically obeys
the Hirota equation.  Moreover, one
can impose the following gauge conditions on \(T\), and see that they
automatically propagate to \(F\) (due to \eqref{eq:126}):
\begin{align}
  \label{eq:127}
  \left\{
    \begin{aligned}
      T_{0,s}&=T_{0,0}^{[-s]}\\
      T_{a,0}&=T_{0,0}^{[+a]}\\
      T_{N,s}&=T_{N,0}^{[-s]}
    \end{aligned}
  \right. \qquad\Rightarrow&\qquad
  \left\{
    \begin{aligned}
      F_{0,s}&=F_{0,0}^{[-s]}\\
      F_{a,0}&=F_{0,0}^{[+a]}\\
      F_{N-1,s}&=F_{N-1,0}^{[-s]}
    \end{aligned}
  \right.\,.
\end{align}

\begin{figure}
  \centering
\includegraphics{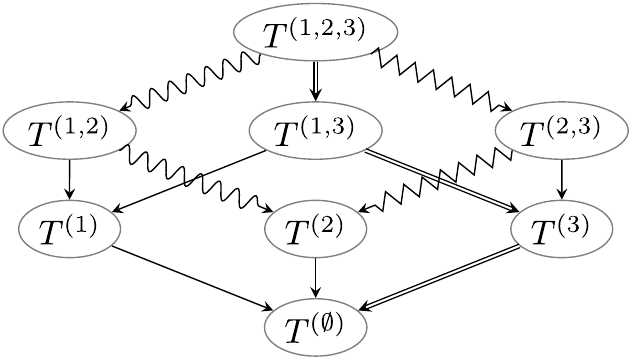}  \caption{Hasse diagram for \(\gl(3)\) T-functions. \(T^{(123)}\) is the original
    T-function out of which a sequence of B\"acklund transforms is
    constructed.
Each arrow corresponds to a
    B\"acklund transformation, reducing by one the number \(|A|\) of indices labeling
  the function \(T^{(A)}\). One should note that different sequences of
  arrows having the same starting and ending point (e.g. the different
  types of wavy arrows) correspond to sequences of B\"acklund
  transformations resulting in the same T-functions. The double
  arrows form a nesting path, i.e. a sequence of arrows from
  \(T^{(\bar\emptyset)}\) to \(T^{(\emptyset)}\).}
  \label{fig:Hasse}
\end{figure}

One can then iterate this procedure: a B\"acklund transform of \(F\)
is a solution of Hirota equation on the \(\gl(N-2)\)-strip. The simplest
example of a sequence of B\"acklund transformations is given by
characters , i.e. for the case when \(T_{a,s}(u)=\chi_{a,s}(\tG)\) for some \(\tG\in \GL(N)\).
We can denote by
\(G^{(b_1,b_2,\dots,b_n)}\in \GL(n)\) a matrix with eigenvalues \(\x_{b_1},
\x_{b_2}, \dots \x_{b_n}\) (where \(\x_1,\x_{2}, \dots \x_{N}\) are the
eigenvalues of \(\tG\)), and for any multi-index \(A\subset \{1,..,N\}\)
denoting the nesting path,
we set \begin{align}
  \label{eq:128}
  T^{(A)}_{a,s}&=\chi_{a,s}\left(\tG^{(A)}\right) \left(\prod_{j\in \{A\}}
    {\x_j}\right)^{-\bi\, u -|A| -\tfrac {s-a-1} 2} \,.
\end{align}
Then each function \(T^{(A)}\) is a B\"acklund transform of \(T^{(Ab)}\)
(for any \(b\notin \{A\}\)). These successive B\"acklund transforms, labeled
by a multi-index \(A\subset \{1,..,N\}\) can be represented by  Hasse
diagram (see figure \ref{fig:Hasse}) \cite{Tsuboi:2009ud}.

From   this example, as well as  from the boundary condition in \((a,s) \) space, we see that each B\"acklund transform can be viewed as a decrease by one of the rank of the symmetry group, as  one might already guess from the ``variation of constants'' method described above which decreases the degree of the finite-difference equation by one at each step as well.

Since the B\"acklund transform of T-functions fits into the same
Hasse diagram as for these characters,  one can define
Q-functions as
\begin{align}
  \label{eq:129}
  Q_A(u)&=T^{(A)}_{0,0}(u+\tfrac \bi 2 |A|)\,.
\end{align}

Let us now call nesting path a sequence of B\"acklund transforms from
\(T^{(\bar\emptyset)}\) to \(T^{(\emptyset)}\) (\textit{e.g.} such as the green sequence of
arrows on figure \ref{fig:Hasse}). Each nesting path is associated
to a sequence of multi-indices \(A_0\), \(A_1\), \(\dots\), \(A_N\), where
\be\label{eq:bosnpath}
\emptyset\equiv\{A_0\}\subset\{A_1\}\subset\dots \subset\{A_N\}\equiv\Bset\,,
\ee
such that  \(|A_n|=n\): For instance
the green nesting path of figure \ref{fig:Hasse} is associated to
\(A_0=\emptyset\), \(A_1=3\), \(A_2=13\), \(A_3=123\). Then one can show \cite{Krichever:1996qd,Zabrodin:1996vm} 
from \eqref{eq:126} that
\begin{align}\label{eq:158}
  \sum_{s\ge 0}
  T_{1,s}^{[+s]}e^{\bi\,s\, \partial_u}&=Q_{\bar\emptyset}^{[1-N]}
W_{A_N;A_{N-1}}W_{A_{N-1};A_{N-2}}\dots W_{A_1;A_0}Q_\emptyset^-\\
\textrm{with}\ \ \ W_{I;J}&=\left(1-\frac{Q_{I}^{[+3-|I|]}}{Q_{I}^{[+1-|I|]}}
\frac{Q_{J}^{[-1-|J|]}}{Q_{J}^{[+1-|J|]}}
e^{\bi\,\partial_u}\right)^{-1}\,,
\end{align}
where \(e^{\bi \partial_u}f(u)=f(u+\bi) e^{\bi \partial_u}\) and where
\((1-f e^{\bi \partial_u})^{-1}=1+f e^{\bi \partial_u}+f e^{\bi \partial_u}f e^{\bi \partial_u}+\dots\).

One can then show (see \cite{Gromov:2010km,Leurent:2012xc}) that the QQ-relation
\eqref{eq:2} arises\footnote{More precisely, this procedure gives the
  QQ-relation up to an \(\bi\)-periodic constant factor which can be viewed as an irrelevant
  normalisation, analogous to the factors \(f_{|A|}\) in
  \eqref{eq:86}.\label{fn:8}} from the constraint \(W_{Aab;Aa}W_{Aa;A}=W_{Aab;Ab}W_{Ab;A}\), i.e. the statement that two sequences of
B\"acklund transformations having the same starting point and the same
endpoint in the Hasse diagram (e.g. the red and blue arrows in figure
\ref{fig:Hasse}) should give rise to the same T-functions.
 Moreover,
one can show that each function \(T^{(A)}\) is then given by
\begin{align}
  \label{eq:124}
T_{a,s}^{(A)}=
  &\star\left(Q_{(a)}^{[+s]}\wedge
  Q_{(|A|-a)}^{[-s-|A|]}\wedge \bg_{\bar A}\right)\,\epsilon^{A\bar A}&\textrm{when }&s\ge 0 \textrm{ and }
  0\le a\le |A|.
\end{align}

Another interesting remark is that  B\"acklund flow suggests a different way to generate the Baxter equation:
\be\label{eq:ONBax}
\big[\ \mathcal{O}^N \cdot Q(u+is/2)\ \big]_{|a=N}=0\,,
\ee
where \(\mathcal{O}(u;a,s)={T_{a,s-1}}\,e^{\frac i2\partial_u-\partial_a}\frac 1{T_{a,s-1}}- T_{a,s}^+ \,e^{-\partial_s-\partial_a}\frac 1{T_{a,s}^+}\). We present the proof in \appref{sec:ONBaxproof}. Note that equations \eqref{eq:ONBax} and \eqref{eq:37h} do not coincide literally. We need to extensively exploit the Hirota equation to show their equivalence.

Although the B\"acklund flow was introduced for the case of Hirota equation on semi-infinite strip, the logic survives if we consider the case of the infinite strip of figure~\ref{fig:InfStrip}. For instance, \eqref{eq:ONBax} holds in either of cases.

\subsection{Solution of Hirota equation on an L- or T-shaped lattice}
\label{sec:Hirota-equation-an}

\subsubsection{Bijection between supersymmetric and non-supersymmetric Q-systems}
In this section we will describe  quite a remarkable fact: one does not need to change the geometric description to accommodate the Q-system for integrable systems with  \(\gl(K|M)\) supersymmetry. One can still use the same Q-system as was used for \(\gl(K+M)\) case. \textit{I.e.} one still considers \(u\)-dependent hyperplanes of \(\mathbb{C}^N\) and imposes the same intersection property \eqref{eq:81D}, however one needs to introduce a different set of coordinates to parameterise it.

Consider a decomposition 
\be\label{eq:dec}
\mathbb{C}^N= \mathbb{C}^{K}\oplus \mathbb{C}^M
\ee  and choose  coordinate vectors \(\xi_1,\ldots,\xi_{N}\) of
\(\mathbb{C}^N\) such that first \(K\) of them span \(\mathbb{C}^{K}\)
and the latter span \(\mathbb{C}^M\). Correspondingly, we introduce a
set of ``bosonic'' indices \({\Bset}=\{1,2,\cdots,K\}\) and a set of
``fermionic'' ones  \({\Fset}=\{K+1,K+2,\dots,N\}\).
Since we will see that in most setups, there is no risk of
confusion\footnote{We should still view the two symbols \(1\in\Bset\) and
  \(1\in\Fset\) as two distinct objects, but the context will allow to
  know which of them is referred to when we use the symbol \(1\).}
between ``bosonic'' and ``fermionic'' indices, one may also label the
latter as \({\Fset}=\{1,2,\dots,M\}\).

The Q-functions which were used in previous paragraphs will be denoted here by small \(q\) to avoid a clash with notations introduced below. The labelling of \(q\)'s is done according to the decomposition \eqref{eq:dec}, i.e. \(q_{AI}\), where \(A\) is a multi-index from \(\Bset\) and \(I\) is a multi-index from \(\Fset\), denote the components of the (p;q)-form
\be\label{eq:ung1}
{\bzdQ}_{(p;q)}=\sum_{|A|=p%
, |I|=q%
} {\bzdQ} _{A I}\xi_{AI} \,,
\ee
where \(\xi_{b_1b_2\dots b_p i_1i_2\dots i_q}\equiv \xi_{b_1}\wedge
\xi_{b_2}\wedge \dots \xi_{b_p}\wedge \xi_{i_1}\wedge
\dots \xi_{i_q}.\) The sum of \((n-k;k)\)-forms 
\be
{\bzdQ}_{(n)}=\sum_{k=0}^n {\bzdQ}_{(n-k;k)}\,
\ee
 is nothing but the \(n\)-form \eqref{eq:7} which defines the hyperplane \(V_{(n)}\) obeying \eqref{eq:81D}.
 
We define the Q-functions \(Q_{A|I}\), which form what will be called the supersymmetric \(\gl(K|M)\) Q-system, by a simple relation\footnote{No summation over \(\bar I\)}
\begin{align}
\label{eq:96}
Q_{A|I} &\equiv \epsilon^{\bar II}\,{\bzdQ}_{A\bar I}\,, &
\textrm{where }\ \ \ \{\bar I\}&={\Fset}\setminus \{I\},
\end{align}  i.e., it is a simple relabeling of the purely bosonic Q-functions. 
In geometric terms, the supersymmetric Q-system is obtained by the partial Hodge transformation along \(\mathbb{C}^M\) direction of \(\mathbb{C}^N\) of the non-supersymmetric Q-system.

This partial Hodge transformation can also be viewed as a rotation of the Hasse diagram, see \secref{sec:polyn-spin-chains} and figure \ref{fig:HasseOr}.

The supersymmetric Q-system of \(\gl(K|M)\)-type was introduced in \cite{Kazakov:2007fy}(see also \cite{Tsuboi:1998ne,Tsuboi:2009ud}. In \cite{Gromov:2010km}, it was observed that the inverse of relation \eqref{eq:96} can be used to map from \(\gl(K|M)\) to \(\gl(N)\) system and it was named ``bosonisation'' (or ``fermionisation'') trick. We extensively rely on this mapping in various places of the paper. 

As should be clear from \eqref{eq:96}, \(Q_{a_1\ldots a_p|i_1\ldots i_q}\) is antisymmetric under a permutation of bosonic \(a\)-indices and under a permutation of fermionic \(i\)-indices. Correspondingly, the graded Q-forms are defined by
\begin{align}
  \label{eq:91}
Q_{(n|p)}&\equiv \sum_{|A|=n,|I|=p} Q_{A|I}\, \xi_{A}\wedge{\zeta}_{I}\,,
\end{align}
where \(\zeta\)'s are some anti-commuting variables independent of \(\xi\) (\(\zeta\)'s and \(\xi's\) are defined to anti-commute between them as well).

For the following discussions, it would be convenient to introduce the Hodge duality map. It is induced from \eqref{eq:52} which can be written more explicitly as 
\(
  \star \bg_{AI} =\epsilon_{AI\bar A\bar I}\,\xi_{\bar A\bar I}\equiv
  (-1)^{|I|\,|\bar
    A|}\epsilon_{A\bar A} \epsilon_{I\bar I} \xi_{\bar A\bar I}\,.
\)
One deduces that the hodge-dual Q-functions should be defined by:
\begin{align}\label{eq:145a}
 \HQ {A|I} &\equiv (-1)^{|A|\,|\bar I|}\epsilon^{\bar A A}\epsilon
  ^{\bar I I} Q_{\bar A|\bar I}.
\end{align}
Finally, for the sake of notational simplicity, we will also sometimes denote  \(Q_A\equiv Q_{A|\emptyset}\) and \(Q_\emptyset\equiv Q_{\emptyset|\emptyset}\) (and the same for Q-functions with upper indices).

\ \\
The bijection between supersymmetric and non-supersymmetric Q-systems is quite a remarkable property; we spend the remainder of this subsection discussing it.

One thing to note is a possibility to rewrite the Hodge transformation as a Grassmannian Fourier transform. Namely, introduce the  sums 
\begin{subequations}
\begin{align}
q[u;\xi_1,\ldots,\xi_N] &\equiv \sum_{n=0}^N q_{(n)}= \sum_{n=0}^{N-M}\sum_{m=0}^M \sum_{|A|=n,|I|=m}q_{AI}\,\xi_A\xi_I\,, \\ 
Q[u;\xi_1,\ldots,\xi_{N-M},\zeta_{\hat 1},\ldots,\zeta_{\hat M}]&\equiv \sum_{n=0}^{N-M}\sum_{m=0}^M (-1)^{M-m+1} \sum_{|A|=n,|I|=m}Q_{A|I}\,\xi_A\zeta_I\,,
\end{align}
\end{subequations}

Then they are related by the Grassmann integral
\be
Q=\int \prod_{i\in\Fset} d\xi_i\, q\,e^{\xi_i \zeta_i} \,.
\ee
This representation suggests adopting a Dirac sea point of view on the bijection transformation. Whereas the description in terms of \(q\)'s corresponds to ``excitations'' of the ``bare vacuum'', description in terms of \(Q\)'s corresponds to ``excitations'' of the ``sea'' created by filling the bare vacuum with all the excitations from the set \(\Fset\). Such an interpretation has a close relation to the Grassmannian construction in the works of Jimbo and Miwa \cite{Jimbo:1983if}.

A similar relation exists between the characters of \(\gl(K|M)\) and \(\gl(N)\) algebras. The characters of compact representations are, correspondingly, the Schur symmetric polynomials \(s_{\lambda}(\x)\) and Schur supersymmetric polynomials  \(s_{\lambda}(\x|\y)\), where \(\lambda\) is a Young diagram, see e.g. \cite{bookDualities}.  Schur polynomials form a ring
\be
s_\lambda\,s_\mu=\sum_{\nu} c_{\lambda\mu}^{\nu}s_{\nu}\,,
\ee
where \(c_{\lambda\mu}^{\nu}\) are the Littlewood-Richardson coefficients which {\it are the same} for the ordinary and supersymmetric cases. Hence, in the limiting case of \(M\to\infty\) and \(K\to\infty\) when none of \(s_\lambda\) is zero due to the bound on a group rank, the rings of ordinary and supersymmetric Schur polynomials are isomorphic. It is not difficult to construct the isomorphism mapping explicitly. We can do this by exploiting the 2\(^{\rm nd}\) Weyl formula from table~\ref{tab:compare}, the reader may also focus on the most important  case of rectangular representations when the Weyl formula reduces to \eqref{GJTcharacter} and can be derived directly from the simplified Hirota equation \eqref{eq:SHE}. The 2\(^{\rm nd}\) Weyl formula expresses all the characters through \(\chi_{1,s}\) -- the characters for the representation \(\lambda=(s,0,0\ldots)\). On the other hand, the generating function for \(\chi_{1,s}(\x|\y)\) of  \(\gl(K|M)\)  is known:\be
\frac{\prod\limits_{i\in\Fset}(1+\epsilon\, \y_i)}{\prod\limits_{a\in\Bset}(1-\epsilon\, \x_a)}=\sum_{s=0}^\infty \epsilon^s\, \chi_{1,s}(\x|\y)\,.
\ee
Hence the map between supersymmetric and non-supersymmetric characters is induced by the replacements
\be\label{eq:yx1}
1+\epsilon\, \y \leftrightarrow \frac 1{1-\epsilon\, \x}\,
\ee
in the generating function. The mapping becomes an isomorphism in the limit when the numbers of \(\y\)'s and \(\x\)'s are infinite. This relation can be thought of as a statement (equivalent to the partial Hodge transformation) that ``adding a fermionic index'' is the same as ``removing a bosonic index''. Indeed, adding a fermionic index or removing a bosonic index is realised by multiplication of the generating function by a factor \(1+\epsilon \alpha\) (where \(\alpha\) is either \(-\y\) or \(\x\)). See also \secref{sec:backl-flow-supersymm} for a motivation of this principle in terms of B\"acklund transforms.

\subsubsection{QQ-relations with a grading}
\label{sec:qq-relations-with}

As explained in the previous section, the supersymmetric Q-system is obtained by  a simple relabelling of ordinary Q-functions: we just use \(Q_{A|I}\) instead of \(q_{A\bar I}\). Therefore, all the QQ-relations in the supersymmetric basis would be, eventually, an algebraic consequence of \eqref{eq:2}. Nevertheless, despite the simplicity of \eqref{eq:2}, the emergent algebraic structure turns out to be very rich.

\ \\
First of all, the original QQ-relation \eqref{eq:2} splits into three
equations due to possibility of multiplying Q-functions with different
gradings
\cite{GohmannSeel,Beisert:2005di,Belitsky:2006cp,Kazakov:2007fy,Zabrodin:2007rq,Bazhanov:2008yc}
\begin{subequations}
\label{eq:42}
  \begin{align}
    \label{eq:41}
       Q_{A|I}Q_{A ab|I} &=Q_{A a|I}^{+} Q_{A b|I}^{-}-
       Q_{A a|I}^{-}
       Q_{A b|I}^{+}\,,\\
\label{eq:49}
       Q_{A a|I}Q_{A|I i} &= Q_{A a|I i}^{+}Q_{A|I}^{-}-
       Q_{A|I}^{+} Q_{A a|I i}^{-} \,,\\
\label{eq:78}
       Q_{A|I}Q_{A|I ij} &=Q_{A|I i}^{+} Q_{A|I j}^{-}-
       Q_{A|I i}^{-} Q_{A|I j}^{+}\,,
     \end{align}
\end{subequations}
It is easy to see how they correspond to \eqref{eq:2}, especially if to note the general rule that adding a fermionic index in \(Q_{A|I}\) is equivalent to removing this %
index from \(q_{A\bar I}\).

Now, we present a handful of algebraic relations which all follow from \eqref{eq:42}. Their derivation is given in \appref{sec:proofs}.

Firstly,  we have the obvious relations
\begin{align}\label{eq:185}
  Q_{(n|0)}
&=\frac{Q_{(1|0)}^{[n-1]}\wedge Q_{(1|0)}^{[n-3]}\wedge\dots \wedge Q_{(1|0)}^{[1-n]}}{
\prod_{1\leq k \leq n-1}{Q_{\emptyset}^{[n-2 k]}}
}\,,&  Q_{(0|p)}
&=\frac{Q_{(0|1)}^{[p-1]}\wedge Q_{(0|1)}^{[p-3]}\wedge\dots \wedge Q_{(0|1)}^{[1-p]}}{
\prod_{1\leq k \leq p-1}{Q_{\emptyset}^{[p-2 k]}}
}\,,
\end{align}
which are identical to the relations of \secref{sec:Hirota-equat-strip} because they do not mix ``bosonic'' and
``fermionic'' indices. Secondly, the following
expressions 
 give all
Q-functions in terms of \(Q_{\emptyset}\), \(Q_{(1|0)}\),
\(Q_{(0|1)}\) and \(Q_{(1|1)}\)\footnote{One should note that
\(Q_{\emptyset}\),
\(Q_{(1|0)}\), \(Q_{(0,1)}\) and \(Q_{(1|1)}\) are not independent: they are
related by (\ref{eq:49}), which states that \(Q_{(1|0)}\wedge
Q_{(0|1)}=Q_{(1|1)}^+Q_{\emptyset}^--Q_{(1|1)}^-Q_{\emptyset}^+\).
}:
\begin{align}
\label{eq:45}  Q_{(n|n)}&=\frac {(-1)^{\frac {n(n-1)}2}} {n!}\frac{Q_{(1|1)}^n}{Q_{\emptyset|\emptyset}^{n-1}}\equiv \frac {(-1)^{\frac {n(n-1)}2}} {n!}\frac{Q_{(1|1)}\wedge
  Q_{(1|1)}\wedge\dots\wedge Q_{(1|1)}}{Q_{\emptyset|\emptyset}^{n-1}}\hspace{-4cm}&&%
\hspace{4.3cm} (n\textrm{ times})&\\
\label{eq:46}Q_{(n|p)}&=
\frac{Q_{(p|p)}^{[t]}}{Q_{\emptyset}^{[t]}}\wedge Q_{(n-p|0)}&&\hspace{1cm}\textrm{where
}n\ge p \nonumber
\\ &&&\textrm{ and }t\in\{n-p,n-p-2,\dots,-n+p\}\\
\label{eq:47}Q_{(n|p)}&= %
   (-1)^{n(p+1)}
\frac{Q_{(n|n)}^{[t]}}{Q_{\emptyset}^{[t]}}\wedge Q_{(0|p-n)}&&\hspace{1cm}\textrm{where
}n\le p \nonumber
\\ &&&\textrm{ and }t\in\{p-n,p-n-2,\dots,-p+n\}\,%
.
\end{align}
These expressions were already implicitly incorporated into \emph{sparce determinants} of \cite{Tsuboi:2009ud}, and rewritten in terms of forms in \cite{Gromov:2014caa} for the \(\psu(2,2|4)\) case without proofs.

These relations can be recast into equations for the components
\(Q_{A|I}\) of these forms: %
the relation (\ref{eq:45}) becomes
\begin{equation}
\label{eq:53}
  Q_{A|I}= \frac{\det\limits_{\substack{a\in A\\i\in I}}
    Q_{a|i}}{(Q_{\emptyset})^{n-1}}\qquad\textrm{or more explicitly}
  \qquad
 Q_{b_1,b_2,\dots,b_n|f_1,f_2,\dots,f_n}= \frac{\det\limits_{1\leq i,j\leq
    n} Q_{b_i|f_j}}{(Q_{\emptyset})^{n-1}}\,,\\
\end{equation}
whereas the equation (\ref{eq:46}) (resp (\ref{eq:47})) states that if
\(|A|=n\), and \(|I|=p\) with \(n\ge p\) (resp \(n\le p\)), then for any
\(t\in\{|n-p|,|n-p|-2,\dots,-|n-p|\}\) we have
\begin{align}
\label{eq:comp46}&Q_{A|I}= \frac{(-1)^{p(n+1)}}{Q_\emptyset^{[t]}}\sum_{\substack {|B|=|I|\\|C|=|A|-|I|}}
{Q_{B|I}^{[t]}\,Q_{C|\emptyset}^{}}\,%
\delta_{A}^{BC}\,,&
  \\ %
\label{eq:comp47}
    \textrm{resp.}\ \ \ %
  &%
Q_{A|I}= \frac{(-1)^{n(p+1)}}{Q_\emptyset^{[t]}}\sum_{\substack {|J|=|A|\\|K|=|I|-|A|}}
{Q_{A|J}^{[t]}\,Q_{\emptyset|K}^{}}\,%
\delta_{I}^{JK}\,,%
\\\label{eq:819}\textrm{where\ \ \ 
}&
\delta_{i_1i_2\dots i_n}^{j_1j_2\dots j_n}\equiv \det\limits_{1\leq a,b\leq n}\delta_{i_a}^{j_b}
\,.
\end{align}
Note that the role of e.g. \(\delta_A^{BC}\) in \eqref{eq:comp46} is to anti-symmetrise the index \(BC\).

Another interesting class of relations is obtained by using both Q-functions and their Hodge duals: 
\begin{align}
\label{eq:67}
Q_{\emptyset|J}&= \frac {%
(-1)^{n%
  |J|%
}%
}{Q_{\bar \emptyset|\bar\emptyset}^{[t]}} \sum_{|A|=n}\HQ{A|\emptyset} Q_{A|J}^{[t]}&&\hspace{1cm}\textrm{when }n=|\Bset|-|\Fset|+|J|\ge 0\nonumber\\[-.5cm]
&&&\textrm{and } t\in \{-n,-n+2,\dots,n-2,n\}\,,
\\\label{eq:147}
Q_{A|\emptyset}&= \frac {(-1)^{n |A|}
}{Q_{\bar \emptyset|\bar\emptyset}^{[t]}} \sum_{|J|=n}\HQ{\emptyset|J} Q_{A|J}^{[t]}&&\hspace{1cm}\textrm{when }n=|\Fset|-|\Bset|+|A|\ge 0\nonumber\\[-.5cm]
&&&\textrm{and } t\in \{-n,-n+2,\dots,n-2,n\}\,,
\\
\label{eq:84}
&\hspace{-1.5cm}
\sum_{|A|=|I|}Q_{A|I}\HQ{A|J}=(-1)^{|I|(|\Fset|+1)}\sum_{\quad\mathclap{|L|=|J|-|I|}\quad}\delta_{LI}^{J} \HQ{\emptyset|L}Q_{\emptyset|\emptyset}\hspace{-1cm}&&\hspace{1cm}\textrm{when }|J|-|I|=|{\Fset}|-|{\Bset}|\ge 0
\,, %
\\\label{eq:93}
&\hspace{-1.5cm}\sum_{a\in \Bset}Q_a^{[+t]}\HQ a=
\begin{cases}
(-1)^{|\Bset|+|\Fset|+1}\HQs \emptyset{\pm} Q_\emptyset^{[t\mp 1]}&\textrm{ if
}t=\pm n \textrm{
  where }n\equiv|\Bset|-|\Fset|>0\\
0&\textrm{ if }t\in \{n-2,n-4,\dots,-n+2\}\,,
\end{cases}\hspace{-5cm}
\\\label{eq:94}
&\hspace{-1.5cm}\sum_{a\in \Bset}Q_a\HQ a=
  \HQs\emptyset - Q_\emptyset^+-\HQs\emptyset + Q_\emptyset^-%
\hspace{-1cm}&&\hspace{1cm}\textrm{if }|\Bset|=|\Fset|\,.
\end{align}
Note that equation \eqref{eq:147} %
is obtained from
\eqref{eq:67} %
by interchanging bosonic and
fermionic indices. Obviously the same can be done for any other
relation. For instance  it
follows from (\ref{eq:93}) that
\begin{align}
\sum_{i\in \Fset}Q_{\emptyset|i}^{[\pm n]}\HQ
  {\emptyset |i}&=(-1)^{|\Bset|+|\Fset|+1}\HQs \emptyset \pm
  Q_\emptyset^{[\pm n \mp 1]}
&&\hspace{1.5cm}\textrm{when }n\equiv|\Fset|-|\Bset|>0
\,.
\end{align}
Similarly, one can take the Hodge dual of each relation, i.e. perform
the substitutions
\begin{align}\label{eq:149}
  Q_{A|I}\mapsto&Q^{A|I}\,,&  Q^{A|I}\mapsto&(-1)^{(\nA+\nI)(|\bar
    A|+|\bar I|)}Q_{A|I}\,,&
\end{align}
which leave the QQ-relations invariant, and are compatible with the
sign in \eqref{eq:145a}.
For instance
\eqref{eq:67} becomes
\begin{align}\label{eq:151}
\HQ {\emptyset|J}&= \frac {
(-1)^{n
  |\Fset|
}
}{\HQs {\bar \emptyset|\bar\emptyset} {[t]}}
 \sum_{|A|=n}Q_{A|\emptyset} \HQs {A|J} {[t]}&&\hspace{1cm}\textrm{when }n=|\Bset|-|\Fset|+|J|\ge 0\nonumber\\[-.5cm]
 &&&\textrm{and } t\in \{-n,-n+2,\dots,n-2,n\}\,,
\end{align}
where the sign \((-1)^{n|\Fset|}\) is obtained by simplifying the
expression \((-1)^{n|J|+|A|\,|\bar A|}\) obtained from the substitution \eqref{eq:149}.

\paragraph{Examples.} It  turned out \cite{Gromov:2014caa} that in the case of
AdS/CFT (where {\THook} in figure~\ref{fig:THook-strips} has \({\Kone}={\Ktwo}={\Mone}={\Mtwo}=2\); and \(Q_{\emptyset|\emptyset}=Q_{\bar\emptyset|\bar\emptyset}=1\)), the above-listed relations are very useful. We give below some of them specified to this particular case:
\begin{itemize}
\item Setting \(|J|=1\)  or  \(|A|=1\) in (\ref{eq:67},\ref{eq:147}), one gets two interesting relations
  \begin{align}\label{eq:141}
    Q_{\emptyset|i}=-\sum_{a}\HQ{a|\emptyset} Q_{a|i}^{\pm}\,, \qquad Q_{a|\emptyset}=-\sum_{i}\HQ{\emptyset|i} Q_{a|i}^{\pm}\,,
  \end{align}
which  correspond to
(4.14a1)-(4.14b1) in \cite{Gromov:2014caa}.  The other relations (4.14) are
obtained by Hodge duality \eqref{eq:149}.
\item Setting \(|I|=|J|=1\) in (\ref{eq:84}), one gets
\(\sum_{a}\HQ{a|i}Q_{a|j}=-\delta^i_j\), which is the relation
(4.15a) in \cite{Gromov:2014caa}.
\item In the AdS/CFT case, the r.h.s. of (\ref{eq:94}) vanishes,
  giving the relation (4.16a) in \cite{Gromov:2014caa}.
\end{itemize}

We did not describe all possible relations in this section. For instance, another interesting class worth mentioning involves equations of finite difference type of order \(\geq 2\). Such kind of Baxter-type relations were exploited for instance in \cite{Gromov2011a}. Very recently, the fourth-order equation having \(Q_{\emptyset|i}\) as four solutions played the decicive role in the derivation of the BFKL equation from the AdS/CFT integrability \cite{Alfimov:2014bwa}. 

We see that the algebra of Q-functions is indeed very rich. We should think about these relations as an opportunity for discovering short-cuts through the Q-system that link the physically most-improtant Q-functions for practical problems to solve.
For each particular problem or calculation, one should look for a specific, most convenient subset of these relations.

\subsubsection{Expression for T-functions in a {\THook}}
\label{sec:expr-t-funct-2}

\begin{figure}
  \centering \includegraphics{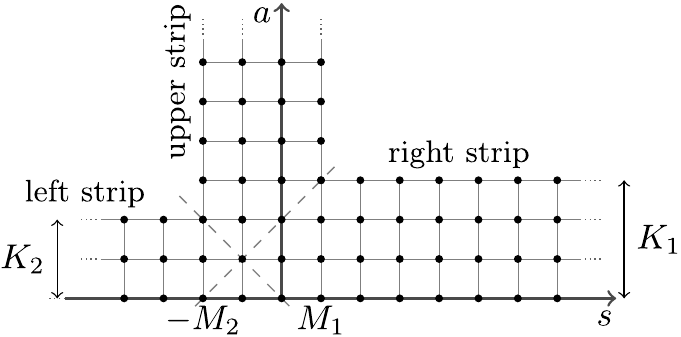}
  \caption{The right, left, and upper strip of the {\THook} are
    delimited by the diagonals (dashed, gray). The figure corresponds
    to \({\Kone}=3\), \({\Ktwo}=2\), \({\Mone}=1\) and \({\Mtwo}=2\).}
  \label{fig:THook-strips}
\end{figure}

At the level of Q-functions, we have seen that it was necessary to
introduce two different sets of indices, which we called ``bosonic'' and
``fermionic'', and which are distinguished in the QQ-relations
(\ref{eq:42}). If we denote by \(K\) (resp \(M\)) the number \(|\Bset|\) (resp \(|\Fset|\)) of bosonic
(resp fermionic) labels, then the Q-functions are related to the
algebra \(\mathfrak{gl}(K|M)\).

At the level of   T-functions which obey
the Hirota equation \eqref{eq:1} on a generic {\THook} fig.\ref{fig:THook-strips}, one should also specify a real
form: It is  \(\su({\Kone},{\Ktwo}|M)\) in the most general case, with \({\Kone}+{\Ktwo}=K\)). As a consequence, the set \(\Bset\) of bosonic indices
should be split into a union of two non-intersecting sets \({\Bset}_1\) and \({\Bset}_2\):
\begin{align}
  {\Bset}&={\Bset}_1\sqcup {\Bset}_2\,,&\textrm{where
  }|{\Bset}_1|&={\Kone}\,,&|{\Bset}_2|&={\Ktwo}\,%
  .
\end{align}
With these two sets, we introduce \emph{graded} and \emph{ungraded} forms in the same way as in, respectively, \eqref{eq:91} and \eqref{eq:ung1}:
\begin{align}\label{nonbzdQ}
   Q_{(n_1,n_2|p)} &\equiv
\sum_{|R|=n_1, |L|=n_2, |I|=p}
Q_{RL|I} ~\xi_R\wedge\xi_L\wedge \bg_I\,,%
\,,\\
\label{bzdQ}
{\bzdQ} &\equiv 
\sum_{|R|=n_1, |L|=n_2, |I|=p}
{\bzdQ}_{RLI} ~\xi_R\wedge\xi_L\wedge \xi_I\,,&&R\subset {\Bset}_1,
L\subset {\Bset}_2, I\subset {\Fset}\,
.
\end{align}

With these notations, the Hirota equation on the \(({\Kone}|{\Mone}+{\Mtwo}|{\Ktwo})\)
{\THook} %
has the following solution %
\begin{subequations}\label{eq:155}
  \begin{align}
    T_{a,s}&=\varepsilon_r(a,s)%
    \star \big(Q_{(a,0|0)}^{[\tilde
      s%
      ]}\wedge Q_{({\Kone}-a,{\Ktwo}|M)}^{[-\tilde
      s%
      ]}\big)%
    &&%
    {\textrm{if }\tilde s\ge \tilde a %
    },%
    \label{eq:60}\\
    T_{a,s}&=\varepsilon_u(a,s)%
    \star \big(Q_{({\Kone},0|{\Mone}-s)}^{[+\tilde
      a%
      ]}\wedge Q_{(0,{\Ktwo}|{\Mtwo}+s)}^{[-\tilde
      a%
      ]}\big)%
    & &%
    {\textrm{if }\tilde a\ge |\tilde
      s|}%
    ,\label{eq:61}\\
    T_{a,s}&=\varepsilon_l(a,s)%
    \star \big(Q_{({\Kone},{\Ktwo}-a|M)}^{[-\tilde
      s%
      ]}\wedge Q_{(0,a|0)}^{[+\tilde
      s%
      ]}\big) %
    &&%
    {\textrm{if }\tilde s\le -\tilde a%
    },\label{eq:62}
  \end{align}
\end{subequations}
where \({\Mone}+{\Mtwo}=M\) and the choice of \({\Mone}\) and \({\Mtwo}\) is
arbitrary and defines the origin of the {\THook}, as in figure
\ref{fig:THook}. In \eqref{eq:155}, we used the notations
{\belowdisplayshortskip=0pt
\belowdisplayskip=0pt
\begin{align*}
\tilde s&=s-s_0\,,&\tilde  a&=a-a_0\,, &
s_0&=\frac{-{\Kone}+{\Ktwo}+{\Mone}-{\Mtwo}} 2\,,&a_0&=\frac{K-M}2\,,
\end{align*}
}{
 \abovedisplayshortskip=0pt
\abovedisplayskip=0pt
\begin{align}\label{eq:157}
\varepsilon_r(a,s)&=\bi^{M(a-s)}(-1)^{a\,{\Ktwo}}\,,&
\varepsilon_l(a,s)&=\bi^{M(a-s)}(-1)^{a({\Kone}+M)}\,,&%
\end{align}
\begin{align*}
\textrm{and}&&
\varepsilon_u(a,s)&=\bi^{M(a-s)}%
(-1)^{(a+K)({\Mtwo}+s)+{\Ktwo}({\Kone}+M)}\,.
\end{align*}
}
The practical meaning of these notations is: \((s_0,a_0)\) is the coordinate of the intersection of the diagonals on figure
\ref{fig:THook-strips}, and \((\tilde s,\tilde a)\) are the coordinates, with
respect to this point, of an arbitrary node on the {\THook}. 

The proof that \eqref{eq:155} indeed solve the Hirota equation is given in \appref{sec:proof-that-Hirota}. There we use, in particular, a possibility to represent the solution in terms of the bosonised functions \eqref{bzdQ}.

The semi-infinite strip of figure \ref{fig:glnstrip} is the case \(K_1=N\), \(K_2=M_1=M_2=0\)  of the {\THook}. In this case, the above expressions of T-functions match the expressions of \secref{sec:Hirota-equat-strip} up to an overall redefinition of the (shift of the) Q-functions:
\begin{align}
  Q_{(n)}&\mapsto Q_{(n)}^{[3N/2]}\,,
\end{align}
which obviously leaves all QQ-relations invariant.

Other interesting special cases of the Wronskian solution \eqref{eq:155} include: the compact real form \(\su(M|K_2)\) corresponding to \(B_1=\emptyset\) and L-hook shape of non-zero T-functions shown in figure~\ref{fig:LHook}; the compact real form \(\su(K_1,|M)\) corresponding to \(B_2=\emptyset\) and a mirror-reflected  L-hook \footnote{For real forms we use notations of \cite{Volin:2010xz}, a more detailed exposition is planned in \cite{MV}.  Although the real form \(\su(K_1,|M)\) is isomorphic to \(\su(K_1|M)\)  and hence comma is usually not written, we should distinguish the case with comma and without when \(\su(K_1,|M)\) and \(\su(M|K_2)\) are simultaneously subalgebras of a bigger non-compact algebra \(\su(K,|M|K)\).}; and, finally the non-compact and non-supersymmetric case \(\su(K_1,K_2)\) which corresponds to \(\Fset=\emptyset\) and the ``slim-hook'' shape first discussed in \cite{Gromov:2010vb}  (see e.g. figure 1b in \cite{Volin:2010xz}).  The slim-hook is solved using purely bosonic Q-system constructed on \(\mathbb{C}^{K_1+K_2}\). We expect that Hirota equation on such a hook will appear in affine Toda integrable models.

\subsubsection{Symmetries}
\label{sec:gauge-transf-t}

Similarly to its bosonic version, the graded Q-system  has rotational and rescaling symmetry.

\paragraph{Gauge transformations.}
It is suitable to parameterise two available rescalings (gauge transformations) by
\begin{align}\label{eq:146}
  Q_{A|I}\mapsto& g_ 1^{[\nA-\nI]}g_2^{[-\nA+\nI]}Q_{A|I}%
\end{align}
which replaces \eqref{eq:79}.

This transformation generates the following two gauge transformation of T-functions:
\be
T_{a,s}\mapsto g_{1}^{[a+s-s_0]}g_2^{[-a+s-s_0]}g_1^{[-a-s+s_0+K-M]}g_2^{[a-s+s_0-K+M]}\,T_{a,s}\,
\ee
Another two gauge degrees of freedom of T-functions (cf. \eqref{eq:73}) are actually fixed for what concerns the solution \eqref{eq:155}. This solution was specially written to satisfy the {\it Wronskian gauge}:
\begin{align}
\label{eq:54}  T_{{\Kone}+1,{\Mone}}&=T_{{\Kone},{\Mone}+1}&\textrm{and}&&T_{{\Ktwo}+1,-{\Mtwo}}&=T_{{\Ktwo},-{\Mtwo}-1}\,,
\end{align}
which immediately implies, by virtue of Hirota equation, 
\begin{align}
T_{{\Kone}+n,{\Mone}}&=T_{{\Kone},{\Mone}+n}&\textrm{and}&&T_{{\Ktwo}+n,-{\Mtwo}}&=T_{{\Ktwo},-{\Mtwo}-n}\,,\ \ \  n\geq 0\,.
\end{align}
and reflects the fact that the corresponding characters are equal: \(\chi_{{\Kone}+n,{\Mone}}=\chi_{{\Kone},{\Mone}+n}\), \(\chi_{{\Ktwo}+n,-{\Mtwo}}=\chi_{{\Ktwo},-{\Mtwo}-n}\,.\)

The signs \(\varepsilon_r\), \(\varepsilon_u\), and \(\varepsilon_l\) in \eqref{eq:155} were chosen, in particular, to ensure the Wronskian gauge condition (\ref{eq:54}). Hence, as in the case of semi-infinite strip, we understand that \eqref{eq:155} is a general solution modulo two gauge transformations.

\paragraph{Rotations.}
The graded \(\gl(N-M|M)\) Q-system is algebraically equivalent to its
bosonised version and hence it is in principle invariant under
\(\GL(N)\) H-transformations originating from
\eqref{eq:QRotate}. However, the obvious explicit rotations are only a
subgroup \(\GL(N-M)\times\GL(M)\) which leaves invariant the
decomposition \eqref{eq:dec}. All other rotations, implicitly there, 
would not preserve the T-functions and hence should not be considered.
 
Furthermore, the T-functions of T-hook are invariant only under unimodular rotations from \(\GL(K_1)\times\GL(K_2)\times \GL(M)\) which preserve the grading of the forms \eqref{nonbzdQ}. It is important to realise that prior to constructing a T-hook, one has to agree how to decompose indices into sets \(\Bset_1\), \(\Bset_2\), and \(\Fset\) and then stick to the bases which respect such a decomposition. Also, it is possible to exchange the role of bosonic and fermionic indices and, in particular, decompose into sets \(\Bset\), \(\Fset_1\), \(\Fset_2\) . The choice of a basis and decomposition into sets depends on a real form one wishes to associate to T-hook and how this real form is related to analytic properties of Q-functions. From the same \(\GL(K|M)\)-system we can construct different T-hooks. It is an additional question to justify which of the T-hooks (maybe several) are physically meaningful in a given explicit problem and what is their physical interpretation.

\subsubsection{B\"acklund  flow in supersymmetric case }
\label{sec:backl-flow-supersymm}
One can also introduce Q-functions from a sequence of B\"acklund transformations. It was demonstrated already for the bosonic case in \secref{sec:backl-transf-qq}, and we saw that QQ-relations can be interpreted as the fact that different paths on the Hasse diagram (see figure \ref{fig:Hasse}) correspond to the same transformation.

This approach can be generalized to the super-symmetric case, i.e. for  
{\LHook} \cite{Kazakov:2007fy} and   {\THook}
\cite{Hegedus:2009ky}.  The relation to Wronskian determinants was shown in \cite{Tsuboi:2011iz}. We remind the arguments for the {\LHook} case
only.
 Consider the Lax pair condition \eqref{eq:126} near the internal boundary of the hook, namely set \((a,s)=(K-1,M)\) in (\ref{eq:125}) and 
\((a,s)=(K,M-1)\))  in (\ref{eq:130}). One can see
that if \(T\) obeys the Hirota equation on a \((K|M)\) {\fHook}, then \(F\)
can obey it on a \((K-1|M)\) or a \((K|M+1)\) {\fHook}. The transformation
from a \((K|M)\) to a \((K-1|M)\) {\fHook} is the exact analog of the
B\"acklund transformation of \secref{sec:backl-transf-qq}, and
corresponds to the removal of a ``bosonic'' index from the Q- and T-functions. By
contrast, it is the inverse of the transformation from a \((K|M)\) to a
\((K|M+1)\) {\fHook} which can be regarded as a B\"acklund transformation
removing a ``fermionic'' index; in this case, the function \(T\) of (\ref{eq:126}) is
the B\"acklund transform of the function \(F\).  Hence we see that the same transformation ``adds a fermionic index'' or ``removes a bosonic index'', justifying the partial Hodge transformation (\ref{eq:96}), and the analogous observation \eqref{eq:yx1} at the level of characters.

Furthermore, one finds out from the linear system (\ref{eq:126}) and the definition (\ref{eq:129}) that  the generating series \eqref{eq:158} can be generalized to the {\LHook}. To this end, we encode a nesting path as a sequence of labels
\be\label{eq:nestingpath}
(A_0|I_0\equiv\emptyset|\emptyset)\subset A_1|I_1\subset A_2|I_2 \subset \dots
\subset (A_{K+M}|I_{K+M}\equiv\Bset|\Fset)\,,
\ee
which are
included into each other and obey \(|A_n|+|I_n|=n\).
Each step \(n\) of the nesting path is a B\"acklund transform
which can be either associated to a ``bosonic'' index (then
\(|A_{n+1}|=|A_n|+1\) and \(I_{n+1}=I_n\)) or a ``fermionic'' index (then
\(A_{n+1}=A_n\) and \(|I_{n+1}|=|I_n|+1\)). Then, the generalization of the generating series \eqref{eq:158} is
 \begin{gather}\label{eq:159}
 (-\bi)^M \sum_{s\ge 0}
  T_{1,s}^{[+s+(M-K)/2]}e^{\bi\,s\, \partial_u}=Q_{\bar\emptyset}^{[1-K+M]}
\,W_{K+M}\,W_{K+M-1}\,\dots\, W_{1}\,Q_\emptyset^-\,,\\
\textrm{with\ \ }W_{n}=
\begin{cases}
\left(1-\frac{Q_{A_n|I_n}^{[+3-|A_n|+|I_n|]}}{Q_{A_n|I_n}^{[+1-|A_n|+|I_n|]}}
\frac{Q_{A_{n-1}|I_{n-1}}^{[-1-|A_{n-1}|+|I_{n-1}|]}}{Q_{A_{n-1}|I_{n-1}}^{[+1-|A_{n-1}|+|I_{n-1}|]}}
e^{\bi\,\partial_u}\right)^{-1}\,,&\textrm{if \(|A_{n}|=|A_{n-1}|+1\)}\\[.4cm]
\,\,\,\,1-\frac{Q_{A_n|I_n}^{[-1-|A_n|+|I_n|]}}{Q_{A_n|I_n}^{[+1-|A_n|+|I_n|]}}
\frac{Q_{A_{n-1}|I_{n-1}}^{[+3-|A_{n-1}|+|I_{n-1}|]}}{Q_{A_{n-1}|I_{n-1}}^{[+1-|A_{n-1}|+|I_{n-1}|]}}
e^{\bi\,\partial_u}\,,&\textrm{if \(|I_{n}|=|I_{n-1}|+1\)}
\end{cases}
\,.
\end{gather}

As an illustration (which will be used in the next section),
the coefficient of \(e^{\bi\, \partial_u}\) gives 
\begin{gather}\label{eq:160}
  T_{1,1}%
  =\bi^MQ_{\bar\emptyset}^{[s_0]} \sum_{n=1}^{K+M} \varepsilon_n %
    \frac{Q_{A_n|I_n}^{[+2\varepsilon_n-|A_n|+|I_n|-s_0]}}{Q_{A_n|I_n}^{[-|A_n|+|I_n|-s_0]}}
\frac{Q_{A_{n-1}|I_{n-1}}^{[-2\varepsilon_n-|A_{n-1}|+|I_{n-1}|-s_0]}}{Q_{A_{n-1}|I_{n-1}}^{[-|A_{n-1}|+|I_{n-1}|-s_0]}}%
Q_{\emptyset}^{[-s_0]}\\
\nonumber \textrm{where
  }\varepsilon_n=(-1)^{|I_n|-|I_{n-1}|}\hspace{2cm}\textrm{ and
  }s_0=\frac{M-K}2\,.
\end{gather}

The QQ-relations (\ref{eq:42}) can  be easily deduced\(^{\textrm{\ref{fn:8}}}\)  from this generating series  \cite{Gromov:2010km} (see also \cite{Leurent:2012xc}). All T-functions can be expressed from (\ref{eq:159}) and (\ref{GJTfunction}), and the result which comes out coincides with the Wronskian expressions (\ref{eq:155}).

\section{Polynomiality and \emph{twist}.}
\label{sec:twisted-case}

In the previous sections, in our study of general Wronskian solutions of Hirota  functional equations with particular ``hook'' boundary conditions,  as well as  the  QQ-relations,   we had no need to precise the  analyticity properties of the functions of  spectral parameter \(u\).  If we now try to do it,  generically it will  impose severe restrictions on the analyticity properties of the whole ensemble of these functions.
For instance if they are assumed to be polynomial or meromorphic functions, or having a given set of singularities we will have rather strong restrictions on the type and position of the singularities and zeros due to the Hirota and QQ functional relations.
It turns out that the analyticity of the T- and Q-functions is an
extremely important ingredient to characterise a given physical model. In this section, we discuss a well-known example of rational spin chains which correspond, in the case of compact representations, to polynomial T-functions, with polynomial Q-functions. In
particular, we  discuss the effect of the twist on polynomiality
conditions.
In the \secref{twistedQSC}, we will consider  the AdS/CFT Q-system which corresponds to
multivalued analytic Q-functions with specific monodromy properties.

\subsection{Polynomiality and spin chains}
\label{sec:polyn-spin-chains}
The spectra of periodic rational spin chains in compact representations of \(\su(K|M)\) with integer fermionic Dynkin label are encoded in the polynomial solutions of the QQ-relations, with certain constraints on the polynomials that precise the details of the spin chain considered (length, representation, inhomogeneities). There are various ways to establish the correspondence between the spectrum of a spin chain and the  QQ-relations, probably the most direct one is to  construct the Q-operators acting in the quantum space of the spin chain (several constructions are available in the literature \cite{Bazhanov:2008yc,Bazhanov:2010ts,Bazhanov:2010jq,Kazakov:2010iu,2011arXiv1112.3310A}) and identify Q-functions with the eigenvalues of these operators.

In \appref{sec:Labels}, we list the required constraints on the polynomials for a generic case\footnote{The discussion of the appendix applies beyond the polynomial case and includes any highest-weight type representations.}. In this section, we discuss one of the most simple and probably the most important examples  -- a homogeneous rational spin chain of length \(L\) in the defining representation. For this spin chain one imposes
\begin{equation}
  \label{eq:226}
    Q_{\emptyset}=1\,,\ \ \ \ \ \ \ \ 
    Q_{\bar\emptyset}^{[s_0]}=T_{0,0}=u^L\,,
\end{equation}
where \(s_0=\frac{M-K}{2}\). 

\begin{figure}
  \centering
\subfigure[Hasse diagram for an \(\su(3)\) spin chain's Q-functions:  The Q-functions of figure \ref{fig:HasseQ} are not written explicitly, only the conditions \(Q_{\emptyset}=1\) and \(Q_{\bar\emptyset}^{[s_0]}=u^L\) are made manifest (the shift \({[s_0]}\) is omitted).
]{\hspace{.6cm}\label{fig:HasseSCb}\includegraphics{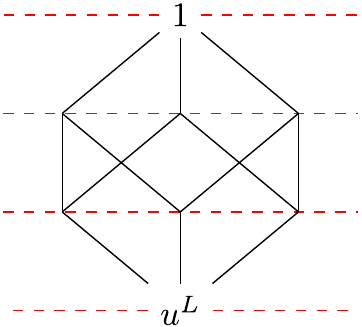}\hspace{.6cm}}\hspace{1cm}
\subfigure[Hasse diagram for an \(\su(2|1)\) spin chain's Q-functions:  Two different equivalent orientations of the Hasse diagram are presented. In the notations of \eqref{eq:96}, the diagram to the left corresponds to functions \({\bzdQ}_{AI}\) while the diagram to the right corresponds to \(Q_{A|I}\).
]{%
  \label{fig:HasseSCf}{\includegraphics{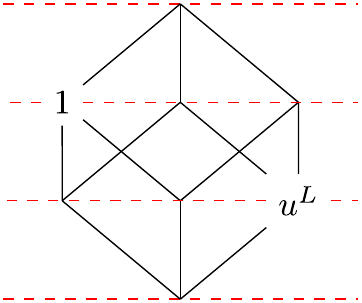}}\raisebox{1.4cm}{=}{\includegraphics{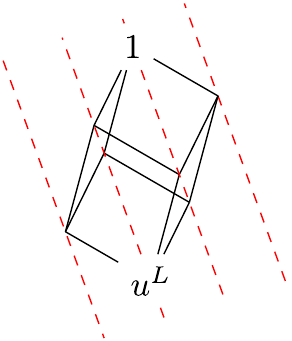}}%
}\\
\subfigure[Notation for the orientation of Hasse diagram: the dashed red lines indicate how to write the QQ-relation for a given facet.]{\hspace{2cm}
  \label{fig:HasseOrient}{\includegraphics{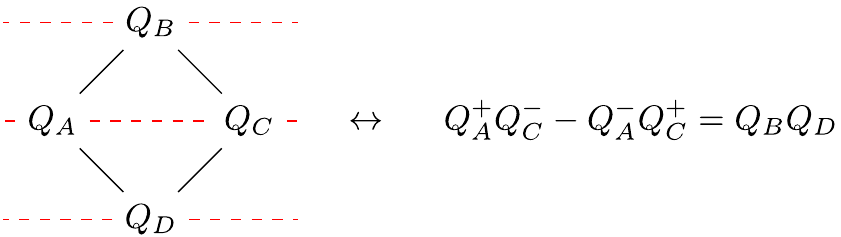}} \hspace{2cm}
}
\caption{\label{fig:HasseOr}Boundary conditions and orientation of the Hasse diagrams.
}
\end{figure}

It is  remarkable that, algebraically, the Q-system is the same for all symmetry algebras \(\su(K|M)\) with given value of \(K+M\). The difference appears only in how the constraints \eqref{eq:226} appear on the Hasse diagram. This phenomenon is illustrated in figure~\ref{fig:HasseOr}, where we see that the ``bosonization trick'' \eqref{eq:96} amounts to a rotation of the Hasse diagram.

Note also how the Hodge duality map acts. It %
flips
the Hasse diagram (upside-down), hence the boundary conditions \eqref{eq:226} change to \(Q_{\es}^{[s_0]}=u^L\,,\ Q_{\bar\es}=1\). These new boundary conditions correspond to the conjugation  of the defining representation. It is not difficult to guess then that the Hodge duality performs  an outer automorphism \(E_{ij}\mapsto -E_{ji}\)  from the point of view of representation theory. 
 
Although there are no other constraints  than \eqref{eq:226}  on the
structure of the polynomial Q-functions, the QQ-relations themselves
strongly constrain  possible polynomials, and one ends up with only
a discrete set of possibilities. All of them can be found by solving Bethe equations for super-symmetric rational spin chains \cite{Kulish1982,Kulish1986} which are a set of algebraic equations for the roots of the polynomials.

The QQ-relations directly imply the Bethe equations as follows \cite{Kazakov:2007fy}: If \(Q_{A a|I}\) has a zero at position \(u=\theta\),
  then equation \eqref{eq:41} implies that \(\theta\) is also a zero of \(Q_{A|I}^+Q_{A ab|I}^+- Q_{A
    a|I}^{++} Q_{A b|I}\) and of \(Q_{A|I}^-Q_{A ab|I}^-+ Q_{A a|I}^{--} Q_{A
    b|I}\). Hence it is a zero of
  the linear combination \(Q_{A|I}^+Q_{A ab|I}^+Q_{A a|I}^{--}+Q_{A|I}^-Q_{A ab|I}^-Q_{A a|I}^{++}\), and  we get the equation
  \begin{align}\label{eq:165}
    -1&=\frac{Q_{A|I}(\theta+\tfrac \bi 2)Q_{A a|I}(\theta- \bi)Q_{A
        ab|I}(\theta+\tfrac \bi 2)}{Q_{A|I}(\theta-\tfrac \bi 2)Q_{A
        a|I}(\theta+\bi)Q_{A ab|I}(\theta-\tfrac \bi
      2)}\,,&\textrm{where }Q_{A a|I}(\theta)=0\,.
  \end{align}
This equation involves the Q-functions corresponding to two
successive ``bosonic'' B\"acklund transformations along the  nesting path \eqref{eq:nestingpath}. 

If one has two subsequent ``fermionic'' B\"acklund transformations, we get analogously
\begin{align}
\label{eq:162}
    -1&=\frac{Q_{A|I}(\theta+\tfrac \bi 2)Q_{A|Ii}(\theta- \bi)Q_{A
        |I i j}(\theta+\tfrac \bi 2)}{Q_{A|I}(\theta-\tfrac \bi 2)Q_{A
        |I i}(\theta+\bi)Q_{A|I i j}(\theta-\tfrac \bi
      2)}\,,&\textrm{where }Q_{A|I i}(\theta)&=0\,.
\end{align}
Finally, one also derives
\begin{align}\label{eq:163}
  1&=\frac{Q_{A a|I i}(\theta+\tfrac \bi 2)Q_{A|I}(\theta-\tfrac \bi
    2)}{Q_{A|I}(\theta+\tfrac \bi 2) Q_{A a|I i}(\theta-\tfrac \bi
    2)}\,,&\textrm{where }Q_{A a|I}(\theta)&=0\,,&\textrm{or
  }Q_{A|Ii}(\theta)&=0\,,
  \end{align}
if a ``fermionic''
B\"acklund transformation is followed by a ``bosonic'' one (case \(Q_{A
  a|I}(\theta)=0\)) or if a ``bosonic''
B\"acklund transformation is followed by a ``fermionic'' one (case \(Q_{A
  |I i}(\theta)=0\)).
  
There is a special case when all there terms of  \eqref{eq:41} (or \eqref{eq:49}, or \eqref{eq:78}) become zero at  some \(u=\theta\). Such zero can be an ``exceptional'' root of Bethe equations which was accidentally trapped into a singular point, we can resolve this singularity by introducing a twist, see \eqref{eq:176}. Another possibility, which is not realised for defining representation but is possible for other cases, is that such zero is not demanded to be a solution of Bethe equation; instead, it belongs to a source term thus specifying the type of a spin chain, see \appref{sec:Labels} for further details.

When exceptional Bethe roots are properly accounted,  a solution of the Bethe equations allows to restore the Q-functions and vice versa, hence the  Q-system and the Bethe equations encode the same information. This is another way to see that the polynomial ansatz with boundary conditions of type \eqref{eq:226} indeed corresponds to a rational spin chain, as there is a handful of ways to derive Bethe equations, including those not relying on construction of Q-operators or even T-operators.

Each eigenspace of the spin chain which forms an irrep  of \(\su(K|M)\) symmetry algebra corresponds to a solution of the QQ-relations. For our particular example of the homogeneous spin chain in the defining representation, the commuting family of operators that act diagonally on the discussed eigenspaces includes the operator with only nearest-neighbour interactions of the spin chain sites: \(\mathcal{H}=\sum_{i=1}^L \mathcal{P}_{i,i+1}\,,\) where \(\mathcal{P}\) is a permutation operator. It is usually interpreted, up to an addition or multiplication by a constant, as the physical Hamiltonian of the system. For an eigenstate characterized by a given solution of the QQ-relations, the eigenvalue (energy) is given by
  \begin{equation}\label{eq:161}
    E=\bi\,\partial_u \log T_{1,1}\raisebox{-0.6em}{\(|_{u=0}\)}=\frac{\bi\,\partial_u T_{1,1}}{T_{1,1}}\raisebox{-0.6em}{\(|_{u=0}\)}\,.
  \end{equation}
In the expression \eqref{eq:160} for \(T_{1,1}\), we see that due to the
 factor \(Q_{\bar\emptyset}^{[s_0]}=u^L\), the terms \(n\neq K+M\) do not
 contribute to \eqref{eq:161} if \(L\ge2\). Thus, we have
\begin{equation}
    E=\bi\,\partial_u \log \left((u \pm
\bi)^L \frac{\mathbf{Q}^{%
    \mp%
          }}{\mathbf{Q}^{\pm%
          }}\right)
\raisebox{-0.5em}{\(|_{u=0}\)}=\pm \left(L-\sum_{k}\frac{1}{\theta_k^2+\frac 14}
\right)\,,
\label{eq:164}
  \end{equation}
where \(\pm\) denotes \(\varepsilon_{K+M}\), the grading of the first B\"acklund transform of the nesting path, and \(\mathbf Q=
{Q_{A_{K+M-1}|I_{K+M-1}}^{[s_0]}}\), while \(\theta_k\)
denote the roots of \(\mathbf
Q\propto\prod\limits_{k}(u-\theta_k)\).

Formulae of type \eqref{eq:164} is an extra information one needs to introduce, apart from finding a solution of a QQ-relations, for computing the spectrum of rational spin chains. By contrast, in the case of the AdS/CFT integrable system, the Hamiltonian is part of   the symmetry algebra charges that define the large-\(u\) asymptotic of Q-functions \eqref{twistedPQ}. One can derive the formulae like \eqref{eq:164}, at least in the asymptotic Bethe Ansatz limit \cite{Gromov:2014caa}, but now as a non-trivial consequence of analytic properties of Q-functions rather than an independent input.

\subsection{Twisted spin chains and Q-system}
\label{sec:twisted-spin-chains}

Spin chains can be deformed by the introduction of a ``twist'', which
changes the periodicity condition\footnote{More explicitely the Hamiltonian becomes \(\mathcal{H}=\sum_{i=1}^{L-1} \mathcal{P}_{i,i+1}+(\mathcal{P}_{1,L}\cdot G^{-1}\otimes\mathbb I^{\otimes L-2}\otimes G)\).}. 
 For rational spin chains, this
twist \(\tG\) can be chosen diagonal without a loss of generality, and we denote
its eigenvalues as \(\x_1,\dots,\x_K,\y_1,\dots,\y_M\).

It is known that in the presence of a twist, the Bethe equations of
the rational spin chain are deformed and become
\begin{subequations}\label{eq:176}
  \begin{align}\label{eq:166}
    -\frac{\tx_b}{\tx_a}&=\frac{\mathcal{Q}_{A|I}(\theta+\tfrac \bi
      2)\mathcal{Q}_{A a|I}(\theta- \bi)\mathcal{Q}_{A
        ab|I}(\theta+\tfrac \bi 2)}{\mathcal{Q}_{A|I}(\theta-\tfrac
      \bi 2)\mathcal{Q}_{A a|I}(\theta+\bi)\mathcal{Q}_{A
        ab|I}(\theta-\tfrac \bi
      2)}\,,&\textrm{where }\mathcal{Q}_{A a|I}(\theta)&=0\,,\\
    \frac{\tx_a}{\ty_i}&=\frac{\mathcal{Q}_{A a|I i}(\theta+\tfrac \bi
      2)\mathcal{Q}_{A|I}(\theta-\tfrac \bi
      2)}{\mathcal{Q}_{A|I}(\theta+\tfrac \bi 2) \mathcal{Q}_{A a|I
        i}(\theta-\tfrac \bi 2)}\,,&\textrm{where }\mathcal{Q}_{A
      a|I}(\theta)&=0\,,&\textrm{or
    }\mathcal{Q}_{A|Ii}(\theta)&=0\,,\\
    -\frac{\ty_j}{\ty_i}&=\frac{\mathcal{Q}_{A|I}(\theta+\tfrac \bi
      2)\mathcal{Q}_{A|Ii}(\theta- \bi)\mathcal{Q}_{A |I i
        j}(\theta+\tfrac \bi 2)}{\mathcal{Q}_{A|I}(\theta-\tfrac \bi
      2)\mathcal{Q}_{A |I i}(\theta+\bi)\mathcal{Q}_{A|I i
        j}(\theta-\tfrac \bi 2)}\,,&\textrm{where }\mathcal{Q}_{A|I
      i}(\theta)&=0\,,\label{eq:167}
  \end{align}
\end{subequations}
which constrains the roots of the polynomials
\(\mathcal{Q}_{A|I}\). Hence these polynomial \(\mathcal{Q}\)-functions do
not obey the same QQ-relations \eqref{eq:42} as in the absence of 
twist. There exist two equivalent ways to describe this situation: one
can either add an exponential prefactor which breaks the polynomiality
of Q-functions, or deform the QQ-relations.

\subsubsection{Twist  as an exponential prefactor}
\label{sec:expon-pref}

  One possibility is to consider  Q-functions which are not
  polynomials anymore. More precisely, \(Q_{A|I}\) is the product of the exponential prefactor \(\left(\frac{\prod_{a\in
        A} \x_a}{\prod_{i\in I} \y_i}\right)^{-\bi \,u}\)
and of a polynomial function denoted by the letter \(\mathcal{Q}\):
  \begin{align}\label{eq:168}
   Q_{A|I}\propto \left(\frac{\prod_{a\in A} \x_a}{\prod_{i\in I}
        \y_i}\right)^{-\bi \,u}  \mathcal{Q}_{A|I}\,,
  \end{align}
then it is immediate to see that (\ref{eq:165}-\ref{eq:163}) for \(Q\) becomes
(\ref{eq:166}-\ref{eq:167}) for \(\mathcal{Q}\), whereas it is a bit less trivial to see that 
\eqref{eq:164}
 is not modified\footnote{In principle, one could expect that in \eqref{eq:164}, the factor \((u \pm\bi)^L\) (from \(Q_{\bar\emptyset}\)) becomes \((\sdet u)^{-\bi \,u}(u \pm\bi)^L\), if \(\sdet \tG\neq 1\). This is not the case as the expression (\ref{eq:161}) holds in a gauge where \(T_{1,1}\) is polynomial, i.e. one has to divide the expression (\ref{eq:160}) by \((\sdet u)^{-\bi \,u}\).}
(in particular, \(\partial_u \log
\frac{\mathbf{Q}^{+}}{\mathbf{Q}^-}\) is invariant under
multiplication of \(\mathbf{Q}\) by \(\x^{-\bi u}\)).

In \eqref{eq:168}, the symbol ``\(\propto\)'' denotes an arbitrary
normalization for the polynomial \(\mathcal{Q}_{A|I}\) (for instance the
coefficient of the leading power can be set to one). This normalization
is not very relevant, as it cancels out in (\ref{eq:165}-\ref{eq:163})
and \eqref{eq:164}.

In this setup, the simplest {\it character} solution of the
QQ-relations \eqref{eq:42} when the twist has pairwise-distinct
eigenvalues is the
following:
\begin{align}\label{eq:169}
  Q^{(\chi)}_{b_1,b_2,\dots, b_n|f_1,f_2,\dots, f_p}&=
\frac{\prod\limits_{i=1}^n\x_{b_i}^{-\bi\, u+\tfrac{p-n-1}2}}
{\prod\limits_{j=1}^n \y_{f_j}^{-\bi\,u+\tfrac{p-n+1}2}}
  \frac{
      \prod\limits_{1\leq i < j \leq n}     \left(
 \x_{b_i}-\x_{b_j}
    \right)  \prod\limits_{1\leq i < j \leq p} \left(
     \y_{f_j}-\y_{f_i}
    \right)}{
 \prod\limits_{i=1}^n \prod\limits_{j=1}^p (\x_{b_i}-\y_{f_j})
}\,.
\end{align}
It is obtained by solving the QQ-relation when \(Q_{b|\es}=%
\x_b%
^{-\bi u}\), \(Q_{\es|f}=\y_f^{\bi u}\). The corresponding T-function, obtained
by plugging \eqref{eq:169} into \eqref{eq:155}, is related to the
characters \(\chi_{a,s}(\tG)\) as follows:
\begin{gather}
  \label{eq:170}
  T_{a,s}^{(\chi)}=\bi^{M\,(a+s+a\,s)} \chi_{a,s}(\tG)\,  Q_{\bar\emptyset}^{[a-s+s_0]}\,.
\end{gather}
Note that if the twist
\(\tG=\mathrm{diag}(\x_1,\dots,\x_K,\y_1,\dots,\y_M)\) belongs to \(\SL(K|M)\),  i.e. if
\(\prod_{a\in\Bset}\x_a=\prod_{i\in\Fset}\y_a\), 
then the factor \(Q_{\bar\emptyset}^{[a-s+s_0]}\) is just a
\(u\)-independent normalization (in particular, it is equal to the Vandermonde determinant in the
bosonic case and the Cauchy double alternant in the \(\SL(M|M)\) case).

In more general situation, one can see that the T-functions are polynomial
functions of \(u\) if \(\tG\in \SL(K|M)\)\footnote{%
 If
 \(\tG\not\in SL(K|M)\), then the T-functions are the product of \((\sdet g)^{-\bi\,u}\)
and of a polynomial, so that they are polynomial up to a gauge.}. By contrast, unlike the untwisted case, their
B\"acklund transforms are in general not polynomial functions of \(u\).
\subsubsection{Twist as a holomorphic connection}
\label{sec:constant-connection}

A more geometric approach consists in adding to the fiber bundle described in \secref{sec:solut-qq-relat} a holomorphic \(\GL(N)\) connection \(A\). In other words, one gauges the global rotational \(\GL(N)\) symmetry making it local.

In this setup, we slightly deform the definition (\ref{eq:3}) of Pl\"ucker coordinates of \(V_{(n)}\): we now introduce the coordinates of \(V_{(n)}\) as forms \(\mathcal{Q}_{(n)}\) such that 
\begin{align}
\label{eq:40}  V_{(1)}&=\left\{\xx\in\mathbb C^N ~;~
\mathcal{Q}_{(1)} \wedge \Pexp{u}{0} {\bf x}=0\right\}\,,\\
  V_{(n)}&=\left\{\xx\in\mathbb C^N ~;~
\mathcal{Q}_{(n)} \wedge \Pexp{u-\bi\tfrac{n-1}2}{0} {\bf x}=0\right\}\,,\label{eq:39}
\end{align}
where the path-ordered integral \Pexp{u}{v} is the parallel transport from spectral parameter \(u\) to \(v\), and the shift \(\bi\tfrac{n-1}2\) was introduced arbitrarily in (\ref{eq:39}) to simplify upcoming expressions. To obtain (\ref{eq:40})-(\ref{eq:39}), one naturally chooses (as a generalization of \eqref{eq:5}):
\begin{align}
  \frac{\mathcal{Q}_{(n)}}{\mathcal{Q}_{\emptyset}^{[-n]}}&=%
  {\left(\Pexp {u+\bi\tfrac{n-1}2}{u-\bi\tfrac{n-1}2}\frac{\mathcal{Q}_{(1)}^{[n-1]}}{\mathcal{Q}_{\emptyset}^{[n-2]}}\right)\wedge 
\left(\Pexp {u+\bi\tfrac{n-3}2}{u-\bi\tfrac{n-1}2}\frac{\mathcal{
Q}_{(1)}^{[n-3]}}{\mathcal{Q}_{\emptyset}^{[n-4]}}\right)\wedge\dots \wedge\frac{ Q_{(1)}^{[1-n]}}{\mathcal{Q}_{\emptyset}^{[-n]}}}\,.
\end{align}

While the \(A=0\) case corresponds to the non-twisted case of the previous sections, the twist corresponds to constant \(A\): indeed, if \(A\) is a constant diagonal matrix and we denote \(\mathrm{diag}(\tx_1,\dots,\tx_K,\ty_1,\dots,\ty_M)=\tG=e^A\) then we get (at the price of repeating the \emph{bosonization trick} in the  supersymmetric case)
\begin{subequations}\label{eq:180}
  \begin{align}\label{xyQQ}
       \mathcal{Q}_{A|I} \mathcal{Q}_{A,a,b|I}&=\tx_a\,
        \mathcal{Q}_{A,a|I}^+ \mathcal{Q}_{A,b|I}^--\tx_b\, \mathcal{Q}_{A,a|I}^-
    \mathcal{Q}_{A,b|I}^+\,,\\
    \mathcal{Q}_{A|I,i}  \mathcal{Q}_{A,a|I} &=\tx_a\,\mathcal{Q}_{A,a|I,i}^+   \mathcal{Q}_{A|I}^- -\ty_i\,\mathcal{Q}_{A,a|I,i}^- \mathcal{Q}_{A|I}^+
\,,\\
\mathcal{Q}_{A|I,i,j} \mathcal{Q}_{A|I}&=\ty_i\,\mathcal{Q}_{A|I,j}^+   \mathcal{Q}_{A|I,i}^- -\ty_j\,\mathcal{Q}_{A|I,j}^-   \mathcal{Q}_{A|I,i}^+\,.
  \end{align}
\end{subequations}
These relations imply the Bethe equations \eqref{eq:176}.

Obviously, this approach is equivalent to the approach of section
\ref{sec:expon-pref}, and the twisted
\(\mathcal{Q}\mathcal{Q}\)-relations \eqref{eq:180} are equivalent to
the standard QQ-relations \eqref{eq:42} up to the change of variables
\begin{align}\label{eq:181}
  \mathcal{Q}_{A|I}&=Q_{A|I}\left(\frac{\prod\limits_{a\in A}\tx_a}{\prod\limits_{i\in
        I} \ty_i}\right)^{\bi \,u+\tfrac{|A|-|I|}2}\,.
\end{align}
This change of variable corresponds to 
\begin{align}
\label{eq:68qq}
  Q_{(n|p)}&=\Pexp{0}{u+\bi\frac{p-n+1}2} \mathcal{Q}_{(n|p)} \,,
\end{align}
which is a simple parallel transport to the origin.

One easily checks (for instance by  use of the mapping \eqref{eq:181} to
the non-twisted case (\ref{eq:185}-\ref{eq:47})) that the \(\mathcal{Q}\mathcal{Q}\)-relations
\eqref{eq:180} are solved  in a gauge where \(\mathcal{Q}_\emptyset=1\)
\footnote{In a gauge where \(\mathcal{Q}_\emptyset\) is not equal to
  one, the relations \eqref{eq:182} still hold up to a denominator, as
  in (\ref{eq:185}-\ref{eq:47}). This denominator can be restored by
  substituting
\(\mathcal{Q}_{(n|p)}\leadsto \mathcal{Q}_{(n|p)}/\mathcal{Q}_{\emptyset}^{[n-p]}\)
into the relations \eqref{eq:182}.}, similarly to the untwisted case, by
\begin{subequations}\label{eq:182}
  \begin{align}
\label{eq:183}    \mathcal{Q}_{(n|0)} &=
\left(%
\tG^{n-1}
  \mathcal{Q}_{(1|0)}^{[n-1]}\right)\wedge
\left(%
\tG^{n-2}
  \mathcal{Q}_{(1|0)}^{[n-3]}\right)
\wedge\dots\wedge
    \mathcal{Q}_{(1|0)}^{[1-n]}
    \,,&\\
 \mathcal{Q}_{(0|p)} &=
\mathcal{Q}_{(0|1)}^{[p-1]}%
\wedge
\left(%
\tG
    \mathcal{Q}_{(0|1)}^{[p-3]}\right)
\wedge
\left(
\tG^2
    \mathcal{Q}_{(0|1)}^{[p-5]}\right)\wedge\dots \wedge
\left(%
\tG^{p-1}
    \mathcal{Q}_{(0|1)}^{[1-p]}\right)%
    \,,\\
    \mathcal{Q}_{\emptyset|i} \mathcal{Q}_{a|\emptyset}
    &=\tx_a\,\mathcal{Q}_{a|i}^+
    -\ty_i\,\mathcal{Q}_{a|i}^-
\label{eq:184}    \\
        \mathcal{Q}_{(n|n)}&=\frac {(-1)^{\frac {n(n-1)}2}} {n!}
    \mathcal{Q}_{(1|1)}^n
    \equiv \frac {(-1)^{\frac {n(n-1)}2}} {n!}
    \mathcal{Q}_{(1|1)}\wedge \mathcal{Q}_{(1|1)}\wedge\dots\wedge
    \mathcal{Q}_{(1|1)}
     \hspace{-4cm}
        \\
    \mathcal{Q}_{(n|p)}&=
        \mathcal{Q}_{(p|p)}^{[p-n]}    \wedge \mathcal{Q}_{(n-p|0)}&&\hspace{-1cm}\textrm{where }n\ge p\,,
    \\
    \mathcal{Q}_{(n|p)}&= (-1)^{n(p+1)}
        \mathcal{Q}_{(n|n)}^{[p-n]}
    \wedge \mathcal{Q}_{(0|p-n)}&&\hspace{-1cm}\textrm{where }n\le p\,,
  \end{align}
\end{subequations}
where the twist appears only in the non-local relations
(\ref{eq:183}-\ref{eq:184}).
Similarly, one can write the T-functions in terms of
\(\mathcal{Q}\)-functions instead of Q-functions as in
\eqref{eq:155}. To this end, one should just substitute \eqref{eq:181}
into \eqref{eq:155}. We do not repeat here these expressions, which become slightly less compact than in the
non-twisted case \eqref{eq:155}~(see
\cite{Tsuboi:2009ud} for similar formulae).
 
Quite curiously, the holomorphic connection point of view allows constructing Q-systems with arbitrary value of \(A\), not only a constant one which we consider in this article.  The non-constant value of \(A\) produces a Q-system which we cannot identify with systems studied in the literature. It would be indeed very interesting to explore this new case.
 
\paragraph{Remark: B\"acklund flow}
If we use these twisted QQ-relations, we can also understand the
\(\mathcal{Q}\)-functions in terms of B\"acklund
transformations~\cite{Kazakov:2007fy,Zabrodin:2007rq}, as in \secref{sec:backl-transf-qq}, with the slight difference that the Lax
Pair \eqref{eq:126} has to be replaced with
\begin{subequations}\label{eq:177}
  \begin{empheq}[left=\empheqlbrace]{align}
    \label{eq:178}
    T_{a+1,s}F_{a,s}^+-T_{a,s}^+F_{a+1,s}&=g_\alpha\,T_{a+1,s-1}^+F_{a,s+1}\\
\label{eq:179}
    T_{a,s+1}F_{a,s}^--T_{a,s}^-F_{a,s+1}&=g_\alpha\,T_{a+1,s}F_{a-1,s+1}^-\,.
  \end{empheq}
\end{subequations}
In this expression, \(g_\alpha\) denotes an eigenvalue of the twist (either \(\tx_\alpha\) if \(\alpha\) is a bosonic index or \(\ty_\alpha\) otherwise) and its index \(\alpha\) is the index which is removed by the B\"acklund transform, in the notations of figure \ref{fig:Hasse}.

\subsection{Dependence on twist and the untwisting limit: illustration on examples}
\label{sec:zero-twist-limit-1}

The dependence of a Q-system on twist can be quite non-trivial. For instance, the behaviour of the Q-functions is  singular when
two eigenvalues of the twist matrix tend to become equal. We can see it already on the example of the 1-st Weyl formula in the table~\ref{tab:compare} where both the numerator and denominator become zero in this limit. If we
focus on the \(\mathcal{Q}\mathcal{Q}\)-relation \eqref{xyQQ}, then we
see that if the \(\mathcal{Q}\)-functions are polynomial (consider the case of compact spin
chains), then their degrees obey
\begin{align}
\label{eq:191}   
\deg \mathcal{Q}_{A,a|I}+ \deg\mathcal{Q}_{A,b|I}
&=
  \begin{cases}
\deg \mathcal{Q}_{A|I}+\deg \mathcal{Q}_{A,a,b|I}
&\textrm{ if }\x_a\neq \x_b
 \\
\deg \mathcal{Q}_{A|I}+\deg \mathcal{Q}_{A,a,b|I}+1&\textrm{ if }\x_a= \x_b\,.
  \end{cases}
\end{align}
For instance, in the case of the \(\su(2)\) Heisenberg spin chain of length \(L\), we have \(\deg
\mathcal{Q}_{12}=L\) and \(\deg \mathcal{Q}_{\emptyset}=0\), which means
that the degree of the polynomial \(\mathcal{Q}_1 \mathcal{Q}_2\)
increases by one in the limit \(\x_1-\x_2\to 0\). This seemingly harmless change in the degree leads, as we shall see, to a significant reorganisation of the Q-system.

Let us consider a more general picture now. The space of all possible diagonal twists is the projective space \(\mathbb{CP}^{N-1}\) parameterised by \([\x_1:\x_2:\ldots : \x_K:\y_1:\y_2\ldots:\y_M]\). We can study how a Q-system changes upon analytic continuation in this space. We then face several different effects when performing such a study:
\begin{itemize}
\item Untwisting limit. The limiting points on hyperplanes \(\x_a=\x_b\), \(\x_a=\y_i\), or \(\y_i=\y_j\) are quite singular as we explained above. This type of limit receives the most of attention in this section, a special emphasis is put on the fully untwisted case when the twist matrix \(\tG\) becomes the identity. In general, the result of the limit \(\tG\to\mathbb{I}\) may depend on how the identity is approached. We discuss only  the limit \(\tG=\mathbb{I}+\epsilon\,g_0\) with \(\epsilon\to 0\) and assume \(g_0\) being in generic position.

\item Degeneration of solutions. Singular points on \(\mathbb{CP}^{N-1}\) of other type live on hyperplanes \(\x_a=0\) or \(\y_i=0\). Space of solutions to QQ-relations degenerates there, and analytic continuation around such hyperplanes has a non-trivial monodromy. 

\item Borel ambiguities. Generic points on \(\mathbb{CP}^{N-1}\) also have certain interest. We included an example of a non-compact rational spin chain into this section. The definition of the associated  Q-system for such a chain suffers from Borel-type ambiguities, with  position of Borel singularities being dependent on the value of the twist.

\item Relation to representation theory of \(\gl(K|M)\). In the
  presence of generic twist, Cartan sub-algebra of \(\gl(K|M)\) is the
  only remaining symmetry\footnote{In the sense that Q- and
    T-operators commute with symmetry generators, see
    \appref{sec:rational-spin-chains}. Also note that the Cartan
    sub-algebra is realised in a standard way, by generators
    \(E_{kk}\), if the twist is dingonal. %
  } of a spin chain. However, the full \(\gl(K|M)\) symmetry is restored in the untwisting limit; this is another way to see why this limit is singular. Certain properties of irreducible representations (irreps) find their counterpart in analytic structure of Q-systems.
\end{itemize}

In this section, we will discuss several explicit examples based on small-rank algebras to illustrate the mentioned effects, the gained experience is then summarised in \secref{sec:untwist summary}. The generalisation from the explicit examples to an arbitrary rank is also done, but only for the question of untwisting limit and for the case of finite-dimensional irreps. We hope to study  other phenomena beyond small-rank cases in future works.

We explore the above-mentioned properties purely assuming existence of a Q-system with certain analytic properties (mostly polynomiality), without questioning its origin. This approach is conceptually important
because there are situations, e.g. the AdS/CFT, where the existence of a Q-system is
known although there is presently no operatorial construction, beyond the leading order in the perturbation theory, of a Hamiltonian
and T- (hence Q-)operators. However, we should note that for the spin chains discussed in this section, all analytic properties follow from  operatorial constructions \cite{Kazakov:2010iu,Bazhanov:2010ts,Bazhanov:2010jq,Bazhanov:1996dr,Derkachov:2008aq}%
. We explicitly demonstrate this link in subsection \ref{sec:su21example}.

\ \\
In the discussion of irreps, we use the following notations (with more details given in \appref{sec:Labels}). The vector of an irrep is characterised by its fundamental weight\footnote{Hats in \(\nu_{\hat i}\) may be omitted when expression is unambiguous.}
\begin{equation}\label{eq:194}
[\lambda_1,\lambda_2,\ldots\,\lambda_K;\nu_{\hat 1},\nu_{\hat 2},\ldots,\nu_{\hat M}]\,,
\end{equation}
where \(\lambda\)'s and \(\nu\)'s are eigenvalues of the corresponding
Cartan generators. In physical jargon, \(\lambda_a\) is called the
``number of spin d.o.f.'', or ``number of spins'' in short\footnote{This
  terminology must not be confused with the name \emph{spin} which is
  related to the value of Casimir operator.}
  in direction \(a\); and \(\nu_{\hat i}\) is the ``number of spins'' in direction \(\hat i\). Indeed, in the case of spin chains with sites in the defining representation, the weight of any eigenstate comprises non-negative integers which sum up to the number of sites, \(\sum\limits_{a}\lambda_a+\sum\limits_{i}\nu_i=L\), each site is being thought of as a spin degree of freedom. 

An irrep can be labelled by the weight of its lowest weight vector. We emphasise that the definition of the latter is not universal as it depends on a total order imposed on the set of indices \(\{1,\ldots,K,\hat 1,\ldots,\hat M\}\), see \appref{sec:Labels}. The corresponding ambiguity finds its counterpart in the untwisting limit of a Q-system. However, quite expectedly, more invariant objects -- T-functions -- do not depend on the choice of  order.

\subsubsection{\texorpdfstring{$\su(2)$}{su(2)}: untwisting should be supplemented with a rotation}
The simplest example to commence with is the \(\su(2)\) XXX spin chain in the defining representation. The twist-related effects were  studied probably the most on this example, quite a detailed  and illuminating analysis was presented in \cite{Bazhanov:2010ts}, including explicit examples of  construction of Q-operators . We will partially repeat the known statements, but also complement this discussion, in the next subsection, with  novel observations about analytic dependence of the Q-functions on the twist. In particular, we remark that the famous umbrella-shaped configurations of Bethe roots are well-approximated by zeros of Laguerre polynomials.

In this example, the only non-trivial QQ-relation is \(Q_1^+Q_2^--Q_1^-Q_2^+=Q_{\es}Q_{12}\). It is explicitly realised as\footnote{We drop the offset \([s_0]\) from the further discussion.}
\be\label{QQXXX}
\fQ_1^+ \fQ_2^--z\, \fQ_2^+ \fQ_1^-\propto u^L\,,
\ee
where \(\fQ_a\) are polynomials defined modulo an overall normalisation, and where we defined \(z\equiv \x_2/\x_1\).

According to \eqref{QQXXX}, the degree of the polynomial
\(\fQ_1\fQ_2\) should be \(L\) if \(z\neq 1\) and \(L+1\) if
\(z=1\). But any limit of a degree \(L\) polynomial cannot have a higher
degree than \(L\)! Hence something non-trivial should happen in the limit \(z\to 1\). Let us find the explicit solutions of \eqref{QQXXX} for \(L=2\) to clarify the situation. One can do this study analytically, but numerical solution already suffices to demonstrate the effect. The Hilbert space is 4-dimensional, hence one should find 4 solutions. Moreover, one expects appearance of spin-1 and spin-0 irreps at the point \(z=1\). We find for \(z=e^{-\frac{2\pi\,\bi}{100}}\) which is sufficiently close to 1:
\begin{itemize}
\item\ Solutions describing the triplet when \(z\to 1 \):
\item[\(M_1\)=0:] \(\fQ_1=1\),\\
                  \(\fQ_2=1+6.28 \times 10^{-2}\,u+1.97\times 10^{-3} u^2\),
\item[\(M_1\)=1:] \(\fQ_1=1-3.14\times 10^{-2}\,u\),\\
                  \(\fQ_2=1+3.14\times 10^{-2}\,u\),
\item[\(M_1\)=2:] \(\fQ_1=1-6.28 \times 10^{-2}\,u+1.97\times 10^{-3} u^2\),\\
                  \(\fQ_2=1\).
\item\ Solution describing the singlet when \(z\to 1\):
\item[\(M_1\)=1:]  \(\fQ_1=u+7.85\times 10^{-3}\),\\
                  \(\fQ_2=u-7.85\times 10^{-3}\),
\end{itemize}
where \(M_1\) is the degree of \(\fQ_1\).

The small in magnitude numbers will become identically zero in the limit \(z=1\). We see that the degree of \(\fQ_1\fQ_2\) actually drops, or it remains the same at most. Furthermore, both Q-functions approach the same value, \(\lim\limits_{z\to 1}\fQ_1=\lim\limits_{z\to 1}\fQ_2=\mathbb{Q}\), so that
\be
\lim_{z\to 1}\left[\fQ_1^+ \fQ_2^--z\, \fQ_2^+ \fQ_1^-\right] =0\,,
\ee
hence there is no way to normalise this Wronskian combination to \(u^L\) in \eqref{QQXXX} without going to its subleading in \((z-1)\) terms.

The \(\su(K)\) generalisation of the observed phenomenon is the following one: In the untwisting limit, all polynomials \(\fQ_A\) with the same value of the number of indices \(|A|\)  tend to the same Q-functions which can be denoted as \(\mathbb{Q}_{\leftarrow|A|}\):
\be\label{dirlimit1}
\mathbb{Q}_{\leftarrow|A|}\propto \lim\limits_{G\to \mathbb I}\fQ_{A}\,.
\ee
We understand \(G\to \mathbb I\) as \(G=\mathbb{I}+\epsilon\,g_0\) with \(\epsilon\to 0\); the equality \eqref{dirlimit1} should hold for all but finite number of the limiting directions  on the group given by \(g_0\). Furthermore we understand that \(\fQ_{A}\) is normalised to be neither infinite nor zero in the \(G\to \mathbb I\) limit. That is we consider some value \(u=u_0\) at which  \(\fQ_A(u_0)\neq 0\) for \(G\) sufficiently close to the identity matrix (e.g.  \(u=0\) for triplet solution in the example above), and normalise \(\fQ_A(u_0)=1\). For example, in this convention \(\fQ_A=1+\frac{u}{z-1}\) would produce \(\propto u\) in the untwisting limit \(z\to 1\).
 
Equation \eqref{dirlimit1} means that, when one takes the direct untwisting limit, one formally obtains only \(N+1\) non-equal Q-functions \(\mathbb{Q}_{\leftarrow|A|}\,,\) with \(|A|=0,1,2,\dots,N,\) out of \(2^N\) Q-functions of the twisted system.

The functions \(\mathbb{Q}_{\leftarrow|A|}\) are quite special. First, we can obtain all these functions from Q-functions on a nesting path \eqref{eq:bosnpath}, hence the corresponding Bethe equations \eqref{eq:165} would be well defined. Second, the energy of a state can be still computed using \eqref{eq:164}, with \(\bQ=\mathbb{Q}_{\leftarrow K-1}\). Hence, in principle, the emerging functions \(\mathbb{Q}_{\leftarrow|A|}\) contain all necessary information. Moreover, it is known (see \appref{sec:Labels}) that
\begin{equation}\label{eqlargeu1}
\mathbb{Q}_{\leftarrow|A|}\sim u^{\,\sum\limits_{b=1}^{|A|}\lambda_b}\,,\ \ u\to\infty\,,
\end{equation}
where \([\lambda_1\lambda_2\ldots \lambda_N]\) is the the lowest weight of an irreducible multiplet associated to the (remnant of) Q-system in the full untwisting point.

On the other hand, it is quite dissatisfying that  other \(2^N-(N+1)\) Q-functions seem to be lost in the untwisting procedure. The art of obtaining these other Q-functions is to take the untwisting limit {\it simultaneously} with rotating the Q-system, e.g. to consider combinations of the type 
\be\label{off-diagonalidea}
\frac{Q_1-Q_2}{\x_1-\x_2}=\frac{\x_1^{-\bi\,u}\fQ_1-\x_2^{-\bi\,u}\fQ_2}{\x_1-\x_2}\,.
\ee
In the limit \(\x_1=\x_2\) for \(\su(2)\) case, this sample combination becomes a polynomial, and its degree can be larger than the degree of \(\fQ_1\) or \(\fQ_2\), due to the expansion 
\be
\x^{-\bi\,u}=1-\bi\,u\log \x+\ldots\,.
\ee
Then, if we define \(Q_{\es}\equiv\mathbb{Q}_{\leftarrow 0}=1\), \(Q_1\equiv \mathbb{Q}_{\leftarrow 1}\), \(Q_{12}\equiv \mathbb{Q}_{\leftarrow 2}=u^L\) to be Q-functions for \(z=1\) case, one still has to introduce \(Q_2\) which would be given by \(Q_2\propto \lim\limits_{z\to 1}\frac{\x_1^{-\bi\,u}\fQ_1-\x_2^{-\bi\,u}\fQ_2}{\x_1-\x_2}\). It is quite clear that  Q-functions defined in such a way satisfy the desired QQ-relation \(Q_1^+Q_2^--Q_2^+Q_1^-=Q_{\es}Q_{12}\).

 In \secref{sec:su21example} we explicit another example of  implementation of rotation of the type  \eqref{off-diagonalidea}, and  we describe a general strategy of defining the untwisting limit alongside with rotation in \appref{sec:constr-rotat}.\\

\subsubsection{\texorpdfstring{$\su(2)$}{su(2)}: analytic continuation in twist meets representation theory}
One can pose a question: what should be known about twisted Q-functions to predict the Q-system emerging in the untwisting limit? For instance, can we predict the values of \(\lambda_b\) in the large-\(u\) behaviour \eqref{eqlargeu1}? The first, naive expectation is that if we know the large-\(u\) behaviour of twisted Q-functions, e.g. in the \(\su(2)\) case
\be
Q_1\sim \x_1^{-\bi\,u}\,u^{M_1}\,,\ \ \ Q_2\sim \x_2^{-\bi\,u}\,u^{M_2}\,,
\ee
with \(M_1+M_2=L\) then we can deduce the large-\(u\) behaviour in the limit \(z=1\). This expectation is wrong as we can see from our explicit numerical example. Generically, {\it both} powers \(M_1\) and \(M_2\) will drop in the direct untwisting limit (without rotations),  and one cannot predict by what amount without a more detailed information about the structure of Q-functions.

\begin{figure}
  \centering
\subfigure[
Zeros of \(\fQ_1\) on example of \(L=34,s=15,m=10\) and twist \(z=e^{-\frac{2\pi\,\bi}{3}}\). Crosses show exact location of Bethe roots and circles -- their approximation using \eqref{Lagapproximation}. \(\fQ_1\) degenerates into a degree-2 polynomial \(\mathbb{Q}=u^2-\frac 14\left(\cot\frac{\pi}{L-1}\right)^2\) in the untwisting limit, if one follows the shortest path from \(z=e^{-\frac{2\pi\,\bi}{3}}\) to \(z=1\).
]{\label{fig:LBetR}\includegraphics[height=0.3\textwidth]{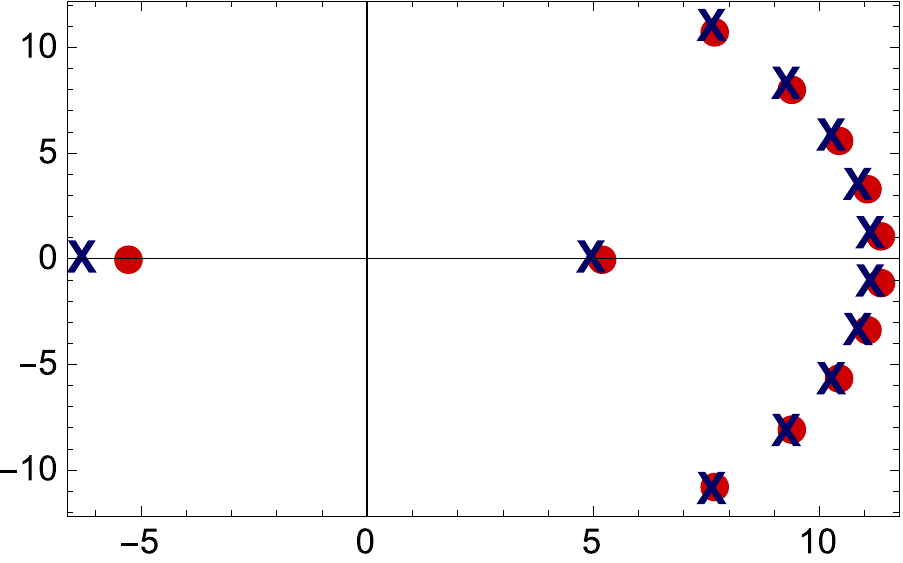}}\hspace{.5cm}
\subfigure[Zeros of polynomial \(\fQ_1^+\fQ_2^-\), an example with \(M_1=3\), \(M_2=5\), and \(z=e^{-4\pi}\). Circles denote zeros of \(\fQ_1^+\) and diamonds -- zeros of \(\fQ_2^-\). The monodromy around \(z=0\) results roughly in rotation of the zeros by the angle \(\frac{2\pi}{L}\) around \(u=0\), thus producing new \(\fQ_a\).  
]{%
  \label{fig:ZeroTwist}{\hspace{1cm}\includegraphics[height=0.3\textwidth]{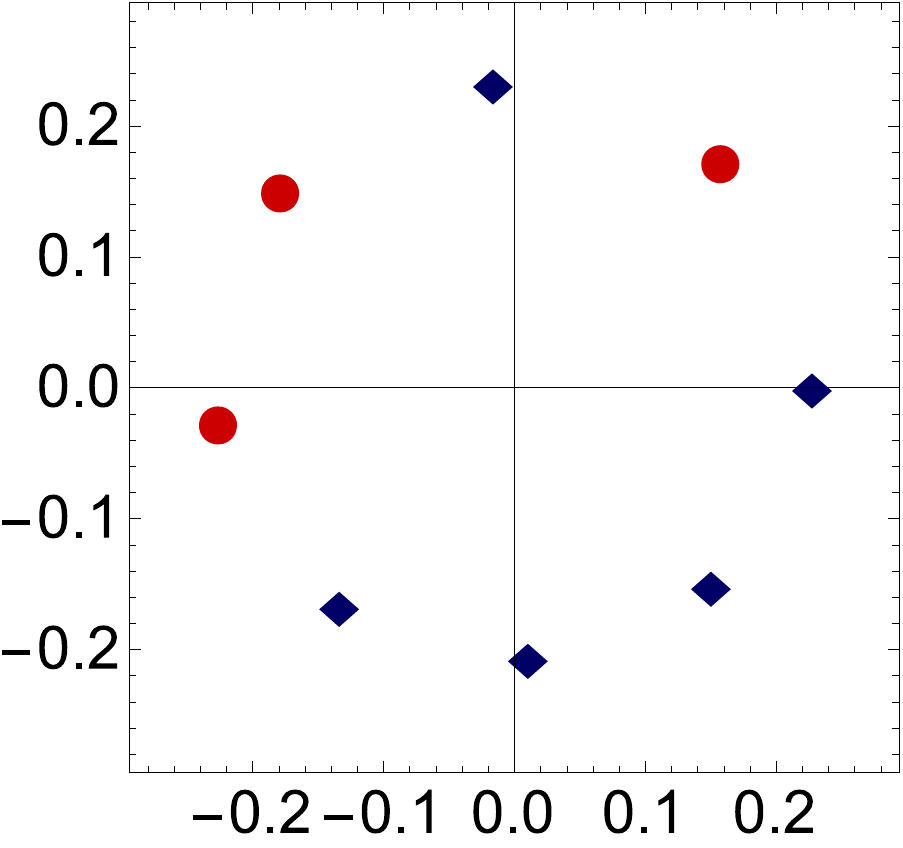}\hspace{1cm}}}%
  \caption{\label{fig:BR}Sample patterns of Bethe roots.}
\end{figure}
One however understands that degree of a polynomial drops if some of its zeros go to \(\infty\). We can find approximate analytic solution describing the structure of these large zeros. As is derived in \appref{sec:LargeBethe}, all twisted Q-functions \(\fQ_1\), \(\fQ_2\) which tend to a polynomial \(\mathbb{Q}\) of degree \(\frac L2-s\), have the following structure when   \(z\to 1\):
\be\label{Lagapproximation}
\fQ_1(u)&\simeq& \mathbb{Q}(u)\,L_{m}^{(-2s-1)}(-\bi\,u\,\log z)\,,
\nonumber\\
\fQ_2(u)&\simeq& \mathbb{Q}(u)\,L_{2s-m}^{(-2s-1)}(+\bi\,u\,\log z)\,,
\ee
for \(0\leq m\leq 2s\), and where \(L_{m}^{(\alpha)}(x)\) are associated Laguerre polynomials, see an example in figure~\ref{fig:LBetR}.
 
From the point of view of representation theory, solution \(\mathbb{Q}\) with \(\deg\mathbb{Q}=\frac L2-s\) corresponds to a spin-\(s\) multiplet. The multiplet consists of \(2s+1\) states and we can observe these states as coming from precisely \(2s+1\) solutions \eqref{Lagapproximation} of the twisted Q-system. Note that \(M_1\equiv\deg\fQ_1=\frac L2-s+m\) and \(M_2\equiv \deg\fQ_2=\frac L2+s-m\) define the weight of an eigenstate:
\be\label{eCg}
\text{ \(M_a\) is the eigenvalue of the Cartan generator \(E_{aa}\) }
\ee 
in the twisted case. The eigenstate remains of the same weight at any value of twist, even at point \(z=1\). However, the relation between the weight and the powers \(M_1=\deg\fQ_1\) and \(M_2=\deg\fQ_2\) does not work at the point \(z=1\), due to the above-discussed power drop and degeneration effects\footnote{When \(z=1\), a link to the representation theory is realised by \eqref{eqlargeu1}.}. Moreover, the powers \(M_a\) are even not uniquely defined when \(z=1\). Indeed, for generic twist, one sees that for every \(a\), the relation between degrees and charges associates \(\mathcal{Q}_a\) with the Cartan generator \(E_{aa}\) and with the eigenvalue \(\x_a\) of the twist -- the labelling of Q-functions is then unambigously identified to the labeling of eigenstates of the twist, cf. (\ref{eq:168}). By contrast, in the \(G\to \mathbb I\) limit, different labelings of Q-function are possible since one can always rotate them. Similarly to the labelling of Q-functions, their asymptotic behavior is ambiguous in  the \(G\to \mathbb I\) limit, since it is not rotation-invariant.

The result about \(2s+1\) being the number of solutions  \eqref{Lagapproximation}  is obtained solely by analytic analysis of the QQ-relation \eqref{QQXXX}, but it is, de-facto, in agreement with representation theory of \(\su(2)\). It would be interesting to generalise this analysis to higher ranks and to derive  in this way a rich set of representation theory properties solely from analytic structure imposed by QQ-relations.\\

Approximation \eqref{Lagapproximation} is valid only in proximity to \(z=1\). If we are far from this point, can we still predict what would happen with Q-functions in the untwisting limit? In fact, the result of the untwisting procedure depends on a path which connects a point \(z\neq 1\) to \(z=1\). Hence, twisted Q-functions are not assigned a-priory to some particular multiplet at \(z=1\). There is a non-trivial monodromy around \(z=0,\infty\) which allows to jump from one multiplet to another. 

Consider for instance the vicinity of \(z=0\). At the leading order of small-\(z\) expansion one  has
\be\label{xzero}
\fQ_1^+\,\fQ_2^-\propto u^L\,,\ \ \ \ {\rm hence}\ \ \ \ \fQ_1=(u^-)^{M_1}\,,\ \ \ \fQ_2=(u^+)^{M_2}\,,
\ee
i.e. all solutions with given weight \([M_1M_2]\) degenerate to the solution \eqref{xzero}. It is then well-expected that these solutions will mix when one performs an analytic continuation around the point \(z=0\).

At the subleading order, the QQ-relation can be written as
\begin{align}
\fQ_1^+\,\fQ_2^-=(1-z)\,u^L+z\,\fQ_2^+\,\fQ_1^-= (1-z)\,u^L+z (u+\bi)^{M_2}\,(u-\bi)^{M_1}+o(z)\,.
\end{align}
The polynomial on the r.h.s. has roots at positions
\be\label{eq:circleBR}
u_k=\sqrt[L]{z}\,e^{i\pi\left(\frac{M_2+1}L-\frac 12\right)}e^{\frac{2\pi i k}{L}}+\mathcal{O}\left(\sqrt[L]z\right)^2\,,\ \  k=0,1,\ldots, L-1\,,
\ee
an example is shown in figure~\ref{fig:ZeroTwist}.

We have to assign these \(L\) distinct zeros to either \(\fQ_1^+\) or \(\fQ_2^-\), and  there are \(L\choose M_1\) ways of doing this. Since \(M_1\) can range from \(0\) to \(L\), we conclude that there are precisely  \(2^L\) solutions of the QQ-relations \eqref{QQXXX}, which is precisely the dimension of the Hilbert space of the length-\(L\) XXX Heisenberg spin chain. We emphasise that  this enumeration result was obtained solely by analysing \eqref{QQXXX}, no connection with XXX spin chain was exploited. Historically, enumeration of solutions to Bethe equations was a non-trivial issue \cite{Bethe:1931hc} which relied on the string hypothesis about the patterns of Bethe roots. This hypothesis is known to be, strictly speaking, wrong. In the above-proposed approach, enumeration becomes indeed simple, at a suitable value of the twist parameter \(z=0\), and it does not require any assumptions. We can look on this result also from the other side: In operatorial derivation of Q-system, we do know that Q-functions -- the eigenvalues of Q-operators -- are polynomials in \(u\) and that QQ-relations are satisfied. However, it requires an extra analysis to show that all polynomial solutions of the QQ-relations are indeed eigenvalues of Q-operators. The obtained enumeration result is a way to resolve this issue.

Analytic continuation produces a cyclic permutation of the Bethe roots \eqref{eq:circleBR} and hence induces nontrivial monodromy on solutions of the Q-system. For instance, the singlet state from our numerical example is exchanged with a vector in the triplet state that has the same weight, upon the analytic continuation.

We saw that analytic continuation in twist is a useful tool allowing one to better control combinatorial and group-theoretical aspects of the Q-system. It can potentially have other interesting applications, one of them is analysing the above-mentioned string hypothesis, see appendix A of \cite{Volin:2010xz}.

\subsubsection{\texorpdfstring{$\gl(1|1)$}{gl(1|1)}: lowest weight depends on a nesting path}
We consider the \(\gl\) case, assuming that \(\tx/\ty\neq 1\), otherwise we won't be able to introduce a non-trivial twist. In higher-rank generalisations one can restrict to \(\sl\) case only.

For a spin chain of length \(L\), one has \(\fQ_{\es|\es}=1\) and \(\fQ_{1|1}=u^L\), thus the only non-trivial QQ-relation \(Q_{1|1}^+Q_{\es|\es}^--Q_{1|1}^-Q_{\es|\es}^+=Q_{1|\es}Q_{\es|1}\)  becomes explicitly
\be
\left(u+\frac \bi2\right)^L-z\left(u-\frac \bi2\right)^L\propto \fQ_{1|\es}\fQ_{\es|1}\,,
\ee
where \(z\equiv \y/\x\).

Since the l.h.s. is a polynomial of degree \(L\), we can distribute zeros of this polynomial between \(Q_{1|\es}\) and \(Q_{\es|1}\) in \(2^L\) ways, thus correctly reproducing the dimension of the Hilbert space.

As before, asymptotics of Q-functions encodes the weight of the state \([\lambda\,;\nu]\), with \(\deg Q_{1|\es}=\lambda\), \(\deg Q_{\es|1}=\nu\).

In the untwisting limit, there is precisely one Bethe root going to infinity, and either from \(Q_{1|\es}\) or from \(Q_{\es|1}\). Hence, if we define the non-twisted Q-functions by
\be
{\mathbb{Q}}_{\leftarrow 0|0}=1\,,\ \  \mathbb{Q}_{\leftarrow 1|0}\propto \lim\limits_{z\to 1} \fQ_{1|\es}\,,\ \  \mathbb{Q}_{\leftarrow 0|1}\propto \lim\limits_{z\to 1} \fQ_{\es|1}\,,\ \  \mathbb{Q}_{\leftarrow 1|1}=u^L\,,
\ee
they can originate from two different twisted Q-systems. And, indeed, all irreps of \(\gl(1|1)\) with non-trivial action of  \(\sl(1|1)\) sub-algebra are 2-dimensional, hence the solution \(\mathbb{Q}\) corresponds to the eigenstates in a certain 2-dimensional representation.

The question is how to label this representation. The representation consists of two vectors, with weights \([\deg\mathbb{Q}_{\leftarrow1|0}+1;\deg\mathbb{Q}_{\leftarrow0|1}]\) and \([\deg\mathbb{Q}_{\leftarrow1|0};\deg\mathbb{Q}_{\leftarrow0|1}+1]\), and we have to pick one of them for the purpose of labelling. In contrast to non-supersymmetric case, there is no distinguished choice of what should be called lowest weight and we can choose either of options.

The situation  becomes clearer with generalisation to higher ranks. In general, one can expect that untwisting without rotations of an \(\su(K|M)\) Q-system  generates a set of  \((K+1)\times (M+1)\) functions \(\mathbb{Q}\) described in \cite{Kazakov:2007fy}
\be\label{untwistedsusy}
\mathbb{Q}_{\leftarrow k|m}\propto \lim_{G\to \mathbb I}\fQ_{A|J}\,,\ \ \ {\rm for}\ \   k=|A|\,,\ m=|J|\,,
\ee
where the limit is understood in the same sense as in \eqref{dirlimit1}.

In supersymmetric algebras, lowest weights are not invariant objects\footnote{To be more strict, we can change the lowest weight vector in non-supersymmetric case as well, by choosing a different Borel decomposition, but its weight would be just the same after we apply the automorphism \(E_{ab}\mapsto E_{\sigma(a)\sigma(b)}\), where \(\sigma\) is a permutation such that \(E_{\sigma(a)\sigma(b)}|{\rm lowest\ weight}\rangle =0\) for \(a>b\). This ``cure'' by an automorphism cannot be done in supersymmetric case.}. We, however, can associate the unique notion of lowest  weight to the choice of the nesting path \eqref{eq:nestingpath}. It is done as follows: For certain \(K+M+1\) functions \(\mathbb{Q}\) which are the untwisting limit \eqref{untwistedsusy} of \(\fQ_{A|J}\) along certain path \eqref{eq:nestingpath}, one parameterises their degree as
\be\label{eq:weightsusy}
\deg\mathbb{Q}_{\leftarrow k|m}=\sum_{a=1}^{k}\lambda_a+\sum_{i=1}^m\nu_i\,.
\ee
On the other hand, each nesting path is in the obvious one-to-one correspondence with the ordering in the set \(\{1,\ldots,K,\hat 1,\ldots,\hat M\}\), see \appref{sec:Labels}. Hence we can say that the choice of the nesting path also defines the notion of the lowest weight. Then one can show that the weight  \([\lambda_1,\ldots,\lambda_K;\nu_1,\ldots,\nu_M]\) is the lowest-weight vector defined by the nesting path used to choose the \(K+M+1\) functions \(\mathbb{Q}\) in \eqref{eq:weightsusy}.

Specifying to the \(\gl(1|1)\) example: If the nesting path is \((\es|\es)\subset (1|\es)\subset (1|1)\), then \([\deg\mathbb{Q}_{1|0},\deg\mathbb{Q}_{0|1}+1]\) would be our lowest weight and if the nesting path is \((\es|\es)\subset (\es|1)\subset (1|1)\), then \([\deg\mathbb{Q}_{1|0}+1,\deg\mathbb{Q}_{0|1}]\) would be our lowest weight.

The formula \eqref{eq:weightsusy} applies without any subtleties if the multiplet in question is long. If the multiplet  is short, and there are plenty of them in \(\su(K|M)\) spin chains, we need to provide a further analysis.

\subsubsection{\texorpdfstring{$\su(2|1)$}{su(2|1)}:  states in short representations involve zero Q-functions}\label{sec:su21example}
The \(K+M+1\) functions \(\mathbb{Q}_{\leftarrow}\) obtained from \(K+M+1\) functions \(\fQ_{A|J}\) along certain nesting path contain, in principle, all the information about the untwisted Q-system. In this sense, the situation is exactly the same as with \(K+1\) functions \(\mathbb{Q}_{\leftarrow}\) in \(\su(K)\) case. However, the direct untwisting limit generates more than \(K+M+1\) distinct functions, see \eqref{untwistedsusy}. In this respect, the situation differs from the \(\su(K)\) case where other Q-functions are accessible only if the untwisting is supplemented with a rotation. It is a priory not obvious that Q-functions generated by \eqref{untwistedsusy} will be consistent with QQ-relations. One can see  that \cite{Kazakov:2007fy}
\be\label{expectedQQ}
\mathbb{Q}_{\leftarrow k+1|m+1}^+\mathbb{Q}_{\leftarrow k|m}^--\mathbb{Q}_{\leftarrow k+1|m+1}^-\mathbb{Q}_{\leftarrow k|m}^+\propto \mathbb{Q}_{\leftarrow k+1|m}\mathbb{Q}_{\leftarrow k|m+1}\,,
\ee
but sometimes it is possible to satisfy this equation only when the coefficient of proportionality is zero, very similarly to untwisting of \eqref{QQXXX}.

The simplest example is the relation \(Q_{a|i}^+-Q_{a|i}^-=Q_{a|0}Q_{0|i}\). If the polynomial \(\fQ_{a|i}=\left(\frac{\x_a}{\y_i}\right)^{\bi\,u}Q_{a|i}\) becomes a constant in the untwisting limit then \(\lim\limits_{\x_a,\y_i\to 1}Q_{a|0}Q_{0|i}=0\), and one should assign, for consistency, \(\mathbb{Q}_{\leftarrow a|0}=0\) or \(\mathbb{Q}_{\leftarrow 0|i}=0\), instead of a non-zero answer stemming from \eqref{untwistedsusy}. We also can spot from \eqref{eq:weightsusy} that \(\deg \fQ_{a|i}=0\) means \(\lambda_a+\nu_i=0\) which is a shortening condition for  representations of \(\su(K|M)\). 

\ \\
We illustrate this issue on a very explicit and relatively simple case of the \(\su(2|1)\) spin chain with two
sites (\(L=2\)) in the defining representation. We will further strengthen our claim that Q-systems can involve zero Q-functions and yet describe physical states by explicitly realising all Q-functions as eigenvalues of the Q-operators. This example is also rich enough to illustrate certain other twist-dependent effects introduced in previous sections.

First, we list below all possible polynomial solutions of the \(\su(2|1)\) QQ-relations without twist with the boundary conditions \(Q_{\es|\es}=1\) and \(Q_{12|1}=u^2\). These solutions can be quickly found by brute force. There is one solution with non-zero Q-functions:
\begin{align}
\label{eq:233}
  Q_{\emptyset}&=1\,,&
 Q_{1|\emptyset}&=1\,,&
 Q_{2|\emptyset}&=\bi\, u+\mathrm{cst}\,,&
Q_{\emptyset|1}&=-2\,,
  \nonumber\\
 Q_{12|\emptyset}&=1\,,&
Q_{1|1}&=2 \bi\, u\,,&
Q_{2|1}&=-u^2%
  -\frac 1 4+2\bi\,\mathrm{cst}\,u\,,&
 Q_{12|1}&=u^2\,,
\end{align} where \(\mathrm{cst}\) denotes an irrelevant constant which originates from the \(\GL(2)\) H-symmetry rotating bosonic indices.

There are also two solutions which contain zero Q-functions:
\begin{subequations}
\label{eq:232}
\begin{align}
\label{eq:232a}
  Q_{\emptyset}&=1\,,&
 Q_{1|\emptyset}&=0\,,&
 Q_{2|\emptyset}&=-u^2,&
Q_{\emptyset|1}&=R\,,
  \nonumber\\
 Q_{12|\emptyset}&=0\,,&
Q_{1|1}&=1\,,&
Q_{2|1}&=\Psi^+(u^2\,R)\,,&
 Q_{12|1}&=u^2\,,
\end{align}
\begin{align}
\label{eq:232b}
  Q_{\emptyset}&=1\,,&
 Q_{1|\emptyset}&=R\,,&
 Q_{2|\emptyset}&=-u^2&
Q_{\emptyset|1}&=0\,,
  \nonumber\\
 Q_{12|\emptyset}&=-\left|\begin{smallmatrix}R^+&R^-\\(u^+)^2&(u^-)^2\end{smallmatrix}\right|\,,&
Q_{1|1}&=1\,,&
Q_{2|1}&=0\,,&
 Q_{12|1}&=u^2\,,
\end{align}
\end{subequations}
where \(R\) is an arbitrary polynomial and \(\Psi(u^2\,R)\) is a polynomial that satisfies \(\Psi(u^2\,R)-\Psi^{++}(u^2\,R)=u^2\,R\).

Second, we assign the irreps in the Hilbert space to the presented solutions. The Hilbert space is 9-dimensional and it decomposes into two irreps of \(\su(2|1)\):\\[0.5em]
super-symmetrisation of two defining representations
\begin{center}
\begin{tabular}{ccccc}
state & weight & \(1<2<\hat1\) & \(1<\hat 1< 2\) & \(\hat 1< 1<2\) \\
\(|\!\uparrow\uparrow\rangle\) & [20;0] & HW & HW & \\
\(|\!\uparrow\downarrow\rangle+|\!\downarrow\uparrow\rangle\) & [11;0] &  &  & \\
\(|\!\downarrow\downarrow\rangle\) & [02;0] &  & LW & LW \\
\(|\!\uparrow\!\theta\rangle+|\theta\!\uparrow\rangle\) & [10;1] &  &  &  HW\\
\(|\!\downarrow\!\theta\rangle+|\theta\!\downarrow\rangle\) & [01;1] & LW &  & 
\end{tabular}\,,
\end{center}
and super-antisymmetrisation 
\begin{center}
\begin{tabular}{ccccc}
state & weight & \(1<2<\hat1\) & \(1<\hat 1<2\) & \(\hat 1<1<2\) \\
\(|\!\uparrow\downarrow\rangle-|\!\downarrow\uparrow\rangle\) & [11;0] & HW &  & LW\\
\(|\!\uparrow\!\theta\rangle-|\theta\!\uparrow\rangle\) & [10;1] &  & HW &  \\
\(|\!\downarrow\!\theta\rangle-|\theta\!\downarrow\rangle\) & [01;1] &  & LW  & \\
\(|\theta\theta\rangle\) & [00;2] & LW &  & HW
\end{tabular}\,.
\end{center}
The choice of the ordering \(1<2<\hat1\), \(1<\hat 1< 2\), or \(\hat
1< 1<2\)  is in one-to-one correspondence with the preferred choice of
the nesting path. For instance \(1 <\hat 1<2\) corresponds to \((\es|\es)\subset (1|\es)\subset (1|1)\subset (12|1)\). 

We can use \eqref{eq:weightsusy} to identify Q-systems with corresponding irreps. We see that the four-dimensional representation corresponds to the Q-system \eqref{eq:233}. A less obvious claim is that {\it both} Q-systems \eqref{eq:232} correspond to the five-dimensional representation. To perform identification of weights, we should choose a nesting path which avoids zero Q-functions: \eqref{eq:232a} is  used with the ordering \(\hat1<1<2\) while \eqref{eq:232b} is used with the orderings \(1<2<\hat1\) or \(1<\hat1<2\).  The reader can check correctness of \eqref{eq:weightsusy} when we choose \(R\propto 1\). 

The five-dimensional representation is an example of short, or atypical representation. Such a representation is characterised by a property that some of states are annihilated by more than a half of fermionic generators. Hence, these states can be highest- or lowest-weight ones for more than one index ordering. The practical output that we rely on is a possibility to realise condition \(\lambda_a+\nu_i=0\)  for a highest or lowest weight if one choose an appropriate ordering for which \(a\) and \(i\) are the neighbours in this ordering sequence. In compact rational spin chains all weights are non-negative integers. Hence \(\lambda_a+\nu_i=0\) implies \(\lambda_a=\nu_i=0\). Then condition of being lowest weight implies \(\lambda_b=0\) for \(b\leq a\) and \(\nu_j=0\) for \(j\leq i\). This significantly restricts the possible Q-systems describing short representations, it also explains why we chose \(R\propto 1\) above\footnote{This last argument is based on the representation theory. It would be nice to observe it solely from analytic properties of a Q-system, we however do not perform this analysis here.}.

The choice \(R\propto 1\) seems to be natural for the purpose of
correct weight counting. And it also stems from the operatorial
construction given below.  However, quite remarkably, such invariant
quantities as T-functions or energy do not depend on the choice of
\(R\). We can even put \(R=0\) and get that the two solutions \eqref{eq:232a} and \eqref{eq:232b} coincide!
To say more, many T-functions computed for the states in short representations are identically 0, the non-zero ones live on a smaller L-hook. The observed phenomena are present in supersymmetric Q- and T-systems of any rank. In fact, in the case of character solution \eqref{eq:169}, we can recognise in these effects one of the defining properties of supersymmetric Schur polynomials. We discuss this question in detail in \appref{app:shortmultiplets}.

\ \\
Finally, we support the observations made above by explicit analysis of  operators \(\hat Q\) acting on the Hilbert space. These operators are constructed according to the procedure of \cite{Kazakov:2010iu}, and the presence of twist is essential, see  \appref{sec:rational-spin-chains} for details. The  explicit expressions obtained in the basis  \(\ket{\uparrow\uparrow}, \ket{\uparrow\downarrow}, \ket{\uparrow\!\theta}, \ket{\downarrow\uparrow}, \ket{\downarrow\downarrow}, \ket{\downarrow\!\theta}, \ket{\theta\!\uparrow}, \ket{\theta\!\downarrow}, \ket{\theta\theta}\) are
\begin{subequations}\label{eq:187}
  \begin{align}
{\hat{\mathcal{Q}}}_\emptyset&=1\,,&{\hat{\mathcal{Q}}}_{12|1}&=
u^2\frac{%
  (\x_1-\x_2)}{(\x_1-\y)(\x_2-\y)}
\,,
\end{align}
\begin{align}%
  {{\hat{\mathcal{Q}}}_{1|\emptyset}}%
  &= \left(
      \begin{smallmatrix}%
u^{2} -\frac{\x_1(\x_2-\y)}{(\x_1-\x_2)(\x_1-\y)}\left(2\bi u+\frac{\x_1+\x_2}{\x_1-\x_2}\right)
&0&0&0&0&0&0&0&0\\
0&u+\frac{\bi\,\x_1}{\x_2-\x_1}+\frac{\bi\,\x_1}{\x_1-\y}&0&-\frac{\bi\,\x_2}{\x_1-\x_2}&0&0&0&0&0\\
0&0&u+\frac{\bi\,\x_1}{\x_2-\x_1}+\frac{\bi\,\x_1}{\x_1-\y}&0&0&0&-\frac{\bi\,\y}{\x_1-\y}&0&0\\
0&-\frac{\bi\,\x_1}{\x_1-\x_2}&0&u+\frac{\bi\,\x_1}{\x_2-\x_1}+\frac{\bi\,\x_1}{\x_1-\y}&0&0&0&0&0\\
0&0&0&0&1&0&0&0&0\\
0&0&0&0&0&1&0&0&0\\
0&0&-\frac{\bi\,\x_1}{\x_1-\y}&0&0&0&u+\frac{\bi\,\x_1}{\x_2-\x_1}+\frac{\bi\,\x_1}{\x_1-\y}&0&0\\
0&0&0&0&0&0&0&1&0\\
0&0&0&0&0&0&0&0&1
\end{smallmatrix}
    \right)\,,
\\  %
\hat{\mathcal{Q}}_{2|\emptyset}%
&=\left(
      \begin{smallmatrix}%
1&0&0&0&0&0&0&0&0\\
0&u+\frac{\bi\,\x_2}{\x_1-\x_2}+\frac{\bi\,\x_2}{\x_2-\y}&0&\frac{\bi\,\x_2}{\x_1-\x_2}&0&0&0&0&0\\
0&0&1&0&0&0&0&0&0\\
0&\frac{\bi\,\x_1}{\x_1-\x_2}&0&u+\frac{\bi\,\x_2}{\x_1-\x_2}+\frac{\bi\,\x_2}{\x_2-\y}&0&0&0&0&0\\
0&0&0&0&u^{2} -\frac{\x_2(\x_1-\y)}{(\x_2-\x_1)(\x_2-\y)}\left(2\bi u+\frac{\x_2+\x_1}{\x_2-\x_1}\right)&0&0&0&0\\
0&0&0&0&0&u+\frac{\bi\,\x_2}{\x_1-\x_2}+\frac{\bi\,\x_2}{\x_2-\y}&0&-\frac{\bi\y}{\x_2-\y}&0\\
0&0&0&0&0&0&1&0&0\\
0&0&0&0&0&\frac{\bi\,\x_2}{\x_2-\y}&0&u+\frac{\bi\,\x_2}{\x_1-\x_2}+\frac{\bi\,\x_2}{\x_2-\y}&0\\
0&0&0&0&0&0&0&0&0
\end{smallmatrix}
    \right)\,,
\\
\hat{\mathcal{Q}}_{\emptyset|1}%
&=\left(
      \begin{smallmatrix}%
1&0&0&0&0&0&0&0&0\\
0&1&0&0&0&0&0&0&0\\
0&0&u+\frac{\bi\,\x_2}{\x_2-\y}+\frac{\bi\,\y}{\x_1-\y}&0&0&0&\frac{\bi\,\y}{\x_1-\y}&0&0\\
0&0&0&1&0&0&0&0&0\\
0&0&0&0&1&0&0&0&0\\
0&0&0&0&0&u+\frac{\bi\,\x_2}{\x_2-\y}+\frac{\bi\,\y}{\x_1-\y}&0&\frac{\bi\,\y}{\x_2-\y}&0\\
0&0&\frac{\bi\,\x_1}{\x_1-\y}&0&0&0&u+\frac{\bi\,\x_2}{\x_2-\y}+\frac{\bi\,\y}{\x_1-\y}&0&0\\
0&0&0&0&0&\frac{\bi\,\x_2}{\x_2-\y}&0&u+\frac{\bi\,\x_2}{\x_2-\y}+\frac{\bi\,\y}{\x_1-\y}&0\\
0&0&0&0&0&0&0&0&u^2+2\bi u\frac{\x_1\x_2-\y^2}{(\x_1-\y)(\x_2-\y)}-\frac{\x_1\x_2+\y^2}{(\x_1-\y)(\x_2-\y)}
\end{smallmatrix}
    \right)\,.
  \end{align}
\end{subequations}
The presented 5 operators mutually commute. The \(\mathcal{Q}\)-functions are their eigenvalues. The operators \({\hat{\mathcal{Q}}}_{1|1}\), \({\hat{\mathcal{Q}}}_{2|1}\), and \({\hat{\mathcal{Q}}}_{2|1}\)  were not shown explicitly. They are also polynomials in \(u\) and rational functions in twist variables and they  can be easily restored using the \(\mathcal{Q}\mathcal{Q}\)-relations.

\paragraph{Eigenstate \(|\!\uparrow\uparrow\rangle\).}The most intriguing is to look on a state which becomes a member of atypical representation in the untwisting limit. We will concentrate on \(\ket{\uparrow\uparrow}\), it is already an eigenstate of \(\hat Q\)-operators. After the change of variables \eqref{eq:181}, we obtain the following eigenvalues:
\be
&&Q_\emptyset=1\,,
\nonumber\\
&&Q_{1|\emptyset}=\x_1^{-\bi u-3/4} \left(u^{2} -\frac{\x_1(\x_2-\y)}{(\x_1-\x_2)(\x_1-\y)}\left(2\bi u+\frac{\x_1+\x_2}{\x_1-\x_2}\right)\right)\,,
\quad
Q_{2|\emptyset}=\x_2^{-\bi\,u-3/4}\,, \quad Q_{\emptyset|1}=\y^{\bi\,u-1/4}\,,
\nonumber\\
&&Q_{12|\emptyset}=(\x_1\x_2)^{-\bi\,u-5/4}\left(u^2+\bi\,u\frac{\x_1+\y}{\x_1-\y}-1/4\right)(\x_1-\x_2)\,, 
\nonumber\\
&&Q_{2|1}=\left(\frac{\x_2}{\y}\right)^{-\bi u-1/4}\frac 1{\x_2-\y}\,,\quad
Q_{1|1}=\left(\frac{\x_1}{\y}\right)^{-\bi u-1/4}\frac 1{\x_1-\y}\left(u^2-\bi u\frac{\x_1+\x_2}{\x_1-\x_2}-\frac{\x_1^2+6\x_1\x_2+\x_2^2}{4(\x_1-\x_2)^2}\right)\,,
\nonumber\\
&&
 Q_{12|1}=
 \left(\frac{\x_1\x_2}{\y}\right)^{-\bi u-3/4}\,u^2\frac{(\x_1-\x_2)}{(\x_1-\y)(\x_2-\y)}\,.
\ee
For instance, one can compute the energy\footnote{We remind that 
the present convention for the Hamiltonian is 
\(\mathcal{H}=\sum_{i=1}^{L-1} \mathcal{P}_{i,i+1}+(\mathcal{P}_{1,L}\cdot G^{-1}\otimes\mathbb I^{\otimes L-2}\otimes G)\).
} of this state which turns out to be  \(2\), as can be seen in (\ref{eq:164}) where \(r=0\) (\(\mathcal{Q}_1\) has no roots).

If we perform a straightforward untwisting limit in the style of \eqref{dirlimit1} and \eqref{untwistedsusy} we will find that all Q-functions are proportional to identity, except for \(\mathbb{Q}_{\leftarrow 2|0}\propto u\) and \(\mathbb{Q}_{\leftarrow 2|1}\propto u^2\). Such set of Q-functions does not satisfy the relation \eqref{expectedQQ}.

A sober way to proceed is to perform a rotation which will produce Q-functions that have an explicitly regular limit \(G\to \mathbb I\) (for almost any direction) but, however, may become also zero. Then we guarantee that QQ-relations would survive the limit.

On one hand, one can use the rotation \eqref{eq:QRotate} with the matrix
\begin{subequations}\label{eq:68}
\begin{align}
\label{eq:194a}  h &=
  \left(
    \begin{array}{ccc}
-\frac{(\x_1-\x_2)^2(\x_1-\y)}{2\alpha}&0&0\\
(\x_1-\y)\left(\frac{-1}{\x_1-\x_2}-\frac 5 4-\frac{41}{32}(\x_1-\x_2)\right)
&
-\frac{2(\x_2-\y)}{(\x_1-\x_2)^3}&0\\
0&0&\alpha
    \end{array}
  \right)\,,
\end{align}
where \(\alpha\) is an arbitrary (but non-zero) constant. More precisely \(\alpha\) is
independent of \(u\), and it may be a function of \(G\) 
such that \(\epsilon\ll \alpha\) when \(G=\mathbb I+\epsilon \,g_0\to\mathbb I\). Then, one  finds that the Q-functions
obtained after the rotation have a \(G\to \mathbb I\) limit given by
\eqref{eq:232a}, with \(R=\lim\limits_{G\to\mathbb I}\alpha\). The choice \(R=0\) can be
obtained in several ways, for instance one can put \(\alpha=\sqrt{\x_2-\y}\).

On the other hand, one can use the rotation \eqref{eq:QRotate} with the matrix

\begin{align}
\label{eq:194b} 
h&=
  \left(
    \begin{array}{ccc}
-\alpha \frac{(\x_1-\x_2)^2(\x_1-\y)}{2(\x_2-\y)}&0&0\\
(%
{\x_1-\y}%
)
\left(\frac {-1}{\x_1-\x_2}-\frac 5 4
  -\frac {41} {32} (\x_1-\x_2)\right)
&-2\frac{\x_2-\y}{(\x_1-\x_2)^3}&0\\0&0&\frac{\x_2-\y}{\alpha}
    \end{array}
  \right)\,,
\end{align}
\end{subequations}
(with the same condition on \(\alpha\) as before), and produce Q-functions
with a \(G\to \mathbb I\) limit given by
\eqref{eq:232b}, with \(R=\lim\limits_{G\to\mathbb I}\alpha\). It is manifest that the two
rotations differ by slight normalisations only, and the choice \(\alpha=\sqrt{\x_2-\y}\) makes them coincide.

The procedure to construct these rotations is the following one: the
diagonal entries are designed to make Q-functions along a chosen
nesting path regular and non-zero in the untwisting limit\footnote{Example: If we choose the nesting path \((\es|\es)\subset
  (\es|1)\subset (1|1) \subset (12|1)\), then we obtain the rotation
  \eqref{eq:194a} with \(\alpha=1\), whereas if we chose the nesting path \((\es|\es)\subset
  (1|\es)\subset (1|1) \subset (12|1)\), then we obtain the rotation
  \eqref{eq:194b} with \(\alpha=1\).
} , i.e. we perform the limit of type \eqref{untwistedsusy} on the chosen \(K+M+1\) Q-functions. The off-diagonal entries  are introduced to reproduce all Q-functions of the untwisted Q-system, they execute the idea \eqref{off-diagonalidea}. If we are interested only in Q-functions of type \(\mathbb{Q}_{\leftarrow}\) then we can skip constructing off-diagonal terms.  Note that in the presence of the off-diagonal terms, e.g. rotated \(Q_{2|\es}\) is not a product of polynomial and exponential prefactor. Such a mixture allows to get polynomials of higher degree in the untwisted limit, the off-diagonal terms are fine-tuned to achieve this goal. A generic algorithm to construct rotations is explained in \appref{sec:constr-rotat}.

\paragraph{Eigenstate \(|\!\downarrow\downarrow\rangle\).} This one also has the energy equal to \(2\). It is analysed in full analogy to \(\ket{\uparrow\uparrow}\). Two rotational matrices yielding \eqref{eq:232} in the untwisting limit are
\begin{subequations}
\begin{align}
  h=&   \left(
    \begin{array}{ccc}
\frac{\x_1-\y}{\alpha}
&0&0\\
(%
{\x_1-\y}%
)
\left(\frac{1/16}{\x_1-\x_2}-\frac {3/2}{(\x_1-\x_2)^2}+\frac{2}{(\x_1-\x_2)^3}\right)
&\frac{\x_2-\y}{\x_1-\x_2}&0\\0&0&\alpha
    \end{array}
  \right)
\end{align}
and
\begin{align}
  h=&   \left(
    \begin{array}{ccc}
\alpha&0&0\\
(%
{\x_1-\y}%
)
\left(\frac{1/16}{\x_1-\x_2}-\frac {3/2}{(\x_1-\x_2)^2}+\frac{2}{(\x_1-\x_2)^3}\right)
&\frac{\x_2-\y}{2(\x_1-\x_2)}&0\\0&0&\frac{\x_1-\y}{\alpha}
    \end{array}
  \right)\,.
\end{align}
\end{subequations}

\paragraph{Eigenstate \(|\theta\theta\rangle.\)} The energy of this state is \(-2\). One can use the rotation
\begin{align}\label{eq:234}
  h=&   \left(
    \begin{array}{ccc}
1&0&0\\
\frac {-1} {\x_1-\x_2}&\frac 1{\x_1-\x_2}&0\\
0&0&(\x_1-\y)(\x_2-\y)
    \end{array}
  \right)\,
\end{align}
to repreduce  \eqref{eq:233} in the untwisting limit.

As this eigenstate is not cluttered with effects related to atypical representation, it is the simplest example to observe how off-diagonal elements of the rotation matrix allow to increase the degree of a polynomial. Indeed \(\fQ_{1|\es}=\fQ_{1|\es}=1\) on this state, however \(Q_{2|\es}=\bi\,u+{\rm cst}\) in \eqref{eq:233}.

\paragraph{Other eigenstates.} The remaining 6 eigenstates are obtained by diagonalizing three \(2\times 2\) blocks in matrices \eqref{eq:187}.  
These states and their energies read as follows
\begin{equation}
    \label{eq:237}
    \begin{array}{|c|c|}
      \hline\textrm{State}&\textrm{twisted energy}\\
      \hline
      \sqrt{\x_1}\ket{\uparrow\downarrow}\pm\sqrt{\x_2}\ket{\downarrow\uparrow} & \pm\frac{\x_1+\x_2}{\sqrt{\x_1\x_2}}\\
      \hline
\sqrt{\x_1} \ket{\uparrow\theta} \pm \sqrt{\y}\ket{\theta\uparrow} 
 &\pm\frac{\x_1+\y}{\sqrt{\x_1\y}}\\
 \hline
\sqrt{\x_2} \ket{\downarrow\theta} \pm \sqrt{\y}\ket{\theta\downarrow}
 &\pm\frac{\x_2+\y}{\sqrt{\x_2\y}}\\
 \hline
     \end{array}
         \raisebox{-2em}{\,,}
  \end{equation}
and the reader can straightforwardly construct rotations which provide a smooth \(G\to\mathbb I\) limit, following the lines of \appref{sec:constr-rotat}. 

 These states are examples demonstrating a non-trivial monodromy around co-dimension one hyperplanes \(\x_1=0\) etc, where the twist matrix \(G\) becomes degenerate, cf. \eqref{xzero}. Going around the degeneration points changes the branch of the corresponding square root. On the branch were \(\sqrt{1}=1\), the  sign ``\(+\)'' in \eqref{eq:237} corresponds to the states which become a part of the atypical (five-dimensional) representation in the untwisting limit (hence Q-functions have the limit (\ref{eq:232})), whereas sign ``\(-\)'' corresponds to the states which become a part of the typical (four-dimensional) representation (i.e. Q-functions have the limit (\ref{eq:233})).

\subsubsection{\texorpdfstring{$\sl(2)$}{sl(2)}: non-compactness leads to Stokes phenomena}
Finally, we will consider the XXX spin chain in a non-compact representation. Such a spin chain is not described by entirely polynomial Q-functions, but it is still based on rational R-matrix, hence it is natural to consider it in the same section. Understanding certain features of such a system is quite important for further study of the AdS/CFT integrability which is also based on a non-compact algebra.

The non-compactness is distinguished by appearance of a certain singularity, a pole in the rational case: \(\fQ_{12}=u^{-L}\), so the QQ-relation to solve is
\begin{equation}\label{eq:6}
\fQ_1^+\,\fQ_2^--z\,\fQ_1^-\,\fQ_2^+\propto u^{-L}\,.
\end{equation}
One further demands that \(\fQ_1\) will be a {\it polynomial}. Let us denote its  degree \(\deg\fQ_1=M_1\). On the other hand, \(\fQ_2\) cannot be a polynomial as \(\fQ_2\sim u^{-L-M_1}\) at large \(u\).

Similarly to the compact case,  the degree of the polynomial function \(\fQ_1\) can drop in untwisting limit, and we can find  the following analytic approximating solution
\be\label{LagapproximationNC}
\fQ_1(u)&\simeq& \mathbb{Q}(u)\,L_{m}^{(-2s-1)}(-\bi\,u\,\log z)\,,\ \ \ {\rm with}\ \  \deg\mathbb{Q}=-\frac L2-s\,. 
\ee
The main difference with  \eqref{Lagapproximation} is the change in sign of \(s\), so now one has \(-2s-1\geq L-1\geq 0\). As a consequence, we have no upper bound on the value of \(m\), i.e. \(m\in\mathbb{Z}_{\geq 0}\). This labelling by \(m\) enumerates the eigenstates of an infinite-dimensional lowest-weight representation of \(\sl(2)\), with spin \(s\). Another consequence of \(-2s-1\geq 0\) is that all zeros of the associated Laguerre polynomial are real. The polynomial \(\mathbb{Q}(u)\) satisfies untwisted Bethe equations, it is known to have real Bethe roots as well.

Unlike  the compact case, the limits of \(\fQ_1\) and \(\fQ_2\) are two
independent functions which we denote as \(\mathbb{Q}\propto
\lim\limits_{z\to 1}\fQ_1\) and \(\mathbb{Q}'\propto \lim\limits_{z\to1}\fQ_2\). The degree\footnote{We define degree of a non-polynomial function the value of its exponent when \(u\to\infty\)} of \(\mathbb{Q}'\), \(\deg\mathbb{Q}'=-\frac L2+s+1\) is negative but it is larger than that of \(\fQ_2\). It is indeed possible that the degree of a non-polynomial function increases in certain limits, a simple example is \(\lim\limits_{z\to 1}\frac{u^{M_2}}{1+(z-1)u}\).\\

Another interesting question to discuss is the interplay between dominant and sub-dominant solutions. Think of \(Q_1,Q_2\), the Q-functions of the untwisted Q-system, as of two solutions of the Baxter equation \eqref{Baxter}. In the compact \(\su(2)\) case, the function \(Q_1=\mathbb{Q}_{\leftarrow 1}\) is obtained by the direct untwisting limit \eqref{dirlimit1}, whereas the function \(Q_2\) is obtained with the help of a rotation. Apart from this difference related to untwisting limit, \(Q_1\) can be also singled out as the sub-dominant solution of the Baxter equation at large \(u\). \(Q_2\) is a dominant solution and hence it is not defined uniquely: any combination of the type \(Q_2+{\rm const}\,Q_1\) would still qualify as a dominant solution. The transformation \(Q_2\mapsto Q_2+{\rm const}\,Q_1\) is a residual H-rotation which respects the ordering in degree of Q-functions.

In the non-compact \(\sl(2)\) case, the situation appears to be contr-intuitive if one uses ordering in degree as a way to select ``distinguished'' Q-functions. The two Q-functions \(Q_1\equiv \mathbb{Q}\) and \(Q_2\equiv \mathbb{Q}'\) also satisfy the Baxter equation \eqref{Baxter}, but now with \(\phi=u^{-L}\). \(Q_1\) seems now to be a dominant one, nevertheless \(Q_1\) is defined uniquely because this is the only polynomial solution. At first glance, \(Q_2\) is sub-dominant and hence it should be defined uniquely. Alas, it is not. We are going to investigate this subtlety.

As we shall see, the subtlety is also present in the twisted case: the function \(\fQ_1\) is uniquely determined from the fact that it is a polynomial, whereas it is less elementary to give a unique prescription for \(\fQ_2\). 

We will reconstruct \(\fQ_2\) from the fact that it satisfies \eqref{eq:6}. We introduce an operator \(\Psi_z\) which satisfies the property
\be\label{psiz}
\Psi_z(f)-z\,\Psi_z^{++}(f)=f\,.
\ee
Then, the most general expression for \(\fQ_2\) can be written in the form
\be\label{Q2generalpos}
\fQ_2=\fQ_1\,\Psi_z^+\left(\frac 1{u^L \fQ_1^+\fQ_1^-}\right)+\CP(u)\,\fQ_1\, z^{{\bi\, u}}\,,
\ee
where \(\CP(u)\) is an \(\bi\)-periodic function.

\(\Psi_z\) is not defined by \eqref{psiz} uniquely. We further narrow the ambiguities in its definition by the requirement that \(\Psi_z(f)\) is regular if \(f\) is regular and that the large-\(u\) asymptotic expansion of \(\Psi_z(f)\) is related to the large-\(u\) expansion of \(f\)  by
\be\label{psizas}
\Psi_z(f)\simeq \frac 1{1-z\,e^{\bi\,\partial_u}}f=\left(\frac 1{1-z}+\frac{\bi\,z}{(1-z)^2}\partial_u+\ldots\right)f\,.
\ee
For instance, these constraints on \(\Psi_z(f)\) imply that  \(\Psi_z(f)\) is a polynomial if \(f\) is a polynomial. Already from \eqref{psizas} we see that \(z=1\) is quite special. Large-\(u\) expansion should be re-summed, we refer to \cite{Marboe:2014gma} for a discussion of properties of \(\Psi\equiv \Psi_{z=1}\) in the case \(z=1\).

As we seek for \(\fQ_2\) with power-like asymptotics, the term with \(z^{{\bi\,u}}\) factor should be discarded. Furthermore, one can always write a solution using the following ansatz (cf. \cite{Marboe:2014gma})
\be\label{Q2ansatz}
\fQ_2=\fQ_1\sum_{k=1}^L c_k\,\Psi_z^+\left(\frac 1{u^k}\right)+R(u)\,,
\ee
where coefficients \(c_k\) are defined by the small-\(u\) expansion \(\frac 1{u^L\fQ_1^+\fQ_1^-}=\sum\limits_{k=1}^L\frac{c_k}{u^k}+\CO(u)^0\), and where \(R(u)\) is a polynomial of degree \(L-1\) which is uniquely fixed by requirement to cancel all positive powers in \(u\) in the large-\(u\) expansion of \eqref{Q2ansatz}.

Hence we can focus on studying \(\Psi_z\left(\frac 1{u^k}\right)\) which coincides, after an appropriate rescaling of parameters, with the Lerch transcendent. We can immediately write down its integral representation by rewriting a symbolic expression \(\frac 1{1-z\,e^{\bi\,\partial_u}}\frac 1{u^k}\) as a Laplace integral
\be\label{Borelint}
\Psi_z\left(\frac 1{u^k}\right)=\frac 1{\Gamma(k)}\int\limits_0^{\infty\times e^{\bi\,\phi}}\frac{t^{k-1}e^{-u\,t}}{1-z\,e^{-\bi\,t}}\,dt\,.
\ee
The announced ambiguity in construction of \(\fQ_2\) can be explicitly seen here: the direction of integration is chosen to make the integral convergent, it is correlated with the direction in which \eqref{psizas} is chosen to hold and the analytic function defined by the integral {\it does depend} on the choice of direction.

A different perspective on this effect is to note that the expansion \eqref{psizas} produces asymptotic non-convergent series. Their Borel resummation leads to \eqref{Borelint} which has Borel ambiguities at the poles 
\be
t_n=-\bi\,\log z+2\pi\,n\,,\ \ \ n\in\mathbb{Z}\,.
\ee 
The resulting ambiguity in the definition of \(\fQ_2\) is \(\delta_n \fQ_2\propto \fQ_1(u)\,z^{\bi\,u}e^{-2\pi\,n\,u}\), which has the form of the second term in \eqref{Q2generalpos} that we attempted to discard. But, as \(n\in\mathbb{Z}\), one can always find Borel ambiguities which are exponentially suppressed in \(u\). Due to this exponential suppression they are sub-dominant compared to the first term in \eqref{Q2generalpos} and we cannot discard such terms based on the large-\(u\) behaviour argument. Hence \(\fQ_2\) cannot be defined uniquely.

We observed here a qualitative distinction between differential and finite-difference equations: In the case of differential equations, the dominance of a solution is decided by analysing its large-\(u\) asymptotic. In the case of finite-difference equations, solutions can be summed with periodic coefficients, not only constants, to produce a new solution; and it can happen that periodic coefficients themselves decide the dominance of different terms in the sum. This effect becomes visible in a non-compact case. Indeed,  we can forbid non-constant periodic functions as coefficients in the compact case, by requiring polynomiality of the solutions. However, in the non-compact case, we cannot forbid periodic functions completely. Although we can require power-like behaviour when \(u\to\infty\), at least in certain directions, periodic functions will still appear as subdominant terms due to Borel ambiguities.

For any value of twist \(z\), except the cases \(0\leq z<1\) and \(z>1\), there are two distinguished choices for \(\Psi_z\).

\begin{figure}[t]
 \centering
\subfigure[Borel complex plane. The position of poles are denoted by crosses. Integration contour \(A\) defines the UHPA solution. Integration contour \(B\) defines another, non-equivalent solution.]{\label{fig:tplane}
\hspace{.4cm}\includegraphics[height=0.3\textwidth]{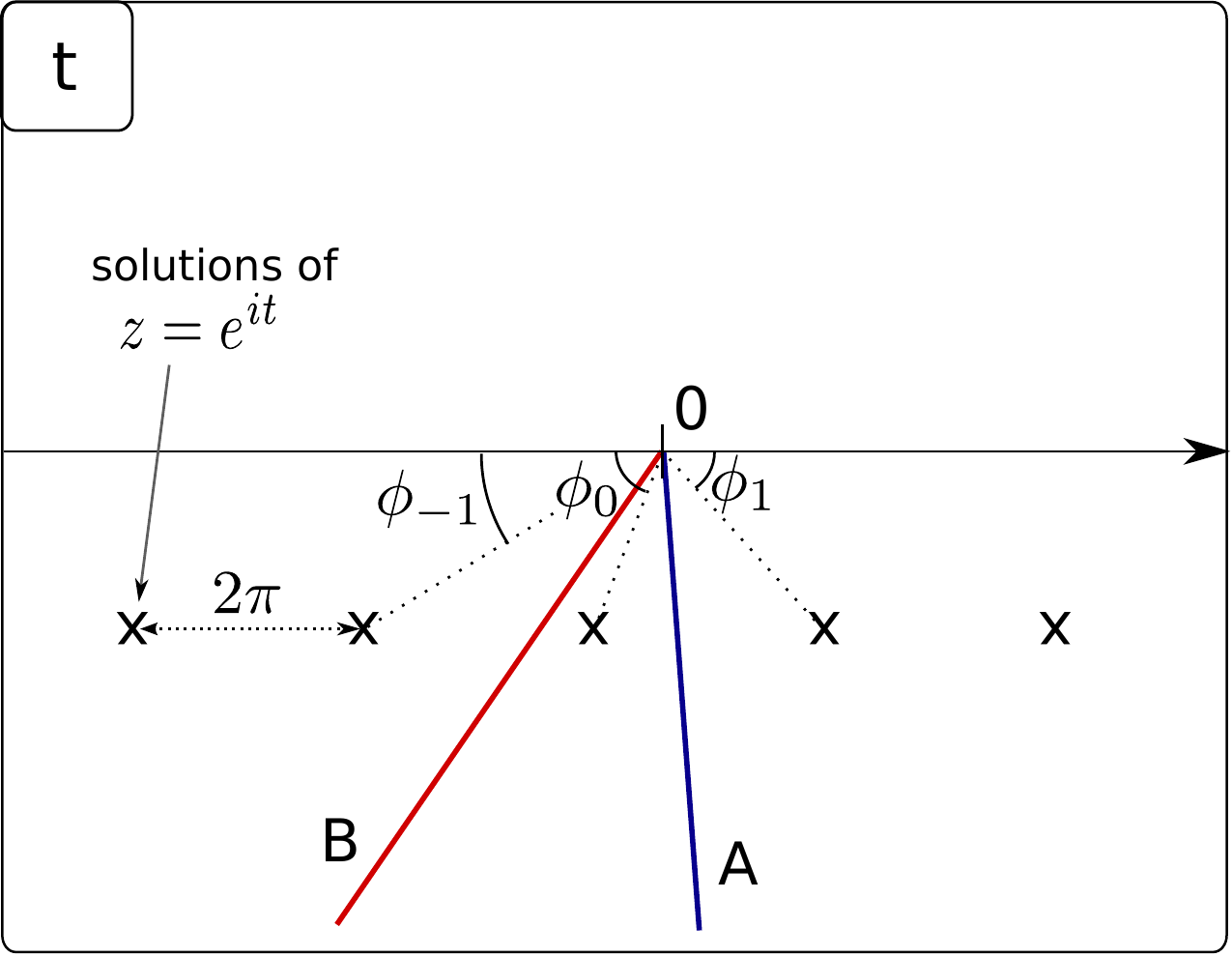}\hspace{.4cm}}\hspace{1.3cm}
\subfigure[Sectors in \(u\)-plane where expansion \eqref{psizas} is applicable. Two sectors are shown, the ones corresponding to \(\Psi_z\) defined by integration contours \(A\) and \(B\).]{\label{fig:uplane}
\hspace{.5cm}\includegraphics[height=0.3\textwidth]{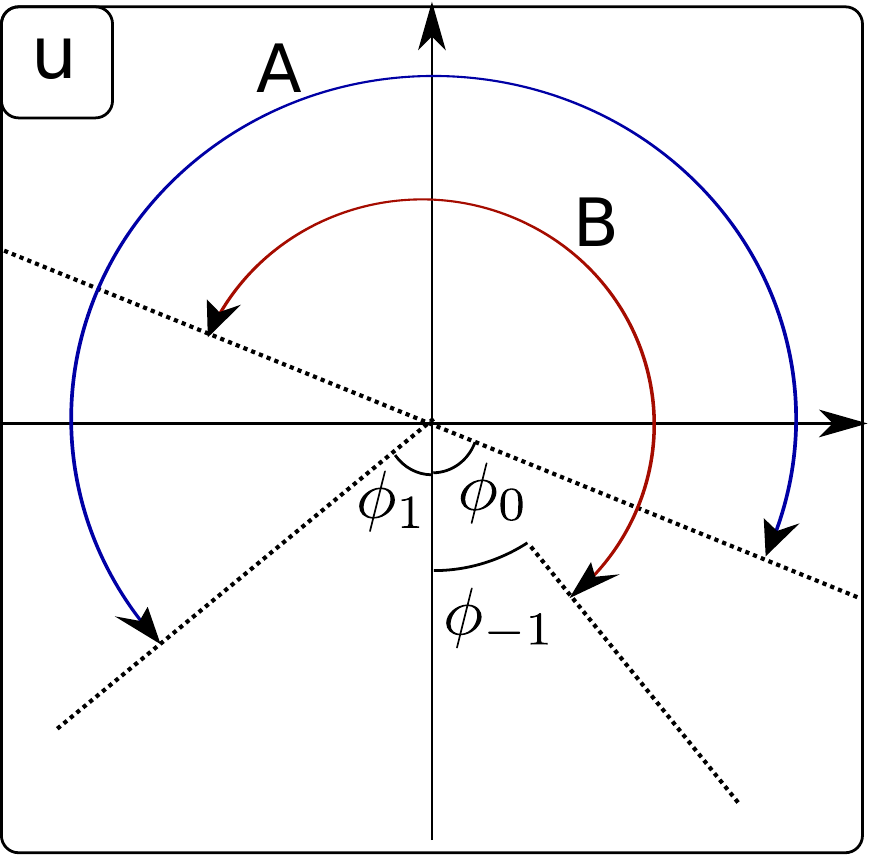}\hspace{.5cm}}

\caption{Illustration of Borel ambiguities (left) and emergence of Stokes sectors (right). A case with \(|z|>1\).
}
 \label{fig:ancones}
\end{figure}

The first one, upper half-plane analytic ({\bf UHPA}), is denoted by \(\Psi_z^{\uparrow}\) and is defined as solution having large-\(u\) expansion \eqref{psizas} valid in the largest possible cone containing  \(u\to+\bi\,\infty\,\). It can be defined by integration over \(t\) in \eqref{Borelint} from \(0\) to \(-\bi\,\infty\) for \(\Im m(u)>0\) and then by analytic continuation. For \(|z|\leq 1\), this solution can be represented as a convergent series\footnote{For \(z=1\), the sum is marginally divergent when \(k=1\). We define such a sum assuming the same regularisation as for the case of digamma function, see \cite{Marboe:2014gma}.}
\be\label{Psiser1}
\Psi_z^{\uparrow}\left(\frac 1{u^k}\right)=\sum_{s=0}^\infty\frac{z^s}{(u+\bi\,s)^k}\,.
\ee
Note that in the case of \(|z|\leq 1\) the sector of applicability of \eqref{psizas} (Stokes sector) is any direction save \(u\to -\bi\,\infty\). The sector becomes smaller if  \(|z|>1\) but it still includes both \(u\to+\infty\) and \(u\to-\infty\) directions, except when \(z>1\); example is shown in figure~\ref{fig:ancones}. For \(z>1\) the UHPA solution is not defined uniquely as there are two solutions which have Stokes sector of equal size.

The second one, the lower half-plane analytic ({\bf LHPA}), is denoted by \(\Psi_z^{\downarrow}\) and is defined as solution having large-\(u\) expansion \eqref{psizas} valid in the largest possible sector containing  \(u\to-\bi\,\infty\,\). Correspondingly, the integration in \eqref{Borelint} is from \(0\) to \(+\bi\,\infty\) for \(\Im m(u)<0\), and \(\Psi_z^{\downarrow}\) is defined by analytic continuation if \(\Im m(u)>0\). The corresponding series
\be\label{Psiser2}
\Psi_z^{\downarrow}\left(\frac 1{u^k}\right)=-\sum_{s=1}^\infty\frac{z^{-s}}{(u-\bi\,s)^k}\,
\ee
is convergent for \(|z|\geq 1\). LHPA is not defined uniquely for \(0\leq z<1\).

The difference between two solutions can be found explicitly by computing the emering integral by residues:
\begin{align}
\Psi_z^{\uparrow}\left(\frac 1{u^k}\right)-\Psi_z^{\downarrow}\left(\frac 1{u^k}\right)=\frac{(-\partial_u)^{k-1}}{\Gamma(k)}\int\limits_{+\bi\,\infty}^{-\bi\,\infty}\frac{e^{-u\,t}}{1-z\,e^{\bi\,t}}dt=\pi\frac{(-\partial_u)^{k-1}}{\Gamma(k)}\frac{(-z)^{\bi\,u}}{\sinh(\pi\,u)}\,.
\end{align}
We see that this difference indeed produces the term of a type \(\CP(u)z^{\bi\,u}\fQ_1(u)\) in \eqref{Q2generalpos}, and this term is exponentially suppressed in \(u\) in the region where large-\(u\) asymptotic expansion is applicable simultaneously for both UHPA and LHPA solutions.

The distinction between UPHA and LHPA Q-functions is paramount for the case of AdS/CFT quantum spectral curve, one could even understand the Riemann-Hilbert conditions of the spectral curve as a way to build UPHA system from LHPA and vice-versa \cite{Gromov:2014caa}.  However, we see now that  the phenomenon appears already in rational non-compact spin chains.

Although the presence of twist is not an absolute requirement for the presented analysis, the twisted case illuminates and enriches the emergent  Stokes effects. First, quite convenient series \eqref{Psiser1} and \eqref{Psiser2} cease to converge simultaneously if \(|z|\neq 1\) and one start to look for a more universal integral definition \eqref{Borelint} which clearly suffers from Borel ambiguities. Second,  the Borel poles depend on twist, and the definition of UPHA and LHPA, which relies on position of these poles, is not smooth in \(z\). Consider for instance UPHA solutions. Crossing the \(z>1\) line in \(z\)-plane requires to pick up a Borel pole in \(t\)-plane thus generating solution which is no longer an UPHA. We hence perceive the line \(1<z<\infty\) as a branch cut. Its branch points are of infinite degree. Indeed, the  discontinuity  \(\Psi_{z+\bi\,0}^{\uparrow}-\Psi_{z-\bi\,0}^{\uparrow}\)  across the cut involves \(\log z\).

\subsection{Dependence on twist and the untwisting limit: general picture}
\label{sec:untwist summary}Back in \secref{sec:solut-qq-relat}, we understood that Q-system realises a maximal flag in \(\mathbb{C}^N\) whose \(u\)-dependence is constrained by the intersection property \eqref{eq:81D}. Since then, we observed  two conceptually different ways to parameterise this flag using \(N\) Q-functions (in the gauge \(Q_{\es}=1\)). We list the essential properties of these two parameterisations (\(\su(N)\) case is kept in mind, but most of the statements can be generalised to supersymmetric and noncompact cases):

\paragraph{Covariant parameterisation:} 
\begin{itemize}
\item[-] Uses functions  
\be\label{eq:parnl}
[Q_1(u):Q_2(u):\ldots Q_N(u)]\,.
\ee 
It can be geometrically thought as a map \(\Sigma\to \mathbb{CP}^{N-1}\), where \(\Sigma\) is the space of spectral parameter \(u\).
\item[-] Is {\bf not invariant} (but co-variant) under H-symmetry transformations. Hence, a choice of a particular basis \eqref{eq:parnl} fixes the H-symmetry freedom.
\item[-] All other Q-functions are restored straightforwardly and uniquely by the determinant relation \eqref{eq:4}.
\item[-] The physical constraint \(Q_{\bar\es}=u^L\)  (see figure~\ref{fig:HasseOr}) is a highly non-local equation in the \(u\)-plane.
\end{itemize}

\paragraph{Nested parameterisation:}
\begin{itemize}
\item[-] Uses Q-functions along a  nesting path, e.g.
\be
\label{eq:parlo}
Q_{\leftarrow 1}=Q_1\,,\ Q_{\leftarrow 2}=Q_{12}\,,\ \ldots\,,\ Q_{\leftarrow N}=Q_{12\ldots N}\equiv Q_{\bar\es}\,.
\ee
\item[-] Is {\bf invariant} under the action of  Borel subgroup of the H-symmetry transformations (lower-triangular matrices with respect to the chosen ordering)\footnote{More accurately, diagonal H-matrices still affect the overall normalisation of Q-functions. The overall normalisation is, however, not essential for our discussion.}. Hence, H-symmetry is only partially broken by the choice of \eqref{eq:parlo}.
\item[-] All other Q-functions are restored  by a systematic usage of the QQ-relation \eqref{eq:2}; the explicit computation requires to solve linear first-order finite-difference equations, and the Q-functions are found not uniquely but modulo the Borel subgroup of H-rotations.
\item[-] The constraint \(Q_{\bar\es}=u^L\) is imposed naturally.
\end{itemize}

The nesting path parameterisation behaves smoothly in the untwisting limit in the sense that its limit can be used to parameterise an untwisted Q-system. In fact, a safe way to realise the untwisting limit of the whole Q-system is to choose the set of nested Q-functions
\be
Q_{\leftarrow |A|}\equiv Q_{A},\ \ \  A\  \text{belongs to a chosen nesting path,}
\ee
normalise them to be non-singular and non-zero in the untwisting limit, take the limit only of these functions\footnote{We use  font \(\mathbb{Q}\) to label a Q-system without twist.},
\be
\mathbb{Q}_{\leftarrow |A|}= \lim\limits_{G\to\mathbb{I}} Q_{\leftarrow |A|}\,,
\ee
 and then restore all other Q-functions using QQ-relations. Note that \(\mathbb{Q}_{\leftarrow |A|}\) are always non-zero by construction, and one can construct a meaningful Q-system from any nested set of non-zero Q-functions.

On the other hand, the covariant parameterisation is quite singular: the set functions \eqref{eq:parnl} degenerates upon direct untwisting limit and we cannot use it anymore to parameterise a Q-system. The problem with covariant parameterization is that it defines a meaningful Q-system only if the basis Q-functions have a non-vanishing determinant: \(Q_{(1)}^{[1-N]}\wedge Q_{(1)}^{[3-N]}\wedge\ldots Q_{(1)}^{[N-1]}\neq 0\). The latter property can be violated when certain limits are taken, in particular this happens to be the case in the untwisting limit.

We can give a symmetry argument why the covariant parameterisation behaves badly in the untwisting limit. Imposing the gauge-fixing condition that the twist is diagonal essentially brakes the H-symmetry. In the untwisting limit,  the gauge-fixing condition is no longer required, hence the (global) H-symmetry is restored. Any choice of a covariant basis would spontaneously brake the restored symmetry, but there is no preferred way to do this choice by performing  the untwisting limit. So, instead of having a meaningful limit, the covariant parameterisation degenerates when \(G\to\mathbb{I}\). 

We can choose, by hands, how to brake the symmetry, by performing twist-dependent H-rotations, e.g. \eqref{eq:194}, and in this way to define a smooth limit of a covariant basis. It can be summarised by a formula
\be\label{eq:Qlimh}
\mathbb{Q}_a=\lim\limits_{G\to\mathbb{I}} \left(h_{a}{}^{a'}\,Q_{a'}\right)\,,
\ee
where \(h\) is a twist-dependent rotation matrix which should be fine-tuned to remove degeneracy from the limiting system. Finding \(h\) is equivalent in complexity to solving QQ-relations. 

Nested parameterisation is invariant enough under H-transformation to have a smooth limit. It is, however, not fully invariant, and the choice of a nesting path plays a certain role which we will now discuss.

We need to precise how the Q-functions are labeled. In the diagonal twist gauge, we follow the following rule: The function \(Q_A\) with index \(A\) is the one with the exponential prefactor \(\prod\limits_{a\in A}x_a^{-\bi\,u}\) in the large-\(u\) asymptotics. In the non-twisted case, the labelling  is more subtle, but there  still exists a natural choice. First, one rotates the basis such that all one-indexed Q-functions have a distinct degree, and then, among all Q-functions \(Q_{A}\) with \(|A|=n\), we assign the label \(Q_{12\ldots n}\) to the polynomial of the lowest degree.

In principle, there are \(N!\) choices of nesting paths, or, equivalently of the total order in the set \(\{1,2,\ldots, N\}\). In the twisted case all of the choices enter indeed on the same footing. However, in the non-twisted case, the order \(1<2<\ldots <N\) in the above-introduced labelling scheme, used as an example in \eqref{eq:parlo},  is distinguished for three reasons:
\begin{itemize}
\item[-] the Q-functions of the distinguished nesting can be uniquely defined as sub-dominant solutions of the corresponding Baxter equations.
\item[-] these nested Q-functions emerge from {\it any} nested set in the untwisting limit. I.e. if \(\mathbb{Q}_{\leftarrow n}=\lim\limits_{G\to\mathbb{I}}Q_{\leftarrow n}\) then \(\mathbb{Q}_{\leftarrow n}=\mathbb{Q}_{12\ldots n}\) independently of how \(Q_{\leftarrow n}=Q_{a_1\ldots a_n}\) was chosen.
\item[-]  The large-\(u\) behaviour correctly reproduces the weights of the representation according to \eqref{eqlargeu1}; note that the statement \eqref{eqlargeu1} is formulated following the distinguished nested path.
\end{itemize}

The present discussion can be generalised to \(\su(K_1,K_2|M)\) case. The full H-symmetry is broken to \(\GL(K_1)\times \GL(K_2)\times \GL(M)\) by {\it analytic} and {\it boundary} requirements on the solution. However the full H-symmetry is not broken on the level of QQ-relations, and it re-emerges partially producing some interesting ambiguities.  

The Q-functions of the twisted Q-system have a natural labelling \(Q_{A_1,A_2|J}\), where \(A_1\subset\Bset=\{1,2,\ldots K_1\}\), \(A_2\subset\Bset=\{\dot 1,\dot 2,\ldots \dot K_2\}\), and \(J\subset\Fset=\{\hat 1,\hat 2,\ldots,\hat M\}\), imposed by the prefactor \(\prod\limits_{a\in A_1}\prod\limits_{b\in A_2}\prod\limits_{j\in J}\left(\frac{\x_a\,\x_b}{\y_j}\right)^{-\bi\,u}\) in the large-\(u\) asymptotic. Due to possible Borel ambiguities, which are remnants of broken H-symmetry, we have to further precise the maximal Stokes sector where one demands such an asymptotic behaviour.  The covariant basis, comprised from one-indexed functions \(Q_{a,\es|\es}\), \(Q_{\es,b|\es}\), \(Q_{\es,\es|j}\), definitely suffers from the Borel ambiguities. However, it is possible to choose a nesting path where all Q-functions are rational functions times the exponential prefactor and hence defined uniquely. Nesting paths with order in which \(a<b\) if \(a\in\Bset_1\) and \(b\in\Bset_2\) have such a property\footnote{The statement was made by inspecting the structure of Bethe equations for arbitrary highest-weight representation written in \appref{sec:Labels}. Intrinsic Q-system study has still to be performed; it might show that a weaker constraint on the order suffices.}.

By analogy, one can define \(\mathbb{Q}_{\leftarrow k_1,k_2\,|\,m}\propto \lim\limits_{G\to\mathbb{I}}Q_{A_1,A_2|J}\), where \(k_i=|A_i|\) and \(j=|J|\). It is expected that due to the degeneracy effect, the function \(\mathbb{Q}_{\leftarrow k_1,k_2\,|\,m}\) does not depend on the choice of sets \(A_1,A_2,J\), but only on the number of indices involved\footnote{This is a conjecture which we believe should be true in a general position. The main argument is that each QQ-relation \eqref{eq:2} has a potentially singular dependence only on one twist ratio \(\x_a/\x_b\), \(\y_i/\y_j\), or \(\x_a/\y_j\). Hence, in general position we can essentially analyse each QQ-relation separately, and hence apply our detailed knowledge of rank-1 examples.}. We then constrain the distinguished orderings in the untwisted Q-system by demanding
\be\label{eq:some3}
\mathbb{Q}_{1\ldots k_1,1\ldots k_2|1\ldots m}=\mathbb{Q}_{\leftarrow k_1,k_2\,|\,m}\,.
\ee
I.e  \((K_1+K_2+M)!\) different nesting paths degenerate into \(\frac{(K_1+K_2+M)!}{K_1!K_2!M!}\) paths by taking the untwisting limit. We further prefer to constrain to the paths which contain only rational Q-functions, hence we limit ourselves to \(\frac{K_1!K_2!}{(K_1+K_2)!}(K_1+K_2+M)!\) possibilities in the twisted case and \(\frac{(K_1+K_2+M)!}{(K_1+K_2)!M!}\) possibilities in the untwisted case.

Defining all Q-functions with distinguished order using \eqref{eq:some3} may lead to contradictions in QQ-relations, e.g. \eqref{expectedQQ} might be violated. Hence we follow the above-outlined strategy: to choose one unique path, perform the limit for Q-functions along this path and then restore the other Q-functions using QQ-relations. Note that the strategy with the use of rotation matrix \eqref{eq:Qlimh} also requires to choose a nesting path to decide which functions to only regularise/make non-zero, and which to  rotate. The result of untwisting can indeed depend on the nesting choice, cf. \eqref{eq:232a} vs. \eqref{eq:232b} (this is another place where the broken H-symmetry re-emerges). However, the ambiguity proves to be unphysical as explained in \appref{app:shortmultiplets}. The untwisted result should also comply with weights expected from representation theory, according to \eqref{eq:weightsusy}, with \(k=k_1+k_2\). To get the agreement, the lowest weight should be defined by the same total order that defines the nesting path.

\section{Twisted Quantum Spectral Curve}
\label{twistedQSC}

The AdS\(_5\)/CFT\(_4\) system represents the most emblematic example of AdS/CFT duality between the  Green-Schwarz-Metsaev-Tseytlin superstring sigma model on \(AdS_5\times S^5\) background on the string side of the duality, and  \(\mathcal{N}=4\) SYM theory on the CFT side. It was realized that, at least in the planar sector, this system is  integrable   \cite{Beisert:2010jr}.  Recently, the integrability equations, originally discovered as the AdS/CFT Y-system \cite{Gromov:2009tv}, and later brought into a  TBA form \cite{Bombardelli:2009ns,Gromov:2009bc,Arutyunov:2009ur,Cavaglia2011}    were recast by the present authors  together with N.Gromov into a  concise and elegant finite system of Riemann-Hilbert equations -- the Quantum Spectral Curve (QSC)  \cite{Gromov2014a,Gromov:2014caa}.  The QSC approach has shown its efficiency and universality in the recent papers \cite{Gromov:2014bva,Alfimov:2014bwa,Marboe:2014gma,Gromov:2015wca,Gromov:2015vua} and has  been then successfully applied to the  AdS\(_4\)/CFT\(_3\) duality \cite{Cavaglia:2014exa,Gromov2014c,Anselmetti:2015mda}.

In this section, we will generalise the AdS\(_5\)/CFT\(_4\) QSC  construction to the case of arbitrary diagonal twist which corresponds, in several subcases, to a handful of  integrable modifications of \({\cal N}=4\) SYM (see e.g. \cite{Beisert:2005if}). In e.g. rational spin chains, the twist is a particular deformation of the spin chain boundary conditions. We saw in \secref{sec:twisted-case} that such a deformation amounts to the introduction of a constant connection \(A\). In  the AdS\(_5\)/CFT\(_4\) case, a description on the level of spin chains or other explicit physical model is not available at arbitrary coupling. Hence the twist of QSC is understood solely as the introduction of a constant connection \(A\), we denote the eigenvalues of \(e^A\) as \(\x_a\) and \(\y_i\). The use of the terminology ``twist'' is justified at weak and strong coupling where the introduced twist can be indeed given its more standard physical meaning.

As we discussed in section \ref{sec:twisted-case}, there is a mapping between the twist parameters and the charges. For spin chain, the twist matrix \(\tG=e^A\) does not only execute H-rotations, but also it  can be thought as an element of the Cartan subgroup of the symmetry group. Such a mapping also manifests itself in the large-\(u\) asymptotics: a multi-index function \(Q_{A|I}\) has an asymptotics  \(Q_{A|I}\sim\prod_a\left(\x_a^{\bi\,u}u^{\lambda_a}\right)^{\alpha_a}\prod_i\left(\y_i^{-\bi\,u}u^{\nu_i}\right)^{\beta_i}u^{n}\), where \([\pmb{\lambda};\pmb{\nu}]\) is the weight of state \eqref{eq:194}, and the numbers \(\alpha_a,\beta_i,n\) depend on the indices \(A|I\) but do not depend on the state considered.

By analogy, we think about twist parameters as group elements also in the case of the AdS\(_5\)/CFT\(_4\). There are actually six twist parameters:
\begin{equation}\label{PSUconstraint} \tG=\{\y_1,...,\y_4|\x_1,\dots,\x_4\}\in \mathsf{PSU}(2,2|4),\quad \text{where}\quad \x_1\x_2\x_3\x_4=1,\qquad   \y_1\y_2\y_3\y_4=1, \end{equation} where the two constraints are due to super-unimodularity and projectivity of \(\mathsf{PSU}(2,2|4)\) group.  

 Explicitly the charges \(\lambda_a\) and \(\nu_i\) associated correspondingly to \(\x_a\) and \(1/\y_i\) read as follows:
\begin{subequations}\label{eq:1001}
  \begin{align}\label{eq:1000}
    \lambda_1&=\frac {J_1+J_2-J_3}2\,,&
    \lambda_2&=\frac {J_1-J_2+J_3}2\,,&
    \lambda_3&=\frac {-J_1+J_2+J_3}2\,,&
    \lambda_4&=\frac {-J_1-J_2-J_3}2\,,\\
    \nu_1&=\frac {-\Delta +S_1+S_2}2\,,&
    \nu_2&=\frac {-\Delta -S_1-S_2}2\,,&
    \nu_3&=\frac {+\Delta +S_1-S_2}2\,,&
    \nu_4&=\frac {+\Delta -S_1+S_2}2\,.\label{eq:1002}\end{align}
\end{subequations}
To avoid confusion, let us stress that twists do not define the value of charges. We just say that twists and (exponentiation of charges) are elements of the same group.

One introduces a freedom of speech and says that we twist a given  symmetry if the value of the twist in the direction of the corresponding Cartan elements is different from one. For instance, we say that \(\x\)-twists realise the twisting of \(\su(4)\simeq \so(6)\) R-symmetry and that  \(\y\)-twists twist the \(\su(2,2)\simeq \so(2,4)\) conformal symmetry of \(\mathcal{N}=4\) SYM. Note that a twisted model is generically no longer invariant under  the original symmetry transformations, only invariance under  the Cartan subalgebra action remains. In particular, unless  \(\x_a=\y_i\) for some \(a,i\), the supersymmetry is fully broken in the presence of twist.

For what concerns twisting the charges, one can make a curious remark. It was noticed in  \cite{Gromov:2007ky}, in the approximation of the twisted ABA equations\footnote{The dictionary between our twist notations and the angles of
 \cite{Gromov:2007ky}  in the \(sl(2)\) favored grading is:\\
\(\y_1=e^{i\phi_1},\quad \x_1=e^{i\phi_2},\quad \x_2=e^{i\phi_3},\quad
\y_2=e^{i\phi_4},\quad \y_3=e^{i\phi_5},\quad \x_3=e^{i\phi_6},\quad
\x_4=e^{i\phi_7},\quad \y_4=e^{i\phi_8}\,.\) In the \(SU(2)\) favored
grading, one has to exchange \(x_j\leftrightarrow y_j\).}, that
 \begin{eqnarray}\label{xyMom}
\frac{\x_1\x_2}{\y_1\y_2}=\frac{\y_3\y_4}{\x_3\x_4}=e^{i{\cal P}}\,,
\end{eqnarray}
where \({\cal\ P}\) is the total momentum of a state. The identities \eqref{xyMom} are also true beyond the ABA approximation The first one is a trivial consequence of \eqref{PSUconstraint} and the second one is a dynamical consequence of the QSC equations. On the other hand, the combination \(\frac{\x_1\x_2}{\y_1\y_2}\) twists the charge \(E=\Delta-J_1\) which is nothing but the energy of a state. Hence the total momentum \({\cal\ P}\) plays a role of twist for the AdS time \(\tau_{AdS}\).\\[-1em]

On the CFT side of duality, e.g. from the point of view of  asymptotic
Bethe ansatz, the deformed theory represents a non-local spin chain
with twisted periodic boundary conditions.  The corresponding SYM
action is not known for a general twisting. However, for some
particular twistings such SYM theories were conjectured and
successfully tested.      Such is the case of  the so-called
\(\gamma\)-deformation, when \(\y\)-twists are absent and
\(\x\)-twists are pure phase factors.    The corresponding
\(\gamma\)-deformed,  \({\cal N}=0\)     SYM action has three exactly
marginal  deformations of scalar-scalar and fermion-scalar
interactions:  the commutators of scalar fields with themselves and
with the fermions should be replaced  by the deformed q-commutators
\cite{Beisert:2005if}.  It is believed that this \(\gamma\)-deformed
theory is non-conformal anymore at finite \(N_c\)  but it still
preserves its conformality  at \(N_c=\infty\).\footnote{At finite
  \(N_c\) the \(\gamma\)-deformed theory is non-conformal, and even at
  infinite \(N_c\)  certain operators of small length, such as
  \(trZ^2\), have a divergent dimension and demand the addition of
  double trace counterterms in the action, leading to the running
  coupling \cite{Fokken2014b,Jin:2013baa}. The theory then runs in  into the \(\beta\)-deformed \(N=1\) SYM. This is always the case at finite \(N_c\). However, in the 't~Hooft limit \(N_c\to\infty\) we expect that these short operators can be decoupled  from OPE's and most of the correlators will take the conformal form.  In that sense, both conformality and integrability of \(\gamma\)-deformed SYM theory are restored in the 't~Hooft limit. }   On the string side, the \(\gamma\)-deformed coset has been known already since long~\cite{Frolov:2003xy} and the corresponding string sigma model appears to be integrable.  This AdS/CFT-correspondence was successfully tested by comparing the results of L\"uscher correction from  TBA on the string side \cite{deLeeuw:2011rw,Ahn:2011xq} with the leading order weak coupling correction on the Yang-Mills side    \cite{Fokken2014b,Fokken2014a} for the BMN vacuum. We will reproduce this result in the next section for testing the twisted QSC.

 For a particular one-parametric case of gamma twisting, the     \(\beta\)-deformation (when all \(\gamma_j=\beta\))  the corresponding CFT dual is identified with a particular case of    Leigh-Strassler \(\mathcal{N}=1\) SYM theory \cite{Leigh1995}.   The   \(\beta\)-deformation of the \(AdS_5\times S^5\) string sigma model  was proposed in \cite{Frolov:2005iq,Arutyunov:2010gb} and it is related to  Lunin-Maldacena background \cite{Lunin:2005jy} on the string side of duality. 

The \(\y\)-twists describe the deformations of \(AdS_5\). They presumably  correspond to the introduction of a non-commutativity of space-time coordinates in the dual deformed SYM theory  \cite{Beisert:2005if}. Generically, three  \(y\)-twists (the condition \eqref{PSUconstraint} always imposed) break the conformal group   \(\SU(2,2)\) to the remaining \(U(1)\times U(1)\times {\mathbb R }\) subgroup, where the line \(\mathbb{R}\) corresponds to the action of the dilatation operator   \({\cal D}\)  related to the non-compact translational isometry of the conformal group.  

 In this section, we first introduce the most general twisting of QSC, and then show how one can reduce the number of twists and consider  some examples of partial twisting which correspond to preserving certain, generically non-abelian subalgefbras of the full \(\mathsf{psu}(2,2|4)\). The untwisting procedure is far from trivial (it is already rather subtle for the spin chains, as we have seen  in the previous section) and it can drastically modify the analyticity conditions of QSC  system. 
 
 In the next section, we will demonstrate how one can work with twisted QSC by reproducing the known result for the anomalous dimension of the BMN vacuum at single wrapping orders.

\subsection{Twisting of Quantum Spectral Curve}

 Now we will give the twisted version of QSC formulation of the  AdS/CFT spectral problem. We will proceed with the generic twisting    keeping all 6 independent twist variables as arbitrary complex numbers.  Recall \cite{Gromov:2014caa} that the QSC is essentially characterized by certain       of Riemann-Hilbert type conditions imposed on a set   of \(2^8=256\) Baxter's Q-functions \(Q_{A|I}(u)\), which depend on the spectral parameter \(u\) and on two sets of indices, \(A\subset\{1,2,3,4\}\), \(\,I\subset\{1,2,3,4\}\), and the dependence is antisymmetric with respect to permutations of the indices inside each of these sets \(A\) and \(I\).
 These Q-functions obey the QQ-relations \eqref{eq:42}. In the AdS/CFT context, and in a specific, most natural  gauge, the following constraints hold: 
  \begin{equation}\label{eq:random1}
Q_{\emptyset|\emptyset}=1,\qquad Q_{\bar\emptyset|\bar\emptyset}:=Q_{1234|1234}=1.
\end{equation} 
The first condition is a simple normalization, whereas the second one is non-trivial and should be interpreted as following from the quantum analog of unimodularity \cite{Gromov:2010km}. Indeed, \(\frac{Q_{\bar\emptyset|\bar\emptyset}^+}{Q_{\bar\emptyset|\bar\emptyset}^-}\) becomes a quantum determinant in the case of rational spin chain with the same symmetry. \(Q_{\bar\emptyset|\bar\emptyset}={\rm const}\) is the only solution of \(\frac{Q_{\bar\emptyset|\bar\emptyset}^+}{Q_{\bar\emptyset|\bar\emptyset}^-}=1\) consistent with analytic properties required below and the projectivity constraint \eqref{PSUconstraint}. Normalisation \eqref{eq:random1} can be achieved by a constant in \(u\) basis rotation \(H\), with \(\det H\neq 1\).

We notice that the twisting of spin chains, as it was done in the previous section by \eqref{eq:168}, only modifies the analyticity of Q-functions, in particular, adding the exponential asymptotics  at \(u\to\infty\) without changing  the Baxter relations and QQ relations. We will follow this inspiration in the case of the AdS/CFT QSC and assume that the QQ-relations and the Riemann-Hilbert relations remain intact after twisting. Hence we will modify the  \(u\to\infty\) asymptotics of Q-functions by exponential factors defined by twists.   The consequent modification of analytic properties  will be however greatly constrained by the structure of  QSC equations.

In what follows, we use the notations of
\cite{Gromov:2014caa}:  namely, the Q-functions  \(\bP_a\) (resp
\(\bQ_{j}\)) denote the functions \(Q_{a|\emptyset}\) (resp
\(Q_{\emptyset|j}\)) in a specific gauge,  discussed in detail in 
\cite{Gromov:2014caa} and used through the whole present section. The functions \(\bP^a\) and \(\bQ^{j}\) are
their Hodge dual: \(\bP^a\equiv\HQ {a|\emptyset}=(-1)^a\fQ_{\bar
  a|1234}\) and \(\bQ^j\equiv\HQ {\emptyset|j}=(-1)^j\fQ_{1234|\bar j}\)
-- where \(\bar n\) denotes the sorted multi-index which forms the
complement of \(n\) in \(\{1,2,3,4\}\) (for instance \(\bar
3=124\)). The Riemann-Hilbert relations, in a particular  form of \(\bP\mu\) and \(\bQ\omega\) systems will be detailed further in \secref{sec:twist-texorpdfstr-mu}.

 A natural ansatz for the large \(u\) asymptotics of  QSC, generalizing the formulae of the paper   \cite{Gromov:2014caa}
(section~3.2.3)     to the twisted case, when all 6 twists are turned on, is
 \begin{equation}
  \begin{aligned}
\bP_a\simeq \mathsf{A}_a\, \x_a^{\bi u}u^{-\lambda_a}\,,\ \qquad \bP^a\simeq \mathsf{A}^a\, \x_a^{-\bi u}u^{\lambda_a},\qquad a=1,2,3,4\,,\\
\bQ_{j}\simeq \mathsf{B}_j\,\y_j^{-\bi u}u^{-\nu_j}\,,\qquad \ \bQ^{j}\simeq \mathsf{B}^j\,\y_j^{\bi u}u^{\nu_j}\,,\qquad j=1,2,3,4\,,   \label{twistedPQ}
\end{aligned}
\end{equation}
  where \(\mathsf{A}_a\), \(\mathsf{A}^a\), \(\mathsf{B}_j\) and
  \(\mathsf{B}^j\) are constant prefactors and  the powers of \(u\) are
  given, for the generic twisting, by equation (\ref{eq:1001}).

This ansatz will be further justified in \secref{sec:turning-twist}, where we will show that in general, setting a  part of twists to zero (or to one, in
 terms of \(\x\) and \(\y\) variables) leads to certain 
  shifts by integers in powers of  certain Q-functions with respect to the fully twisted case \eqref{twistedPQ}. %
We will compute the coefficients of
asymptotics of various Q-functions for various
configurations of partial twisting. In particular, we will
express in terms of Cartan charges and twist variables the following eight invariant products of single-indexed Q-functions: \(\mathsf{A}_1\mathsf{A}^1\), \(\mathsf{A}_2\mathsf{A}^2\), \(\dots\), \(\mathsf{A}_4\mathsf{A}^4\), \(\mathsf{B}_1\mathsf{B}^1\), \(\dots\), \(\mathsf{B}_4\mathsf{B}^4\).
These formulae appeared to be extremely useful for various applications of the QSC, for example for recovering the weak coupling limit  \cite{Gromov2014a} or the BFKL limit \cite{Alfimov:2014bwa} of twist-2 operators. We will generalize here these results to an arbitrary configurations of twists.

The amount of conserved supersymmetry depends on the number of  pairs \((a,j)\) of equal twists \(\x_a=\y_j\) (each of twists enters only once into this counting). 

To motivate a bit our prescription \eqref{twistedPQ}, we note that using the quasiclassical correspondence    \cite{Gromov:2014caa}
\begin{eqnarray}
&&{\bf P}_a\sim \exp\left(-\int^u \hat p_{a}(v)dv\right)\;\;,\;\;
{\bf P}^a\sim \exp\left(+\int^u \hat p_{a}(v)dv\right)\,,\\
&&{\bf Q}^j\sim \exp\left(-\int^u \check p_{i}(v)dv\right)\;\;,\;\;
{\bf Q}_j\sim \exp\left(+\int^u \check p_{i}(v)dv\right)\,.
\end{eqnarray}we obtain from here and \eqref{twistedPQ} the large \(u\) asymptotics of twisted quasimomenta for generic twist:
\begin{eqnarray}
\left(\begin{array}{c}
e^{i\hat p_1}\\
e^{i\hat p_2}\\
e^{i\hat p_3}\\
e^{i\hat p_4}
\end{array}
\right)
\simeq
\left(\begin{array}{c}
\x_1e^{i(+J_1+J_2-J_3)/u}\\
\x_2e^{i(+J_1-J_2+J_3)/u}\\
\x_3e^{i(-J_1+J_2+J_3)/u}\\
\x_4e^{i(-J_1-J_2-J_3)/u}
\end{array}
\right)
\;\;,\;\;
\left(\begin{array}{c}
e^{i\check p_1}\\
e^{i\check p_2}\\
e^{i\check p_3}\\
e^{i\check p_4}
\end{array}
\right)
\simeq
\left(\begin{array}{c}
\y_1e^{i(+\Delta-S_1+S_2)/u}\\
\y_2e^{i(+\Delta+S_1-S_2)/u}\\
\y_3e^{i(-\Delta-S_1-S_2)/u}\\
\y_4e^{i(-\Delta+S_1+S_2)/u}
\end{array}
\right).
\end{eqnarray}
Generalizing what was noticed in \secref{sec:twisted-case}
(equation \eqref{eq:170}) for an \(\mathbb L\)-hook, one can show that
if we neglect, at large \(u\), the exponentials in \eqref{twistedPQ}
and keep only the twists, then we can insert  \eqref{twistedPQ} into  \eqref{eq:60}-\eqref{eq:62},
 and reproduce the \(\SU(2,2|4)\) characters of rectangular irreps
 given by  (2.19) of \cite{Gromov:2010vb}.
 As an even stronger motivation of our twisting  ansatz, we could also reproduce the twisted asymptotic Bethe ansatz of  \cite{Beisert:2005if,Gromov:2007ky} following the guidelines of a similar calculation for the untwisted case done in \cite{Gromov:2014caa}.

\subsection{Twisted \texorpdfstring{$\bP\mu$}{P-mu} and \texorpdfstring{$\bQ\omega$}{Q-omega} systems} 
\label{sec:twist-texorpdfstr-mu}
\begin{figure}
\centering{\includegraphics{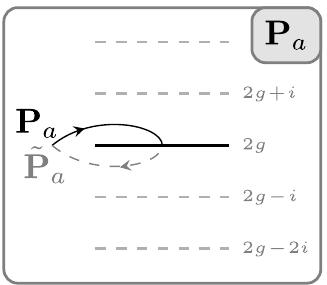}~\includegraphics{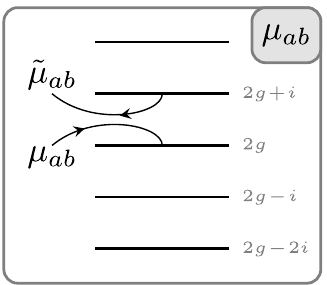}~\includegraphics{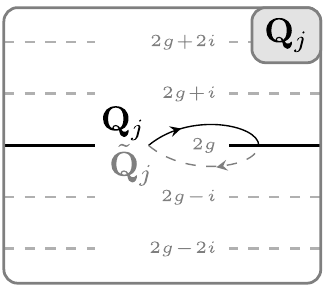}~\includegraphics{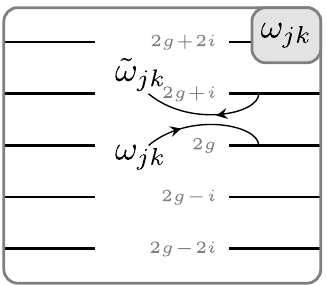}}
 \caption{Cut structure of \(\bP\) and \(\mu\), \(\bQ\) and \(\omega\) and their analytic continuations \(\tilde\bP\) and \(\tilde\mu\), \(\tilde{\bQ}\) and \(\tilde{\omega}\)  \cite{Gromov2014a,Gromov:2014caa}. These pictures are in a choice of sheet where the functions \(\mu_{ab}\) and \(\omega_{ij}\) are not \(\bi\)-periodic, giving the periodicity conditions \(\mu^{++}=\tilde\mu\) and \(\omega^{++}=\tilde\omega\).}
\label{fig:Cuts}
\end{figure}

      The   \(\bP\mu\) and \(\bQ\omega\) systems were
  formulated in \cite{Gromov2014a,Gromov:2014caa} as a particular, and currently intensively used in the literature, incarnation of the QSC.\footnote{These systems of RH relations are only two of many possible types of RH conditions on the full Q-system; the convenient choice of RH conditions can be specific to a particular computation. } Recall that
  \(\bP_a\) and \(\bP^a\) have a single ``short'' Zhukovsky cut along
  the interval \([-2g,2g]\) at the real axis on their defining (physical)
  sheets, whereas \(\bQ_j\) and \(\bQ^j\) have a single ``long''
  Zhukovsky cut along the infinite ``interval''
  \([2g,+\infty[\cup]-\infty,-2g]\) with the same branch points, as
  shown in Fig.~\ref{fig:Cuts}. The main ingredient of the
  \(\bP\mu\) and \(\bQ\omega\) systems is the relations describing the
  monodromy around  these branch points.
Using the notation \(\tilde f\) to denote the analytic continuation of
a function \(f(u)\) around the branch point at \(u=\pm 2 g\), we can formulate the     \(\bP\mu\) and \(\bQ\omega\) systems as  relations  between the original   \(\bP\)
and \(\bQ\) functions and their analytic continuations    \(\tilde
\bP\) and \(\tilde\bQ\). These relations essentially follow from the
equivalence of  choosing functions analytic either in the upper-half plane
\(\mathrm{Im}(u)>0\) (a standard choice for \(\bP\)
and \(\bQ\) functions), or  in the lower-half plane when \(\mathrm{Im}(u)<0\) (then corresponding 
to  \(\tilde \bP\) and \(\tilde\bQ\)). This equivalence means there is
an \(H\)-rotation transforming  \(\tilde
\bP_a\) and \(\tilde\bQ_i\) into the Hodge duals \(\bP^a\) and
\(\bQ^j\)%
.          In the twisted case, these main \(\bP\mu\) and \(\bQ\omega\) relations remain unchanged with respect to the untwisted case of   \cite{Gromov:2014caa}:
 \begin{eqnarray}\tilde \bQ_j=\omega_{jk}\bQ^k\label{QtildeQ}\,,\\   \tilde \bP_a=\mu_{ab}\bP^b \label{PtildeP}\,,\end{eqnarray}  where \(\mu_{ab }\) is \(\bi\)-periodic on a sheet with long cuts, i.e. it has an infinite sequence of long Zhukovsky cuts at  \(u\in([2g,+\infty[\,\cup\,]-\infty,-2g])+\bi \mathbb{Z}\}\), whereas  \(\omega_{jk }\) is \(\bi\)-periodic on a sheet with short cuts \(u\in\{[-2g,-2g]+\bi \mathbb{Z}\}\)  (see  Fig.~\ref{fig:Cuts} and
\cite{Gromov:2014caa}   for more details).  Both matrices  \(\mu_{ab }\)  and  \(\omega_{jk }\) turn out to be antisymmetric and one can consistently normalise them to have unit Pfaffian, hence 
\begin{equation} \mu^{ab }=-\frac{1}{2}\epsilon^{abcd}\mu_{cd }\,,\qquad \omega^{ab }=-\frac{1}{2}\epsilon^{abcd}\omega_{cd }\,.\,\end{equation}  They are  related in the same way as in the untwisted case
\cite{Gromov:2014caa}:  
 \begin{equation}\label{muQomega}
\hat \mu_{ab}=\frac{1}{2}Q_{ab|jk}^-\,\hat \omega^{jk}\,,\qquad \check \omega_{jk}
=\frac{1}{2}Q_{ab|jk}^-\,\check\mu^{ab}
\end{equation}
so that, on the sheet with short cuts (denoted by ``hat''), \(\mu_{ab}\) can be viewed as a linear combination of 4-index functions \(Q_{ab|jk}\) with \(\bi\)-periodic coefficients \(\omega^{jk}\), and vice versa for \(\omega\) with long cuts (denoted by ``check'').
 Let us also remind the obvious quasi-periodicity relations: 
 \begin{equation}\label{quasiPer}
\tilde{\hat\mu}_{ab}=\hat \mu_{ab}^{++}\,,\qquad  \tilde{\check\omega}_{jk}=\check\omega_{jk}
\end{equation}

Another important set of equations defining the monodromies of \(\hat\mu_{ab}\)
follows from directly from certain QQ-relations \cite{Gromov:2014caa}. Namely, we have (for short cuts)\footnote{Throughout the text, the ``hat'' and ``check'' symbols will be removed, and the choice of cuts (usually ``short'' cuts) will be specified in the context.}
    \begin{eqnarray}\mu_{ab}^{++}-\mu_{ab}&&=\bP_a\tilde\bP_b-\bP_b\tilde
      \bP_a=\\&&=(\delta_b^c\bP_a\bP^d -\delta_a^c\bP_b\bP^d)\mu_{cd
      },\end{eqnarray}
 and a similar equation for \(\omega\) in terms of \(\bQ\) (for long cuts):
  \begin{eqnarray}\omega_{jk}^{++}-\omega_{jk}&&=\bQ_j\tilde\bQ_k-
  \bQ_k\tilde \bQ_j=\\&&=(\delta_k^i\bQ_j\bQ^l-\delta_j^i\bQ_k\bQ^l)\omega_{il},\end{eqnarray}

Another set of useful  relations between Q-functions which can be obtained from the QQ relations \eqref{eq:42} and \eqref{eq:141}
 reads\footnote{The present
  Q-functions have a non-polynomial asymptotic behavior and
  correspond to the functions denoted as \(Q\) in \secref{sec:twisted-case}, hence the obey the QQ-relation
  \eqref{eq:42}. By contrast, \(\mathcal{Q}\)-functions obeying the modified
  \(\mathcal{Q}\mathcal{Q}\)-relation \eqref{eq:180} would have a polynomial asymptotics.}~ %
\begin{eqnarray}
&&{\fQ}_{a|j}^+ -{\fQ}_{a|j}^-= \bP_a\bQ_j\,,
\label{QPrel1}
\\
&&  \bQ_j=-\sum_a \bP^a {\fQ}^\pm_{a|j}\,,
\label{QPrel2}
\\
&&    \bP_a=-\sum_j\bQ^j{\fQ}_{a|j}^\pm\,.
\label{QPrel3}
\end{eqnarray}

We also have various orthogonality relations following from the algebraic properties of this Q-system (see \secref{sec:qq-relations-with}):
\begin{align}
{\fQ}_{a|i} {\fQ}^{b|i}&= -\delta_{a}^b\,,& {\fQ}_{a|i} {\fQ}^{a|j}&= -\delta_{i}^j\,,
\label{QQort}
\\
 \bP_a\bP^a&=0\,,&\bQ_j\bQ^j&=0
\label{QQPPort}
\end{align}
The only difference between untwisted   \(\bP\mu\) and \(\bQ\omega\) system  and the twisted ones, with  various  full or partial twistings,  resides in the large \(u\) asymptotics of functions entering these   \(\bP\mu\) and \(\bQ\omega\) systems. We know already from \eqref{twistedPQ}  the asymptotics of twisted \(\bP_a\) and \(\bQ_j\)~.  For the efficient applications of   \(\bP\mu\) and \(\bQ\omega\) systems, we can also calculate the leading asymptotics for \(\mu_{ab}\) on the sheet with short cuts assuming that  \(\omega^{jk}\) is a finite \(\bi\)-periodic function on that sheet.

 In what follows, we will work out  the asymptotics of  \(\bP_a\) and \(\bQ_j\), as well as of some other Q-functions, in various cases of particular twisting.

\subsection{Asymptotics of Q-functions for full and partial twistings %
}
\label{sec:turning-twist}

In the degenerate case when some eigenvalues are equal whereas others
are distinct, it is also possible to express the asymptotics of the
different Q-functions. One can define \(\x_a\), \(\y_i\), \(\lL\) and
\(\lM\) such that \begin{align}\label{eq:220}
    \bP_a &\simeq \mathsf{A}_a\,
    \x_a^{\bi u}u^{-\lL}%
  \,,&
\bQ_{j} &\simeq \mathsf{B}_j\,\y_j^{-\bi u}u^{-\lM%
[j]}\,,
  \end{align}
where we assume without loss of generality that when \(\tx_a=\tx_b\) then \(\lL\neq
\lL[b]\). Indeed, if \(\tx_a=\tx_b\) then we can \(H\)-rotate these 4-vectors (take linear
  combinations of \(Q_a\) and \(Q_b\)) so as to ensure that \(\lL\neq \lL[b]\).
Similarly, we can assume without loss of generality that if \(\ty_i=\ty_j\) then we can always choose a basis with  \(\nu_j\neq \nu_i\). Finally, if \(\tx_a=\ty_i\) then we also assume a generic situation when \(\lL+\lM-1\neq 0\). The equality \(\lL+\lM-1= 0\) corresponds to the multiplet shortening effect. If it holds at arbitrary coupling then the energy would be a protected quantity, but QSC is precisely devised to consider the non-protected case.

From the QQ-relations \eqref{eq:42}, it follows that all functions \(\left(\frac{\prod_{a\in A} \x_a}{\prod_{i\in I}
        \y_i}\right)^{-\bi \,u} Q_{A|I}\) have a power-like asymptotics;
    more precisely \eqref{eq:42} gives
     \begin{subequations}\label{eq:213}
      \begin{align}
        \label{eq:212}
        Q_{Aab|I}\simeq&
        \frac{Q_{Aa|I}Q_{Ab|I}}{Q_{A|I}}\frac{f_{ab}}{\sqrt{\x_a\x_b}}&\textrm{where~} f_{ab}&=
  \begin{cases}\x_b-\x_a&\textrm{ if }\x_a\neq \x_b\\\bi \x_a\tfrac{\lL[b]-\lL}{u}&\textrm{ if }\x_a=\x_b\,,
  \end{cases}\\
        Q_{A|Iij}\simeq&
        \frac{Q_{A|Ii}Q_{A|Ij}}{Q_{A|I}}\frac{f_{ij}}{\sqrt{\y_i\y_j}}&\textrm{where~} f_{ij}&=
  \begin{cases}\y_i-\y_j&\textrm{ if }\y_i\neq \y_j\\\bi \y_i\tfrac{\lM[j]-\lM}{u}&\textrm{ if }\y_i= \y_j\,,
  \end{cases}\\
  \label{eq:213c}
        Q_{Aa|Ii}\simeq&\left.
        \frac{Q_{Aa|I}Q_{A|Ii}}{Q_{A|I}}\right/\frac{f_{ai}}{\sqrt{\x_a \y_i}}&\textrm{where~} f_{ai}&=
  \begin{cases}\y_i-\x_a&\textrm{ if }\x_a\neq \y_i\\\bi \x_a\tfrac{1-\lL-\lM}{u}&\textrm{ if }\x_a= \y_i\,,
  \end{cases}
\end{align}
    \end{subequations}
where the \(\simeq\) symbol denotes the equivalent of \(u\to\infty\) asymptotics (see the end of \secref{sec:asymptotics}). Indeed, we deduce \eqref{eq:213} using that if   \(f\simeq \mathsf{A}_f
{\x}_f^{-\bi u} u^{p_f}\) and \(g\simeq  \mathsf{A}_g \x_g^{-\bi u} u^{p_g}\) then
\begin{align}
\label{eq:222}
\begin{vmatrix}
  f^+&g^+\\f^-&g^-
\end{vmatrix}\simeq&
\begin{cases}
  f\,g\frac{\x_f-\x_g}{\sqrt{\x_f \x_g}}&\textrm{ if }\x_f=\x_g\\
  f\,g\frac{\bi (p_f-p_g)}{u}&\textrm{ if }\x_f\neq \x_g
\end{cases}\,.
\end{align}
Applying the recurrence over the number of indices, starting from
\eqref{eq:220} and \(Q_{\emptyset}=1\), we derive the large-\(u\) asymptotics for an arbitrary Q-function:
\begin{align}
  \label{eq:214}
  Q_{A|J}&\simeq\prod_{a\in A}%
    \frac{\mathsf{A}_a \x_a^{\bi u}
    u^{-\lL}}{\x_a^{\frac{|A|-|I|-1}2}}
  \prod_{i\in J}%
    \frac{\mathsf{B}_i \y_i^{-\bi u}
    u^{-\lM}}{\y_i^{\frac{|I|-|A|-1}2}}%
  \frac{\prod_{\substack{a,b\in A\\a <  b}} f_{a,b}
    \prod_{\substack{i,j\in I\\i <  j}} f_{i,j}}{\prod_{a\in A}\prod_{i\in I} f_{a,i}}\,.
\end{align}
In addition, one should consider the constraint \(Q_{\bar
  \emptyset}=1\), which implies
\begin{align}
  \label{eq:216}
\sum_{1\le a\le 4} \lL+\sum_{1\le i\le 4} \lM
=\sum_{1\le a,i
    \le 4} \delta_{\x_a,\y_i}-\sum_{1\le a < b
    \le 4} \delta_{\x_a,\x_b}-\sum_{1\le i < j
    \le 4} \delta_{\y_i,\y_j}\,.
\end{align}

As particular cases of \eqref{eq:214},  the Hodge duals of \(\bP_a\)
and \(\bQ_i\), given by \(\bP^a=\bP^a/Q_{\bar\emptyset}\simeq\frac{
  {1/\x_a}}{\mathsf{A}_a\x_a^{\bi
    u}u^{-\lL}}\frac{\prod_{1\le i \le 4}f_{a,i}}{\prod_{b\ne
    a}f_{b,a}}\) and \(\bQ^i=\bQ^i/Q_{\bar\emptyset}\simeq\frac{ 1/{\y_i}}{\mathsf{B}_i\y_i^{-\bi
    u}u^{-\lM}}\frac{\prod_{1\le a \le 4}f_{a,i}}{\prod_{j\ne
    i}f_{j,i}}\),  can be rewritten as
  \begin{align}\label{eq:203}
    \bP^a&\simeq \mathsf{A}^a \x_a^{-\bi u}
    u^{\lL-\sum_{i}\delta_{\x_a\y_i}+\sum_{b\ne a} \delta_{\x_a\x_b}}\,,&
    \bQ^i&\simeq \mathsf{B}^i \y_i^{\bi u}
    u^{\lM-\sum_{a}\delta_{\x_a\y_i}+\sum_{j\ne i} \delta_{\y_i\y_j}}\,,
  \end{align}
where we introduce \(\mathsf{A}^a\) and \(\mathsf{B}^i\) defined by
\begin{align}
  \label{eq:204}
  \mathsf{A}_a\mathsf{A}^a&=\frac 1 {\x_a} \frac{\prod_{1\le i \le 4}z_{a,i}}{\prod_{b\ne
    a}z_{b,a}}\,,&
  \mathsf{B}_i\mathsf{B}^i&=\frac 1 {\y_i} \frac{\prod_{1\le b \le 4}z_{b,i}}{\prod_{j\ne
    i}z_{j,i}}\,,\qquad (\text{no sum over}\,\,a,i) 
\end{align}
 and use the
notation
\begin{align}\label{eq:225}
  z_{ab}&=
  \begin{cases}\x_b-\x_a&\textrm{ if }\x_a\neq \x_b\\\bi \x_a({\lL[b]-\lL}){}&\textrm{ if }\x_a=\x_b\,,
  \end{cases}\,,&\textrm{ so that}\ \ \ &f_{ab}=z_{ab}\, u^{-\delta_{\x_a\x_b}}\,.
\end{align}
Obviously, \(z_{ai}\) and \(z_{ij}\) are defined similarly, by
\(f_{ai}=z_{ai}\, u^{-\delta_{\x_a\y_i}}\) and \(f_{ij}=z_{ij} u^{-\delta_{\y_i\y_j}}\).

\paragraph{Relation to Cartan charges and Dynkin labels}
One can now understand how the powers \(\lL\) and \(\lM\) are related to
the \(SO(6)\times SO(2,4)\) Cartan charges \(\{J_1,J_2,J_3|\Delta,S_1,S_2\}\) or, equivalently,  to the corresponding \(SU(4)\times SU(2,2)\) weights
\begin{align}\lambda _1&= \frac{J_1+J_2-J_3}{2},&  \lambda _2&= \frac{J_1-J_2+J_3}{2},& \lambda _3&= \frac{-J_1+J_2+J_3}{2},&  \lambda _4&= \frac{-J_1-J_2-J_3}{2},\nonumber\\
\nu _1&= \frac{-\Delta +S_1+S_2}{2},&  \nu _2&= \frac{-\Delta -S_1-S_2}{2} ,& \nu _3&= \frac{\Delta +S_1-S_2}{2},&  \nu_4&= \frac{\Delta -S_1+S_2}{2}\,,\label{weightstocharges}\end{align}
written for the
Kac-Dynkin-Vogan diagram
\begin{eqnarray}
\{\fQ_{\emptyset|1},\fQ_{1|1},\fQ_{12|1},\fQ_{12|12},\fQ_{12|123},\fQ_{123|123},\fQ_{1234|123}\}
   \label{QABA}
\end{eqnarray}
which corresponds to the \(sl_2\) ABA diagram. We associate to this diagram the ordering
\begin{align}
\label{eq:211}  \hat 1 \prec 1 \prec 2 \prec \hat 2 \prec \hat 3 \prec 3 \prec 4 \prec \hat 4\,,
\end{align}
where the ``hat'' symbol is used to recognize fermionic indices -- denoted by the letters \(i\), \(j\), etc, as opposed to the ``bosonic'' indices denoted by the letters \(a\), \(b\), etc. The ordering \eqref{eq:211} simply corresponds to the order in which the indices are added to the Q-functions in \eqref{QABA}.

As explained in \appref{sec:Labels}, the asymptotics of the Q-functions along this diagram are given by (omitting the constant prefactor)
\begin{gather}
\label{eq:224}  \fQ_{\emptyset|1}\sim \y_1^{-\bi u} u^{-\nu_1}\,, \hspace{1cm} \fQ_{1|1}\sim \left(\tfrac{\x_1}{\y_1}\right)^{\bi u} u^{-\nu_1-\lambda_1} \,,\hspace{1cm} \fQ_{12|1}\sim\left(\tfrac{\x_1\x_2}{\y_1}\right)^{\bi u} u^{-\nu_1-\lambda_1-\lambda_2}
\nonumber \,,\\\fQ_{12|12}\sim\left(\tfrac{\x_1\x_2}{\y_1\y_2}\right)^{\bi u} u^{-\nu_1-\nu_2-\lambda_1-\lambda_2}\sim \left(\tfrac{\y_3\y_4}{\x_3\x_4}\right)^{\bi u} u^{\nu_3+\nu_4+\lambda_3+\lambda_4}\,,,\\
\fQ_{12|123}\sim\left(\tfrac{\y_4}{\x_3\x_4}\right)^{\bi u} u^{\nu_4+\lambda_3+\lambda_4}
\,, \hspace{1cm}
\fQ_{123|123}\sim\left(\tfrac{\y_4}{\x_4}\right)^{\bi u} u^{\nu_4+\lambda_4}
\,, \hspace{1cm}
\fQ_{1234|123}\sim \y_4^{\bi u} u^{\nu_4}\,.
\nonumber
\end{gather}
By comparison with \eqref{eq:214}, we get\footnote{We notice that, as a
  consequence of \(\sum_a \lambda_a=0=\sum_i \nu_i\), the condition
  \eqref{eq:216} is satisfied by the expression \eqref{eq:221}.}
\begin{align}
\label{eq:221}  \lL&=\lambda_a -\sum_{b\prec a} \delta_{x_ax_b}+\sum_{i\prec a}\delta_{x_ay_i}\,,& \lM&=\nu_i
-\sum_{j\prec i}\delta_{y_iy_j}+\sum_{a\prec i} \delta_{x_ay_i}.
\end{align}
Inserting these expressions into the equations \eqref{eq:220} and \eqref{eq:203} we obtain
\begin{subequations}\label{eq:218}
  \begin{equation}
  \begin{aligned}
\label{eq:219}
    \bP_a&\simeq\mathsf{A}_a \x_a^{\bi u}
    u^{-\lambda_a+\sum_{b\prec a} \delta_{\x_a\x_b} -\sum_{i\prec a}\delta_{\x_a\y_i}}\,,&\hspace{.5cm}
    \bP^a&\simeq\mathsf{A}^a \x_a^{-\bi u}
    u^{\lambda_a+\sum_{b\succ a} \delta_{\x_a\x_b} -\sum_{i\succ a}\delta_{\x_a\y_i}}\\
    \bQ_i&\simeq\mathsf{B}_i \y_i^{-\bi u}
    u^{-\nu_i-\sum_{a\prec i}\delta_{\x_a\y_i}+\sum_{j\prec i} \delta_{\y_i\y_j}}\,,&
    \bQ^i&\simeq\mathsf{B}^i \y_i^{\bi u}
    u^{\nu_i-\sum_{a\succ i}\delta_{\x_a\y_i}+\sum_{j\succ i} \delta_{\y_i\y_j}}\,,
  \end{aligned}
\end{equation}
where
\begin{align}\label{AABBcoef}
    \mathsf{A}_a\mathsf{A}^a&=\frac 1 {\x_a} \frac{\prod_{1\le i \le 4}z_{a,i}}{\prod_{b\ne
    a}z_{b,a}}\,,&
  \mathsf{B}_i\mathsf{B}^i&=\frac 1 {\y_i} \frac{\prod_{1\le a \le 4}z_{a,i}}{\prod_{j\ne
    i}z_{j,i}}\,,
\end{align}
with
\begin{equation}
  \begin{aligned}
    z_{ab}=-z_{ba}&=
  \begin{cases}\x_b-\x_a&\textrm{ if }\x_a\neq \x_b\\
    \bi \x_a(\lambda_b-\lambda_a -\sum_{a\prec c\prec b}\delta_{\x_c\x_a}+\sum_{a\prec i\prec b}\delta_{\x_a\y_i}-1)&\textrm{ if }\x_a=\x_b\textrm{ and }a\prec b\,,
  \end{cases}\\
z_{ij}=-z_{ji}&=
  \begin{cases}\y_i-\y_j&\textrm{ if }\y_i\neq \y_j\\
\bi \y_i(\nu_j-\nu_i-\sum_{i\prec k\prec j}\delta_{\y_i\y_k}+\sum_{i\prec a\prec j} \delta_{\x_a\y_i}-1
)&\textrm{ if }\y_i= \y_j\textrm{ and }i\prec j\,,
\end{cases}\\
z_{ai}=-z_{ia}&=
  \begin{cases}\y_i-\x_a&\textrm{ if }\x_a\neq \y_i\\
\bi \x_a (-\lambda_a-\nu_i-\sum_{a\prec b\prec i}\delta_{\x_a\x_b}+\sum_{a\prec j\prec i}\delta_{\x_a\y_j})&\textrm{ if }\x_a= \y_i\textrm{ and }a\prec i\,,\\
\bi \x_a (-\lambda_a-\nu_i+\sum_{i\prec b\prec a}\delta_{\x_a\x_b}-\sum_{i\prec j\prec a}\delta_{\x_a\y_j})&\textrm{ if }\x_a= \y_i\textrm{ and }i\prec a\,.\\
  \end{cases}
\end{aligned}
\end{equation}
\end{subequations}

For future sections, the asymptotics (\ref{eq:218}) will be summarized as 
  \begin{equation}\label{eq:143}
  \begin{aligned}
    \bP_a&\simeq\mathsf{A}_a \x_a^{\bi u}
    u^{-\lL}\,,&\hspace{.5cm}
    \bP^a&\simeq\mathsf{A}^a \x_a^{-\bi u}
    u^{\uL}\\
    \bQ_i&\simeq\mathsf{B}_i \y_i^{-\bi u}
    u^{-\lM}\,,&
    \bQ^i&\simeq\mathsf{B}^i \y_i^{\bi u}
    u^{\uM}\,,
  \end{aligned}
\end{equation}
where \(\lL\) and \(\lM\) are given by (\ref{eq:221}), whereas \(\uL\) and \(\uM\) are given by
\begin{align}
  \label{eq:145}
    \uL&=\lambda_a+\sum_{b\succ a} \delta_{\x_a\x_b} -\sum_{i\succ a}\delta_{\x_a\y_i}&
    \uM&=\nu_i-\sum_{a\succ i}\delta_{\x_a\y_i}+\sum_{j\succ i} \delta_{\y_i\y_j}\,.
  \end{align}

This terminates the description of the calculation of asymptotics of Q-functions, including the constant factors, powers in terms of Cartan charges and exponential factors defined by  twists.  Let us consider now some particular cases of the full or partial untwisting.

\subsection{Particular cases of twisting}

In this subsection we will consider some simplest and/or physically most interesting cases of full and partial twisting and give the results for the leading asymptotics of the most important Q-functions.

As particular important examples, we  give the results for the fully twisted case, as well as for the \(\beta\)- and \(\gamma\)-deformations. The latter case will be  used and tested in the next section for the computation of energy of the BMN vacuum in the weak coupling appropximation. The asymptotics of some other cases of twisting can be found in \appref{app:TwistedAss}. In addition, \appref{sec:comp-impl} provides a computer implementation of the formulae of \secref{sec:turning-twist} which can be used in particular to obtain the formulae of the present subsection.

\subsubsection{Leading asymptotics for fully twisted case}
\label{sec:lead-asympt-fully}
As we already mentioned, the supersymmetry in this case is completely broken leaving only a bosonic \(\mathsf{U}(1)^5\times\mathbb{R}\) subgroup of the full \(\mathsf{PSU}(2,2|4)\).

As was mentioned in \eqref{twistedPQ} in this case we have the asymptotics  \eqref{eq:143} with the powers given by
\begin{align}
    \lL&=\lambda_a,&\lM&=\nu_i,&
    \uL&=\lambda_a,&\uM&=\nu_i,&&& a,i=1,2,3,4.
  \end{align}
  We can also express it through charges by the use of \eqref{weightstocharges}.

For  the 8 products  \(\mathsf{A}_a\mathsf{A}^a\) and  \(\mathsf{B}_j\mathsf{B}^j\) of the asymptotic factors we obtain from the general formula \eqref{AABBcoef}:
\begin{align}
\forall a\textrm{, }& \mathsf{A}_a\mathsf{A}^a=\frac{\displaystyle \prod_{j}(\x_a-\y_j)}{\displaystyle \x_a\prod_{b\ne
    a}(\x_a-\x_b)}\,,&
\forall j\textrm{, }& \mathsf{B}_j\mathsf{B}^j=\frac{\displaystyle \prod_{a}(\x_a-\y_j)}{\displaystyle \y_j\prod_{i\ne j}(\y_i-\y_j)}\,.
\label{AABB9}\end{align}

 Notice also that from the general formula \eqref{eq:214} we get
   \begin{align}\label{dQdP}
\fQ_{a|j}&\simeq \frac{ \mathsf{A}_a\,\mathsf{B}_j}{\sqrt{\y_j/\x_a}-\sqrt{\x_a/\y_j}}u^{-\lL-\lM[j]}\,
\left(\frac{\x_a}{\y_j}\right)^{\bi u}\,,
\intertext{and}
\label{Qabjk}
Q_{ab|jk}&\simeq \frac{\mathsf{A}_a\,\mathsf{A}_b\,\mathsf{B}_j\mathsf{B}_k\left(\sqrt{\frac{\x_a}{\x_b}}-\sqrt{\frac{\x_b}{\x_a}}\right)\left(\sqrt{\frac{\y_k}{\y_j}}-\sqrt{\frac{\y_j}{\y_k}}\right) u^{-\lL-\lL[b]-\lM[j]-\lM[k]}}
{\left(\sqrt{\frac{\y_j}{\x_a}}-\sqrt{\frac{\x_a}{\x_j}}\right)
\left(\sqrt{\frac{\y_j}{\x_b}}-\sqrt{\frac{\x_b}{\x_j}}\right)
\left(\sqrt{\frac{\y_k}{\x_a}}-\sqrt{\frac{\x_a}{\x_k}}\right)
\left(\sqrt{\frac{\y_k}{\x_b}}-\sqrt{\frac{\x_b}{\x_k}}\right)}
\left(\frac{x_ax_b}{y_jy_k}\right)^{\bi u}
\,,
\end{align}
where the coefficients \(\mathsf{A}_a\) and \(\mathsf{B}_j\) are given by (\ref{AABB9}).

In view of equation \eqref{muQomega}, this allows to control the asymptotics of \(\hat\mu_{ab}\): indeed \(\hat\mu_{ab}\) is the linear combination of \(Q_{ab|jk}\) with  coefficients \(\hat\omega^{jk}\), and these coefficients are \(\bi\)-periodic with constant asymptotics at large \(u\), hence they are constant.

\subsubsection{\texorpdfstring{$\gamma$}{gamma}-deformation}
\label{sec:gamma-deformation}  If we denote all three angles of \(S^5\) corresponding to the generators \(\{J_1,J_2,J_3\}\) by \(\{e^{\bi\Phi_1}= \x_1\x_2,\,e^{\bi\Phi_2}= \x_1\x_3,\,e^{\bi\Phi_3}= \x_2\x_3\}\) the \(\gamma\)-{\it deformation} is given by  the following choice of twists\footnote{our \(\gamma_j\) as well as \(\beta\) of the next subsection, coincide with  \cite{Fokken2014b,Fokken2014a} but they are \(2\pi\)   times bigger than those of   \cite{Tongeren2013}.} 
 \begin{eqnarray}\label{S5-twist}e^{\bi\Phi_a}=e^{\bi\,\epsilon_{abc}\gamma_b J_c}\,,\qquad a=1,2,3\,,    \end{eqnarray} or   
\begin{align}\label{gamma-twist} 
\x_1&=\ e^{\frac{\bi}{2}  \left(\left(\gamma _2-\gamma _1\right) J_3-\left(\gamma _1+\gamma _3\right) J_2+\left(\gamma _2+\gamma _3\right) J_1\right)},&
\x_2&= e^{\frac{\bi}{2} \left(\left(\gamma _1+\gamma _2\right) J_3-\left(\gamma _2+\gamma _3\right) J_1+\left(\gamma _1-\gamma _3\right) J_2\right)},\notag\\
\x_3&= e^{\frac{\bi}{2}  \left(-\left(\gamma _1+\gamma _2\right) J_3+\left(\gamma _1+\gamma _3\right) J_2+\left(\gamma _3-\gamma _2\right) J_1\right)},&
\x_4&= e^{\frac{\bi}{2} \left(\left(\gamma _3-\gamma _1\right) J_2+\left(\gamma _1-\gamma _2\right) J_3+\left(\gamma _2-\gamma _3\right) J_1\right)}\,.
\end{align}
   with real \(\gamma\)'s and \(\y_1=\y_2=\y_3=\y_4=1\). This means a choice of the  \(S^5\) twists which obeys the two conditions
\begin{equation}
  \label{eq:77}
  \begin{cases}
    \x_1\x_2\x_3\x_4=1\,,\\
    \left(\x_{1} \x_{2}\right)^{J_{1}} \left(\x_{1} \x_{3}\right)^{J_{2}}
\left(\x_{2} \x_{3}\right)^{J_{3}}=1\,.
  \end{cases}
\end{equation}

Again, using the general formulae\footnote{To use the equations of \secref{sec:turning-twist}, one should note that if the charges \(J_a\) are non-zero then for generic \(\gamma\), one has \(\forall a\neq b,~ \x_a\neq\x_b\).} eqs.\eqref{eq:220}-\eqref{eq:221} we get for it the asymptotics  \eqref{eq:143} with the powers given by
\begin{align}\label{eq:90}
  \lL&=\lambda_a\,,&
  \uL&=\lambda_a\,,&
  \lM&=\nu_i+1-i\,,&
  \uM&=\nu_i+4-i\,.&
\end{align}
The existence of shifts in some \(\lL\),  \(\lM\), \(\uL\) and/or \(\uM\) arises as soon as several eigenvalues are equal. It comes from \(\delta\) symbols in \eqref{eq:221} and \eqref{eq:145} and can be seen as originating from the shift  \(+1\) in the r.h.s. of \eqref{eq:191}.

For  the 8 products  \(\mathsf{A}_a\mathsf{A}^a\) and  \(\mathsf{B}_j\mathsf{B}^j\) of the asymptotic factors in \(\gamma\) deformed theory we obtain from the general formula \eqref{AABBcoef}:
\begin{align}\label{eq:217}
  \mathsf{A}_a\mathsf{A}^a&=\frac{\left(\x_a-1\right)^4}{\x_a\prod_{b\neq a}\left(\x_a-\x_b\right)}\,,&
  \mathsf{B}_i\mathsf{B}^i&=\bi\frac{\prod_a(\x_a-1)}{\prod_{j\neq i}\left(\lM-\lM[j]\right)}\,.
\end{align}

The supersymmetry is completely broken  for the generic \(\gamma\)'s.
 In \secref{sec:BMNvac} we will use these results for the study of a particular state -- the \(\gamma\)-deformed BMN vacuum -- and calculate its energy in the weak coupling approximation.
\subsubsection{\texorpdfstring{$\beta$}{beta}-deformation}
\label{sec:beta-deformation}
The \(\beta\)-{\it deformation} is a particular case of the \(\gamma\)-deformation, with all three \(\gamma\)-twists equal  \(\gamma_1=\gamma_2 =\gamma_3=\beta\), or 
\begin{align}\label{beta-twist} \x_1&= e^{\bi \beta  (J_1- J_2)},& \x_2&=  e^{\bi\beta(J_3- J_1)},& \x_3&= e^{\bi\beta(J_2- J_3)},&  \x_4&= 1\,.
\end{align}
  Another  possible choice of twists,  corresponding to a coset background,  is obtained by  changing the sign of all \(\x_a\)'s.  This coset corresponds to Lunin-Maldacena background~\cite{Lunin:2005jy} and it is dual to a particular case of Leigh-Strassler \(\mathcal{N}=1\) deformation of \(\mathcal{N}=4\) SYM~\cite{Leigh1995}. 

 The asymptotics of Q-functions can be again obtained using the general formulae of \secref{sec:turning-twist}: in particular the asymptotics of single-indexed \(\bP\) and \(\bQ\)-functions is given by \eqref{eq:219}, where\footnote{For instance, the relation \(\lL-\lambda_a=(0,0,0,3)\) means \(\lL[1]=\lambda_1\), \(\dots\), \(\lL[3]=\lambda_3\), \(\lL[4]=\lambda_4+3\).}
 \begin{align}
\lL-\lambda_a&=(0,0,0,3)\,,&\lM-\nu_i&=(0,-1,-2,-2)\,,\nonumber\\
\uL-\lambda_a&=(0,0,0,-1)\,,&\uM-\nu_i&=(2,1,0,0)\,,\\
\mathsf{A}_a\mathsf{A}^a&=
\begin{cases}
  \frac{(\x_a-1)^3}{\x_a\prod_{\substack{b\le3\\b\ne a}}(\x_a-\x_b)}&\textrm{ if }a\le 3\\
\frac{\prod_{i}(\uL[4]+\lM)}{\prod_{b\ne4}(1-\x_b)} &\textrm{ if }a=4
\end{cases}\,,
&\mathsf{B}_i\mathsf{B}^i&=\frac{(\uL[4]+\lM)\prod_{a\le3}(1-\x_a)}{\prod_{j\ne i}(\lM-\lM[j])}\,.
 \end{align}

The residual supersymmetry is \(\mathcal{N}=1\) and the full symmetry of the coset is \(\mathsf{U}(1)\times \mathsf{U}(1)\times \mathsf{PSU}(2,2|2)\).

\section{BMN vacuum in gamma-deformed case, weak coupling expansion}
\label{sec:BMNvac}

In this section, we will study by means of the  twisted quantum
spectral curve   a particular, simplest possible operator -- BMN
vacuum \({\rm Tr}\, Z^L\) in the gamma-deformed theory. Supersymmetry
is fully broken in the presence of the gamma-deformation, and the
conformal dimension of the BMN vacuum is no longer protected.  At the
same time, one does not need Bethe roots to describe this state since
the whole contribution to its dimension comes entirely from wrapping
effects. Hence this is probably the  simplest example of twisted
object to perform computation with. Due to its simplicity, the
dimension of this operator was computed perturbatively, directly from
the SYM, to the leading single-wrapping orders by QFT methods
\cite{Fokken2014a}, confirming integrability-based predictions of \cite{deLeeuw:2011rw}.

We will show how to compute the conformal dimension of the BMN vacuum at weak coupling at the single-wrapping order using the twisted QSC. The result is already known in the literature, even the double-wrapping orders have been computed \cite{Ahn:2011xq} using L\"uscher-type approach. We do not aim so far to improve these results, rather we initiate a computation to demonstrate how the twisted QSC works and hope that it will be boosted in future to an efficient computation up to very high orders, similarly to as it already happened in non-twisted case \cite{Marboe:2014gma}. In the process of  our computation, we pave a new, more transparent way, compared to \cite{Gromov:2014caa},  of deriving the asymptotic Bethe Ansatz approximation to QSC solution and make first steps towards deriving L\"uscher-type formulae directly from the QSC.
 \subsection{Input data, notations, and symmetries.}
The BMN vacuum is characterized by the following set of charges
\be
J_1 =L\,,\  J_2=J_3=S_1=S_2=0\,,\  \Delta(g,\gamma_{\pm})\leq L\,,
\ee
where equality \(\Delta=L\) is reached  at \(g=0\) or  \(\gamma_+=0\) or \(\gamma_-=0\). Correspondingly, the weights \eqref{eq:1001} are given by
\begin{align}
\lambda_1&=\lambda_2=\frac L2\,, &   \lambda_3&=\lambda_4=-\frac L2\,,&
\nu_1 &=\nu_2=-\frac{\Delta}{2}\,, & \nu_3=\nu_4&=\frac \Delta 2\,,
\end{align}

For this choice of charges, only \(\gamma_2\) and \(\gamma_3\) out of three parameters of \(\gamma\)-deformation are relevant, c.f. \eqref{S5-twist}, which enter in combinations \(\gamma_{\pm}\equiv \frac 12(\gamma_3\pm\gamma_2)L\). Consequently, the twists are  identified as follows
\begin{align}\label{eq:229}
q\equiv \x_1 &=e^{\bi \gamma_+}\,,  & q^{-1}=\x_2=e^{-\bi\gamma_+}\,,
\notag\\
\dot q\equiv \x_3 &=e^{\bi\gamma_-}\,, &\dot q^{-1}=\x_4=e^{-\bi\gamma_-}\,.
\end{align}
and \(\y_i=1.\) We will also use the notation \(\x_{ab}\equiv \x_a\,\x_b\,\). In particular,  \(\x_{12}=\x_{34}=1\).

The large-\(u\) asymptotics of Q-functions are deduced, following the analysis of \secref{sec:gamma-deformation}, to be
\begin{subequations}
\label{eq:asbPbQ}
\begin{align}
\bP_a&\simeq \mathsf{A}_a\,\x_a^{\bi u}\,u^{-\lambda_a}\,,
&
\bP^a&\simeq \mathsf{A}^a\,\x_a^{-\bi u}\,u^{+\lambda_a}\,,
\\
\label{eq:asbQ}
\bQ_i&\simeq \mathsf{B}_i\,u^{-\hat\nu_i}\,,
&
\bQ^i&\simeq \mathsf{B}^i\,u^{+\hat\nu_i+3}\,,
\end{align}
\end{subequations}
with
\begin{align}\label{eq:normAABB}
\mathsf{A}_a\,\mathsf{A}^a&=\frac 1{\x_a}\frac{(1-\x_a)^4}{\prod\limits_{b\neq a}(\x_a-\x_b)}\,,
&
 \mathsf{B}_i\mathsf{B}^i&=- \bi\frac{\prod\limits_{a=1}^4(1-\x_a)}{\prod\limits_{j\neq i}(\hat\nu_j-\hat\nu_i)}=
 \begin{cases}
   (-1)^i\frac{-\bi \prod_{a=1}^4(1-\x_a)}{(\Delta-2)(\Delta-3)}&\textrm{ if }i\in\{1,4\}
\\   (-1)^i\frac{-\bi \prod_{a=1}^4(1-\x_a)}{(\Delta-1)(\Delta-2)}&\textrm{ if }i\in\{2,3\}\,,
 \end{cases}
\end{align}
and \(\lM=\nu_i+1-i\).

One can note that if \(L=3\) then \(B_1B^1\) and \(B_4B^4\) develop a pole as \(g\to 0\). Also note that at \(L=2\) all the four products \(B_iB^i\) develop such a pole. For \(L=2\) this singular behaviour persists on the final formula for energy and it corresponds to rearrangements in comparative large-\(u\) magnitudes of \(\bQ_i\) given by \eqref{eq:asbQ}.  Hence this  case should be treated separately. As for the case \(L=3\), we will see from the result at the end of the section that the formula for energy predicts the non-singular correct value. Hence the pole \(L=3\) in the formula is probably not physical.  In what follows, we stick only to the  regular case  \(L\geq 4\), but see the comments and references at the very end of this section.
From (\ref{eq:214}) one can also deduce that \(Q_{ab|ij}\sim \x_{ab}^{\bi\,u}\,u^{-\lL[a]-\lL[b]-\lM-\lM[j]-1}\) where \(\lM[3]>\lM[4]>\lM[1]>\lM[2]\). Hence, as \(\omega^{ij}\sim 1\), the term with \(ij=12\) dominates in \(\mu_{ab}=\frac 12 Q_{ab|ij}^-\,\omega^{ij}\) (see \eqref{muQomega}). Therefore, we  obtain
\begin{align}\label{eq:muscaling}
\mu_{12}&\sim u^{\Delta-L}\,,
&
\mu_{34}&\sim u^{\Delta+L}\,,
&
\mu_{\alpha\dot\alpha}&\sim u^{\Delta}\,\x_{\alpha\dot\alpha}^{\bi\,u}\,,
\end{align}
where
\be
\alpha\in\{1,2\}\ \ \ {\rm and}\ \ \  \dot\alpha\in\{3,4\}.
\ee
We now introduce normalised variables \(\bp\) and \(\tm\) suitable for further analysis.  
\begin{align}\label{bPmu}
\bP_a &= \frac{\mathsf{A}_a\, \x_a^{\bi u}}{(g\,x)^{L/2}}\bp_a\,,
&
\bP^a &= \frac{\mathsf{A}^a\, \x_a^{-\bi u}}{(g\,x)^{L/2}}\bp^a\,,
&
\mu_{ab}&=\frac{{\x_{ab}^{\bi u}\,\mathsf{A}_a\, \mathsf{A}_{b}}}{g^L}\,\tm_{ab}\,.
\end{align}
Zhukovsky variable \(x\) is defined, as usually, by the relation \(\frac{u}{g}=x+\frac{1}{x}\). We always  consider it, as well as any other functions in this section, as a function in physical kinematics (with short cuts): \(x(u)=\frac{u}{2g}\left(1+\sqrt{1-\frac{4g^2}{u^2}}\right)\). For the purpose of weak coupling expansion, one should remember that \(x= \frac ug-\frac gu+\ldots\) at either small \(g\) or large \(u\), so that the expansion goes in powers of \(\frac{g^2}{u^2}\).

Note in particular that the large-\(u\) behaviour of \(\bp\) is given by
\begin{align}\label{eq:231}
\bp_\alpha &\simeq 1\simeq \bp^{\dot\alpha}\,,&
\bp_{\dot\alpha}&\simeq 1\cdot u^L\simeq \bp^{\alpha}\,.
\end{align}
We also denote by \(M_{ab}\) the prefactors in the large-\(u\) asymptotic of \(\tm\) (these prefactors will be explicitly determined later):
\begin{align}\label{eq:230}
  \tm_{12}&\simeq M_{12}\cdot u^{\Delta-L}\,,& 
  \tm_{34}&\simeq M_{34}\cdot u^{\Delta+L}\,,& 
  \tm_{\alpha\dot\alpha}&\simeq M_{\alpha\dot\alpha}\cdot u^{\Delta}\,.
\end{align}

The bosonic H-symmetry of QSC \cite{Gromov:2014caa} is mostly destroyed by the introduction of twists, only the diagonal rescaling remains:
\begin{align}\label{eq:228}
\bP_a&\to \alpha_a\, \bP_a& \bP^a&\to  \frac 1{\alpha_a}\, \bP^a\,,
\end{align}
provided that  \(\alpha_1\alpha_2\alpha_3\alpha_4=1\) to preserve the \(\Pf(\mu)=1\) property. Hence the values of \(\mathsf{A}_a\) and \(\mathsf{A}^a\) are not fixed universally, only the product
\be
\PR \equiv \textsf{A}_1\textsf{A}_2\textsf{A}_3\textsf{A}_4
\ee
is fixed, and we will eventually determine it explicitly. However, the normalised quantities \(\bp\) and \(\tm\) do not depend on  rescalings of \(\mathsf{A}\)'s.

The input data is highly symmetric, with the consequence that   the following transformations map a solution to itself, up to an appropriate rescaling (\ref{eq:228}):
\begin{itemize}
\item Exchange \(1\leftrightarrow 2\):
\begin{align}
q &\to 1/q\,,\ \dot q\to \dot q\,,
\\
\nonumber
&&\bP_1&\to\bP_2\,,&\bP_2&\to-\bP_1\,,& \bP^1&\to\bP^2\,,&\bP^2&\to-\bP^1\,,&&\hphantom{\hspace{5em}}
\\
\nonumber
&& \hspace{-8em}\text{other \(\bP\) unchanged,}\hspace{-7.5em}
\\ 
\nonumber
&&\mu_{12}&\to\mu_{12}\,,& \mu_{1\dot\alpha}&\to \mu_{2\dot\alpha}\,,&\ \mu_{2\dot\alpha}&\to-\mu_{2\dot\alpha}\,,&\mu_{34}&\to\mu_{34}\,;
\end{align}
\item Exchange \(3\leftrightarrow 4\):
\begin{align}
q&\to q\,,\  \dot q\to 1/{\dot q}\,,
\\
\nonumber
&&\bP_3&\to\bP_4\,,&\bP_4&\to-\bP_3\,,& \bP^3&\to\bP^4\,,&\bP^4&\to-\bP^3\,,&&\hphantom{\hspace{5em}}
\\
\nonumber
&& \hspace{-8em}\text{other \(\bP\) unchanged,}\hspace{-7em}
\\ 
\nonumber
&&\mu_{12}&\to\mu_{12}\,,& \mu_{\alpha3}&\to \mu_{\alpha4}\,,&\ \mu_{\alpha4}&\to-\mu_{\alpha3}\,,&\mu_{34}&\to\mu_{34}\,;
\end{align}
\item Exchange \(\{1,2\}\leftrightarrow\{4,3\}\) (analog of LR-symmetry in \cite{Gromov:2014caa}):
\begin{align}
\label{eq:LR}
q &\leftrightarrow \dot q\,,
\\
\nonumber
&&\bP_1&\leftrightarrow+\bP^4\,,& \bP_2&\leftrightarrow-\bP^3\,,& \bP_3&\leftrightarrow+\bP^2\,,&\bP_4&\leftrightarrow-\bP^1&&\hphantom{\hspace{5em}}\,,
\\
\nonumber
&&\bP^1&\leftrightarrow-\bP_4\,,& \bP^2&\leftrightarrow+\bP_3\,,& \bP^3&\leftrightarrow-\bP_2\,,&\bP^4&\leftrightarrow+\bP_1\,,
\\
\nonumber
&&\mu_{14}&\leftrightarrow\mu_{23}\,,&&\hspace{-2em}\text{other \(\mu\) unchanged.}\hspace{-5em}
\end{align}
\end{itemize}
For instance, the answer for the conformal dimension should be invariant under replacements \(q\leftrightarrow 1/q\), \(\dot q\leftrightarrow 1/\dot q\), and \(q\leftrightarrow \dot q\). 

In the computations of this section, we will also routinely use the following properties, which are consequences of (\ref{eq:normAABB}) and (\ref{eq:229}):
{\setlength{\belowdisplayskip}{0pt} \setlength{\belowdisplayshortskip}{0pt}\begin{align}
\mathsf{A}_1\mathsf{A}^1&=-\mathsf{A}_2\mathsf{A}^2\,,&\mathsf{A}_3\mathsf{A}^3&=-\mathsf{A}_4\mathsf{A}^4\,,
\end{align}\setlength{\abovedisplayskip}{0pt} \setlength{\abovedisplayshortskip}{0pt}\begin{align}
\mathsf{A}_3\mathsf{A}^3\frac{1+\dot q}{1-\dot q}+\mathsf{A}_1\mathsf{A}^1\frac{1+q}{1-q}+1&=0\,,
\end{align}}{\setlength{\abovedisplayskip}{0pt} \setlength{\abovedisplayshortskip}{0pt}
\begin{align}
\left[\frac 1{\x_{\alpha3}-1}-\frac 1{\x_{\alpha4}-1}\right] &=(\mathsf{A}^3\mathsf{A}_3)\frac{(\dot q+1)^2}{(\dot q-1)^2}\,,
&
\left[\frac 1{\x_{1\dot\alpha}-1}-\frac 1{\x_{2\dot\alpha}-1}\right] &= (\mathsf{A}^1\mathsf{A}_1)\frac{(q+1)^2}{(q-1)^2}\,.
\end{align}}
Finally, we introduce a handy notation to work with indices: \(\alpha,\beta,\ldots \in\{1,2\}\), \(\dot\alpha,\dot\beta,\ldots\in \{3,4\}\). Setting  the normalisation  of Levi-Civita symbols as \(\epsilon_{12}=\epsilon_{34}=+1\,, \) \(\epsilon^{12}=\epsilon^{34}=+1\,,\) one defines
\begin{align}
(\epsilon\bP)_{\alpha} &\equiv \epsilon_{\alpha\beta}\bP^\beta\,, &  (\epsilon\bP)_{\dot\alpha}&\equiv\epsilon_{\dot\alpha\dot\beta}\bP^{\dot\beta}\,,
\\
(\epsilon\bP)^{\alpha} &\equiv \epsilon^{\alpha\beta}\bP_\beta\,, &  (\epsilon\bP)^{\dot\alpha}&\equiv\epsilon^{\dot\alpha\dot\beta}\bP_{\dot\beta}\,,
\end{align}
and we use this convention for other functions with the same index structure. One property which uses this notation is
\begin{align}
\frac{(\epsilon\,\mathsf{A})_{\alpha}}{\mathsf{A}_{\alpha}}&=-\frac{\mathsf{A}_1\mathsf{A}^1}{\mathsf{A}_1\mathsf{A}_2}\,,&
\frac{(\epsilon\,\mathsf{A})_{\dot\alpha}}{\mathsf{A}_{\dot\alpha}}&=-\frac{\mathsf{A}_3\mathsf{A}^3}{\mathsf{A}_3\mathsf{A}_4}\,,
\end{align}
note that the r.h.s. does not depend on \(\alpha\) or \(\dot\alpha\).

\subsection{Asymptotic \texorpdfstring{$\bP\mu$}{P-mu}-system}
We will use the following terminology: ``pre-wrapping'' orders signify a
collection of  perturbative corrections in
\(g^2\) 
from \(g^2\) to \(g^{2L-2}\) with respect to the leading order approximation\footnote{In the literature, some authors also call pre-wrapping order the order \(g^{2L-2}\), i.e. the last of the pre-wrapping orders.}.
Similarly,
 ``single-wrapping'' orders
means a collection of orders from \(g^{2L}\) to \(g^{4L-2}\), while ``\(n\)-wrapping'' orders means all orders
 from \(g^{2nL}\) to \(g^{2(n+1)L-2}\). 

In this subsection we will find the explicit solution in all single-wrapping orders. One should note that at any perturbative order in \(g\), the Zhukovsky cuts at \([-2\,g\,,\,2\,g]+\bi\,\mathbb Z\) degenerate into isolated poles at \(u=\bi\,\mathbb{Z}\). The success of the perturbative expansion relies on our ability to control the functions at these poles.

\paragraph{Leading order.} We start by identifying the value of \(\bp\)'s and \(\tm\)'s at the leading order of the perturbative expansion.

\ \\
First, we note that all \(\tm_{ab}\) should be polynomials at the leading order. The proof of this property was given in section 3.2.1. of \cite{Marboe:2014gma}. We repeat it here because we will recursively apply it at higher perturbative orders: Represent \(\mu(u)\) and \(\mu(u+\bi)\) as follows
\begin{subequations}
\label{eq:singmuu}
\be
\mu(u)&=&\frac 12\left(\mu+\mu^{[2]}\right)+\sqrt{u^2-4g^2}\left(\frac{\mu-\mu^{[2]}}{\sqrt{u^2-4g^2}}\right)\,,
\\
\mu(u+\bi)&=&\frac 12\left(\mu+\mu^{[2]}\right)-\sqrt{u^2-4g^2}\left(\frac{\mu-\mu^{[2]}}{\sqrt{u^2-4g^2}}\right)\,.
\ee
\end{subequations}
On the r.h.s. of (\ref{eq:singmuu}), all combinations in brackets  are regular at \(u=0\) at any order of perturbative expansion. Indeed, they do not have branch points on the real axis at finite coupling, e.g. \(\mu+\mu^{[2]}=\mu+\tilde\mu\), so the singularities at \(u=0\) simply cannot develop. Hence \(\mu\) is regular at \(u=0,\bi\) at the leading order, and the singularity at subleading orders  can arise only from expansion of \(\sqrt{u^2-4g^2}\) in front of the second bracket.

Now we use the relation \(\mu^{[2]}_{ab}=\mu_{ab}+\bP^c\bP_b\,\mu_{ac}-\bP^c\bP_a\,\mu_{bc}\) and regularity of \(\bP\)'s outside the real axis to recursively prove that \(\mu\) has no poles at \(u=\bi\,\mathbb{Z}_{>0}\) provided that \(\mu\) is regular at \(u=\bi\). Similarly, we use the relation \(\mu_{ab}=\mu_{ab}^{[2]}-\bP^c\bP_b\,\mu_{ac}^{[2]}+\bP^c\bP_a\,\mu_{bc}^{[2]}\) to recursively deduce regularity of \(\mu\) in the lower half-plane from its regularity at \(u=0\). Note that all 6 \(\mu_{ab}\) should be regular at \(u=0\) and \(u=\bi\) simultaneously because they are intertwined in the recursive procedure.

Hence we proved that \(\mu\), and therefore \(\tm\), are entire functions at the leading order. Hence \(\tm\)'s should be polynomials as they have power-like asymptotics.

 Since \(\tm_{12}\sim u^{\Delta-L}\) and \(|\Delta-L|<1\) at small \(g\), \(\tm_{12}\) is forced to be simply a constant at all perturbative orders in which it is still a polynomial. Hence, at the leading order we have for sure \(\tm_{12}=M_{12}\)  where the constant \(M_{12}\) was defined in (\ref{eq:230}).

\ \\
Second, from the relation \(\tilde\bP_a=\mu_{ab}\bP^b\) and polynomiality of \(\tm\) we conclude that \(\tilde\bP_a\) is free of singularities everywhere except probably at the origin, and this property propagates to \(\tilde\bp_a\) and similarly we have the same absence of singularities for \(\tilde\bp^a\). On the other hand, consider the expansion of \(\bp\) (where \(\bp\) without index would denote in this section any of different functions \(\bp_1,\bp_2,\bp^3\) or \(\bp^4\) ) and \(\tilde\bp\) into the convergent series \cite{Marboe:2014gma}:
\be\label{eq:bpsums}
\bp=1+\sum_{k=1}^\infty \frac{c_k}{x^k}\,,\qquad   \tilde\bp=1+\sum_{k=1}^\infty c_k\,x^k\,,
\ee
which shows, in particular, that \(\tilde\bp\) is regular at \(u=0\). Given the mentioned analytic properties of \(\tilde\bp\) and its power-like large-\(u\) behaviour, we deduce that \(\tilde\bp\) is simply a polynomial in \(u\) and hence the infinite sums \eqref{eq:bpsums} are truncated at some finite number.

\ \\
Third, consider the following exact relation \cite{Gromov:2014caa}~\footnote{
Derivation:
\begin{align*}
\tilde\bP_\alpha(\epsilon\tilde\bP)_{\dot\alpha}-\bP_\alpha(\epsilon\bP)_{\dot\alpha}&=\epsilon_{\dot\alpha\dot\beta}\mu^{\dot\beta\beta}\left(\tilde\bP_{\alpha}\bP_{\beta}-\bP_{\alpha}\tilde\bP_{\beta}\right)+\epsilon_{\dot\alpha\dot\beta}\e^{\dot\beta\dot\gamma}\mu^{34}\left(\tilde\bP_{\alpha}\bP_{\dot\gamma}-\bP_{\alpha}\tilde\bP_{\dot\gamma}\right)
\\
{}&=\epsilon_{\dot\alpha\dot\beta}\,\epsilon_{\alpha\beta}\,\mu^{\dot\beta\beta}(\mu_{12}-\tilde\mu_{12})+\mu_{12}(\mu_{\alpha\dot\alpha}-\tilde\mu_{\alpha\dot\alpha})
\\
{}&=-\mu_{\alpha\dot\alpha}(\mu_{12}-\tilde\mu_{12})+\mu_{12}(\mu_{\alpha\dot\alpha}-\tilde\mu_{\alpha\dot\alpha})=\mu_{\alpha\dot\alpha}\,\tilde\mu_{12}-\mu_{12}\,\tilde\mu_{\alpha\dot\alpha}\,,
\end{align*}
where we used \(\epsilon_{\vphantom{\dot\beta}\alpha\beta}\,\epsilon_{\dot\alpha\dot\beta}\,\mu^{\beta\dot\beta}=\mu_{\vphantom{\dot\alpha}\alpha\dot\alpha}\,.\)
}
\begin{align}\label{eq:mu12tmu}
\mu_{\alpha\dot\alpha}\,\mu_{12}^{[2]}-\mu_{12}\,\mu_{\alpha\dot\alpha}^{[2]}
&=
\tilde\bP_\alpha(\epsilon\tilde\bP)_{\dot\alpha}-\bP_\alpha(\epsilon\bP)_{\dot\alpha}\,.
\end{align}
At weak coupling, a significant simplification happens on the r.h.s.: following from the definition (\ref{bPmu}), we have 
\(
\tilde\bP_\alpha(\epsilon\tilde\bP)_{\dot\alpha}-\bP_\alpha(\epsilon\bP)_{\dot\alpha}\propto x^{+L}\tilde\bp_\alpha(\epsilon\tilde\bp)_{\dot\alpha}-x^{-L}\bp_\alpha(\epsilon\bp)_{\dot\alpha}
\). On the other hand, we see from the truncated series \eqref{eq:bpsums} that 
\(x^{-L}\bp_\alpha(\epsilon\bp)_{\dot\alpha}/(x^{+L}\tilde\bp_\alpha(\epsilon\tilde\bp)_{\dot\alpha})=\CO(g^{2L})\). Hence \(\bP_\alpha(\epsilon\bP)_{\dot\alpha}\) is suppressed compared to \(\tilde\bP_{\alpha}(\epsilon\tilde\bP)_{\dot\alpha}\) by a factor \(g^{2L}\), so that it does not contribute to the perturbative expansion of \eqref{eq:mu12tmu} until the first wrapping order! This is precisely the simplification which validates the asymptotic Bethe Ansatz approximation.

Apart from dropping the \(\bP_\alpha(\epsilon\bP)_{\dot\alpha}\) term from \eqref{eq:mu12tmu}, we also use that \(\tm_{12}\) is constant  and derive
\be\label{eq:m12tm}
{\tm_{\alpha\dot\alpha}^{\vphantom{[]}}}-\frac 1{z_{\alpha\dot\alpha}}{\tm_{\alpha\dot\alpha}^{[2]}}=-\frac{\textsf{A}^{\dot\alpha}\textsf{A}_{\dot\alpha}}{M_{12}\,\PR} {(g\,x)^{+L}\tilde\bp_\alpha(\epsilon\tilde\bp)_{\dot\alpha}}\,,
\ee
the equation which is valid at least at the leading order. However, we will extend its validity to all pre-wrapping orders in a moment.

The leading order of \(\tm_{\alpha\dot\alpha}\) is a polynomial of degree \(L\), as one can deduce from the large-\(u\) asymptotics \eqref{eq:muscaling}. Hence the r.h.s. of \eqref{eq:m12tm} is also a polynomial of degree \(L\). But the factor \((g\,x)^L\simeq u^L\) is already a polynomial of such  degree. Hence \(\tilde\bp=1\) in this approximation.

\paragraph{All pre-wrapping orders.} One can prove that \(\bp=\tilde\bp=1\) and \(\tm_{ab}\) are polynomials at all single-wrapping orders. The proof is done by induction. Assume that the statement holds at the order \(n\). Then one should perform the following steps.

\ \\
First, one observes that \(\tilde\bp=1\) at order \(n\) implies that \(\tilde\bp\) is regular at \(u=0\) at the order \(n+1\). One can draw this conclusion by an elementary analysis of the second expansion in \eqref{eq:bpsums}. 

\ \\
Second, prove that all \(\tm_{ab}\) have no poles at \(u=0\,,\bi\) at order \(n+1\). For this, one uses that \(\tm_{12}={\rm const}\) and hence \(\tm_{12}-\tm_{12}^{[2]}=0\) at order \(n\), then equations \eqref{eq:singmuu} tell us that \(\tm_{12}\) cannot have singularities at \(u=0,\,\bi\) at order \(n+1\). Then, in general, the singularity of any \(\mu_{ab}\) at \(u=0,\bi\) is a singularity of the combination \(\mu_{ab}-\mu_{ab}^{[2]}\) at \(u=0\). But this combination, up to the   factors inessential for the issue,  appears precisely on the l.h.s. of \eqref{eq:m12tm}. At the same time, the r.h.s. is regular at the order \(n+1\) because \(\tilde\bp\) is regular\footnote{Strictly speaking, equation \eqref{eq:m12tm} uses approximation \(\tm_{12}={\rm const}\) which has not been proven yet at the order \(n+1\). A more careful approach is to deduce  that \(\tm_{\alpha\dot\alpha}/\tm_{12}\) is regular at \(u=0,\bi\) and hence \(\tm_{\alpha\dot\alpha}\) is regular from \eqref{eq:mu12tmu} and already proven regularity of \(\tm_{12}\) at these points.}.

\ \\
Finally, applying the same logic as was used after equation \eqref{eq:singmuu} one concludes that \(\tm_{ab}\) are entire functions and hence, again, polynomials. Thus, again, \(\tilde\bp=1\) from power-counting in \eqref{eq:m12tm}.

These recursive arguments can be repeated until the moment when the r.h.s. of \eqref{eq:m12tm} develops a singularity for the first time. As we saw, it cannot originate from \(\tilde\bp\). Hence, it originates from the perturbative expansion of \(x^L\). One has
\begin{align}
(g\,x)^L=g^L\left(x^L+\frac 1{x^L}\right)-\frac {g^L}{x^L}=\big[ \text{a polynomial in \(u\)}\big]-\frac {g^{2L}}{u^L}+\CO(g^{2L+2})\,,
\end{align}
hence the singularity does not emerge until the leading wrapping order \(g^{2L}\) (this is also the order when \(\bP_{\alpha}(\epsilon\bP)_{\dot\alpha}\) starts to contribute).

As a conclusion, at all orders up to \(g^{2L-2}\), one has
\be
\tilde\bp_1=\tilde\bp_2=\tilde\bp^3=\tilde\bp^4=1\,,
\,,\ \ \ \ \ \bp_1=\bp_2=\bp^3=\bp^4=1\,.
\ee
and \(\tm_{\alpha\dot\alpha}\) is the polynomial solution of equation \eqref{eq:m12tm}. 

We introduce an operator \(\Psi_z\) which satisfies the property
\be
\Psi_z(f)-\frac 1{z}\Psi_z(f)^{++}=f\,,
\ee
We also require that \(\Psi_z(f)\) is a polynomial if \(f\) is a polynomial, to uniquely define the action of \(\Psi\) on polynomials in the case \(z\neq 1\).

Then one can write
\be
\tm_{\alpha\dot\alpha}=-\frac{A^{\dot\alpha}A_{\dot\alpha}}{M_{12}\,\PR} \Psi_{z_{\alpha\dot\alpha}}\left[(g\,x)^{L}\right]\,.
\ee
The remaining \(\bP\)'s are found by  elementary algebra from the equations of \(\bP\mu\)-system:
\be
(\epsilon\,\bP)_{\alpha}=\frac{\tilde\bP_\alpha-\mu_{\alpha\dot\alpha}\bP^{\dot\alpha}}{\mu_{12}}\,,\ \ \  \bP_{\dot\alpha}=\frac{(\epsilon\tilde\bP)_{\dot\alpha}+\mu_{\alpha\dot\alpha}(\epsilon\bP)^{\alpha}}{\mu_{12}}\,.
\ee
By considering these expressions at infinity and using that \(A_aA^a\) for \(a=1,\ldots,4\) are the quantities fixed by \eqref{eq:normAABB}, one finds explicit expressions for \(M_{12}\) and \(\PR\):
\be\label{eq:MPi}
 \tm_{12}=M_{12}=\frac{1+q}{1-q}\,,\ \ \  \PR=\frac{q-1}{q+1}\frac{\dot q+1}{\dot q-1}(A^3A_3)^2\,.
\ee
Then the explicit expressions for \(M_{\alpha\dot\alpha}\) follow
\begin{align}
M_{\alpha\dot\alpha}=\frac{\tx_{\alpha}-\tx_{\dot\alpha}}{2-[2]_{\dot q}}\,.\  \ 
\end{align}
One finds \(\tm_{34}\) from \(\Pf(\tm)=\tm_{12}\tm_{34}-\tm_{13}\tm_{24}+\tm_{14}\tm_{23}\propto g^{2L}=0\), and in particular
\be
M_{34}=\frac{1+\dot q}{1-\dot q}\,\frac{2-[2]_q}{2-[2]_{\dot q}}\,.
\ee 
Finally, the explicit expressions for nontrivial \(\bp\)'s become
\begin{align}
\bp_{\dot\alpha}&=\frac 1{M_{12}}(\tm_{1\dot\alpha}-\tm_{2\dot\alpha})-(gx)^L\frac{[2]_q+[2]_{\dot q}}{2-[2]_{\dot q}}+\CO(g^{2L})\,,
\\
\tilde\bp_{\dot\alpha}&=\frac 1{M_{12}}(\tm_{1\dot\alpha}-\tm_{2\dot\alpha})+\CO(g^{2L})\,,
\\
(-1)^{\alpha}(\epsilon\bp)_{\alpha}&=\frac{1}{M_{34}}(\tm_{\alpha3}-\tm_{\alpha4})+(gx)^L\frac{[2]_q+[2]_{\dot q}}{2-[2]_{q}}+\CO(g^{2L})\,,
\\ 
(-1)^{\alpha} (\epsilon\tilde\bp)_{\alpha}&=\frac{1}{M_{34}}(\tm_{\alpha3}-\tm_{\alpha4})+\CO(g^{2L})
\end{align}
{Note the large-\(u\) behaviour:} \(\bp_{\dot\alpha}\simeq u^{L}+\ldots\) but  \(\tilde\bp_{\dot\alpha}\simeq \frac{2-[2]_q}{2-[2]_{\dot q}}u^L+\ldots\).

\subsection{Asymptotic Q-system}
In subsequent section we will reduce the computation of energy to the L\"uscher-type formula which requires T-functions \(\bT_{a,\pm 1}\) in the physical gauge \(\bT\) \cite{Gromov:2014caa,Gromov2011a} as an input. In this section we compute the necessary Q-functions to reconstruct \(\bT_{a,\pm 1}\). As it will be clear, these are the functions \(Q_{12|\tau}\) with \(\tau\in\{1,2\}\) and \(Q^{34|\dot\tau}\) with \(\dot\tau\in\{3,4\}\). The functions \(Q^{34|\dot\tau}\) are deduced from the LR-symmetry \eqref{eq:LR}, hence we will not spell them explicitly.

The departing point is the generalisation of \eqref{eq:mu12tmu} to an arbitrary set of indices:
\be
\mu_{ab}^{[2]}\mu_{ac}-\mu_{ac}^{[2]}\mu_{ab}=\epsilon_{abcd}(\tilde\bP_a\tilde\bP^d-\bP_a\bP^d)\,.
\ee
From the results of previous section we know that \(\tilde\bP/\bP \propto g^{-L}\) for all \(\bP\)'s,
hence one always has the approximation:
\be
\mu_{ab}^{[2]}\mu_{ac}-\mu_{ac}^{[2]}\mu_{ab}=\tilde\bP_a(\epsilon_{abcd}\tilde\bP^d)+\ldots\,,
\ee
at all  pre-wrapping orders. But this equation looks precisely like the QQ-relation \eqref{eq:41} with \(A=\{a,b\},\,\,I=\{1,2\}\)! It is hence tempting to identify \(\tilde\bP\)'s and \(\mu\)'s with certain Q-functions. We perform the following identification
\be\label{eq:identification}
Q_{ab|12}^-\propto\mu_{ab}\,,\ \  Q_{a|12}\propto\,\tilde\bP_{a}\,,\ \   Q^{a|34}=\frac 16\epsilon^{bcda}Q_{bcd|12}\propto\tilde\bP^{a}\,.
\ee
One can think of \eqref{eq:identification} as a definition of some Q-functions, by introduction of a formal labelling. But in fact, it is not difficult to show that these are indeed Q-functions of the quantum spectral curve. Indeed, the large-\(u\) asymptotics is correct and given \eqref{eq:identification} one derives
\be\label{eq:identification2}
Q_{a|\es}\,Q_{a|12}=Q_{a|1}^+Q_{a|2}^--Q_{a|1}^-Q_{a|2}^+=\bP_a\bQ_1Q_{a|2}^--(1\leftrightarrow 2)=\bP_a Q_{ac|12}^-\bP^c\simeq \bP_a\,\tilde\bP_a\,,
\ee
so the conjectured Q-functions \eqref{eq:identification} are properly linked with \(\bP_a\) (and \(\bP^a\)). We know that having \(\bP_a\) and \(\bP^a\) is sufficient, in principle, to derive  all the Q-functions of QSC \cite{Gromov:2014caa}. Hence if we found a Q-system which contains \(\bP_a\) and \(\bP^a\) with standard identification \(\bP_a=Q_{a|\emptyset}\) and \(\bP^a=Q^{a|\emptyset}\), and we did so  by \eqref{eq:identification} and \eqref{eq:identification2} indeed, this Q-system should be the one of QSC.

The normalisation factors in \eqref{eq:identification} are restored easily, as we know the normalised large-\(u\) asymptotics of Q-functions \eqref{eq:220},\eqref{eq:217} and of \(\mu\)'s and \(\bP\)'s. We adapt the same strategy of restoring prefactors in the following, and to make things more precise, we define \(\nQ_{A|I}\) as
\be
Q_{A|I}={\rm X}_{A}^{\bi u}\, \Upsilon_{A|I}\, \nQ_{A|I}\,,
\ee
where \({\rm X}_{A}^{\bi u}\, \Upsilon_{A|I}\) are chosen in a way that \(\nQ_{A|I}\simeq 1\cdot u^{n_{A|I}}\) at large-\(u\).
Then we can fix the following Q-functions:
\begin{description}
\item[\(Q_{a|\emptyset}\)] \(=\bP_a\):
\begin{align}
\nQ_{\alpha|\es}&=(g\,x)^{-L/2}\,,&  \nQ_{\dot\alpha|\es}&=(g\,x)^{-L/2}\,\bp_{\dot\alpha}\,.
\end{align}
\item[\(Q_{a|\tau}\,,\)] where \(\tau\in \{1,2\}\):

Consider the equation \(\nQ_{\alpha|1}^+\nQ_{\alpha|2}^--\nQ_{\alpha|1}^-\nQ_{\alpha|2}^+\propto 1\). It is not difficult to solve it:
\be
\nQ_{\alpha|1}= 1\,,\ \ \  \nQ_{\alpha|2}= u+c_{\alpha}\,,\ \ \ \  c_{\alpha}=-\frac {\bi}2\frac{1+x_{\alpha}}{1-x_{\alpha}}\,,
\ee
where the constant \(c_{\alpha}\) is in principle arbitrary, but we have chosen it to get a particularly simple expression for \(\nQ_{\es|2}\) (see below).

To find \(Q_{\dot\alpha|\tau}\), one applies a Pl\"ucker identity
\be
Q_{ab|12}\,Q_{c|\tau}+Q_{bc|12}\,Q_{a|\tau}+Q_{ca|12}\,Q_{b|\tau}=0
\ee
for the case \(ab=12\) and recalls that \(\mu_{ab}^+\propto Q_{ab|12}\), getting \(Q_{\dot\alpha|\tau}\propto -\epsilon^{\alpha\beta}\,\mu_{\alpha\dot\alpha}^+\,Q_{\beta|\tau}\,\). The corresponding normalised expression is
\be
\nQ_{\dot\alpha|\tau}=\nQ_{1|\tau}\frac{1-1/{x_{\dot\alpha}}}{1+q}\left(q\,\tm_{1\dot\alpha}^++ \tm_{2\dot\alpha}^+\right)\,.
\ee

\item[\(Q_{\es|\tau}\)]\(=\bQ_{\tau}\,,\)  from the relation \(Q_{\alpha|\tau}^+-Q_{\alpha|\tau}^-=\bP_{\alpha}\bQ_{\tau}\):
\be
\nQ_{\es|1}=(g\,x)^{L/2}\,,\ \ \  \nQ_{\es|2}=(g\,x)^{L/2}\,u\,.
\ee
\item[\(Q_{12|\tau}\,,\)] the QQ-relation \(Q_{ab|i}Q_{a|12}={Q_{ab|12}^{+}Q_{a|i}^--Q_{ab|12}^{-}Q_{a|i}^+}{}\) becomes the most attractive for the case \(ab=12\), when it reduces in the asymptotic limit to
\be
Q_{12|\tau}&\propto\frac{\bP_a}{\tilde\bP_a}\,\bQ_\tau\propto x^{-L} \bQ_\tau\,.
\ee
One has then explicitly 
\be
 \nQ_{12|1}=(gx)^{-L/2}\,,\ \ \  \nQ_{12|2}=(gx)^{-L/2}\,u\,.
\ee
\item[Other \(Q\)'s:] We briefly comment on how to find all other Q-functions. Although it won't be used in this paper, it would be a necessary step for performing higher-loop computations in the future.

First, one uses \(Q_{12|12}^+Q_{12|\tau\dot\tau}^--Q_{12|12}^-Q_{12|\tau\dot\tau}^+\propto Q_{12|\tau}Q_{12|12\dot\tau}\), which is specified in the asymptotic limit as
\be
q_{12|\tau\dot\tau}^+-q_{12|\tau\dot\tau}^-\propto \epsilon_{\dot\tau\dot\tau'}\,q_{12|\tau}q^{34|\dot\tau'}\,,
\ee
to compute \(Q_{12|\tau\dot\tau}\):
\begin{subequations}
\label{q12ij}
\begin{align}
q_{12|13}&=\bi(L-1)\Psi^+\left[\frac 1{(gx)^L}\right]
\,,\\
q_{12|14}=q_{12|23}&=\bi(L-2)\Psi^+\left[\frac u{(gx)^L}\right]
\,,\\
q_{12|24}&=\bi(L-3)\Psi^+\left[\frac {u^2}{(gx)^L}\right]\,,
\end{align}
\end{subequations}
The action of \(\Psi\) is unambiguous in the case of \eqref{q12ij} and \(L\geq 4\) as we require that the result is a function which decreases at \(u\to\infty\) and which is analytic in the upper half-plane.

From \(\text{Pf}_{ij}(Q_{12|ij})=Q_{12|12}Q_{12|34}-Q_{12|13}Q_{12|24}+Q_{12|14}Q_{12|23}=0\)  one finds \(Q_{12|34}\), and hence all \(Q_{12|ij}\) are known by now.

The last steps are to restore \(Q_{\alpha|\dot\tau}\) from
\be
Q_{\alpha|\dot\tau}Q_{12|12}+Q_{\alpha|1}Q_{12|2\dot\tau}-Q_{\alpha|2}Q_{12|1\dot\tau}=0\,,
\ee
\(Q_{\dot\alpha|\dot\tau}\) from
\be
Q_{\dot\alpha|1}Q_{12|2\dot\tau}-Q_{\dot\alpha|2}Q_{12|1\dot\tau}+Q_{\dot\alpha|\dot\tau}Q_{12|12}=Q_{12|12}Q_{12\dot\alpha|12\dot\tau}\,,
\ee
and, finally \(\bQ_{\dot\tau}\) from \(Q_{\alpha|\dot\tau}^+-Q_{\alpha|\dot\tau}^-=\bP_\alpha\bQ_{\dot\tau}\).

The function \(Q_{12\dot\alpha|12\dot\tau}=-\epsilon_{\dot\alpha\dot\alpha'}\epsilon_{\dot\tau\dot\tau'}Q^{\dot\alpha'|\dot\tau'}\) is known due to LR-symmetry \eqref{eq:LR}, and all Hodge-dual functions are most easily found using LR-symmetry.
\end{description}
\subsection{Asymptotic T-system and energy}

An obvious way to extract the energy of a state is to solve the RH equations of the QSC and then to read off \(\Delta\) from the powers of asymptotics of appropriate functions. For example, we know that for the BMN vacuum in question \(\mu_{12}\sim u^{\gamma}\), where \(\gamma=\Delta-J_1\) is the anomalous dimension. But in practice this method may be sometimes not very convenient because it requires certain information about Q-functions at the same order at which we want to compute the anomalous dimension, even at one order more if to be precise. In our example, the anomalous dimension starts to be non-trivial only at wrapping orders, \(\gamma={\cal\ O}(g^{2L})\), hence one would have to analyse the Q-system at \(L\) orders more compared to what was done in previous sections in order to find all one-wrapping orders of \(\gamma\). To avoid this kind of difficulties we can always use the good old TBA formula  which would allow us to compute the energy at the one-wrapping orders knowing  some particular Q-functions only asymptotically. This formula, exact for any coupling, states \cite{Gromov:2009tv,Gromov:2009bc,Bombardelli:2009ns,Arutyunov:2009ur}
\begin{equation}\label{EnergyTBA}
\Delta-J_1=\sum_{a=1}^\infty \int_{-\infty}^\infty\frac{dv}{2\pi\bi}\p_v\check p_a(u)\log\left(1+ Y_{a,0}(u)\right)\,,
\end{equation}   where \(\frac{1}{\bi g}\check p_a(u)=\frac{1}{\bi g}a+\check  x^{[-a]}-\frac{1}{\check x^{[-a]}}-\check x^{[a]}+\frac{1}{\check x^{[a]}}\) is the ``mirror''momentum and \(Y_{a,s}\) are the Y-functions on the ``mirror'' sheet with long cuts.  This formula was used in \cite{Ahn:2011xq} to compute the energy of the \(\gamma\)-deformed  BMN vacuum up to two wrappings, using the direct solution of TBA equations.

  The purpose of this section is to demonstrate the twisted QSC at work on the BMN vacuum at one wrapping. Hence we  rederive in \appref{app:E2}  the TBA  formula \eqref{EnergyTBA} in a somewhat shorter way compared to a relatively cumbersome way of reversing  the historical derivation of QSC from TBA in \cite{Gromov2011a,Gromov:2014caa}. Then, at the end of this section, we evaluate \(\log(1+Y_{a,0})\) with single-wrapping precision as
\be
\log(1+Y_{a,0})\simeq \label{Y=1over}
 Y_{a,0} = \frac{1}{ \frac{{\bT_{a,0}}^{+} {\bT_{a,0}}^{-}}{\bT_{a,1} {\bT_{a,-1}}}-1}\simeq \frac{\bT_{a,1} {\bT_{a,-1}}}{{\bT_{a,0}}^{+} {\bT_{a,0}}^{-}}\,,
\ee  
where T-functions are computed as special combinations of the Q-functions originating from the Wronskian formulae  \eqref{eq:155}. The approximations made in \eqref{Y=1over} are valid under assumption that \(Y_{a,0}\) is small; we will confirm below that, indeed, \(Y_{a,0}=\CO(g^{2L})\).

The T-functions \(\bT_{a,0}\) and \(\bT_{a,1}\) are the elements of the mirror {\THook}. Construction of its Wronskian solution \eqref{eq:155} requires to choose a particular basis in the Q-system, by means of symmetry transformations, such that splitting of bosonic indices into two sets \(\Bset_1 =\{1,2\}\,,\,\,\Bset_2=\{3,4\}\), and further usage of \eqref{eq:155} would produce T-functions with correct analytic properties identified in \cite{Gromov2011a,Gromov:2014caa}. The appropriate basis for the mirror {\THook} construction was given in appendix B of \cite{,Gromov:2014caa}. This basis is not the same as the one used in QSC, but, of course, it is related to the QSC basis in a certain  way. As a result, we can express T-functions in terms of Q-functions of the QSC basis only, but after several non-trivial steps, the details are given in \appref{sec:TfunctionsfromThook}. The resulting explicit formula we will operate with is
\be\label{eq:bTa1}
\bT_{a,1}=\hat Q_{12|i}^{[a]}\hat Q_{12|j}^{[-a]}(\hat\omega^{ij})^{[a]},
\ee
which is valid slightly above the real axis;  \(\hat Q\) notation means that the expression \(\hat Q_{12|j}^{[-a]}\) is computed by analytic continuation from the upper half-plane to the lower half-plane using the physical kinematics.

Correspondingly,
\be\label{eq:bTa0}
\bT_{a,0}=(-1)^a\frac 12\hat Q_{12|ij}^{[a]}\hat Q_{12|i'j'}^{[-a]}(\hat\omega^{ii'}\hat\omega^{jj'})^{[a-1]}\,.
\ee
These expressions for \(\bT_{a,0}\) and \(\bT_{a,1}\) do not have the structure of the Wronskian ansatz \eqref{eq:155} precisely for the reason that we are using a basis which is related to the Wronskian ansatz basis by a transformation which is a symmetry of QQ-relations but not of relations \eqref{eq:155}. The expressions equivalent to \eqref{eq:bTa1} and \eqref{eq:bTa0} were already suggested in appendix D of \cite{Gromov2011a} where the function \(\omega\) appeared for the first time. 

To accomplish the computations, we need to determine \(\omega^{ij}\). In the asymptotic approximation, one finds
\be
 \omega^{ij}=-\frac 12(Q^{ab|ij})^-\mu_{ab}= \bi\frac{2-[2]_q}{\,g^L}\frac 1{B_1B_2}\frac 12 (Q^{ab|ij}Q_{ab|12})^-=\bi\frac{2-[2]_q}{g^L}\frac 1{B_1B_2}\delta^{ij}_{12}\,.
\ee
Now one can derive the explicit expression for the required T-functions:
\begin{subequations}
\label{TfromQ}
\begin{eqnarray}
\label{Ta1fromQ}
\bT_{a,1}&=&\bi\,\frac{2-[2]_q}{g^L}\frac 1{B_1B_2}\epsilon^{\tau\tau'}\hat Q_{12|\tau}^{[+a]}\hat Q_{12|\tau'}^{[-a]}=- (A_1A_2)^2\frac{2+[2]_q}{g^L}\frac {a}{(g^2x^{[+a]}x^{[-a]})^{L/2}}\,,
\\
\bT_{a,0}&=&-(-1)^a\left(\frac{2-[2]_q}{g^L}\frac 1{B_1B_2}\right)^2 \hat Q_{12|12}^{[+a]}Q_{12|12}^{[-a]}=(-1)^a(A_1A_2)^2\frac 1{g^{2L}}\left(\frac{1+q}{1-q}\right)^2\,,
\\
\bT_{a,-1}&=&- (A_1A_2)^2\left(\frac{q+1}{q-1}\frac{\dot q-1}{\dot q+1}\right)^2\frac{2+[2]_{\dot q}}{g^L}\frac {a}{(g^2x^{[+a]}x^{[-a]})^{L/2}}\,,
\end{eqnarray}
\end{subequations}
where the expression for \(\bT_{a,-1}\) was obtain by applying LR-symmetry transformation \eqref{eq:LR} on \eqref{Ta1fromQ} and using \(\frac{A^3A^4}{A_1A_2}=\frac{A_3A^3A_4A^4}{\Pi}=-\frac{q+1}{q-1}\frac{\dot q-1}{\dot q+1}\).

Finally, we compute \(Y_{a,0}\) in the approximation \eqref{Y=1over}
\begin{eqnarray}
 Y_{a,0} \simeq \frac{\bT_{a,1} {\bT_{a,-1}}}{ \mu_{12}^2} 
=g^{2L}\,  \,(\dot q^{1/2}-\dot q^{-1/2})}{^2(q^{1/2}-q^{-1/2})^2\,\,a^2\left(g^2x^{[a]}x^{[-a]}\right)^{-L}. \end{eqnarray}
 Let us remind that \(x\) is  here the function in the physical kinematics (short cuts). To compute the mirror momentum, we have to substitute \(x^{ [+a]}\to \check x^{ [+a]},\quad x^{ [-a]}\to1/ \check x^{ [-a]} \).

It remains only to perform integration in \eqref{EnergyTBA} to reproduce  the energy for the \(\gamma\)-twisted BMN vacuum in the single-wrapping approximation: 
\begin{eqnarray*}
&&E_0^{(1-wrap)}=-(q^{1/2}-q^{-1/2})^2\,\,(\dot q^{1/2}-\dot q^{-1/2})^2g\sum_{a=1}^\infty a^2\int\frac{dv}{2\pi}\p_v(\check x^{[-a]}-\frac{1}{\check x^{[-a]}}-\check x^{[a]}+\frac{1}{\check x^{[a]}})\left( \frac{\check x^{[a]}}{\check x^{[-a]}}\right)^L \\
&&= -(q^{1/2}-q^{-1/2})^2\,\,(\dot q^{1/2}-\dot q^{-1/2})^2\sum_{a=-\infty}^\infty  a^2\int \frac{dv}{2\pi}\frac{x^{[a]}+\frac{1}{x^{[a]}}}{  x^{[a]}-\frac{1}{x^{[a]}}}\left( \frac{1}{x^{[-a]}x^{[a]}}\right)^L\,,
\end{eqnarray*}
where we used the identity \(g\p_u\left(x-\frac{1}{x}\right)=\frac{x+\frac{1}{x}}{  x-\frac{1}{x}}=\frac{u}{  \sqrt{4g^2-u^2}}\). It is true up to the order \(g^{4L-2}\), just before the second wrapping appears.   It coincides of course with the result computed before from TBA or from the direct perturbation theory    \cite{deLeeuw:2011rw,Ahn:2011xq,Fokken2014a}.  For the leading single-wrapping order \(g^{2L}\), the formula becomes more explicit:  \begin{eqnarray*}
E_0^{(leading)}&=&
-(q^{1/2}-q^{-1/2})^2\,\,(\dot q^{1/2}-\dot q^{-1/2})^2\,\, g^{2L}\sum_{a=-\infty}^\infty a^2\int _{-\infty}^{\infty}\frac{dv}{2\pi}\frac{1}{(v^2+\frac{a^2}{4})^L}=
\\
&=&-2(q^{1/2}-q^{-1/2})^2\,\,(\dot q^{1/2}-\dot q^{-1/2})^2\,\, g^{2L}\binom{2L-2}{L-1}\,\zeta_{2L-3}\,.
\end{eqnarray*}

Notice that this formula is non-singular at \(L=3\) and it predicts the right value of energy, in spite of the presence of singularity at this value of \(L\) in some Q-functions, see \eqref{eq:normAABB}.  At the contrary, as was observed in  \cite{Ahn:2011xq}, at \(L=2\) the formula is singular and it ceases to predict the right energy. The reason for it is probably related to the phenomenon pointed out in \cite{Fokken2014b}: this operator leads to a new counter-term in the action of twisted \({\cal N}=4\) SYM which breaks down its conformal symmetry. In the 't~Hooft limit this operator/state can be self-consistently removed from the spectrum of the theory, but at finite \(N_c\) this is not possible.

\section{Conclusion}
\label{sec:conclusion}

In this paper,  we gave a general description of grassmannian
structure emerging from fusion relations in integrable rational
Heisenberg super-spin chains. The general solution
\cite{Tsuboi:2011iz} of Hirota equations for transfer-matrices in a
{\THook}, corresponding to arbitrary highest weight irreps of
\(sl(K_1,K_2|M)\) superalgebra, and its proof \cite{Tsuboi:2011iz}, are presented in an elegant way in terms of exterior forms  built out of a finite number of Baxter's Q-functions. A particular attention is payed to the case of twisted spin chains and to  subtleties  of partial or full untwisting limit.

Then we used  our observations to construct the  twisted version of the Quantum Spectral Curve (QSC) of the   AdS\(_5\)/CFT\(_4\) duality, thus extending the QSC proposal formulated  \cite{Gromov2014a,Gromov:2014caa} in  the untwisted cased to the full or partial twisting of the superstring sigma model on \(AdS_5\times S^5\) background.      Via AdS/CFT duality, this twisted QSC describes  exact solutions for the spectra of anomalous dimensions   of an extended range of   interesting super-Yang-Mills gauge theories in the planar limit with the number of supersymmetries \({\cal\ N}<4\).   For  generic   configurations of twists, the  actions of such gauge duals are unknown, though they are established  in some particular cases, such as the beta-deformation corresponding to the so-called Leigh-Strassler deformation of \(\mathcal{N}=4\) SYM, and a more general \(\gamma\) deformation for the fully twisted R-symmetry where the corresponding SYM action (see e.g. \cite{Fokken2014}) is explicitly non-supersymmetric.   We presented the construction of QSC not only for twisted string sigma model in the case of generic twisting (6 arbitrary twist parameters) but also  for  an arbitrary partial twisting, representing a subtle limit when some twists become equal to each other. In particular, we computed the asymptotics of large spectral parameter for arbitrary Q-functions entering the Q-system  describing of twisted QSC. Since  the results seem to be as meaningful as in the untwisted case it poses an interesting question of construction of the gauge duals for each of configurations of twists. 

Further on, we  checked our twisted QSC formalism on the computation in the weak coupling approximation for the single-wrapping energy of a peculiar state - the  \(\gamma\)-deformed BMN vacuum corresponding to  \({\cal\ N}=0\) deformation  \({\cal\ N}=4\) SYM in the 't~Hooft limit, successfully reproducing the results of TBA computation \cite{deLeeuw:2011rw,Ahn:2011xq}.

 It would be interesting to perform a systematic weak and strong coupling expansion for various twisted cases similarly to \cite{Marboe:2014gma}, as well as to study the subtle limit of small twisting -- the intermediate regime between particular configurations of distinct and coinciding twist parameters.   Another interesting problem could be the BFKL limit for various twisted SYM actions. The twisting of the conformal group -- the isometry of  \(AdS_5\)   -- is believed to describe certain non-commutative YM theories\cite{Beisert:2005if,Tongeren2013}, though their classical actions and the renormalization properties are not yet established.  It would be interesting to use the QSC formalism to get more of the physical information about these exotic theories and to understand the consequences of the breakdown of conformal invariance. 

Other interesting theories to consider by our QSC method are the orbifold SYM models and their AdS duals, obtained from the general twisted case by choosing some twists as equal to \(\exp[i\text{(rational number)}]\) (see \cite{Tongeren2013} for description). 

The twisted quantum spin chains appeared to be a good starting point for the construction of operatorial formalism for  T-systems and Q-systems in terms of the so-called co-derivative formalism \cite{Kazakov:2007na,Kazakov:2010iu,2011arXiv1112.3310A}. It is conceivable that such a method could provide us with the possibility to recover the operatorial formulation of various sigma-models at finite volume, including the AdS/CFT integrability, in the physical space. After all, the sigma models are not that different from the quantum spin chains: the former could be often represented as a specific continuous limit of the latter. 

Our method of twisting of QSC is certainly generalizable to  other interesting sigma models, such as the principal chiral field where the twist are introduced in a similar way into the asymptotics of Q-functions \cite{Kazakov:2010kf}.
  It would be good to  extend the QSC metods to these cases and to perform the numerical calculations of their energy spectrum.

\section*{Nota added}  Recently, we were informed by N.Gromov and F.Levkovich-Maslyuk about their forthcoming paper where they  used a similar construction of twisted QSC  for the study of cusped Wilson loop where two twist angles are introduced, one on \(S^5\) and one on \(AdS_5\).     We  agreed to synchronyze the publications of our works in the HEP Arxiv.

\section*{Acknowledgments}
\label{sec:acknowledgments}

We thank     N.Gromov, I.Kostov, Ch.Sieg,  Z.Tsuboi, M.Wilhelm,  and K.Zarembo for useful discussions.
Our work  was supported
by the People Programme (Marie Curie Actions)
of the European Union�s Seventh Framework Programme
FP7/2007-2013/ under REA Grant Agreement
No 317089 (GATIS). The work of V.K.  has received funding
from the European Research Council (Programme
Ideas ERC-2012-AdG 320769 AdS-CFT-solvable), from
the ANR grant StrongInt (BLANC- SIMI- 4-2011) and
from the ESF grant HOLOGRAV-09- RNP- 092.

 V.K. is also very grateful to Princeton Advanced Study Institute where a part of this work was done,
 for hospitality and to  the Ambrose Monell Foundation for the generous support during his visit in 2014. V.K. is also very grateful to the  Humboldt University of Berlin  for their hospitality during a part of this work and to the partial funding of this research through  the  Humboldt Kosmos 2015-2017 Programme of the Excellence Initiative.

\pagebreak
\appendix

\part*{Appendices}
\label{part:appendices}

\addcontentsline{toc}{part}{Appendices}
\section{Further details and proofs}

\subsection{Derivation of \texorpdfstring{\eqref{eq:86}}{(\ref*{eq:86})} via Pl\"ucker
identities.}
\label{sec:proof-qq-relation}
In this subsection we prove that (\ref{eq:86}) follows from (\ref{eq:82}). Consider the Pl\"ucker
identity \eqref{Plucker} and set 
\(\xx_N=\xx\), \(\yy_N=\yy\) and \(\xx_i =Q_{(1)}^{[N+1-2i]}\), \(\yy_i
=Q_{(1)}^{[N-1-2i]}\) for \(1\le i\le N-1\). Then one gets
\begin{multline}
  \star(Q_{(N-1)}^+ \wedge \xx)\,\,\star (Q_{(N-1)}^-\wedge
   \yy)%
 =%
   \star (Q_{(N-1)}^+ \wedge \yy)\,\,\star (Q_{(N-1)}^-\wedge
   \xx)%
 \\+\,\frac{f_{N-1}^+f_{N-1}^-}{f_{N}f_{N-2}}\star\,Q_{(N)}\,\,\star(Q_{(N-2)} \wedge \xx\wedge
   \yy)  \,.
\end{multline}
Note that in this example, only the terms \(a=N-1\) and \(a=N\) give a non-vanishing
contribution to the right side of \eqref{Plucker} (other terms vanish
because \(\yy_a=\xx_{a+1}\)).
For \(\xx,\yy\in\{\zeta_1, \dots, \zeta_N\}\), this gives the QQ-relation
\eqref{eq:86} when \(\nA=N-2\)%
.

To show the QQ relation \eqref{eq:2} when the multi-index \(I\) has an
arbitrary number \(l\) of elements, we write another
obvious consequence of \eqref{Plucker}:
   \begin{multline}\label{bosonicQQforms}
\star(Q_{(l+1)}^+ \wedge \xx_{1}\wedge\dots\wedge\xx_{{N}-l-2} \wedge \xx)\,\,\star(Q_{(l+1)}^-\wedge \xx_{1}\wedge\dots\wedge\xx_{{N}-l-2} \wedge \yy)=\\
\star(Q_{(l+1)}^+ \wedge \xx_{1}\wedge\dots\wedge\xx_{{N}-l-2} \wedge \yy)\,\,\star(Q_{(l+1)}^-\wedge \xx_{1}\wedge\dots\wedge\xx_{{N}-l-2} \wedge \xx)\\
+\frac{f_{l+1}^+f_{l+1}^-}{f_{l}f_{l+2}}\star(Q_{(l+2)} \wedge \xx_{1}\wedge\dots\wedge\xx_{{N}-l-2} )\,\,\star(Q_{(l)}\wedge \xx_{1}\wedge\dots\wedge\xx_{{N}-l-2} \wedge \xx\wedge \yy)
\end{multline}
with only three terms surviving there.
If we choose \(\xx_1,\dots,\xx_{{N}-l-2},\xx,\yy\in\{\zeta_1, \dots,
\zeta_N\}\), \eqref{bosonicQQforms} reduces to the
QQ-relation \eqref{eq:86}.

\subsection{Q-functions for the semi-infinite strip}
\label{sec:q-functions-semi}

Starting from the results of \secref{sec:ident-q-funct}, let us
show that generic  solutions of the Hirota equation
(\ref{eq:1}) on the semi-infinite strip of figure \ref{fig:glnstrip}
are given by (\ref{eq:34a}).

First, one should note that by repeating the arguments of the
\secref{sec:ident-q-funct}, the generic solution of Hirota equation
on this semi-infinite strip obeys
\begin{align}\label{eq:95}
  T_{a,s}&=\star\left(Q_{(a)}^{[+s]}\wedge
    P_{(N-a)}^{[-s]}\right)\,&\textrm{when}
  \begin{cases}
    s\ge 0\\
\textrm{or }s=-1 \textrm{ and } a<N\,.
  \end{cases}
\end{align}
As compared to the \secref{sec:ident-q-funct}, this expression does not hold
for \(T_{N,-1}\) because the Hirota equation at \((a,s)=(N,0)\) is
modified (\(T_{N+1,0}=0\) does not hold anymore), and it does not hold
for arbitrarily negative \(s\) because the Hirota equation does not allow to
recursively express
\(T_{a,s}=\frac{T_{a,s+1}^+T_{a,s+1}^--T_{a+1,s+1}T_{a-1,s+1}}{T_{a,s+2}}\)
when the denominator is equal to zero.

The requirement \(T_{a,-s}=0\) for \(a=1,2,\dots,N-1\) can be plugged
into (\ref{eq:95}), allowing to conclude that \(P_{(1)}\propto
Q_{(1)}^{[-N]}\). Indeed, we can deduce from \(T_{N-1,-1}=0\) that
\(P_{(1)}\) is a linear combination \(P_{(1)}=\sum_{k=2}^N \alpha_k Q_{(1)}^{[N-2\,k]}\,.\)
Hence, we have
\begin{align}
0=T_{N-2,-1}&=\star\frac{Q_{(N-2)}^-\wedge\sum_{k=2}^N \alpha_k^{[+2]}
  Q_{(1)}^{[N-2\,k+2]}\wedge P_{(1)}}{P_{\emptyset}^+}\\
&=\alpha_2^{[+2]} \star\frac{Q_{(N-1)}\wedge
  P_{(1)}}{P_{\emptyset}^+/Q_{\emptyset}^{[N-3]}}=
\alpha_2^{[+2]}  \frac {Q_{\emptyset}^{[N-3]}}{P_{\emptyset}^+} T_{N-1,0}\,,
\end{align}
which allows to deduce\footnote{To conclude that \(\alpha_2=0\) we use that the T-functions are non-zero on the dots of the
  lattice in figure~\ref{fig:glnstrip}, so that \(\frac
  {Q_{\emptyset}^{[N-3]}}{P_{\emptyset}^+}  T_{N-1,0}\) is non-zero.} that \(\alpha_2=0\). Reproducing the argument
for \(T_{N-3,-1}\), we obtain \(\alpha_3=0\), and at the last step
(\(T_{1,-1}=0\)) we obtain \(\alpha_{N-1}=0\), which gives \(P_{(1)}=\alpha_{N}
Q_{(1)}^{[-N]}\). Inserting this into the relation \(P_{(n)}
=\frac{P_{(1)}^{[n-1]}\wedge P_{(1)}^{[n-3]}\wedge\dots \wedge P_{(1)}^{[1-n]}}{
\prod_{1\leq k \leq n-1}{P_{\emptyset}^{[n-2 k]}}}
\), one gets \(P_{(n)}= g_{(+-)}^{[N-n]}g_{(--)}^{[n-N]} Q_{(n)}\), where
\(g_{(+-)}\) and \(g_{(--)}\) are two functions such that
\(\frac{g_{(+-)}^+}{g_{(+-)}^-}=\frac{P_\emptyset^{[1-N]}}{\alpha_N
  Q_\emptyset^{[1-2N]}}\) and
\(g_{(--)}=\frac{P_\emptyset^{[+N]}}{Q_\emptyset g_{(+-)}^{[+2N]}}\).

At this point we have shown that
\begin{align}\label{eq:89}
  T_{a,s}&=g_{(+-)}^{[a-s]} g_{(--)}^{[-a-s]} \star\left(Q_{(a)}^{[+s]}\wedge Q_{(N-a)}^{[-s-N]}\right)&\textrm{when}\ \ \ &s\ge 0
  \textrm{ and } 0\le a\le N\,,
\end{align}
which coincides with (\ref{eq:34}) up to a gauge transformation. If we introduce functions \(f_1\), \(f_2\), \(g_1\) and \(g_2\)
defined by \(f_1^2=g_{(--)}\), \(g_1=1/f_1\), \(f_2^2=g_{(+-)}^{[+N]}\) and
\(g_2=f_2^{[+N]}\), then the relation (\ref{eq:89}) becomes \(T_{a,s}=
\frac{f_1^{[a+s]}f_2^{[a-s-N]}}{f_1^{[-a-s]}f_2^{[-a+s-N]}}
g_1^{[a+s]}g_1^{[-a-s]} g_2^{[-a+s]}g_2^{[+a-s-2N]}
{\star\left(Q_{(a)}^{[+s]}\wedge Q_{(N-a)}^{[-s-N]}\right)}\). If we
redefine \(Q_A \mapsto g_ 1^{[\nA]}g_2^{[-\nA]}Q_A\), which still obeys the
QQ-relations and the relation (\ref{eq:5}), then we obtain
\begin{align}
  T_{a,s}&=\frac{f_1^{[a+s]}f_2^{[a-s-N]}}{f_1^{[-a-s]}f_2^{[-a+s-N]}} \star\left(
    Q_{(a)}^{[+s]}\wedge  Q_{(N-a)}^{[-s-N]}\right)&\textrm{when }&s\ge 0
  \textrm{ and } 0\le a\le N\,,
\end{align}
as we wished to prove.

\subsection{Index splitting QQ-relations}\label{sec:index-splitting}
Among the numerous relations between the Q-functions, implied by the
relations (\ref{eq:2},\ref{eq:5}), there are such that arise
if we
split the set \(\Bset=\{1,2,\dots,N\}\) into the disjoint union of three
subsets: \(\Bset=\Sset_1\sqcup\Sset_2\sqcup\Sset_3\). If \(A_1\), \(A_2\) and \(A_3\) are
multi-indices with \(\{A_1\}\subset \Sset_1\) (and \(\{A_2\}\subset
\Sset_2\), \(\{A_3\}\subset
\Sset_3\)), then we denote \(Q_{A_1;A_2;A_3}\equiv Q_{A_1A_2A_3}\). Then
we have
\begin{gather}
  \sum_{|A|=n} Q_{A;\emptyset;\Sset_3}^{[t]}
  \HQ{A;\emptyset;\Sset_3}=(-1)^n\sum_{|B|=n}
  Q_{\emptyset;B;\Sset_3}^{[t]}
  \HQ{\emptyset;B;\Sset_3}\,,\quad\qquad
\textrm{ where}\ \ \  n\le \frac{|\Sset_1|+|\Sset_2|-|\Sset_3|}2
\nonumber\\\textrm{and}\ \ \ t\in\{p-2n,p-2n-2,\dots,-p+2n\} \textrm{ with }p=|\Sset_1|+|\Sset_2|-|\Sset_3|\,,
\label{eq:103}
\end{gather}
where the sums run over sorted multi-indices \(A\subset \Sset_1\) and \(B\subset \Sset_2\).

At the level of forms, if we denote
\(Q_{(n;p;q)}\equiv{\displaystyle \sum_{|A|=n,\,|B|=p,|C|=q}}Q_{A;B;C}\,\,\bg_A\wedge\bg_B\wedge\bg_C\),
then the relation (\ref{eq:103}) reads
 \begin{align}\label{eq:115}
Q_{(n;0;|\Sset_3|)}^{[t]}    \wedge Q_{(|\Sset_1|-n;|\Sset_2|;0)}
    =(-1)^n
 Q_{(0;n;|\Sset_3|)}^{[t]}  \wedge Q_{(|\Sset_1|;|\Sset_2|-n;0)}\,.
 \end{align}

A generalization of \eqref{eq:103} arises when we relax the condition
\(t\in\{p-2n,p-2n-2,\dots,-p+2n\}\) and allow \(t=\pm (p-2n+2)\); This
generalization reads
\begin{gather}
\sum_{|A|=n} \HQ{A;\emptyset;\Sset_3} Q_{A;\emptyset;\Sset_3}^{[t]}=(-1)^n\sum_{|B|=n}
  Q_{\emptyset;B;\Sset_3}^{[t]} \HQ{\emptyset;B;\Sset_3}+\varepsilon_\pm\sum_{|B|=n-1}
  Q_{\emptyset;B;\Sset_3}^{[t\mp 1]} (\HQ{\emptyset;B;\Sset_3})^\pm\,,\nonumber\\
\textrm{ where}\ \ \ p=|\Sset_1|+|\Sset_2|-|\Sset_3|\,,\quad n\le \frac{p+1}2\,,\quad
t=\pm(p-2n+2)\label{eq:111}\\\nonumber
\textrm{and}\ \ \ \varepsilon_+=(-1)^{|\Sset_1|+|\Sset_2|+1}\,\qquad\qquad\varepsilon_-=(-1)^{|\Sset_3|}\,.
\end{gather}
Obviously, it follows from (\ref{eq:103}) that the relation
(\ref{eq:111}) can also be written as
\begin{gather}
\sum_{|A|=n} \HQ{A;\emptyset;\Sset_3} Q_{A;\emptyset;\Sset_3}^{[t]}=(-1)^n\left(\sum_{|B|=n}
  Q_{\emptyset;B;\Sset_3}^{[t]} \HQ{\emptyset;B;\Sset_3}-\varepsilon_\pm\sum_{|A|=n-1}
  Q_{A;\emptyset;\Sset_3}^{[t\mp 1]} (\HQ{A;\emptyset;\Sset_3})^\pm\right)\,,\nonumber\\
\textrm{ where}\ \ \ p=|\Sset_1|+|\Sset_2|-|\Sset_3|\,,\quad n\le \frac{p+1}2\,,\quad
t=\pm(p-2n+2)\label{eq:116}\\\nonumber
\textrm{and}\ \ \ \varepsilon_+=(-1)^{|\Sset_1|+|\Sset_2|}\,\qquad\qquad\varepsilon_-=(-1)^{|\Sset_3|}\,.
\end{gather}

\subsection*{%
Proof}
\label{sec:index-splitting-1}
Let us now show that if Q-functions obey (\ref{eq:5},\ref{eq:4}), then the relation (\ref{eq:103},\ref{eq:115}) holds.
First, one can note that \eqref{eq:103} is invariant under the gauge
transformation \eqref{eq:142}, hence it is sufficient to show that it holds
when \(Q_\emptyset=1\), i.e. when the Q-functions obey
\eqref{eq:143}. In order to simplify the notations for equations like
\eqref{eq:143}, we will use the notation \(F^{[a]\dots[b]}\equiv
 F^{[a]}\wedge F^{[a-2]} \wedge \dots \wedge F^{[b]}\) (for
 \(a-b\in2\mathbb{N}\)), with the convention%
\(F^{[a]\dots[a+2]}\equiv
 1\). %
Then, we obtain
\begin{align}\label{eq:112}
\star\left( \sum_{|A|=n}  \HQ{A;\emptyset;\Sset_3}  Q_{A;\emptyset;\Sset_3}^{[t]}
\right)&=
  \sum_{|A|=n}  Q_{\bar A ;\Sset_2;\emptyset} \epsilon^{\bar A \Sset_2A\Sset_3} Q_{A;\emptyset;\Sset_3}^{[t]}
\bg_\Bset=
  Q_{(|\Sset_1|-n;|\Sset_2|;0)} \wedge Q_{(n;0;|\Sset_3|)}^{[t]} \\&=\label{eq:113}
  \sum_{k=1}^{|\Sset_1|}  Q_{(k;|\Sset_1|+|\Sset_2|-n-k;0)} \wedge
 \sum_{k'=1}^{|\Sset_1|}
  Q_{(k;0;|\Sset_3|+n-k)}^{[t]}\\&=%
  {(Q_{(1;0;0)}+Q_{(0;1;0)})^{[|\Sset_1|+|\Sset_2|-n-1]\dots[-|\Sset_1|-|\Sset_2|+n+1]}}%
\nonumber\\&\qquad\qquad \wedge %
{(Q_{(1;0;0)}+Q_{(0;0;1)})^{[t+|\Sset_3|+n-1]\dots[t-|\Sset_3|-n+1]}}{
\label{eq:114}
}%
\end{align}
where (\ref{eq:112}) uses (\ref{eq:75}) to rewrite the l.h.s. in terms
of forms, and (\ref{eq:113}) is a key argument that the sums in the
r.h.s. vanish if \(k\neq |\Sset_1|-n\) or \(k'\neq n\),
because an expression of
degree \(|\Bset|\) in the \(\bg_a\)'s must contain each \(\bg_a\) exactly
once. Finally the expression (\ref{eq:114}) is obtained by using
(\ref{eq:5}) (with the substitution \(\Bset\leadsto \Sset_1\sqcup\Sset_2\),
hence \(Q_{(1)}\leadsto Q_{(1;0;0)}+Q_{(0;1;0)}\)) to express
\(\sum_{k=1}^{|\Sset_1|}  Q_{(k;|\Sset_1|+|\Sset_2|-n-k;0)}\) (and
analogously for \(\sum_{k'=1}^{|\Sset_1|}
  Q_{(k;0;|\Sset_3|+n-k)}^{[t]}\)).

\begin{figure}
  \centering
  \includegraphics{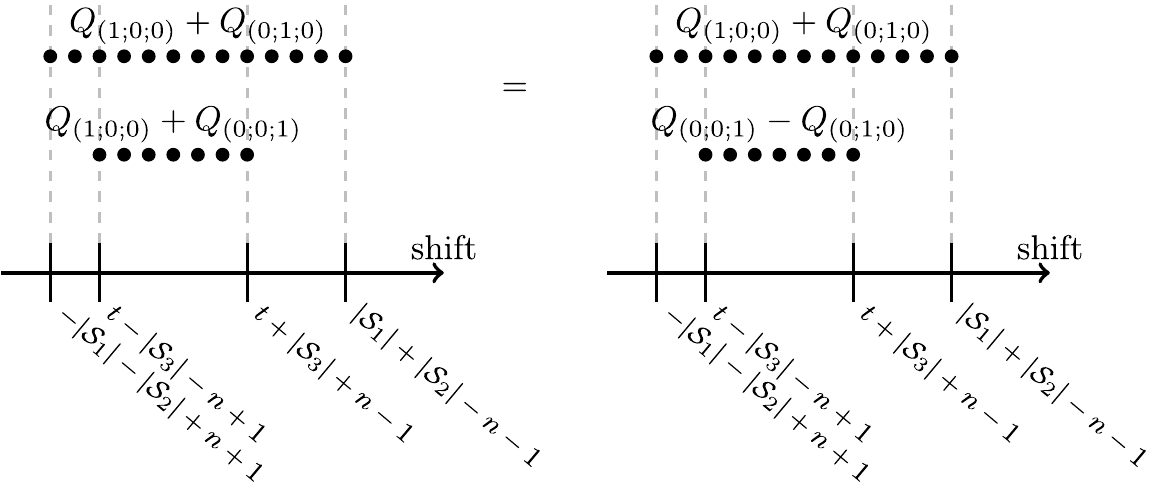}
  \caption{Graphical representation of the equality of the expressions
    \eqref{eq:114} (left) and \eqref{eq:144} (right).
Each product \(F^{[a]\dots[b]}\equiv
 F^{[a]}\wedge F^{[a-2]} \wedge \dots \wedge F^{[b]}\) is represented
 by the expression of \(F\) above a series of dots with horizontal
 positions \(a\), \(a-2\), \(\dots\), \(b\). The equality reduces to the
 statement that for a fixed shift \(s\), one has
 \((Q_{(1;0;0)}^{[s]}+Q_{(0;1;0)}^{[s]})\wedge
 (Q_{(1;0;0)}^{[s]}+Q_{(0;0;1)}^{[s]})=
(Q_{(1;0;0)}^{[s]}+Q_{(0;1;0)}^{[s]})\wedge (Q_{(0;0;1)}^{[s]}-Q_{(0;1;0)}^{[s]})\).
}
  \label{fig:Splitting}
\end{figure}

The r.h.s. of \eqref{eq:114} can be graphically represented as the
l.h.s. of figure \ref{fig:Splitting}.

 At this point, we can notice that if
 \(t\in\{p-2n,p-2n-2,\dots,-p+2n\}\), then
the set of the shifts
\(\{t+|\Sset_3|+n-1,t+|\Sset_3|+n-3,\dots,t-|\Sset_3|-n+1\}\) of
\(Q_{(1;0;0)}+Q_{(0;0;1)}\) in the second factor of (\ref{eq:114})
  is a subset of the set of the shifts
\(\{|\Sset_1|+|\Sset_2|-n-1,|\Sset_1|+|\Sset_2|-n-3,\dots,-|\Sset_1|-|\Sset_2|+n+1\}\) of %
the first factor%
.
 The antisymmetry of the ``wedge'' product %
 hence allows to subtract \(Q_{(1;0;0)}+Q_{(0;1;0)}\) to each \(Q_{(1;0;0)}+Q_{(0;0;1)}\) in the second factor, so that we get
\begin{align}
\label{eq:144}\star\left( \sum_{|A|=n}  \HQ{A;\emptyset;\Sset_3}  Q_{A;\emptyset;\Sset_3}^{[t]}
\right)&=
{(Q_{(1;0;0)}+Q_{(0;1;0)})^{[|\Sset_1|+|\Sset_2|-n-1]\dots[-|\Sset_1|-|\Sset_2|+n+1]}}{
}\nonumber\\&\qquad\qquad \wedge %
{(Q_{(0;0;1)}-Q_{(0;1;0)})^{[t+|\Sset_3|+n-1]\dots[t-|\Sset_3|-n+1]}}%
\\&=
  \sum_{k=1}^{|\Sset_1|}  Q_{(k;|\Sset_1|+|\Sset_2|-n-k;0)} \wedge
 \sum_{k'=1}^{|\Sset_2|} (-1)^{k'}
  Q_{(0;k';|\Sset_3|+n-k')}^{[t]}
\\&=(-1)^n  \star\left(  \sum_{|B|=n}
  Q_{\emptyset;B;\Sset_3}^{[t]}
  \HQ{\emptyset;B;\Sset_3} \right),
\end{align}
which proves (\ref{eq:103}).%

Let us now show that if
  \(t=\pm(2+p-2n)\)%
, the same arguments lead to the three-terms relation (\ref{eq:111}).
Let us show this for the \(t=-(2+p-2n)\) case:
\begin{align}
 \star\left(   \sum_{|A|=n} \HQ{A;\emptyset;\Sset_3} Q_{A;\emptyset;\Sset_3}^{[t]}
  \right)&=Q_{(|\Sset_1|-n;|\Sset_2|;0)}\wedge Q_{(n;0;|\Sset_3|)}^{[t]}\\&=%
  {(Q_{(1;0;0)}+Q_{(0;1;0)})^{[|\Sset_1|+|\Sset_2|-n-1]\dots[-|\Sset_1|-|\Sset_2|+n+1]}}{
}\nonumber\\&\qquad\qquad \wedge %
{(Q_{(1;0;0)}+Q_{(0;0;1)})^{[t+|\Sset_3|+n-1]\dots[t-|\Sset_3|-n+3]}}{
}\nonumber\\&\qquad\qquad \qquad\wedge %
{(Q_{(1;0;0)}+Q_{(0;0;1)})^{[t-|\Sset_3|-n+1]}}%
\,.
\intertext{In this expression, both the second and the third factor originate
from \(Q_{(n;0;|\Sset_3|)}^{[t]}\), but these factors are such that
(when \(t=-(2+p-2n)\)) the
shifts of \(Q_{(1;0;0)}+Q_{(0;0;1)}\) in the second factor is a subset
of the shifts of \(Q_{(1;0;0)}+Q_{(0;1;0)}\) in the first
factor. Consequently, one can subtract \(Q_{(1;0;0)}+Q_{(0;1;0)}\) to each
\(Q_{(1;0;0)}+Q_{(0;0;1)}\) in the second factor -- exactly like we did
when showing the equality of \eqref{eq:114} and \eqref{eq:144}.
Moreover, the
condition \(t=-(2+p-2n)\) implies
\(t-|\Sset_3|-n+1=-|\Sset_1|-|\Sset_2|+n-1\), and we can rewrite the
last term as
\(%
{(Q_{(1;0;0)}+Q_{(0;1;0)})^{[-|\Sset_1|-|\Sset_2|+n-1]}}%
+
{(Q_{(0;0;1)}-Q_{(0;1;0)})^{[t-|\Sset_3|-n+1]}}%
\):
}
&=%
{(Q_{(1;0;0)}+Q_{(0;1;0)})^{[|\Sset_1|+|\Sset_2|-n-1]\dots[-|\Sset_1|-|\Sset_2|+n+1]}}{
}\nonumber\\&\qquad\qquad \wedge %
{(Q_{(0;0;1)}-Q_{(0;1;0)})^{[t+|\Sset_3|+n-1]\dots[t-|\Sset_3|-n+3]}}{
}\label{eq:110}\\&%
\hspace{-1cm}\wedge \left(%
  {(Q_{(1;0;0)}+Q_{(0;1;0)})^{[-|\Sset_1|-|\Sset_2|+n-1]}}%
  +
{(Q_{(0;0;1)}-Q_{(0;1;0)})^{[t-|\Sset_3|-n+1]}}%
\right)\nonumber\\&=(-1)^{|\Sset_3|}Q_{(|\Sset_1|;|\Sset_2|-n+1;0)}^-\wedge Q_{(0;n-1;|\Sset_3|)}^{[t+1]}\nonumber\\&\qquad\qquad
+(-1)^n Q_{(|\Sset_1|;|\Sset_2|-n;0)}\wedge Q_{(0;n;|\Sset_3|)}^{[t]}\,,\label{eq:109}
\end{align}
where the first (resp. second) term of (\ref{eq:109}) is obtained by keeping the
first (resp. second) term of the last factor in (\ref{eq:110}). Finally, (\ref{eq:109}) proves
\begin{align}
\sum_{|A|=n} \HQ{A;\emptyset;\Sset_3} Q_{A;\emptyset;\Sset_3}^{[t]}=(-1)^n\sum_{|B|=n}
  Q_{\emptyset;B;\Sset_3}^{[t]} \HQ{\emptyset;B;\Sset_3}+(-1)^{|\Sset_3|}\sum_{|B|=n-1}
  Q_{\emptyset;B;\Sset_3}^{[t+1]} (\HQ{\emptyset;B;\Sset_3})^-\,,
\end{align}
which is the \(t=-(2+p-2n)\) case of the relation (\ref{eq:111}). The
proof of the  \(t=2+p-2n\) case is identical.

\subsection{Example of a non-Wronskian solution to the Hirota equation}
\label{sec:example-non-wronsk}

In the main text, we wrote the generic solution to Hirota equation,
but one should keep in mind that there exist degenerate solutions of
Hirota equations which do not have a Wronskian form.

As an example, for any \(G_1,G_2\in \SU(3)\), the function
\begin{align}
  T_{a,s}(u)&=
  \begin{cases}
    1&\textrm{ if }a=0\textrm{ or }a=3\\
    \chi_{a,s}(G_1)&\textrm{ if } s\ge 0\textrm{ and }0\le a\le 3\\
    \chi_{a,-4-s}(G_2)&\textrm{ if } s\le -4\textrm{ and }0\le a\le 3\\
    0&\textrm{ otherwise}
  \end{cases}
\end{align}
obeys the Hirota equation on the infinite strip of figure
\ref{fig:InfStrip} (with size \(N=3\)), but it cannot be written in the
form \eqref{eq:87}. Indeed, if \(T\) could be written in the
form \eqref{eq:87}, then by applying the arguments of \secref{sec:ident-q-funct} at \(s_0=-3\), the functions \(Q_1\), \(Q_2\) and \(Q_3\)
would obey
\begin{align}
  \label{eq:70}
    \begin{vmatrix}
    Q_i & Q_i^{[+2]}& Q_i^{[+4]}& Q_i^{[+6]}\\
0&0&0&1\\
0&0&1&\mathrm{tr}(g_1)\\
0&1&\mathrm{tr}(g_1)&\chi_{1,2}(g_1)
  \end{vmatrix}
&=0.
\end{align}
This would imply that \(Q_1=Q_2=Q_3=0\), which contradicts the fact that
\(T_{1,0}\neq 0\).

\subsection{\label{sec:ONBaxproof}Derivation of the Baxter equation in the form \texorpdfstring{\eqref{eq:ONBax}}{(\ref*{eq:ONBax})}}
Consider first the operator \(\tilde{\mathcal{O}}(u;a,s)={T_{a,s}}\,e^{\frac i2\partial_u-\partial_a}\frac 1{T_{a,s}}- T_{a,s-1}^+ \,e^{+\partial_s-\partial_a}\frac 1{T_{a,s-1}^+}\). It was constructed to have the property 
\be
F_{a,s}(u)=\tilde{\mathcal{O}}(u;a,s) F_{a,s}(u)\,,
\ee
which can be checked using \eqref{eq:125}\,. First, we know that \(F_{N,s}=0\). On the other hand, \(\tilde{\mathcal{O}}^N F_{a,s}\) evaluated at \(a=N\) is a certain linear combination of \(F_{0,s+k}^{[N-k]}\) with \(k=0,1,\ldots,N\). Using \(F_{0,s}=\pm Q_{\emptyset}^{[s]}(Q^{b})^{[-s-N+1]}\), one can write
\be
0=\big(\tilde{\mathcal{O}}^N F_{a,s}\big)_{|a=N}=Q_{\emptyset}^{[s+N]}\,\tilde{\mathcal{O}}^N (Q^b)^{[-s-N+1]}{}_{|a=N}\,.\ee
Thus we derived the Baxter equation \(\tilde{\mathcal{O}}^N W(u-\frac {\bi\,s}2)_{|a=N}=0\) which is solved by  \(W=(Q^b)^{[-N+1]}\). To get from here the Baxter equation \eqref{eq:ONBax} which is solved by \(Q_b\), we simply note that, according to \eqref{eq:92}, substitution \(T_{a,s}\to T_{a,-N-s}\) changes the role of \(Q_b\) and \(Q^b\).

\subsection{Proofs of [derivative] QQ-relations in a supersymmetric Q-system}
\label{sec:proofs}
Here we prove the relations
(\ref{eq:45}-\ref{eq:94}).

\paragraph{Proof of (\ref{eq:45})}%
\label{sec:proof-}

We will prove the recurrence relation
\begin{align}\label{eq:100}
 Q_{(n|n)}\wedge Q_{(1|1)}&=(-1)^n(n+1)Q_{\emptyset}Q_{(n+1|n+1)}\,,
\end{align}
which is equivalent to (\ref{eq:45}). For simplicity, we first
assume that \(|\Bset|=|\Fset|=n+1\). In this case, we obtain
(\ref{eq:100}) as follows:
\begin{equation}
  \begin{aligned}
  \star\left(Q_{(n|n)}\wedge
    Q_{(1|1)}\right)&=(-1)^n\sum_{\substack{a\in\Bset\\i\in\Fset}}
  \epsilon_{\bar a a} \epsilon_{\bar i i} Q_{\bar a|\bar i} Q_{a|i}=
(-1)^n\sum_{\substack{a\in\Bset\\i\in\Fset}}
  \epsilon_{\bar a a} \epsilon_{i \bar i} {\bzdQ}_{\bar a|i} {\bzdQ}_{a|\bar i}\\&=-\sum_{\substack{a\in\Bset\\i\in\Fset}}
  \epsilon_{ a \bar a} \epsilon_{\bar i i} \star\left({\bzdQ}_{(n+1)}\wedge\xi_a\wedge\xi_{\bar i}\right) \star\left({\bzdQ}_{(n+1)}\wedge\xi_{\bar a}\wedge\xi_{i}\right)
\\&=n\,\sum_{\substack{a\in\Bset\\j\in\Fset}}
  \epsilon_{ a \bar a} \epsilon_{\bar j j} \star\left({\bzdQ}_{(n+1)}\wedge\xi_a\wedge\xi_{\bar j}\right) \star\left({\bzdQ}_{(n+1)}\wedge\xi_{\bar a}\wedge\xi_{j}\right)\\&\qquad-
\sum_{\substack{a\in\Bset\\i\in\Fset}}
  \epsilon_{a \bar a} \epsilon_{\bar i i} \star\left({\bzdQ}_{(n+1)}\wedge\xi_i\wedge\xi_{\bar i}\right) \star\left({\bzdQ}_{(n+1)}\wedge\xi_{\bar a}\wedge\xi_{a}\right)\,.
  \end{aligned}
\end{equation}
The last equality is the Pl\"ucker identity\footnote{In order to
write this Pl\"ucker identity, it is important to note that
\({\bzdQ}_{(n+1)}=\frac{{\bzdQ}_{(1)}^{[n]}\wedge {\bzdQ}_{(1)}^{[n-2]}\wedge
\dots \wedge {\bzdQ}_{(1)}^{[-n]}}{\prod_{k=1}^{n-1}{\bzdQ}_{\emptyset}^{[-n-1+2k]}}\).
}: the last term
in the r.h.s. corresponds to the exchange
\(\xi_{i}\leftrightarrow\xi_{a}\) and the other term corresponds to the
exchange \(\xi_i\leftrightarrow\xi_j\) with \(j\in\bar i\) (\(\xi_j\)
appears in the product \(\xi_{\bar i}=\xi_{j_1}\wedge\xi_{j_2}\wedge\dots\)
where \(\bar i=j_1,j_2 \dots\)). Noticing that in this last equality,
the first term of the r.h.s. is equal to the l.h.s. (up to a factor
\(-n\)), we obtain\footnote{One can note that on the first line of
  \eqref{eq:152}, all terms of the sum in the r.h.s. are equal, hence
  the second equality.}
\begin{equation}
  \begin{aligned}\label{eq:152}
  \star\left(Q_{(n|n)}\wedge
    Q_{(1|1)}\right)&=-\frac 1 {n+1}
\sum_{\substack{a\in\Bset\\i\in\Fset}}
  \epsilon_{a \bar a} \epsilon_{\bar i i} \star\left({\bzdQ}_{(n+1)}\wedge\xi_i\wedge\xi_{\bar i}\right) \star\left({\bzdQ}_{(n+1)}\wedge\xi_{\bar a}\wedge\xi_{a}\right)\,\\
&=\frac {(-1)^{n}} {n+1}(n+1)^2{\bzdQ}_{\Bset}{\bzdQ}_{\Fset}=%
{(-1)^{n}} {(n+1)} Q_{\emptyset} Q_{\bar \emptyset}\\&=%
{(-1)^{n}} {(n+1)} \star\left(Q_{\emptyset}  Q_{(n+1|n+1)}\right)\,,
  \end{aligned}
\end{equation}
which proves (\ref{eq:45}) when \(|\Bset|=|\Fset|=n+1\).

In order to show that (\ref{eq:45}) holds
also when \(|\Bset|>=n+1\) or \(|\Fset|>=n+1\), we simply use the fact
that the QQ-relations are not sensible to the numbers \(|\Bset|\)
and \(|\Fset|\) of indices.

First, %
we rewrite (\ref{eq:45}) %
in terms of
coordinates as
\begin{align} \label{eq:101}
  Q_{A|I}&=\frac 1 {n+1} \sum_{\substack{a\in A\\i\in I}}
  Q_{(A\setminus a)|(I\setminus i)} Q_{a|i} \epsilon_{(A\setminus a)
    a} \epsilon_{(I\setminus I) i}\,,&\textrm{ where }&|A|=|I|=n
\end{align}
where \((A\setminus a)\) (resp \((I\setminus i)\)) denotes a multi-index
corresponding to the set \(\{A\}\setminus\{a\}\) (resp
\(\{I\}\setminus\{i\}\)). Equation \eqref{eq:101} holds without
condition on \(|\Bset|\) and \(|\Fset|\), because for any \(A\) and \(I\), we
can restrict QQ-relations to the subset of the Q-functions \(Q_{B,J}\)
where \(B\subset A\) and \(J\subset I\) -- in other words we set
\(\Bset=A\) and \(\Fset=I\). For this subset of Q-functions, the above
proof holds, because we have artificially enforced
\(|\Bset|=|\Fset|=n+1\).

Hence we have proven
(\ref{eq:100}) for arbitrary \(n\) -- without any condition on \(|\Bset|\)
and \(|\Fset|\). This means that we have proven
(\ref{eq:45}).

\paragraph{Proof of (\ref{eq:46}-\ref{eq:47})}

By the same argument as above, we can assume that
\(|\Bset|=n\) and
\(|\Fset|=p\) without loss of generality (using the fact that the
relation (\ref{eq:comp46}) is  not sensible to \(|\Bset|\) and \(|\Fset|\)).

Then, we have %
\begin{align}
    {Q_{(p|p)}^{[t]}}%
  \wedge Q_{(n-p|0)}&=(-1)^{p(n-p)} {\bzdQ}_{(p;0)}^{[t]}  \wedge {\bzdQ}_{(n-p;p)}=(-1)^{n\,p} {\bzdQ}_{(0;p)}^{[t]}  \wedge {\bzdQ}_{(n;0)}=Q_{(n|p)} Q_\emptyset^{[t]}\,,
\end{align}
where the first equality uses the bosonization trick (\ref{eq:96}) and the
second equality is the relation (\ref{eq:115}) with \(\Sset_1=\Bset\),
\(\Sset_2=\Fset\) and \(\Sset_3=\emptyset\) and \(n=p\).
This
proves (\ref{eq:46}).

Obviously
,
(\ref{eq:47}) is %
obtained by exchanging the roles of bosonic and
fermionic indices (see (\ref{eq:comp46}-\ref{eq:comp47})).

\paragraph{Proof of (\ref{eq:67})}
If we set \(A=\Bset\) and \(I=\bar J\) in (\ref{eq:comp46}), then we get
\begin{align}
  \HQ{\emptyset|J}=\sum_{|C|=n}\frac{\HQs{C|J}{[t]}}{Q_\emptyset^{[t]}}Q_{C|\emptyset}\,.
\end{align}
Remembering that the functions \(\HQ{A|I}\) obey exactly the same
QQ-relation as \(Q_{A|I}\), we can substitute \(Q_{A|I}\to \HQ{A|I}\)
(and hence \(\HQ{A|I}\to (-1)^{|A|\,|\bar I|}\epsilon^{\bar A A}\epsilon^{\bar I I} \HQ{\bar A|\bar I}=(-1)^{(|A|+|I|)(|\bar A|+|\bar I|)}Q_{A|I}\)) and get exactly (\ref{eq:67}).

\paragraph{Proof of (\ref{eq:84})}\label{sec:proof-refeq:84}
Let us see how the equation (\ref{eq:84}) arises from (\ref{eq:45}):
let \(|I|\) and \(|J|\) be fermionic multi-indices such that
\(|J|-|I|=|{\Fset}|-|{\Bset}|\ge 0\)%
, then
\begin{align}
\sum_{|A|=|I|}Q_{A|I}\HQ{A|J}&=\sum_{|A|=|I|} (-1)^{|A|\,|\bar A|}
\epsilon^{\bar A A} \epsilon^{\bar J J} Q_{A|I}Q_{\bar A|\bar J}\intertext{where one can note that, on the r.h.s., \(|A|=|I|\) and \(|\bar A|=|\bar
J|\), which allows to use (\ref{eq:53}) to express the Q-functions
of the r.h.s, and get
}&=\frac{\epsilon_{\bar J J}}{(Q_\emptyset)^{|\Bset|-2}}\Det_{\substack{a\in\Bset\\i\in I\bar J}}Q_{a,i}\,.
\label{eq:98}
\end{align}
By comparison, we can now study the r.h.s. of (\ref{eq:84}): there is
at most one non-vanishing term in the sum in the r.h.s: the term where
\(\{L\}=\{J\}\setminus\{I\}\), if \(\{I\}\subset\{J\}\). If
\(\{I\}\not\subset\{J\}\),  then the r.h.s. is zero, while the expression
(\ref{eq:98}) of the l.h.s. is also zero because two columns of the
determinant are equal. If \(\{I\}\subset\{J\}\), let us denote by \(L\)
the sorted multi-index such that \(\{L\}=\{J\}\setminus\{I\}\). Then we
have:
\begin{align}
  \HQ{\emptyset|L}Q_{\emptyset|\emptyset}&=\epsilon^{I\bar
    JL}Q_{\overline \emptyset|I\bar J}Q_{\emptyset|\emptyset}=(-1)^{|I|\,|\bar J L|}\epsilon^{\bar
    JLI}Q_{\overline \emptyset|I\bar J}Q_{\emptyset|\emptyset}\nonumber\\&=(-1)^{|I|\,|\bar J
    L|}\delta^{LI}_{J}\epsilon^{\bar J J}Q_{\overline \emptyset|I\bar
    J}Q_{\emptyset|\emptyset}=(-1)^{|I|(|\Fset|+1)} \delta^{LI}_{J}\epsilon^{\bar J J}Q_{\overline \emptyset|I\bar
    J}Q_{\emptyset|\emptyset}\label{eq:99}
\end{align}
Then, from  \eqref{eq:99}, \eqref{eq:53} and \eqref{eq:98} we get (\ref{eq:84}).

\paragraph{Proof of (\ref{eq:93})}\label{sec:proof-refeq:93}

Denoting \(n\equiv|\Bset|-|\Fset|\) and assuming that \(n>0\), %
one gets
\begin{align}
  \sum_{a\in \Bset}Q_a^{[+t]}\HQ a&=
  \sum_{a\in \Bset} Q_a^{[+t]} (-1)^{|\Fset|} \epsilon^{\bar a a}
  Q_{\bar a|\bar \emptyset}=(-1)^{|\Fset|+|\Bset|-1}
  Q_{(1|0)}^{[+t]}\wedge
  Q_{\left(|\Bset|-1\middle||\Fset|\vphantom{\bar J}\right)}\nonumber\\&=(-1)^{n+1}
  Q_{(1|0)}^{[+t]}\wedge
Q_{\left(n-1\middle|0\right)}\wedge \frac{Q_{\left(\vphantom{\bar J}|\Fset|\middle||\Fset|\right)}^{[n-1]}}{Q_{\emptyset}^{[n-1]}}\,.\label{eq:105}
\end{align}
One can note that since \(Q_{(n-1|0)}\) does not involve fermionic
indices, it is given by \(Q_{(n-1|0)}=\frac{Q_{(1|0)}^{[n-2]}\wedge
  Q_{(1|0)}^{[n-4]}\wedge \dots\wedge
  Q_{(1|0)}^{[2-n]}}{\prod_{k=1}^{n-2}Q_{(\emptyset)}^{[-n+1+2k]}}\),
as in (\ref{eq:5}). Hence, t%
he r.h.s. of (\ref{eq:105}) vanishes if \(t\in \{n-2,n-4,\dots,-n+2\}\) because \(Q_{(1|0)}^{[+t]}\wedge
Q_{\left(n-1\middle|0\right)}=0\). By contrast, if \(t=\pm n\) then we
get
\begin{align}
  \sum_{a\in \Bset}Q_a^{[\pm n]}\HQ a&=(-1)^{n+1}
  Q_{\emptyset}^{[\pm (n- 1)]}
Q_{\left(n\middle|0\right)}^{\pm}\wedge \frac{Q_{\left(\vphantom{\bar
        J}|\Fset|\middle||\Fset|\right)}^{[n-1]}}{Q_{\emptyset}^{[n-1]}}\nonumber\\
&=(-1)^{n+1}Q_{\emptyset}^{[\pm (n- 1)]} \HQs\emptyset\pm\,.
\end{align}

\paragraph{Proof of (\ref{eq:94})}\label{sec:proof-refeq:94}
The relation (\ref{eq:94}) is a particular case of (\ref{eq:111}): if
we assume that \(|\Bset|=|\Fset|\) then we have
\begin{align}
 \HQs\emptyset - Q_\emptyset^+ &=
{\bzdQ}_{\Bset;\emptyset}^-  {\bzdQ}_{\emptyset;\Fset}^+
=(-1)^{|\Fset|} {\bzdQ}_{\Bset;\emptyset}^-  \hHQs {\Bset;\emptyset} +
\nonumber \\&= {\bzdQ}_{\emptyset;\Fset}^-  \hHQs {\emptyset;\Fset} +
- \sum_{b\in\Bset} {\bzdQ}_{\bar b;0} \hHQ {\bar b;0}
\nonumber\\&= Q_{\emptyset}^-  \HQs {\emptyset} +
+ \sum_{b\in\Bset} Q_{b|\emptyset} \HQ {b|\emptyset}\,,
\end{align}
where the third equality is the relation (\ref{eq:116}) with
\(\Sset_1=\Bset\), \(\Sset_2=\Fset\), \(\Sset_3=\emptyset\),
\(n=|\Bset|=|\Fset|\) and \(t=-2\), and in the fourth equality we substitute
\(\hHQ{\bar b;0}=\epsilon^{b\bar b}Q_{b|\emptyset}\) and \({\bzdQ}_{\bar
  b;0}=(-1)^{|\Fset|}\epsilon^{\bar b b}\HQ {b|\emptyset}\) to get (\ref{eq:94}).

\subsection{Proof of the Wronskian solution of Hirota on the {\THook}}
\label{sec:proof-that-Hirota}

We will now show that (\ref{eq:60}-\ref{eq:62}) solves the Hirota
equation. This means that we will have to prove the five following statements:
{\renewcommand{\theenumi}{\alph{enumi}}
\renewcommand{\labelenumi}{(\theenumi)}
\begin{enumerate}
\item \label{item:1}The expressions (\ref{eq:60}) and (\ref{eq:62})
  coincide when
\(\tilde a\le\tilde s\le -\tilde a\)
  , i.e. on the intersection of the
  ``right strip'' and the ``left strip'' of figure
  \ref{fig:THook-strips}.
\item \label{item:2}The expressions (\ref{eq:60}) and (\ref{eq:61})
  (resp. (\ref{eq:61}) and (\ref{eq:62})) coincide when
\(\tilde a=\tilde s\ge 0\) (resp. \(\tilde a=-\tilde s\ge 0\))%
  , i.e. on the diagonal joining the
  ``right strip''  (resp the ``left strip'') to the ``upper strip''of figure \ref{fig:THook-strips}.
\item \label{item:5}The Hirota equation is satisfied outside the
  diagonals, i.e. when \(|\tilde s|\neq \tilde a\).
\item \label{item:4}The Hirota equation is satisfied when
  \(\tilde a=\tilde s> 0\) and when \(\tilde a=-\tilde s> 0\),
   i.e. on all the diagonals defining
  the border of the ``upper strip''of figure \ref{fig:THook-strips}
  except their intersection.
\item \label{item:3}If the point \((s_0,a_0)\) (where the diagonals intersect) belongs to
  the {\THook} (i.e. if \({\Kone}+{\Ktwo}\ge m\) and \(k+m\) is odd), the Hirota equation is
  satisfied at this point as well.
\end{enumerate}}
 The statements (\ref{item:1}) and (\ref{item:2}) mean that
 (\ref{eq:60}-\ref{eq:62}) defines indeed a function \(T_{a,s}\),
 whereas the statements (\ref{item:5}-\ref{item:3}) mean that this function
 obeys the Hirota equation.

 \paragraph{\ref{item:1}): Intersection of the right and left strips}
As we will detail now, the relation (\ref{eq:115}) immediately implies
that the expressions
(\ref{eq:60}) and (\ref{eq:62}) coincide on the intersection of the right and left strips:
Assuming that %
\(\tilde a\le\tilde s\le -\tilde a\), i.e. \(a-\frac{K-M}2\le\tilde s\le-a+\frac{K-M}2\)%
,
we have
\begin{align}
  Q_{(a,0|0)}^{[\tilde s]}\wedge
Q_{({\Kone}-a,{\Ktwo}|M)}^{[-\tilde s]}&=(-1)^{M(K-a)}{\bzdQ}_{(a;0;M)}^{[\tilde s]}\wedge {\bzdQ}_{({\Kone}-a,{\Ktwo}|0)}^{[-\tilde s]}
\\
&=(-1)^{M(K-a)+a}{\bzdQ}_{(0;a;M)}^{[\tilde s]}\wedge {\bzdQ}_{({\Kone},{\Ktwo}-a|0)}^{[-\tilde s]}
\\&=(-1)^{a%
}
Q_{(0,a|0)}^{[+\tilde s  ]}\wedge
Q_{({\Kone},{\Ktwo}-a|M)}^{[-\tilde s ]}\\&=(-1)^{a(K+M)}
Q_{({\Kone},{\Ktwo}-a|M)}^{[-\tilde s ]}\wedge
Q_{(0,a|0)}^{[+\tilde s  ]}\,,
\label{eq:107}
\end{align}
where the first and third equalities are the bosonization trick, and
the second equality is \eqref{eq:115}, with \(\Sset_1=\Bset_2\),
\(\Sset_2=\Fset\), \(\Sset_3=\Bset_1\),  \(n={\Ktwo}-a\), and \(t=-2\tilde s\); in this
equality, the condition \(t\in\{p-2n,p-2n-2,\dots,-p+2n\}\) is exactly
the condition \(\tilde a\le\tilde s\le -\tilde a\) which defines the
intersection of the left and write strip.

From (\ref{eq:107}) we see that the expressions
(\ref{eq:60}) and (\ref{eq:62}) coincide on the
intersection of the left and write strip, since
\(\varepsilon_l(a,s) = (-1)^{a(K+M)}\varepsilon_u(a,s)\).

 \paragraph{\ref{item:2}): Diagonals delimiting the upper strip}
For the simplicity we will focus on the diagonal between the upper
strip and the right strip
(i.e. the case \(a= s+{\Kone}-{\Mone}\ge
  -s+{\Ktwo}-{\Mtwo}\)) while the result for other diagonal
(i.e. \(a=
  -s+{\Ktwo}-{\Mtwo} \ge s+{\Kone}-{\Mone}\))
has an identical proof. Assuming that \(a= s+{\Kone}-{\Mone}\ge
  -s+{\Ktwo}-{\Mtwo}\), we have
  \begin{align}
    Q_{(a,0|0)}^{[\tilde s  ]}\wedge Q_{({\Kone}-a;{\Ktwo}|M)}^{[-\tilde s  ]}
    &=(-1)^{M(k+a)}{\bzdQ}_{(a;0;M)}^{[\tilde s  ]}\wedge{\bzdQ}_{({\Kone}-a;{\Ktwo};0)}^{[-\tilde s  ]}
\\&= (-1)^{M(k+a)+{\Kone}-a}{\bzdQ}_{({\Kone};0;M+a-{\Kone})}^{[\tilde s  ]}\wedge{\bzdQ}_{(0;{\Ktwo};{\Kone}-a)}^{[-\tilde s  ]}
\\&= (-1)^{M(k+a)+{\Kone}-a+M\,{\Ktwo}}Q_{({\Kone};0|{\Kone}-a)}^{[\tilde s  ]}\wedge{\bzdQ}_{(0;{\Ktwo}|M+a-{\Kone})}^{[-\tilde s  ]}\label{eq:106}
\\&= (-1)^{(M+1)({\Kone}+a)}Q_{({\Kone};0|{\Mone}-s)}^{[\tilde a]}\wedge{\bzdQ}_{(0;{\Ktwo}|{\Mtwo}+s)}^{[-\tilde a]}\,,
  \end{align}
where the first equality is the bosonization trick (\ref{eq:96}), the
second equality is the relation (\ref{eq:115}) where we choose
\(\Sset_1=\Bset_1\), \(\Sset_2=\Bset_2\), \(\Sset_3=\Fset\), \(n=a\) and
\(t=2\tilde s\). From  the condition \(a= s+{\Kone}-{\Mone}\ge
  -s+{\Ktwo}-{\Mtwo}\) characterizing the diagonal between the upper
strip and the right strip, one sees that  \(n\le
\frac{|\Sset_1|+|\Sset_2|-|\Sset_3|}2\) and
\(s=-|\Sset_1|+|\Sset_2|-|\Sset_3|+2n\), so that the conditions of
relation (\ref{eq:115}) do hold. Finally, the third equality
(\ref{eq:106}) is the bosonization trick (\ref{eq:96}) while the last
equality is a rewriting of the result (using the condition \(a= s+{\Kone}-{\Mone}\ge
  -s+{\Ktwo}-{\Mtwo}\)) to match the expression (\ref{eq:61}) of the
  T-functions in the upper-strip (one can note that on this diagonal \(\varepsilon_u(a,s)=(-1)^{(M+1)({\Kone}+a)}\varepsilon_r(a,s)\)).

 \paragraph{\ref{item:5}): Hirota equation outside the diagonals}
On each strip, the T-functions (\ref{eq:60}-\ref{eq:62}) have the form
(\ref{eq:87}), up to an irrelevant sign. Hence it is  clear that they
obey the Hirota equation (as long as the nodes \((a,s)\), \((a+1,s)\),
\((a-1,s)\), \((a,s+1)\) and \((a,s-1)\) lie on the same strip of the
{\THook}, i.e. as long as \((a,s)\) does not lie on a diagonal).

 \paragraph{\ref{item:4}): Hirota equation on the diagonals}

We will focus on the diagonal between the upper
strip and the right strip
(i.e. the case \(a= s+{\Kone}-{\Mone}\ge
  -s+{\Ktwo}-{\Mtwo}\)) while the result for other diagonal
(i.e. \(a=
  -s+{\Ktwo}-{\Mtwo} \ge s+{\Kone}-{\Mone}\))
has an identical proof. In the present proof of the Hirota equation,
we denote\footnote{The factor \(\varepsilon_r(a,s)\) appearing in
  equation \eqref{eq:156} was defined in \eqref{eq:157}.}
\begin{align}\label{eq:156}
  \tilde T_{a,s}&=\varepsilon_r(a,s)
\star \big(Q_{(a,0|0)}^{[\tilde s
  ]}\wedge
Q_{({\Kone}-a,{\Ktwo}|m)}^{[-\tilde s
  ]}\big)\,,
\end{align}
so that \(\tilde T_{a,s}=T_{a,s}\) if \(\tilde s\ge \tilde a\). It has the
form (\ref{eq:87}) (up to an irrelevant sign) hence it obeys the
Hirota equation \(\tilde T_{a,s}^+\tilde T_{a,s}^-=\tilde
T_{a+1,s}\tilde T_{a-1,s}+\tilde T_{a,s+1}\tilde T_{a,s-1}\). When
\((a,s)\) lies on the diagonal, we have \(\tilde T_{a,s}=T_{a,s}\),
\(\tilde T_{a,s+1}=T_{a,s+1}\) and \(\tilde T_{a-1,s}=T_{a-1,s}\), hence
\begin{align}
   T_{a,s}^+ T_{a,s}^-&=\tilde
T_{a+1,s} T_{a-1,s}+ T_{a,s+1}\tilde T_{a,s-1}\,,&\textrm{when
}\tilde a&=\tilde s\,.
\end{align}
Therefore, the Hirota equation \(T_{a,s}^+ T_{a,s}^-=
T_{a+1,s} T_{a-1,s}+ T_{a,s+1}T_{a,s-1}\) is equivalent to the
statement
\begin{align}
  \label{eq:28}
\left(\tilde T_{a+1,s}-T_{a+1,s}\right)T_{a-1,s} + T_{a,s+1}\left(\tilde
  T_{a,s-1}   -  T_{a,s-1}  \right)&=0\,,
\end{align}
which %
actually follows from
the relation\footnote{Suspicious readers may wonder whether the
  relations (\ref{eq:108}) mean that we are only describing very
  specific (somehow degenerate) solutions of the Hirota equation. This
  is actually not the case: any solution of the Hirota equation obeying the
  Wronskian gauge condition (\ref{eq:54}) has to obey the relations (\ref{eq:108}).}
\begin{align}
  \label{eq:108}
  \tilde T_{a+1,s}-T_{a+1,s} &=-(-1)^{{\Kone}-a} T_{a,s+1}&
\tilde
  T_{a,s-1}   -  T_{a,s-1} &=(-1)^{{\Kone}-a}T_{a-1,s} &\textrm{when }\tilde
  a&=\tilde s\,%
.
\end{align}
This relation \eqref{eq:108} can be proven as follows
\begin{align}\label{eq:117}
 \star T_{a,s+1}&=\varepsilon_r(a,s+1)
Q_{(a,0|0)}^{[\tilde s+1]}\wedge
Q_{({\Kone}-a,{\Ktwo}|M)}^{[-\tilde s-1]}%
\\&=\varepsilon_r(a,s+1)(-1)^{M(K-a)}
{\bzdQ}_{(a;0;M)}^{[\tilde s+1]}\wedge
{\bzdQ}_{({\Kone}-a;{\Ktwo};0)}^{[-\tilde s-1]}%
\\&=\varepsilon_r(a,s+1)(-1)^{M(K-a)}\big(
(-1)^{{\Kone}-a}
{\bzdQ}_{({\Kone};0;M-{\Kone}+a)}^{[\tilde s+1]}\wedge
{\bzdQ}_{(0;{\Ktwo};{\Kone}-a)}^{[-\tilde s-1]}\nonumber\\&\hspace{4cm}
+(-1)^{K-a-1}
{\bzdQ}_{(a+1;0;M)}^{[\tilde s]}\wedge
{\bzdQ}_{({\Kone}-a-1;{\Ktwo};0)}^{[-\tilde s]}
\big)
\\&=\varepsilon_r(a,s+1)(-1)^{M(K-a)}\big(
(-1)^{{\Kone}-a+M{\Ktwo}}
Q_{({\Kone},0|{\Kone}-a)}^{[\tilde s+1]}\wedge
Q_{(0,{\Ktwo}|M-{\Kone}+a)}^{[-\tilde s-1]}\nonumber\\&\hspace{4cm}
+(-1)^{(K-a-1)(M+1)}
Q_{(a+1,0|0)}^{[\tilde s]}\wedge
Q_{({\Kone}-a-1,{\Ktwo}|M)}^{[-\tilde s]}
\big)
\\&=\varepsilon_r(a,s+1)\big((-1)^{(M+1)({\Kone}-a)}
\star T_{a+1,s}/\varepsilon_u(a+1,s)
\nonumber\\&\hspace{4cm}
+(-1)^{M+K-a+1} \star \tilde T_{a+1,s}/\varepsilon_r(a+1,s)
\big)
\\&=(-1)^{{\Kone}-a}\star \left(
T_{a+1,s}
-\tilde T_{a+1,s}
\right),\label{eq:118}
\end{align}
where the key point is the third equality, which is the relation
(\ref{eq:116}) with \(n={\Kone}-a\), \(\Sset_1=\Bset_1\), \(\Sset_2=\Fset\),
\(\Sset_3=\Bset_2\) and \(t=-2\tilde s-2=-(p-2n+2)\).
 The relation \({\tilde
  T_{a,s-1}   -  T_{a,s-1} = (-1)^{{\Kone}-a}T_{a-1,s}}\) follows from
\(\tilde T_{a+1,s}-T_{a+1,s} =- (-1)^{{\Kone}-a} T_{a,s+1}\) (by setting
\(a\to a-1\) and \(s\to s-1\)), and it concludes the proof of
(\ref{eq:108}), showing %
  that the Hirota equation holds on the diagonal delimiting
  the upper and the right strip.

 \paragraph{\ref{item:3}): Hirota equation at the intersection of the diagonals}

The proof follows the same lines as the proof of case \ref{item:4})
above: One should prove that (\ref{eq:108}) holds at the node \(\tilde
a=\tilde s=0\). The previous proof already shows that \(\tilde
T_{a+1,s}-T_{a+1,s} =- (-1)^{{\Kone}-a} T_{a,s+1}\) holds at the node \(\tilde
a=\tilde s=0\). But for the relation \({\tilde
  T_{a,s-1}   -  T_{a,s-1} = (-1)^{{\Kone}-a}T_{a-1,s}}\), we cannot
repeat exactly the steps (\ref{eq:117}-\ref{eq:118}) because  the
condition \(n\le \frac{p-1}2\)
(for the relation (\ref{eq:116}))
would not be satisfied. Instead, we proceed as follows:
\begin{align}
\star  T_{a,s-1}&=
\varepsilon_l(a,s-1)\,
Q_{({\Kone},{\Ktwo}-a|M)}^+\wedge
Q_{(0,a|0)}^-\\&=\varepsilon_l(a,s-1)(-1)^{M\,a}
{\bzdQ}_{({\Kone};{\Ktwo}-a;0)}^+\wedge
{\bzdQ}_{(0;a;M)}^-
\\&=\varepsilon_l(a,s-1)(-1)^{(M+1)a}\big(
{\bzdQ}_{({\Kone}-a;{\Ktwo};0)}^+\wedge
{\bzdQ}_{(a;0;M)}^- \nonumber\\&\hspace{4.5cm}-(-1)^M
{\bzdQ}_{({\Kone};{\Ktwo}-a+1;0)}\wedge
{\bzdQ}_{(0;a-1;M)}\big)
\\&=\varepsilon_l(a,s-1)(-1)^{(M+1)a}\big(
(-1)^{M\,a}Q_{({\Kone}-a,{\Ktwo}|M)}^+\wedge Q_{(a,0|0)}^- \nonumber\\&\hspace{4.5cm}-(-1)^{M\,a}
Q_{({\Kone},{\Ktwo}-a+1|M)}\wedge
Q_{(0,a-1|0)}\big)
\\&=\varepsilon_l(a,s-1)\big(
(-1)^{a(K+M)}\star \tilde T_{a,s-1} /\varepsilon_r(a,s-1)
-(-1)^{a}
\star T_{a-1,s}/\varepsilon_l(a-1,s)
\\&=\star \left(\tilde T_{a,s-1}-(-1)^{{\Kone}-a}T_{a-1,s}\right)
\end{align}
where the third equality is the relation \eqref{eq:116} with
\(\Sset_1=\Bset_2\), \(\Sset_2=\Bset_1\), \(\Sset_3=\Fset\), \(n=a\) and
\(t=-2\), and the last equality uses the condition \(K-M = 0\,\,({\rm mod }
2)\) to simplify signs. This proves the relation \({\tilde
  T_{a,s-1}   -  T_{a,s-1} = (-1)^{{\Kone}-a}T_{a-1,s}}\) and completes
the proof that any T-functions obeying \eqref{eq:155}
satisfy the Hirota equation.

\paragraph{Generic solution}

At this point, we have shown that \eqref{eq:155} provides a solution
of the Hirota equation, but one can ask whether all solutions of the
Hirota equation fit the ansatz  \eqref{eq:155}.
Reproducing either the arguments of \secref{sec:ident-q-funct} or of \secref{sec:backl-transf-qq} (see
e.g. \cite{Kazakov:2007fy}), one can associate to a
solution\footnote{This will be possible under mild 
  assumption of existence of generic solutions   for  T-functions. For instance, in \secref{sec:ident-q-funct}, we had to assume that the Baxter equation
  had \(N\) independent solutions, and that the functions \(1/T_{a,s}\)
  were well defined.} of Hirota
equation a set of Q-functions such that T is expressed as
\eqref{eq:155}.

Without entering into the details of such proof, one can already
convince oneself that \eqref{eq:155} is the generic solution by
counting the number of independent functions:
\begin{itemize}
\item The solution \eqref{eq:155} is characterized by
  \(|\Bset|+|\Fset|+1\) independent Q-functions, for instance the functions
  \({\bzdQ}_\emptyset\), \(({\bzdQ}_a)_{a\in\Bset}\),  \(({\bzdQ}_i)_{i\in\Fset}\).
\item Under the gauge constraint \eqref{eq:54}, the solution to Hirota
  equation on the {\THook} of figure \ref{fig:THook} is characterized  by \(K+M+1\) independent
  T-functions. For instance, if \(K_1=K_2\), one can chose the functions
  \((T_{a,M_1})_{0\le a\le K_1}\), \((T_{a,M_1+1})_{0\le a\le K_1}\) and
  \((T_{K_1+1,s})_{-M_2<s<M_1}\). Similarly if\footnote{The number of
    independent functions is invariant under the \(s\mapsto -s\)
    transformation which maps a \((K_1|M|K_2)\) {\THook} into a
    \((K_2|M|K_1)\) {\THook}. Hence if \(K_1\neq K_2\) we can choose
    \(K_1<K_2\) without loss of generality.
  } \(K_1<K_2\) , one can chose the functions
  \((T_{a,M_1})_{0\le a\le K_1}\), \((T_{a,M_1+1})_{0\le a\le K_1}\),
  \((T_{K_1+1,s})_{-M_2<s<M_1}\) and \((T_{a,-M_2})_{K_1+1\le a\le K_2}\).
\end{itemize}

\section{More details of the zero-twist limit in spin chains}\label{sec:more-details-zero}

\subsection{Large Bethe roots and zeros of Laguerre polynomials}
\label{sec:LargeBethe}
The goal of this subsection is to derive approximate expressions \eqref{Lagapproximation} valid near the point \(z=1\). We will focus on the case of \(\fQ_1\). Zeros of this polynomial, \(\fQ_1(u_i)=0\), will be  called Bethe roots. They can be found from the Bethe equations
\be\label{bom}
{\prod_{j=1}^{M_1}}\frac{u_i-u_j+\bi}{u_i-u_j-\bi}=-z\left(\frac{u_i+\bi/2}{u_i-\bi/2}\right)^L\,,\qquad i=1,2,\ldots M_1\,,
\ee
which follow from \eqref{QQXXX}, cf. \eqref{eq:165}.

In the untwisting limit, certain Bethe roots approach infinity, and we denote such Bethe roots as \(\theta_{\alpha}\). We will make an assumption that is justified a-posteriori that to-be-infinite Bethe roots are far not only from the to-stay-finite Bethe roots but also far from each other when \(z\to 1\). By taking  \(\log\) of Bethe equations \eqref{bom} and performing large-\(\theta_{\alpha}\) expansion one gets
\be\label{largebethe}
-\log z+\sum_{\beta,\beta\neq \alpha}\frac {2\bi}{\theta_\alpha-\theta_\beta}-s\frac{2\bi}{\theta_\alpha}=0\,,
\ee
where \(s=\frac L2-M_0\), and \(M_0\) is number of to-stay-finite Bethe roots.

Introduce \(x=\frac{\bi\,\theta\,\log z}{2}\). Then we have
\be
1+\sum_{j,j\neq i}\frac 1{x_i-x_j}-{s}\frac{1}{x_i}=0\,.
\ee
Using the standard matrix model trick of multiplying with \(\sum \frac 1{x-x_i}\) we can write a Riccati equation on the resolvent \(R(x)\equiv\sum\limits_{i}\frac 1{x-x_i}\):
\be
\frac 12(R^2+R')+R\left(1-\frac {S}x\right)=\frac {m}{x}\,,
\ee
where \(m=M-M_0\) is number of to-be-infinite Bethe roots.

This Ricatti equation is mapped to the linear second-order ODE by \(R=\frac{\psi'}{\psi}\), by noticing that \(R^2+R'=\frac{\psi''}{\psi}\):
\be
\psi''+2\left(1-\frac {S}x\right)\psi'-\frac{2\,m}{x}\psi=0\,.
\ee
We are looking for solution with the polynomial large-\(x\) asymptotics \(\psi\sim x^{m}\). The equation above is almost precisely the one for the associated Laguerre polynomials \(L_n^{(\alpha)}(x):\)
\be
x\, y''+(\alpha+1-x)\,y'+n\,y=0\,.
\ee
We derive the solution:
\be
\psi(x)=L_{m}^{(-2S-1)}(-2x)\,.
\ee
Zeros of \(\psi(x)\) are precisely \(x_\alpha\), and now we just recall that
\be
 x_\alpha=\frac{\bi\,\theta_\alpha\,\log z}{2}\,.
\ee
Note that Laguerre polynomials have degenerate zeros \(x=0\) if \(m\geq 2s+1\), with  degree of degeneration \(2s+1\). Hence we consider only solutions with \(m<2s+1\) when there is no degenerate zeros.

The polynomial \(\mathbb{Q}\) in  \eqref{Lagapproximation} is a polynomial with zeros at Bethe roots that remain finite in the \(z\to 1\) limit. From \eqref{bom}, it is easy to see that these finite Bethe roots satisfy
\be\label{boma}
{\prod_{j=1}^{M_0}}\frac{u_i-u_j+\bi}{u_i-u_j-\bi}=-\left(\frac{u_i+\bi/2}{u_i-\bi/2}\right)^L\,,\qquad i=1,2,\ldots,M_0\,,
\ee
when \(z=1\).
\subsection{Construction of the rotation}
\label{sec:constr-rotat}

In \secref{sec:zero-twist-limit-1}, we announced the existence of
a rotation of the \(Q\)-functions which allows to take the limit \(G\to\mathbb
I\) in a style \eqref{eq:Qlimh}. We provided several explicit examples of rotations, in particular \eqref{eq:194a} and \eqref{eq:194b}. In this appendix, we explain how a rotation matrix is constructed. First, we show this on an explicit example \eqref{eq:194a} and then generalise the logic to arbitrary case.

\paragraph{Example of the rotation \eqref{eq:194a}}
Many different rotation matrices can provide Q-functions with a \(G\to\mathbb I\) limit. As explained in \secref{sec:zero-twist-limit-1}, a way to chose a particular rotation is to choose a nesting path and then demand that the rotation leaves the Q-functions on this nesting path invariant (up to a normalisation). Then the rotation matrix obeys the property: \(h_{\alpha,\beta}=0\) if \(\alpha<\beta\) according to the order dictated by the nesting path. The rotation \eqref{eq:194a} is obtained from the nesting path \((\es|\es)\subset (\es|1)\subset (1|1) \subset (12|1)\).

The effect of the rotation is to multiply \(Q_{\emptyset|1}\) by \(h_{3,3}\),
  \(Q_{1|1}\) by \(h_{1,1}h_{3,3}\), etc. Hence the diagonal coefficients will be fixed by asking what normalisation provides the nesting path Q-functions with a \(G\to\mathbb{I}\) limit.
 In the present example 
 \(Q_{\emptyset|1}=\y^{\bi\,u-1/4}\) goes to \(1\) when \(G\to\mathbb{I}\), thus we set \(h_{3,3}=1\). By contrast 
\(Q_{1|1}=\left(\frac{\x_1}{\y}\right)^{-\bi u-1/4}\frac 1{\x_1-\y}\left(u^2-\bi u\frac{\x_1+\x_2}{\x_1-\x_2}-\frac{\x_1^2+6\x_1\x_2+\x_2^2}{4(\x_1-\x_2)^2}\right)\) has to be multiplied by (e.g.) \(-(\x_1-\x_2)^2(\x_1-\y)/2\) to get a smooth limit -- and the limit is then one. Hence we set 
\(h_{1,1}=-(\x_1-\x_2)^2(\x_1-\y)/2\), and get \(\lim\limits_{G\to\mathbb I} h_{1,1}h_{3,3} Q_{1|1}=1\). Similarly,
\(Q_{12|1}=\left(\frac{\x_1\x_2}{\y}\right)^{-\bi u-3/4}\,u^2\frac{(\x_1-\x_2)}{(\x_1-\y)(\x_2-\y)}\) should be multiplied by \(\frac{(\x_1-\y)(\x_2-\y)}{(\x_1-\x_2)}\) to get a limit, hence we set \(h_{1,1}\,h_{2,2}\,h_{3,3}=\frac{(\x_1-\y)(\x_2-\y)}{(\x_1-\x_2)}\), i.e. \(h_{2,2}=-\frac{2(\x_2-\y)}{(\x_1-\x_2)^3}\).

As we consider the example of an \(su(2|1)\) spin chain, the rotation
\(h\) cannot be an arbitrary \(\GL(3)\) element as it has to preserve the
decomposition \eqref{eq:dec}; we hence have
\(h_{1,3}=h_{2,3}=h_{3,1}=h_{3,2}=0\). The only coefficient which
remains to fix is thus \(h_{2,1}\), and it has to be chosen in such a way that 
\(h_{2,1}Q_{1|\emptyset}+h_{2,2}Q_{2|\emptyset}\) acquires a smooth
\(G\to\mathbb I\) limit. To do this we will iteratively add counter-terms
to \(h_{2,2}Q_{2|\emptyset}\) until the limit becomes smooth. Since
\(h_{2,2}Q_{2|\emptyset}=-2\frac{(\x_2-\y)}{(\x_1-\x_2)^3}\x_2^{-\bi u
  -3/4}\) has a second-order pole the simplest way to cancel this pole
is by considering the difference
\(A=h_{2,2}Q_{2|\emptyset}
- \frac{\x_1-\y}{\x_1-\x_2} Q_{1|\emptyset}
\), i.e. by substrating the multiple of \(Q_{1|\emptyset}\) which precisely cancels the pole of order two.
This would correspond to setting \(h_{2,1}=- \frac{\x_1-\y}{\x_1-\x_2}\).
 But when one expands this combination \(A\) in
the \(G\to\mathbb I\) limit, it turns out to have a pole of order one:
\(A\simeq -\frac 5 2 \frac{\x_2-\y}{(\x_1-\x_2)^2}\), hence one should
subtract one more term, and consider the combination \(B=A-\frac 5 4
({\x_1-\y})Q_{1|\emptyset}\), which would
correspond to setting \(h_{2,1}=
- (\x_1-\y)\left(\frac 1{\x_1-\x_2}+5/4\right)\). When \(G\to\mathbb I\), this
combination \(B\) is equal to \(-u^2-\frac {41}{16}\,\frac{\x_2-\y}{\x_1-\x_2}
+\mathcal{O}(G)\), which still
doesn't really have a unique limit when \(G\to\mathbb I\). To give it a limit,
one can subtract \(\frac{41}{32}(\x_1-\y)(\x_1-\x_2)
Q_{1|\emptyset}\). Hence we finally get
 \(h_{2,1}=-(\x_1-\y)\left(\frac 1{\x_1-\x_2}+5/4+\frac{41}{32}(\x_1-\x_2)\right)
\), and \(\lim\limits_{G\to\mathbb
  I} h_{2,1}Q_{1|\emptyset}+h_{1,1}Q_{2|\emptyset}=-
u^2\). We have hence obtained all the coefficients of the matrix \(h\) in
\eqref{eq:194a}, in the \(\alpha=1\) case.

\paragraph{Generalization}
\label{sec:generalization}

In higher rank, the procedure is the same: given an arbitrary nesting
path, we can still relabel all Q-functions to turn the nesting path
into \(\emptyset\subset1\subset12\subset\dots\), so that if we require the functions of the nesting path to be preserved (up to a normalisation), we impose the rotation
matrix to be lower-triangular when the matrix entries are order according the order dictated by the nesting path.

The diagonal coefficients are fixed by requiring the functions 
\(Q_1\), \(Q_{12}\) to have a smooth \(G\to \mathbb I\) limit in a generic position.
Then in the bosonic case, for each line \(i\), the coefficients
\(h_{i,j}\) are chosen as
being the necessary counter-terms to give the sum \(\sum_{j} Q_j
h_{i,j}\) a smooth limit when \(G\to\mathbb
I\). This is always possible because all diverging terms cancel from
\(Q_{12\dots i}\), which means that as functions of \(u\), they are
linear combinations of \(Q_{1}\), \(Q_{2}\), \(\dots\), \(Q_{i-1}\). In
addition this procedure ensures that the limits of the function
\(Q_{i}\) are linearly independent (as functions of \(u\)), because we
enforced the condition that \(Q_{123\dots N}\) has non-vanishing limit.

In the super-symmetric case, the procedure is the same except that some
coefficients of the matrix \(h\) are forced to be equal to zero, to
preserve the decomposition \eqref{eq:dec}. This means we have a too
little number of counter-terms to be sure we can make the
single-indexed Q-functions linearly independent, which may result in
some Q-functions having a vanishing \(G\to\mathbb I\) limit, as in
\eqref{eq:232}. As we can see in \eqref{eq:232}, the vanishing of a
Q-function still allows other Q-functions to be non-trivial, and in
particular we obtain (by construction) non-vanishing \(Q_{\bar\emptyset}\).
\subsection{Rational spin chain's Q-operators}
\label{sec:rational-spin-chains}
Q-operators can be constructed very explicitly for rational spin
chain in the defining representations, they are operators which commute with
each other, and their Wronskians give the transfer matrices of the spin
chain. Explicit expression of these operators are given for instance
in \cite{Kazakov:2010iu}, for a length \(L\) twisted rational \(su(K|M)\) spin chain in the defining representation. Their expression reads
(in the present notation for shifts):
\begin{multline}
\label{eq:136}
\mathcal{Q}_{A|I}=\prod_{\substack{a<b\\a,b \in A\cup I}}\left((-1)^{p_ap_b}\left(%
    {\x_a-\x_b}%
  \right)^{(-1)^{p_a+p_b}}\right)\lim_{\substack{\z_k\to1/\x_k\\k\in\bar A\cup \bar I}}\left[
  \left(\prod_{k\in \bar A\cup\bar I} \frac{((-1)^{p_k+1}\bi(1-\x_k\,\z_k))^{J_k}}{J_k ! ~
      w(z_k)^{(-1)^{p_k}}} \right) \right.\\\left.
\bigotimes_{i=1}^L\left(u_i^{[-s_0+|\bar A|-|\bar I|]}+\bi \hat
  D\right) %
\left(\prod_{{k\in \bar A\cup\bar I}}w(\z_k)^{(-1)^{p_k}}\right)\right]\,,
\end{multline}
where we use the notation \(\Bset=\{1,2,\ldots,K\}\), \(\Fset=\{K+1,K+2,\ldots,K+M\}\) and we abusively denote \(\x_{K+i}\equiv\y_i\). We also denote \(p_a=0\) for \(a\in \Bset\) and \(p_i=1\) for \(a\in \Fset\).
The two factors in the first line are normalisations: the first factor
is responsible for the antisymmetry of Q-functions, while the second
in necessary to make the limit \(z_k\to1/\x_k\) smooth (it amounts to
taking a pole) and to ensure that \(Q_{\emptyset=1}\). In this second
factor, the operators \(J_k\equiv E_{kk}\) count the numbers of
spins
in direction \(k\), see discussion after \eqref{eq:194} and \appref{sec:Labels}, and the function \(w(\z)\) is defined by
\begin{align}
w(\z)&=\det \frac 1 {1-G\,\z}=\prod_{k=1}^N  \frac 1 {1-\x_k\,\z}\,.
\end{align}
In the second line of (\ref{eq:136}), 
the \(u_i\)'s are related to the inhomogeneities \(\theta_i\) by
\begin{align}
  u_i&=u-\theta_i\,,
\end{align}
\(|\bar A|=K-|A|\) is the number of indices in \(A\), and the operator \(\hat D\) is a derivative operator with respect to the 
twist \(G\) which obeys
\begin{multline}\label{eq:135}
\frac 1 {\prod_{k=1}^n w(z_k)^{(-1)^{p_k}}}\left[\bigotimes_{i=1}^L\left(\alpha_i+\bi \hat
  D\right) \left(\prod_{k=1}^n w(z_k)\right)\right]_{i_1,i_2,\dots,i_L}^{j_1,j_2,\dots,j_L}=\\
\sum_{\sigma \in \mathcal{S}^L} \left[
\hspace{.3cm}\big(\hspace{-.3cm}\prod_{\substack{1\le k<l\le L\\\sigma(k)>\sigma(l)}}(-1)^{p_{i_k}\,p_{i_l}}\big)
\prod_{c\in \sigma}
\left(\alpha_m\delta_m^{\sigma(n)}\delta_{i_{m}}^{j_m}+
\sum_{k=1}^n(-1)^{p_k}\prod_{m\in
  c}\left[
  \bi\frac{(g\,z_k)^{\Theta{m,\sigma(m)}}}{1-g\,z_m}
\right]_{i_{\sigma(m)}}^{j_m}\right)\right]
\,,
\end{multline}
where the super-script and subscript on the l.h.s. denote
tensor indices of the operator, which is an operator on the Hilbert
space \((\mathbb C^{K|M})^{\otimes L}\) (where \(\mathbb C^{K|M}\) denotes the defining representation of \(\su(K|M)\)). In the r.h.s. of \eqref{eq:135}, a
sum runs over permutations \(\sigma\) belonging to the cyclic group
\(\mathcal{S}^L\), and a product runs over the ``cycles'' \(c\) appearing
in the decomposition of \(\sigma\) into a product of cyclic
permutations -- and over the sites \(m\) on which the cycles \(c\) acts. The notation \(\Theta_{i,j}\) is defined by
\eqref{WeylCharacter}. For instance if \(L=3\) and \(\sigma\) is the
permutation exchanging \(1\) and \(2\), then  the corresponding term in
the r.h.s. of \eqref{eq:135} reads
\(\sum_{k=1}^n \left((-1)^{p_k}\left[\frac{\bi}{1-G\,z_k}\right]^{j_1}_{i_2}
\left[\frac{\bi\,G\,z_k}{1-G\,z_k}\right]^{j_2}_{i_1}\right)%
\left[\alpha_3
\mathbb I+\bi\sum_{k=1}^n \frac{G\,z_k}{1-G\,z_k}\right]^{j_3}_{i_3}%
\)
because the decomposition of \(\sigma\) into cyclic permutations is
\(\sigma=(1,2)(3)\).

Given the explicit operatorial construction, it is easy to deduce commutation of T- and Q-operators with the symmetry generators: 
\begin{subequations}\label{eq:138}
  \begin{align}\label{eq:139}
&\textrm{if }G=\mathrm{diag}(\x_1,\dots,\x_N):&  \forall
\lambda,\,\forall k,\, [T_\lambda,J_k\equiv E_{kk}]&=0\,,\\\label{eq:140}
&\textrm{zero-twist limit: }&
 \forall
    \lambda,\,\forall k,j,\, [\lim_{g\to \mathbb
      I}T_\lambda,E_{kj}]&=0\,.
  \end{align}
\end{subequations}
Both relations \eqref{eq:138} follow from the statement that the
functions \(T_\lambda(u)\) can be written in the form (see e.g. \cite{Kazakov:2007na})
\begin{align}
  T_\lambda=\sum_{\sigma \in \mathcal{S}^L} c_\sigma(G)\, \mathcal{P}_\sigma\,,
\end{align}
where the sum runs over all possible permutations of spin chain sites. The operator  \(\mathcal{P}_\sigma\) realises these permutations:  \(\mathcal{P}_\sigma \left|e_{j_1}\cdots e_{j_L}\right\rangle =
\left|e_{j_{\sigma(1)}}\cdots e_{j_{\sigma(L)}}\right\rangle\). The coefficients \(c_\sigma\) depend only on the twist matrix \(\tG\). They are diagonal operators if \(\tG\) is
diagonal (whence \eqref{eq:139} follows, cf.~\cite{Kazakov:2010iu}), and become proportional to an identity operator in the limit
\(\tG\to\mathbb I\) (whence \eqref{eq:140} follows).

Hence the eigenstates of the T- and Q-operators organise in the irreps of the symmetry algebra. These are one-dimensional representations in the fully twisted case (hence the spectrum is generically non-degenerate, unless some bonus symmetry is present) and the representations labeled by Young diagrams with \(L\) boxes when the twist is absent.

\subsection{L-hook reduction in the case of short multiplets}
\label{app:shortmultiplets}

In \secref{sec:su21example}, we demonstrated that a physically relevant supersymmetric Q-system may have zero Q-functions, i.e. the following situation may emerge:
\begin{align}\label{apropos1}
&&Q_{Aa|I}Q_{A|Ii}&=0  &\textrm{i.e.} && Q_{A|I}&=Q_{Aa|Ii}\neq 0   \,,&&
\end{align}
for some special multi-indices \(A,I\) and indices \(a,i\). The equivalence in \eqref{apropos1} is due to the  QQ-relation \eqref{eq:49}.

Such a situation seems to be problematic: If we choose to set \(Q_{Aa|I}=0\) then \(Q_{A|Ii}\) is completely undefined, and vice versa. For instance, examples (\ref{eq:232a}) and \eqref{eq:232b} contain an unconstrained function \(R\).

In this appendix we demonstrate that this arbitrariness in Q-system does not lead to any ambiguity in physical quantities. Choose for instance \(Q_{Aa|I}=0\) (the choice \(Q_{A|Ii}=0\) shall be processed in full analogy).  Our main statement is that by the use of symmetry transformations which leave T-functions invariant, we can enforce to have
\begin{align}\label{eq:235}
\forall B\not\ni a, \forall J\not\ni i,&&
  Q_{Ba|J}&=0\,,\ \ \text{and hence}&Q_{Ba|Ji}&=Q_{B|J}\,.
\end{align}
We prove this result at the end of this appendix, whereas now we discuss its implications.

\begin{figure}
  \begin{minipage}{.5\linewidth}
    \includegraphics{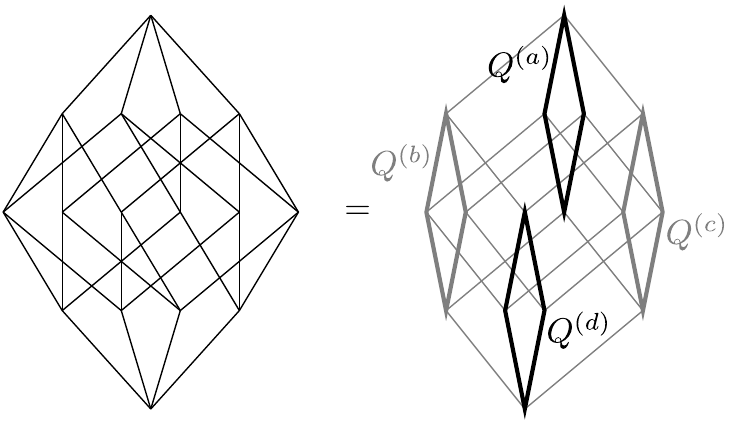}
  \end{minipage}
\hfill
  \begin{minipage}{.5\linewidth}
    \textbf{Hasse diagram splitting:}
    The full Hasse diagram in the l.h.s. splits into four Hasse
    sub-diagrams \(Q^{(a)}\), \(Q^{(b)}\), \(Q^{(c)}\) and \(Q^{(d)}\) defined in
    equation (\ref{eq:240}).

    Out of the four Hasse sub-diagrams on the r.h.s., the diagrams
    \(Q^{(a)}\) and \(Q^{(d)}\) (on
    top and bottom) are equal (see \eqref{eq:235}), while the two
    grayed-out sub-diagrams are unphysical and can be set to zero (see
    \eqref{eq:235}, \eqref{eq:857}).

    Hence the original \((K|M)\)
    Hasse-diagram reduces to only one 
    sub-diagram, which has size \((K-1|M-1)\).
  \end{minipage}
  \caption{Splitting of the Hasse diagram into sub-diagrams,
    illustrating equations \eqref{eq:235}, 
\eqref{eq:857}.}
  \label{fig:degenerasplit}
\end{figure}

Denote by \(B\subset \Bset\) and \(J\subset \Fset\)  arbitrary multi-indices such that \(B\not\ni a\,,\ J\not\ni i\). It is handy to decompose the original \((K|M)\) Q-system into the four subsystems of type \((K-1|M-1)\) each which are defined by 
\begin{align}\label{eq:240}
Q^{(a)}_{B|J}&\equiv Q_{B|J}\,,&
Q^{(b)}_{B|J}&\equiv Q_{Ba|J}=0\,,&
Q^{(c)}_{B|J}&\equiv Q_{B|Ji}\,,&
Q^{(d)}_{B|J}&\equiv Q_{Ba|Ji}\,,
\end{align}
see figure~\ref{fig:degenerasplit}.
From \eqref{eq:235} we see that \(Q^{(d)}=Q^{(a)}\). The system \(Q^{(a)}\) does not contain zero Q-functions. It is the one that posses all physical information. Note for instance that \(Q^{(a)}_{\bar\es}=Q_{\bar\es\backslash a|\bar\es\backslash i}=Q_{\bar\es|\bar\es}=u^L\).

The system \(Q^{(c)}\) has no physical relevance since its Q-functions do not appear in T-functions. Indeed, expression (\ref{eq:155}) for T-functions only involves products of two, Hodge-dual,
Q-functions. But the Hodge-dual of any function in \(Q^{(c)}\) is  \(Q^{(b)}_{\dots}=0\).

We also notice that many T-functions completely vanish:  When \eqref{eq:235} applies, \(T_{K_1,s\geq M_1}=0\) and \(T_{a\geq K_1,M_1}=0\) in \eqref{eq:155}, i.e. both sizes \(K_1\)
and \(M_1\) of the {\THook} decrease by one; on this smaller hook,
the T-functions are given by expression (\ref{eq:155}) in terms of the
Q-functions of the smaller Hasse sub-diagram \(Q^{(a)}\) of size \((K-1|M-1)\).

The reduction of L-hook has a clear counterpart in the theory of characters. The characters of an irrep \(\mu\) of \(\su(K|M)\) is given by the supersymmetric Schur polynomial \(s_{\mu}(\x_1,\x_2,\ldots\,, \x_K\slash \y_1,\y_2,\ldots \y_M)\). It is a defining property of supersymmetric polynomials that they become independent on \(\x_a\) and \(\y_i\) if \(\x_a=\y_i\). Hence, Schur functions become effectively the characters of \(\su(K-1|M-1)\) when \(\x_a=\y_i\).

The system \(Q^{(c)}\) contains exactly those Q-functions that are ambiguous due to the arbitrariness of \(Q_{A|Ii}\). The Q-functions of \(Q^{(c)}\) are still constrainted: by internal to \(Q^{(c)}\) QQ-relations and by the entanglement with \(Q^{(a)}=Q^{(d)}\) through
\begin{align}\label{eq:242}
Q^{(d)}_{Bb|J}Q^{(c)}_{B|J}&=
\begin{vmatrix}
{  Q^{(d)}_{B|J}}^+{Q^{(c)}_{Bb|J}}^+\\
 { Q^{(d)}_{B|J}}^-{Q^{(c)}_{Bb|J}}^-
\end{vmatrix}
\,,&Q^{(c)}_{B|Jj}Q^{(a)}_{B|J}&=
\begin{vmatrix}
{  Q^{(c)}_{B|J}}^+{Q^{(a)}_{B|Jj}}^+\\
  {Q^{(c)}_{B|J}}^-{Q^{(a)}_{B|Jj}}^-
\end{vmatrix}\,.
 \end{align}

One can notice that if we have an arbitrary
solution of the full original \(Q\)-system, we obtain another solution
by setting \(Q^{(c)}_{B|J}=0\), and this solution produces the same
T-functions as the original one, and also the same Q-functions on the
nesting path, except unphysical \(Q_{A|Ii}\). Hence, it is admissible to set the ambiguous Q-functions to zero:
\begin{equation}\label{eq:857}
Q_{B|Ji}=0\,.
\end{equation}

\paragraph{Proof of \eqref{eq:235}}
\label{sec:proof-q_baj=0}

We will proceed by steps to show that \(Q_{Ba|J}=0\) for arbitrary
\(B\not\ni a\) and \(J\not\ni i\), starting from the
case \((B|J)=(A|I)\) where we already know the result by assumption.
\begin{description}
\item [case \((B|J)=(A|Ij)\)] where \(j\neq i\). From \(Q_{Aa|I}=0\) it
  follows that \(\left|
    \begin{smallmatrix}
      Q_{A|I}^+&Q_{Aa|Ij}^+\\Q_{A|I}^-&Q_{Aa|Ij}^-
    \end{smallmatrix}\right|=0\), i.e. \(\left|
    \begin{smallmatrix}
      Q_{Aa|Ii}^+&Q_{Aa|Ij}^+\\Q_{Aa|Ii}^-&Q_{Aa|Ij}^-
    \end{smallmatrix}\right|=0\), i.e. there exists an \bi-periodic
  function \(\alpha\) such that \(Q_{Aa|Ii}=\alpha Q_{Aa|Ij}\).

  If we redefine the \(Q\)-functions by the rotation 
  \begin{align}\label{eq:243}
    h&=
    \left(
      \begin{smallmatrix}
        1\\&\ddots\\&&1&-\alpha\\&&&1\\&&&&\ddots
      \end{smallmatrix}
    \right)
    \,,
  \end{align}
where the coefficient \(-\alpha\) is at position \((j,i)\), then we
obtain \(Q_{Aa|Ij}=0\). One should note that the rotation \eqref{eq:243}
leaves all the Q-functions of the nesting path invariant (as well as
the T-functions, obviously).

\item [case \((B|J)=(A\bar b|I)\)], where we denote as \(A\bar b\) a
  multi-index\footnote{This definition defines \(A\bar b\) up to the ordering,
  which is irrelevant because it only changes the sign of \(Q_{A\bar b|I}\),
  without impacting its zeroness.} corresponding to the set \(\{A\}\setminus\{b\}\), where
  \(b\in A\). 

This case is exactly like above: there exists \(\alpha\) such that
\(Q_{Aa\bar b|I}=\alpha Q_{A|I}\), and we rotate using the same rotation
\eqref{eq:243} where \(\alpha\) stands at position \((a,b)\).

\item[general case] In the general case, on can write \(Q_{Ba|J}\) as a
  determinant where a full line vanishes due to the cases shown
  previously. More precisely, one can write \((B|J)\) as %
  \(A\bar{b_1}\bar{b_2}\dots\bar{b_p}c_1c_2\dots
  c_l|I\bar{j_1}\bar{j_2}\dots\bar{j_m}k_1k_2\dots k_n\) where \(b_h\in
  A\), \(c_h\not\in Aa\), \(j_h\in I\), \(k_h\not \in Ii\). Then a slight
  generalization of (\ref{eq:46}) reads
for instance
\begin{align}\label{eq:239}
  Q_{Ba|J}^{[\dots]}\left(Q_{A|I}^{[\dots]}\right)^{\dots}&=
  \left|
  \begin{smallmatrix}
    Q_{Aa\bar{b_1}|I}& \dots&  Q_{Aa\bar{b_p}|I}& Q_{Aa|Ik_1}& \dots & Q_{Aa|Ik_m}&Q_{Aa|I}^{[\dots]}&Q_{Aa|I}^{[\dots]}&\dots\\
    Q_{Ac_1\bar{b_1}|I}& \dots&  Q_{Ac_1\bar{b_p}|I}& Q_{Ac_1|Ik_1}&
    \dots &
    Q_{Ac_1|Ik_m}&Q_{Ac_1|I}^{[\dots]}&Q_{Ac_1|I}^{[\dots]}&\dots\\
\vdots&&\vdots&\vdots&&\vdots&\vdots&\vdots\\
    Q_{Ac_l\bar{b_1}|I}& \dots&  Q_{Ac_l\bar{b_p}|I}& Q_{Ac_l|Ik_1}&
    \dots &
    Q_{Ac_l|Ik_m}&Q_{Ac_l|I}^{[\dots]}&Q_{Ac_l|I}^{[\dots]}&\dots\\
    Q_{A\bar{b_1}|I\bar{j_1}}& \dots&  Q_{A\bar{b_p}|I\bar{j_1}}& Q_{A|Ik_1\bar{j_1}}&
    \dots &
    Q_{A|Ik_m\bar{j_1}}&Q_{A|I\bar{j_1}}^{[\dots]}&Q_{A|I\bar{j_1}}^{[\dots]}&\dots\\
\vdots&&\vdots&\vdots&&\vdots&\vdots&\vdots\\
    Q_{A\bar{b_1}|I\bar{j_m}}& \dots&  Q_{A\bar{b_p}|I\bar{j_m}}& Q_{A|Ik_1\bar{j_m}}&
    \dots &
    Q_{A|Ik_m\bar{j_m}}&Q_{A|I\bar{j_m}}^{[\dots]}&Q_{A|I\bar{j_m}}^{[\dots]}&\dots\\
  \end{smallmatrix}
  \right|
\end{align}
if \(l+m+1\ge p+k\)\footnote{Otherwise  one gets a similar determinant
  where the last columns are dropped and extra rows are added.}.

The result then follows from the fact that the first line of the
determinant vanishes.

Finally note that \(Q_{Ba|J}=0\) implies that \(Q_{B|J}\propto Q_{Ba|Ji}\). If the coefficient of proportionality is not zero, then we can set it to one at the price of changing the normalization of
T-functions. The case when it is zero for some \(B|J\) means that instead of reduction to a hook of total size \((K-1|M-1)\) one has the size \((K-2|M-2)\) or even smaller, in particular the physically-relevant subsystem \(Q^{(a)}\) would be smaller. We skip discussion of this case as it is done in full analogy to the presented analysis.

\end{description}

\paragraph{Remark:}
\label{sec:remark}
In \secref{sec:qq-relations-with}, we saw that the
derivation of determinant expressions like (\ref{eq:239}) involves
divisions by some Q-functions, and one may expect it fails when some
Q-functions vanish.

In the present proof, we actually assume that a Q-system where some
Q-functions vanish is the limit of a generic Q-system -- where all
Q-functions are non-zero. This assumption is obviously sufficient to
obtain the determinant expressions, and it holds in examples like
(\ref{eq:232}), as the twisted Q-system only has non-zero
Q-functions -- and the functions vanish only in the twistless limit.

\section{Twisted asymptotics and weight of the representation}\label{sec:Labels}

Consider a \(\gl(K|M)\) algebra with generators \(E_{\alpha\beta}\) obeying the commutation relation
\be
[E_{\alpha\beta},E_{\gamma\delta}\}= \delta_{\beta\gamma}E_{\delta\alpha}-(-1)^{(p_{\alpha}+p_{\beta})(p_{\gamma}+p_{\delta})}\delta_{\delta\alpha}E_{\gamma\alpha}\,,
\ee
where indices \(\alpha,\beta,\ldots\) belong either to the set \(\Bset=\{1,2,\ldots,K\}\) or to the set \(\Fset=\{\hat 1,\hat 2,\ldots,\hat M\}\). Correspondingly, \(p\) is a \(\mathbb{Z}_{2}\)-valued function with \(p_{\alpha}=0\) if \(\alpha\in\Bset\) and \(p_{\alpha}=1\) if \(\alpha\in \Fset\). The weight of a state \(|v\rangle\) of a \(\gl(K|M)\) irreducible representation will be labeled by the set 
\be\label{lowwe}
[\lambda_1,\ldots,\lambda_K;\nu_1,\ldots,\nu_M]
\ee
defined by
\be
E_{\alpha\alpha}|v\rangle &=m_{\alpha}|v\rangle\,,\hspace{3em}  {\rm where}\ \ \   m_{a}=\lambda_{a}\,, \ \ \ a\in \Bset\,;  
\,\hspace{2em}
m_i= \nu_{i}\,,\ \ \ i\in \Fset\,.
\ee
Furthermore, \(|v\rangle\) is called the highest-weight state if
\be\label{eq:HWapp}
E_{\alpha\beta}|v\rangle = 0\,\ \ \  \alpha<\beta\,
\ee
and the lowest-weight state if
\be\label{eq:LWapp}
E_{\alpha\beta}|v\rangle = 0\,\ \ \  \alpha>\beta\,.
\ee
The choice of ordering \(<\) between bosonic and fermionic indices affects the choice of the highest- and lowest-weight vectors.

The ordering is typically encoded by the Kac-Dynkin diagram. It is handy to represent it as a two-dimensional path on a \(K\times M\) lattice with crossed nodes corresponding to turning points~\cite{Kazakov:2007fy}. Two examples are shown below:
\begin{align}\label{fig:KDexamples}
\includegraphics[width=0.4\textwidth]{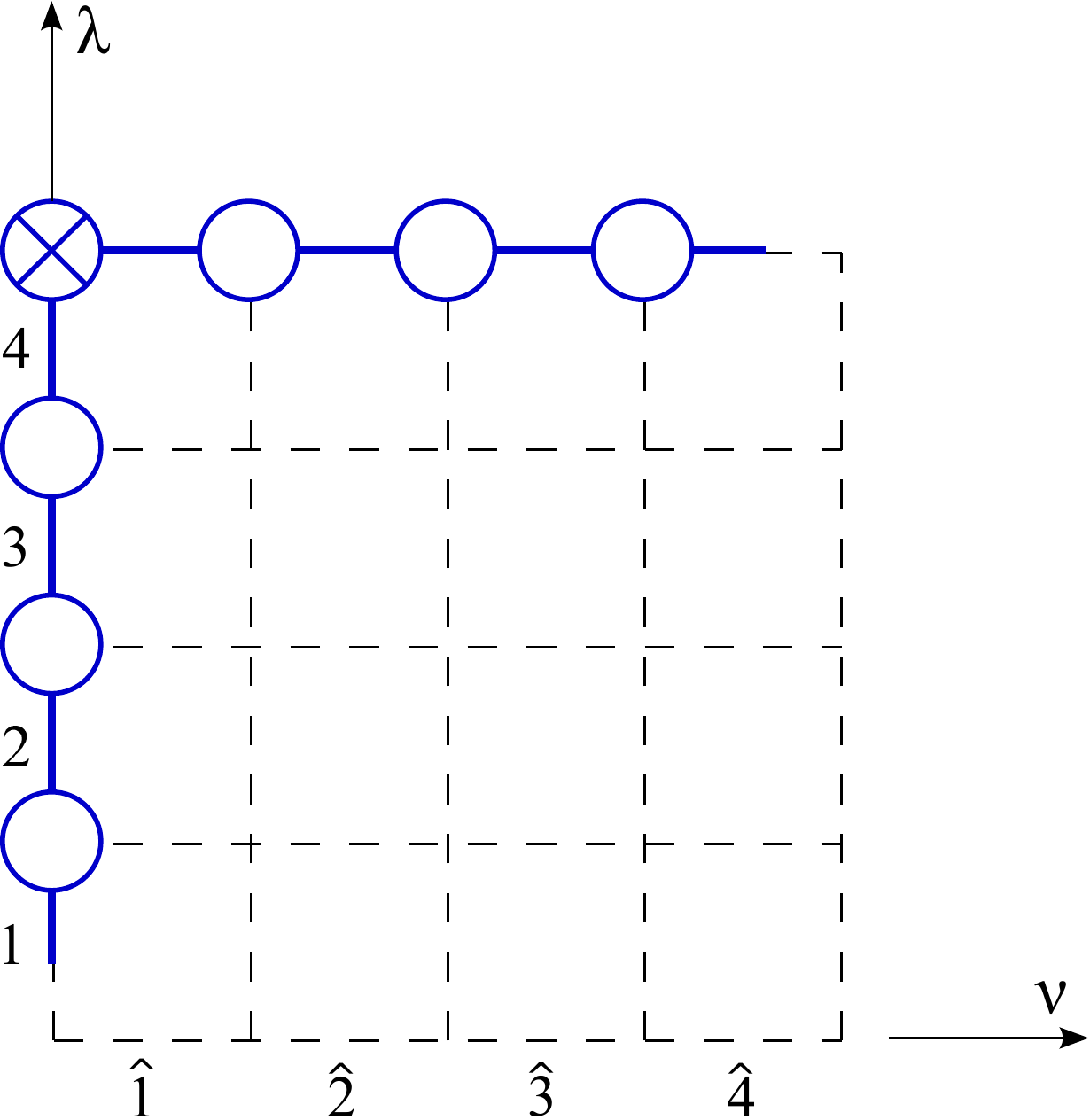}\hspace{2em} \includegraphics[width=0.4\textwidth]{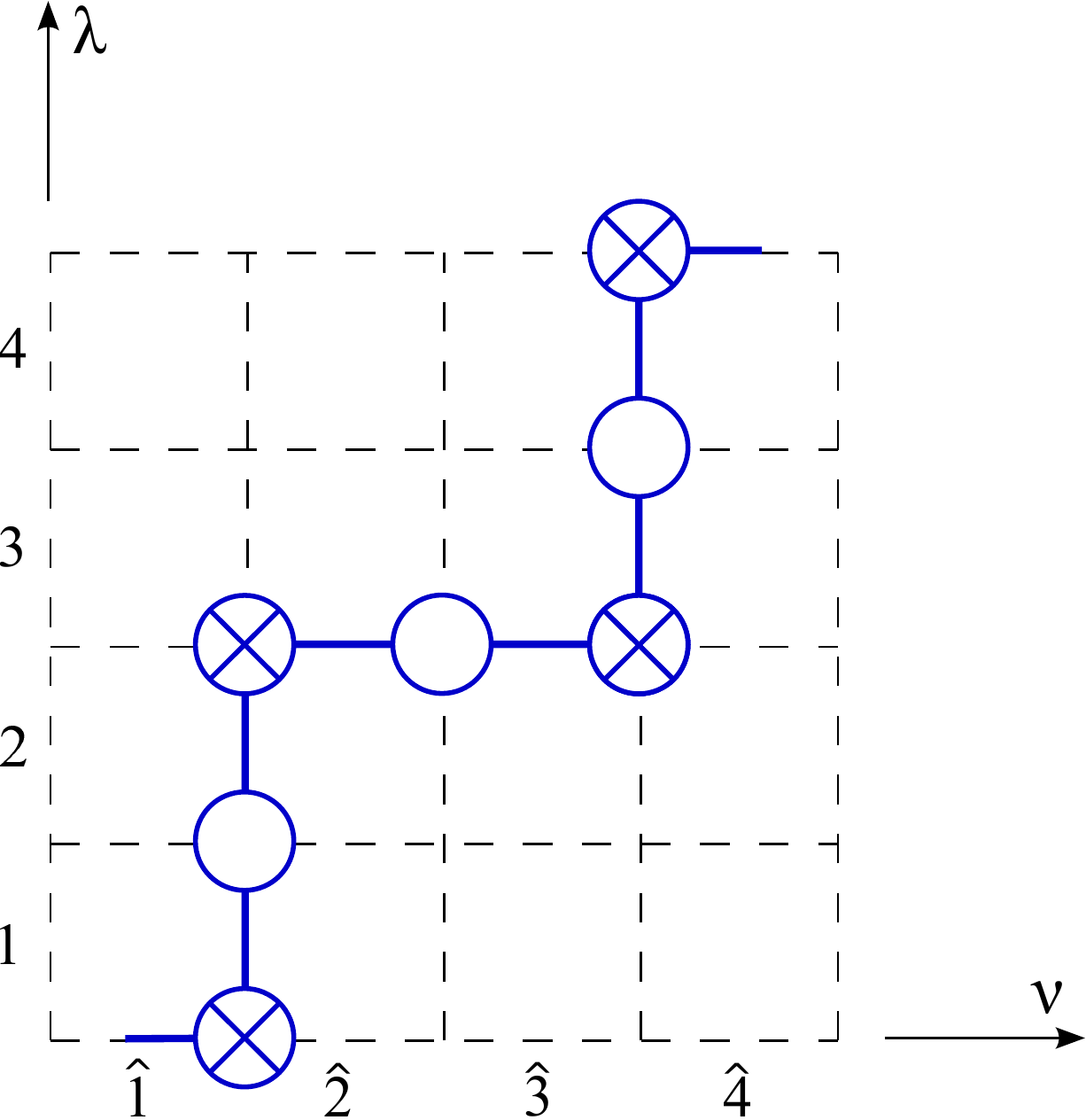}
\end{align}
The global ordering is introduced by the rule: \(\alpha<\beta\) if \(\alpha\) appears before \(\beta\) when one follows the path of the diagram. So the left figure (usually called distinguished diagram) has the ordering \(1<2<3<4<\hat 1<\hat 2<\hat 3<\hat 4\) while the right figure (used in the AdS/CFT asymptotic Bethe Ansatz) has the ordering \(\hat 1<1<2<\hat 2<\hat 3<3<4\).

The highest/lowest weight is transformed following the rule\footnote{The weight of a given vector in a representation module does not depend on the ordering. It is the choice of the highest/lowest weight vector which is responsible for the transformation rule.} (see e.g. \cite{Volin:2010xz} for derivation):
\begin{subequations}
\label{eq:trrule}
\be
\label{eq:trrule1}
\raisebox{-2em}{\includegraphics[width=0.4\textwidth]{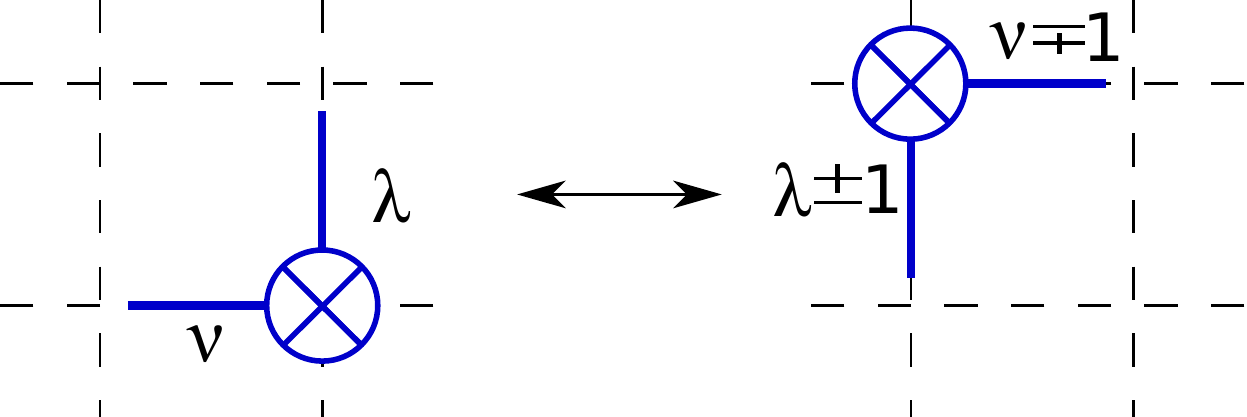}}\,,\ \ \   {\rm if}\ \  \lambda+\nu\neq 0\,,
\\[1em]
\label{eq:trrule2}
\raisebox{-2em}{\includegraphics[width=0.4\textwidth]{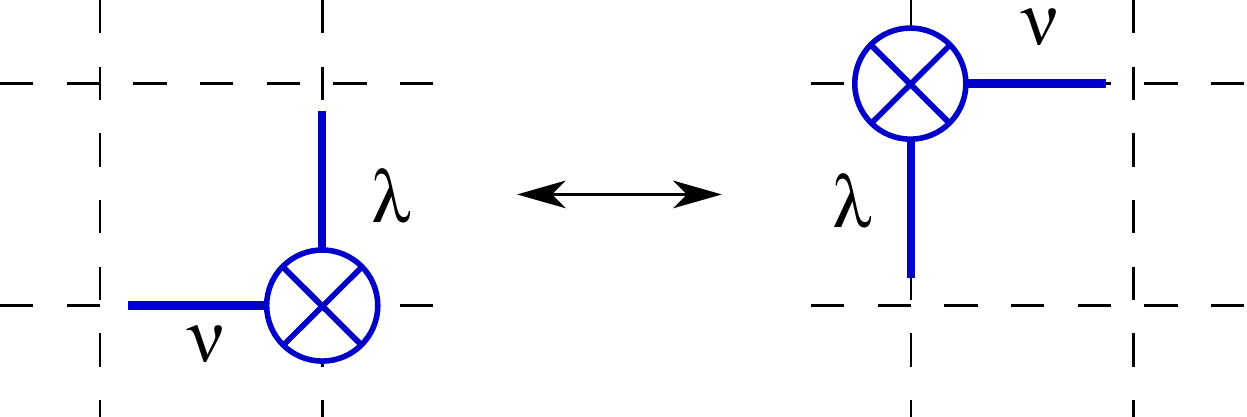}}\,,\ \ \   {\rm if}\ \  \lambda+\nu= 0\,,
\ee
\end{subequations}
where the upper choice of a sign corresponds to the highest weight and the lower choice corresponds to the lowest weight.

\ \\
It would be more convenient, partially for historical reasons, to use the lowest-weight terminology to describe a generic rational integrable spin chain with diagonal twist. The following data defines such a chain: its length \(L\); inhomogeneity parameter \(\theta_k\) and the weight \(\{\pmb{\lambda}^{(k)},\pmb{\nu}^{(k)}\}\) of a lowest-weight representation at a spin chain site \(k\), for \(k=1,2,\ldots,L\); the value of a twist \(G={\rm diag}(\x_1,\x_2,\ldots,\x_N,\y_1,\y_2,\ldots,\y_M)\). 

Such a spin chain is solved by Bethe ansatz techniques, its spectrum is described by solutions of Bethe equations \cite{Kulish:1983rd,Ragoucy:2007kg}. It is not difficult to determine the Q-system providing such equations \cite{Kazakov:2007fy,Tsuboi:2009ud,Gromov:2010km}. To this end, one should introduce a couple of notations. First, one will need a function \(\CP_m(u)\) with the property\footnote{If \(m\) is a non-negative integer then this function is simply a polynomial \(\CP_m=u\,(u+\bi)(u+2\bi)\ldots (u+(m-1)\bi)\,.\) This particular case is enough to cover all spin chains in finite-dimensional representations. However, we allow arbitrary \(m\) to include all the highest-weight representations, not only the finite-dimensional ones.}
\be
\frac{\CP_m^{++}}{\CP_m}=\frac{u+\bi\,m}{u}\,.
\ee
Then define  \(Q_{\alpha_1\ldots\alpha_k}\equiv Q_{A|I}\), where \(A\) is the projection of the set \(\alpha_1\ldots\alpha_k\) on the \(\Bset\)-set and \(I\) the projection on the \(\Fset\)-set,  and denote by ``\(\dbless{\beta}\)'' the set of all indices smaller or equal to \(\beta\) according to the given choice of a Kac-Dynkin diagram. For instance, for the right figure of \eqref{fig:KDexamples},  \(Q_{\dbless{\hat 2}}=Q_{\hat{1}12\hat{2}}=Q_{12|\hat{1}\hat{2}}\). 

Note that the functions \(Q_{\leftarrow\alpha}\) are the functions on a certain nesting path \eqref{eq:nestingpath}. In this way we establish a one-to-one correspondence between the nesting paths and the choice of the global ordering.

Finally, define
\be
s_{\gamma}\equiv (-1)^{p_{\gamma}}\ \ \ {\rm and}\ \ \ \rho_{\beta}\equiv\sum_{\gamma\leq \beta}s_\gamma\,.%
\ee
Then the Q-system for the  spin chain described above is defined by the Q-functions along the chosen Kac-Dynkin diagram, these Q-functions should fit the following ansatz
\begin{align}\label{eq:Qdbless}
Q_{\dbless{\alpha}}=q_{\alpha}\cdot x_{\dbless{\alpha}}^{-\bi\,u}\,\cdot\prod_{k=1}^L\prod_{\beta\leq\alpha}\left(\CP_{s_{\beta}m_{\beta}^{(k)}}^{\vphantom{[]}}\left(u-\theta_k+\frac{\bi}2(\rho_{\alpha}-2\rho_\beta+1+s_{\beta})\right)\right)^{s_{\beta}}\,,\ \ \ \  {1\leq \alpha\leq N+M-1}\,,
\end{align}
with \(q_{\alpha}\) being a polynomial in \(u\); denote its degree as \(K_{\alpha}\). One furthermore demands \(Q_{\es}=1\) and \(K_{N+M}=0\) (so that \(q_{\bar\es}=1\)). For the twist factor, one obviously has \(x_{\dbless{\alpha}}=\prod\limits_{\beta\leq\alpha}x_{\alpha}\), with identification 
\be
x_{a}=\x_a\ \ \ \ {\rm and}\ \ \ \  x_i=1/\y_i\,.
\ee
With such an ansatz for Q-functions, the bosonic Bethe equations would read\footnote{Note that offset of Bethe roots may be different in different literature sources.  The ambiguity arises, in particular, when only the \(\mathsf{sl}\) symmetry instead of \(\gl\) symmetry is present,  hence the physical quantities would depend only on the Dynkin labels \(\omega_{\alpha}=m_{\alpha}-(-1)^{p_{\alpha}+p_{\alpha+1}}m_{\alpha+1}\). For instance, for ``rectangular'' representations, when only \(\alpha\)'s Dynkin label is different from zero, and when representation is the same for all nodes of the spin chain,  one typically uses \(u\to u-\frac{\bi}2s_{\alpha}m_{\alpha}^{(k)}\), to make equations invariant under complex conjugation. Another common overall shift, more suitable for representation theory, is \(Q_{\dbless{\alpha}}\mapsto Q_{\dbless{\alpha}}^{[\rho_{\alpha}]}\), which changes the convention of how QQ-relation \eqref{eq:2} is represented. }
\be
-\frac{q_\alpha^{++}q_{\alpha-1}^-q_{\alpha+1}^-}{q_\alpha^{--}q_{\alpha-1}^+q_{\alpha+1}^+}
=\frac{x_{\alpha+1}}{x_{\alpha}}\left[\prod_{k=1}^{L}
\frac
{u-\theta_k+\bi\,s_{\alpha+1}m_{\alpha+1}^{(k)}-\frac {\bi}2\rho_{\alpha}}
{u-\theta_k+\bi\,s_{\alpha}m_{\alpha}^{(k)}-\frac {\bi}2\rho_{\alpha}\ \ \ }\right]^{s_{\alpha}}\,,
\ee
at zeros of \(q_{\alpha}\);
and fermionic Bethe equations would read
\be
\frac{q_{\alpha-1}^+q_{\alpha+1}^-}{q_{\alpha-1}^-q_{\alpha+1}^+}
={x_{\alpha}\,x_{\alpha+1}}\left[\prod_{k=1}^{L}
\frac
{u-\theta_k+\bi\,s_{\alpha+1}m_{\alpha+1}^{(k)}-\frac {\bi}2\rho_{\alpha}}
{u-\theta_k+\bi\,s_{\alpha}m_{\alpha}^{(k)}-\frac {\bi}2\rho_{\alpha}\ \ \ }\right]^{s_{\alpha+1}}\,,
\ee
at zeros of \(q_{\alpha}\). These are indeed the correct equations stemming from Bethe Ansatz.

Each solution of the Bethe equations corresponds to an irrep of the sub-algebra of \(\gl(K|M)\)  that commutes with the twist matrix \(\tG\).  The lowest weight of the irrep is given by
\be\label{eq:reptheoryresult}
m_{\alpha}=K_{\alpha}-K_{\alpha-1}+\sum_{k=1}^{L} m_{\alpha}^{(k)}\,.
\ee
Now note that \(P_{m}\sim u^{m}\) when \(u\to\infty\). By performing comparison of \eqref{eq:reptheoryresult} with large-\(u\) asymptotic of \eqref{eq:Qdbless}, we conclude that
\be\label{eq:rep}
Q_{\dbless{\alpha}}\sim x_{\dbless\alpha}^{-\bi\,u}\cdot u^{\sum\limits_{\beta\leq\alpha}m_{\beta}}\,.
\ee
We see that power of Q-functions at large-\(u\) is dictated by representation theory, this statement is valid for any rational spin chain. 

The property \eqref{eq:rep} should hold for any choice of the total order (equivalently, nesting paths). One should check then that the large-\(u\) behaviour of Q-functions given by \eqref{eq:rep} for all possible nesting paths is in agreement with QQ-relations which obviously constrain this behaviour, see \eqref{eq:213}. There are three cases to consider: First, in the presence of twist, only Cartan generators remain the symmetry. Hence the irreps are all one-dimensional. Therefore the lowest weight \eqref{lowwe} is the unique weight present in a given irrep: it obviously does not depend on a chosen order, so that we can operate by the rule \eqref{eq:trrule2} when changing from one order to another. The choice \(\x_a\neq \y_i\)  in \eqref{eq:213c} is in full agreement with this rule. Second, if one has \(\x_a=\y_i\) then the symmetry is enhanced and we should operate according to the rule \eqref{eq:trrule1}, unless \(\lambda_a+\nu_i=0\). And indeed, the degrees of Q-functions also involve appropriate \(\pm 1\) factors as it follows from \eqref{eq:213c}, case \(\x_a=\y_i\). Finally, if \(\x_a=\y_i\), \(\lambda_a+\nu_i=0\) and \(a,i\) are neighbours in the chosen order sequence, then one should use again \eqref{eq:trrule2}. We discuss the features of a corresponding Q-system in \secref{sec:su21example} and \appref{app:shortmultiplets}. Such a Q-system is ambiguous, in particular, the functions \(Q_{\pmb{\alpha}a}\) and \(Q_{\pmb{\alpha}i}\), where \({\pmb{\alpha}}\) is the set of all indices that precede both \(a\) and \(i\) in the chosen order, are not uniquely defined. However, these functions prove to be not relevant for physical quantities. If we want, we can always assign them a value that  complies with \eqref{eq:rep} (but such an assignment would be different for different order choices). Note also that the condition \(\deg Q_{\pmb{\alpha}}=\deg Q_{\pmb{\alpha}ai}\) is in a perfect agreement with \eqref{eq:rep} and the property \(\lambda_a+\nu_i=0\) .

Consider now the counterpart of \eqref{eq:rep} in the Hodge-dual basis. Denote
\be
M\equiv \sum_{\alpha} m_{\alpha}=\sum_{k=1}^L m_{\alpha}^{(k)}\,.
\ee
Note that this number depends only on the definition of the spin chain, but not on a particular state that we consider. For a spin chain in  fundamental representation, \(M=L\).

Along the nesting path, Hodge dual basis is related to the original one by the relation \(Q^{[\alpha]\leftarrow}\propto Q_{\dbless\alpha}\), where ``\(^{[\alpha]\leftarrow}\)'' denotes all indices which are larger than \(\alpha\) (and do not include \(\alpha\)). Then it is easy to see the relation
\be\label{eq:repHW}
\frac{\left(\prod\limits_{\beta}\x_{\beta}\right)^{\bi\,u}}{u^M}Q^{{\leftarrow}\alpha}\sim x_{{\leftarrow}\alpha}^{\bi\,u}\cdot u^{-\sum\limits_{\beta\leq\alpha}m_{\beta}}\,,
\ee
with the total order used in \eqref{eq:repHW}  precisely {\bf reverse} of that in \eqref{eq:rep}.

Reversing the order means swapping between lowest-weight and highest-weight description, cf. \eqref{eq:LWapp} vs \eqref{eq:HWapp}. Hence, we see that one relates large-\(u\) asymptotic \eqref{eq:repHW}  in the Hodge-dual description to the highest weight of an irrep (note also the change in signs compared to \eqref{eq:rep}).

In Hodge-dual description, \(Q^{\emptyset}\neq 1\). One can achieve equality \(Q^{\emptyset}=1\) be performing the gauge transformation \(Q_{A|I}\to {\left(\prod\limits_{\beta}\x_{\beta}\right)^{-\bi\,u}}u^M\,Q_{A|I}\), at the price that Q-functions loose their polynomiality if it was present. The lowest-weight description for compact rational spin chains was chosen to allow \(Q_{\es}=1\) and have polynomial Q-functions at the same time.

The AdS/CFT integrable model becomes a rational spin chain at weak coupling \(g\to0\). Hence \eqref{eq:rep} and \eqref{eq:repHW} should hold. In this special case \(Q_{\es}=Q^{\es}=1\), hence there is no actual preference between highest- and lowest-weight descriptions. For historical reasons \cite{Gromov:2014caa}, the highest-weight notation was adopted, with the QSC coinciding with the Hodge-dual basis of this appendix. We expect that \eqref{eq:repHW} should hold at arbitrary coupling \(g\), when rational spin chain description is no longer applicable. The argument we use is the same as in \cite{Gromov:2014caa}: all quantized charges are coupling-independent, while for conformal dimension, the only continuous charge, one performs comparison with TBA, cf. \appref{app:E2}. Of course, we believe that QSC is a fundamental object and a deviation from the property \eqref{eq:repHW} cannot be expected.

\section{Leading QSC asymptotics for some particular cases of twisting}
\label{app:TwistedAss}

In this appendix, we will present the computation of the leading large \(u\) asymptotics of one-indexed Q-functions for some particular configurations of twists of AdS/CFT QSC.

The subsections gives the implementation of the formulae of \secref{sec:turning-twist} in the widespread \verb+Mathematica+ langage and its free, open-source competitor \verb+SageMath+. The further sections  give several examples for different configurations of the twists.

\subsection{Computer implementation of leading asymptotics of QSC Q-functions}
\label{sec:comp-impl}

Let us illustrate how to use the code provided in sections \ref{sec:sagem-impl} and \ref{sec:math-impl}:
we will show, as an example, how to obtain the formulae of \secref{sec:gamma-deformation}.

To implement the equations of \secref{sec:turning-twist} in either \verb+SageMath+ or \verb+Mathematica+, one should first copy-paste the code in \secref{sec:sagem-impl} or \ref{sec:math-impl}, and execute it in a notebook, and then follow the instructions below (people using  \verb+SageMath+ should not forget to activate typesetting to obtain human readable results -- they may also have to restore indentation by hand if copy-pasting removes it). In order to reproduce for instance the formulae of \secref{sec:gamma-deformation}, one should first specify that \(\forall i:~\y_i=1\).
This is done by defining a substitution\footnote{One could also specify values of \(\x_a\) from (\ref{gamma-twist}) but this is not necessary as the eigenvalues \(\x_a\) are pairwise-distinct in the case we consider (i.e. generic parameters \(\gamma_a\)).} rule \verb+Spec={+\(\y_1\)\verb+->1, +\(\y_2\)\verb+->1, +\(\dots\)\verb+}+ (in Mathematica syntax) or a dictionary \verb+Spec={+\(\y_1\)\verb+:1, +\(\y_2\)\verb+:1, +\(\dots\)\verb+}+ (in sage syntax). In what follows, instead of repeating statements in both syntax, we will show commands and their output in two columns: the left column for \verb+SageMath+, and the right column for \verb+Mathematica+.

For instance, let us show how to obtain the equation (\ref{eq:90}), using the functions \verb+hL+, \verb+hN+, \verb+hsL+ and \verb+hsN+, which respectively correspond to ``hat \(\lambda\)'', ``hat \(\nu\)'', ``hat star \(\lambda\)'' and ``hat star \(\nu\)'':

\noindent{\footnotesize\hspace{-.2cm}\begin{tabular}{c|c}
\emph{SageMath}&\emph{Mathematica}\\
    \begin{minipage}{.6\linewidth}
\begin{verbatim}
Spec={y(i):1 for i in [1,..,4]}
matrix([[hL(a),hN(a),hsL(a),hsN(a)] for a in [1,..,4]])
\end{verbatim}
    \end{minipage}
&    \begin{minipage}{.4\linewidth}
\begin{verbatim}
Spec=Table[y[i]->1,{i,1,4}];
Table[{hL[a],hN[a],hsL[a],hsN[a]},
  {a,1,4}]//MatrixForm
\end{verbatim}
    \end{minipage}
  \end{tabular}
  \begin{center}
    \(\left(\begin{array}{cccc}{\hat\lambda_{1}} = \lambda_{1} ~~~~&
{\hat\nu_{1}} = \nu_{1} ~~~~& {\hat\lambda^\star_{1}} =
\lambda_{1} ~~~~& {\hat\nu^\star_{1}} = \nu_{1} + 3
\\{\hat\lambda_{2}} = \lambda_{2} ~~~~& {\hat\nu_{2}} =
\nu_{2} - 1 ~~~~& {\hat\lambda^\star_{2}} = \lambda_{2} ~~~~&
{\hat\nu^\star_{2}} = \nu_{2} + 2 \\{\hat\lambda_{3}} =
\lambda_{3} ~~~~& {\hat\nu_{3}} = \nu_{3} - 2 ~~~~&
{\hat\lambda^\star_{3}} = \lambda_{3} ~~~~& {\hat\nu^\star_{3}} =
\nu_{3} + 1 \\{\hat\lambda_{4}} = \lambda_{4} ~~~~&
{\hat\nu_{4}} = \nu_{4} - 3 ~~~~& {\hat\lambda^\star_{4}} =
\lambda_{4} ~~~~& {\hat\nu^\star_{4}} =
\nu_{4}\end{array}\right)\)
  \end{center}
}
The matrix bellow the two columns is the computer's output when these lines are evaluated (the output is the same\footnote{Up to minor typographic differences, such as the order of terms.} with \verb+Mathematica+ as with \verb+SageMath+).

Similarly, we obtain the equation (\ref{eq:217}) using the functions \verb+AA+ and \verb+BB+ (for \(A_aA^a\) and \(B_iB^i\)):

\noindent{\footnotesize\hspace{-.2cm}\begin{tabular}{c|c}
\emph{SageMath}&\emph{Mathematica}\\
    \begin{minipage}{.5\linewidth}
\begin{verbatim}
matrix([[AA(a),BB(a)] for a in [1,..,4]])
\end{verbatim}
    \end{minipage}
&    \begin{minipage}{.5\linewidth}
\begin{verbatim}
Table[{AA[a],BB[a]},{a,1,4}]//MatrixForm
\end{verbatim}
    \end{minipage}
  \end{tabular}
  \begin{center}
\(\left(\begin{array}{ll}{A^{1}} A_{1} = \frac{{\left(x_{1} -
1\right)}^{4}}{{\left(x_{1} - x_{2}\right)} {\left(x_{1} -
x_{3}\right)} {\left(x_{1} - x_{4}\right)} x_{1}} ~~~~& {B^{1}} B_{1}
= \frac{i \, {\left(x_{1} - 1\right)} {\left(x_{2} - 1\right)}
{\left(x_{3} - 1\right)} {\left(x_{4} -
1\right)}}{{\left({\hat\nu_{1}} - {\hat\nu_{2}}\right)}
{\left({\hat\nu_{1}} - {\hat\nu_{3}}\right)}
{\left({\hat\nu_{1}} - {\hat\nu_{4}}\right)}} \\{A^{2}} A_{2}
= -\frac{{\left(x_{2} - 1\right)}^{4}}{{\left(x_{1} - x_{2}\right)}
{\left(x_{2} - x_{3}\right)} {\left(x_{2} - x_{4}\right)} x_{2}}
~~~~& {B^{2}} B_{2} = -\frac{i \, {\left(x_{1} - 1\right)}
{\left(x_{2} - 1\right)} {\left(x_{3} - 1\right)} {\left(x_{4} -
1\right)}}{{\left({\hat\nu_{1}} - {\hat\nu_{2}}\right)}
{\left({\hat\nu_{2}} - {\hat\nu_{3}}\right)}
{\left({\hat\nu_{2}} - {\hat\nu_{4}}\right)}} \\{A^{3}} A_{3}
= \frac{{\left(x_{3} - 1\right)}^{4}}{{\left(x_{1} - x_{3}\right)}
{\left(x_{2} - x_{3}\right)} {\left(x_{3} - x_{4}\right)} x_{3}}
~~~~& {B^{3}} B_{3} = \frac{i \, {\left(x_{1} - 1\right)}
{\left(x_{2} - 1\right)} {\left(x_{3} - 1\right)} {\left(x_{4} -
1\right)}}{{\left({\hat\nu_{1}} - {\hat\nu_{3}}\right)}
{\left({\hat\nu_{2}} - {\hat\nu_{3}}\right)}
{\left({\hat\nu_{3}} - {\hat\nu_{4}}\right)}} \\{A^{4}} A_{4}
= -\frac{{\left(x_{4} - 1\right)}^{4}}{{\left(x_{1} - x_{4}\right)}
{\left(x_{2} - x_{4}\right)} {\left(x_{3} - x_{4}\right)} x_{4}}
~~~~& {B^{4}} B_{4} = -\frac{i \, {\left(x_{1} - 1\right)}
{\left(x_{2} - 1\right)} {\left(x_{3} - 1\right)} {\left(x_{4} -
1\right)}}{{\left({\hat\nu_{1}} - {\hat\nu_{4}}\right)}
{\left({\hat\nu_{2}} - {\hat\nu_{4}}\right)}
{\left({\hat\nu_{3}} -
{\hat\nu_{4}}\right)}}\end{array}\right)\)  \end{center}
}

Moreover, if one wants this output to be expressed in terms of the charges \(\lambda_a\) and \(\nu_i\) instead of \(\lL\) and \(\lM\), one can use the function \verb+subHat+ to substitute the expression of \(\lL\) and \(\lM\). And if one wants to further substitute \(\lambda_a\) and \(\nu_i\) using expression (\ref{eq:1001}), then one can use the function \verb+subln+, as in the following example:

\noindent{\footnotesize\hspace{-.2cm}\begin{tabular}{c|c}
\emph{SageMath}&\emph{Mathematica}\\
    \begin{minipage}{.53\linewidth}
\begin{verbatim}
def f(i,j):return subln(subHat([AA,BB][j](i+1)))
matrix(4,2,f)
\end{verbatim}
    \end{minipage}
&    \begin{minipage}{.47\linewidth}
\begin{verbatim}
MatrixForm[Table[{AA[a],BB[a]},{a,1,4}
        ]//subHat//subln//Factor]
\end{verbatim}
    \end{minipage}
  \end{tabular}
  \begin{center}
\(\left(\begin{array}{ll}
{A^{1}} A_{1} = \frac{{\left(x_{1} - 1\right)}^{4}}{{\left(x_{1} -
x_{2}\right)} {\left(x_{1} - x_{3}\right)} {\left(x_{1} - x_{4}\right)}
x_{1}} ~~~~& {B^{1}} B_{1} = \frac{i \, {\left(x_{1} - 1\right)}
{\left(x_{2} - 1\right)} {\left(x_{3} - 1\right)} {\left(x_{4} -
1\right)}}{{\left(\Delta - S_{1} - 3\right)} {\left(\Delta - S_{2} -
2\right)} {\left(S_{1} + S_{2} + 1\right)}} \\
{A^{2}} A_{2} = -\frac{{\left(x_{2} - 1\right)}^{4}}{{\left(x_{1} -
x_{2}\right)} {\left(x_{2} - x_{3}\right)} {\left(x_{2} - x_{4}\right)}
x_{2}} ~~~~& {B^{2}} B_{2} = -\frac{i \, {\left(x_{1} - 1\right)}
{\left(x_{2} - 1\right)} {\left(x_{3} - 1\right)} {\left(x_{4} -
1\right)}}{{\left(\Delta + S_{1} - 1\right)} {\left(\Delta + S_{2} -
2\right)} {\left(S_{1} + S_{2} + 1\right)}} \\
{A^{3}} A_{3} = \frac{{\left(x_{3} - 1\right)}^{4}}{{\left(x_{1} -
x_{3}\right)} {\left(x_{2} - x_{3}\right)} {\left(x_{3} - x_{4}\right)}
x_{3}} ~~~~& {B^{3}} B_{3} = \frac{i \, {\left(x_{1} - 1\right)}
{\left(x_{2} - 1\right)} {\left(x_{3} - 1\right)} {\left(x_{4} -
1\right)}}{{\left(\Delta + S_{1} - 1\right)} {\left(\Delta - S_{2} -
2\right)} {\left(S_{1} - S_{2} + 1\right)}} \\
{A^{4}} A_{4} = -\frac{{\left(x_{4} - 1\right)}^{4}}{{\left(x_{1} -
x_{4}\right)} {\left(x_{2} - x_{4}\right)} {\left(x_{3} - x_{4}\right)}
x_{4}} ~~~~& {B^{4}} B_{4} = -\frac{i \, {\left(x_{1} - 1\right)}
{\left(x_{2} - 1\right)} {\left(x_{3} - 1\right)} {\left(x_{4} -
1\right)}}{{\left(\Delta - S_{1} - 3\right)} {\left(\Delta + S_{2} -
2\right)} {\left(S_{1} - S_{2} + 1\right)}}
\end{array}\right)\)
  \end{center}
}

For other cases of twisting than the \(\gamma\)-deformation of \secref{sec:gamma-deformation}, we can still use the same functions, but they should be preceded with another definition of the substitution \verb+Spec+: for instance \secref{sec:lead-asympt-fully} corresponds to \verb+Spec={}+, while \secref{sec:beta-deformation} corresponds to \verb+Spec=Join[Table[y[i]->1,{i,1,4}],{x[4]->1}];+ (assuming the considered state does not have two charges equal).

\subsubsection{Sagemath implementation}
\label{sec:sagem-impl}
{
\scriptsize{
\begin{Verbatim}[commandchars=\@7\&]
#The definitions below allow for a nicely formated output; for instance x(1) returns @m7x_1&
def sb(s):return lambda i:var("%
x=sb("x");y=sb("y");lam=sb("lambda");nu=sb("nu");J=sb("J");S=sb("S");var("Delta_")
def hl(a):return var("hl_%
def hn(a):return var("hn_%
def hsl(a):return var("hsl_%
def hsn(a):return var("hsn_%
idx=lambda s:Integer(str(s)[-1])  @hfill # converts x(a) into the label a
#The definition below specifies the nesting path according to @eqref7eq:211&
NP=[y(1),x(1),x(2),y(2),y(3),x(3),x(4),y(4)]
def delta(i,j):return bool((i==j).subs(Spec)) @hfill # delta(x(a),x(b)) computes \delta_{x_a,x_b}
def bosQ(i):return str(i)[0]=="x" @hfill #checks ``bosonicness'': bosQ(x(a))=1 whereas bosQ(y(i))=0
def hL(a): @hfill #computes \hat @m7@lambda&_a from @eqref7eq:221&
    S=sum([(-1)^(bosQ(b))*delta(x(a),b) for b in NP if NP.index(b)<NP.index(x(a))])
    return hl(a)==lam(a)+S
def hN(i):@hfill #computes \hat @m7@nu&_i from @eqref7eq:221&
    S=sum([-(-1)^(bosQ(b))*delta(y(i),b) for b in NP if NP.index(b)<NP.index(y(i))])
    return hn(i)==nu(i)+S
def hsL(a):@hfill #computes \hat @m7@lambda&_a^\star from @eqref7eq:145&
    S=sum([-(-1)^(bosQ(b))*delta(x(a),b) for b in NP if NP.index(b)>NP.index(x(a))])
    return hsl(a)==lam(a)+S
def hsN(i):@hfill #computes \hat @m7@nu&_i^\star from @eqref7eq:145&
    S=sum([(-1)^(bosQ(b))*delta(y(i),b) for b in NP if NP.index(b)>NP.index(y(i))])
    return hsn(i)==nu(i)+S
def z(a,b): @hfill #From inputs of the form x(a) or y(i), computes z_{a,b} defined by @eqref7eq:225&
    if delta(a,b):
        if bosQ(a) and bosQ(b): return I*a*(hl(idx(b))-hl(idx(a)))
        elif bosQ(a): return I*a*(1-hn(idx(b))-hl(idx(a)))
        elif bosQ(b): return -I*a*(1-hn(idx(a))-hl(idx(b)))
        else: return I*a*(hn(idx(b))-hn(idx(a)))
    else:
        if bosQ(a): return b-a
        else: return a-b
def AA(a): @hfill #computes A_aA^a from @eqref7eq:204&
    LHS=sb("A")(a)*var("A%
    return LHS==(1/x(a)*prod([z(b,x(a))^((-1)^bosQ(b)) for b in NP if b!=x(a)])).subs(Spec)
def BB(i): @hfill #computes B_iB^i from @eqref7eq:204&
    LHS=sb("B")(i)*var("B%
    return LHS==(1/y(i)*prod([z(j,y(i))^(-(-1)^bosQ(j)) for j in NP if j!=y(i)])).subs(Spec)
def subHat(e): @hfill #substitutes the expression @eqref7eq:221& of \hat @m7@lambda&_a and \hat @m7@nu&_i into the expression e
    for i in [1,..,4]:
        e=e.subs(hL(i)).subs(hN(i))
    return e
def l(a): @hfill #returns the expression @eqref7eq:1000& of @m7@lambda&_a
    return lam(a)==1/2*((-1)^(a>2)*J(1)-(-1)^a*J(2)+(-1)^(abs(a-2.5)>1)*J(3))
def n(i): @hfill #returns the expression @eqref7eq:1002& of @m7@nu&_i
    return nu(i)==1/2*(-(-1)^(i>2)*Delta_-(-1)^i*S(1)-(-1)^(abs(i-2.5)>1)*S(2))
def subln(e):@hfill #substitutes the expressions @eqref7eq:1001& of @m7@lambda&_a and @m7@nu&_i into the expression e
    for i in [1,..,4]:
        e=e.subs(l(i)).subs(n(i))
    return e
\end{Verbatim}
}
}

\subsubsection{Mathematica implementation}
\label{sec:math-impl}
{\scriptsize{
\begin{Verbatim}[commandchars=678]
(*The definitions below allow for a nicely formated output; e.g.  hsl[2] returns \hat 6m76lambda8_2^\star  *)
sb[s_]:=Subscript[s,#]&;lam=sb[\[Lambda]];nu=sb[\[Nu]];x[a_]=sb[x][a];y[i_]=sb[y][i];
J[a_]=sb[J][a];S[i_]=sb[S][i];hn=sb[OverHat[\[Nu]]];hl=sb[OverHat[\[Lambda]]];
hsn=sb[SuperStar[OverHat[\[Nu]]]];hsl=sb[SuperStar[OverHat[\[Lambda]]]];
(*The definition below specifies the nesting path according to 6eqref7eq:2118*)
NP={y[1],x[1],x[2],y[2],y[3],x[3],x[4],y[4]};
delta[i_,j_]:=Boole[(i/.Spec)===(j/.Spec)];6hfill(* delta(x(a),x(b)) computes \delta_{x_a,x_b} *)
bosQ[i_]:=Boole[x===i[[1]]]; 6hfill(*checks ``bosonicness'': bosQ(x(a))=1 whereas bosQ(y(i))=0*)
idx=Last; 6hfill(*converts x(a) into the label a*)
hL[a_]:=Block[{S},6hfill(*computes \hat 6m76lambda8_a from 6eqref7eq:2218*)
6hfill S=Sum[(-1)^bosQ[b]*delta[x[a],b]*Boole[Position[NP,b][[1,1]]<Position[NP,x[a]][[1,1]]],{b,NP}];
      hl[a]==lam[a]+S]
hN[i_]:=Block[{S},6hfill(*computes \hat 6m76nu8_i from 6eqref7eq:2218*)
6hfill S=Sum[-(-1)^bosQ[b]*delta[y[i],b]*Boole[Position[NP,b][[1,1]]<Position[NP,y[i]][[1,1]]],{b,NP}];
     hn[i]==nu[i]+S]
hsL[a_]:=Block[{S},6hfill(*computes \hat 6m76lambda8_a^\star from 6eqref7eq:1458*)
6hfill S=Sum[-(-1)^bosQ[b]*delta[x[a],b]*Boole[Position[NP,b][[1,1]]>Position[NP,x[a]][[1,1]]],{b,NP}];
     hsl[a]==lam[a]+S]
hsN[i_]:=Block[{S},6hfill(*computes \hat 6m76nu8_i^\star from 6eqref7eq:1458*)
6hfill S=Sum[(-1)^bosQ[b]*delta[y[i],b]*Boole[Position[NP,b][[1,1]]>Position[NP,y[i]][[1,1]]],{b,NP}];
       hsn[i]==nu[i]+S]
AA[a_]:=sb[A][a]*Superscript[A,a]==1/x[a]*Product[If[b===x[a],1, 6hfill (*computes A_aA^a from 6eqref7eq:2048*)
        z[b,x[a]]^((-1)^bosQ[b])],{b,NP}]/.Spec
BB[i_]:=sb[B][i]*Superscript[B,i]==1/y[i]*Product[If[b===y[i],1, 6hfill (*computes B_iB^i from 6eqref7eq:2048*)
        z[b,y[i]]^(-(-1)^bosQ[b])],{b,NP}]/.Spec
z[a_,b_]:=If[delta[a,b]==1,6hfill (*From inputs of the form x(a) or y(i), computes z_{ab} defined by 6eqref7eq:2258*)
        Which[bosQ[a]==1==bosQ[b],  I a (hl[idx[b]]-hl[idx[a]]),
        bosQ[a]==1,  I a (1-hn[idx[b]]-hl[idx[a]]),
        bosQ[b]==1, -I a (1-hn[idx[a]]-hl[idx[b]]),
        True,       I a (hn[idx[b]]-hn[idx[a]])],
        If[bosQ[a]==1,b-a,a-b]]
l[a_]:=lam[a]==1/2*((-1)^Boole[a>2]*J[1]-(-1)^a*J[2]+(-1)^Boole[Abs[a-2.5]>1]*J[3])6hfill(*equation 6eqref7eq:10008*)
n[a_]:=nu[a]==1/2*(-(-1)^Boole[a>2]*\[CapitalDelta]-(-1)^a*S[1]-(-1)^Boole[Abs[a-2.5]>1]*S[2])
subln[e_]:=e/.Flatten[Table[{Rule@@l[a],Rule@@n[a]},{a,1,4}]]
subHat[e_]:=e/.Flatten[Table[{Rule@@hL[a],Rule@@hN[a]},{a,1,4}]]6hfill(*substitutes 6eqref7eq:10018 into e*)
\end{Verbatim}
}
}

\subsection{Fully untwisted case}

 In this case we have  \(\x_a=\y_j=1,\quad a,j=1,2,3,4 \), and the equations \eqref{eq:218}, give the 
asymptotics 
{\setlength{\belowdisplayskip}{0pt} \setlength{\belowdisplayshortskip}{0pt}
  \begin{align}
  \label{eq:192}
    \lL&=\lambda_a+(a~\mathrm{mod}~2)\,,&\lM&=\nu_i+(a+1~\mathrm{mod}~2)\,,&
    \uL&=\lambda_a-(a+1~\mathrm{mod}~2)\,,&\uM&=\nu_i-(i~\mathrm{mod}~2)\,,
  \end{align}}{\setlength{\abovedisplayskip}{0pt} \setlength{\abovedisplayshortskip}{0pt}\begin{align}
  \label{eq:192a}
  A_aA^a&=\frac{\bi\prod_{i}(\lL+\lM-1)}{\prod_{b\ne a}(\lL-\lL[b])}\,,&
  B_iB^i&=\frac{\bi\prod_{a}(\lL+\lM-1)}{\prod_{j\ne i}(\lM-\lM[j])}\,,&  \end{align}}
which matches the result already  obtained in
\cite{Gromov:2014caa} (eq.(3.64) and (3.68)).

They can also easily be expressed in terms of the charges \(J_1\), \(J_2\), \(\dots\) by executing he code of \secref{sec:comp-impl}. The only difference with the example of \secref{sec:gamma-deformation} is that one should start with setting\footnote{The definition is written in \emph{Mathematica syntax}. In \emph{SageMath} syntax, the same constraint can (for instance) be specified as \verb+Spec={f(i):1 for f in [x,y] for i in [1,..,4]}+.} \verb+Spec={x[_]->1,y[_]->1}+.

Notice that in the non-twisted case, the T-functions   \eqref{eq:60}--\eqref{eq:62} will turn to the dimensions of corresponding rectangular representations  in the large \(u\) limit.

The untwisted coset possesses the full \(\mathcal{N}=4\) supersymmetry.

\subsection{The case \texorpdfstring{$\x_1=\y_4$}{x\_1=y\_4}}    First we consider a simple case of partial twisting and put $\x_1=\y_4=\z$, with other twists being arbitrary.
We can use the formulae of \secref{sec:turning-twist}, or the computer implementation of section\footnote{To specify that \(\x_1=\y_4\), one can start by setting \verb+Spec={x(1):SR.var("z"),y(4):SR.var("z")}+ (in \emph{SageMath} syntax) or \verb+Spec={x[1]->z,y[4]->z}+ (in \emph{Mathematica} syntax).}, obtaining the following coefficients for the asymptotics \eqref{eq:143}:
{\setlength{\belowdisplayskip}{0pt}%
  \setlength{\belowdisplayshortskip}{0pt}%
  \begin{align}
  \label{eq:192b}
    \lL&=\lambda_a\,,&\lM&=\nu_i+\delta_{i,4}\,,&
    \uL&=\lambda_a-\delta_{a,1}\,,&\uM&=\nu_i\,,
  \end{align}}%
{\setlength{\abovedisplayskip}{0pt}
  \setlength{\abovedisplayshortskip}{0pt}%
\begin{align}
  A_aA^a&=
  \begin{cases}
    \frac{\bi(\lambda_1+\nu_4)\prod_{i\le3}(\z-\y_i)}{\prod_{b\neq 1}(\x_b-\z)}&\textrm{ if }a=1\\
    \frac{\prod_{i\le 3}(\x_a-\y_i)}{\x_a\prod_{\substack{b\neq a\\b\ge 2}}(\x_a-\x_b)}&\textrm{ if }a>1
  \end{cases}&
  B_iB^i&= 
  \begin{cases}
    \frac{\prod_{a\ge2}(\x_a-\y_i)}{\y_i\prod_{\substack{j\le3\\j\ne i}}(\y_j-\y_i)}&\textrm{ if }i\le3\\
    \frac{\bi(\lambda_1+\nu_4)\prod_{a\ge2}(\x_a-\z)}{\prod_{j\le3}(\y_j-\z)}&\textrm{ if }i=4\,.
  \end{cases}
 \end{align}}

 The residual supersymmetry is here \(\mathcal{N}=1\). . 
A way to see it for this for arbitrary cases of twisting is as follows:
\begin{itemize}
\item   Take the twist matrix in diagonal form \(g=\text{diag}\{x_1,\dots,x_4|y_1,\dots,y_4\}\);   
\item Take a matrix \(M\)  of \((4|4)\) generators of \(SU(4|4)\) algebra. Find a general solution of equation \([M,g]=0\) where \(M_{i,j}\) is otherwise an arbitrary \((4|4\times(4|4)\) supermatrix, with the labels running the values e.g.  \(i,j=1,2,3,4|-1,-2,-3,-4\). The  matrix elements of \(M\) which remain non-zero correspond to the generators of the remaining symmetry subalgebra; 

\item Bring \(M\) into a block-diagonal form by reshuffling its rows and columns. The sizes of  blocks together with the collection of indices in each grading within each block will indicate the leftover (super)symmetry.
\end{itemize}
For instance, for the case of this subsection we have the twist matrix
 \begin{equation}
g=\text{diag}\{\z,\x_2,\x_3,\x_4|\y_1,\y_2,\y_3,\z\}.
\end{equation} 
After reordering indices (by putting fermionic indices before bosonic ones) to get a block-diagonal matrix, the solution
of the equation  \([M,g]=0\) gives the matrix \(M\) in the form
 {\renewcommand{\arraycolsep}{.15cm}\begin{equation}
M=\left(
\begin{array}{cccc|cccc}
 * & 0 & 0 & 0 & 0 & 0 & 0 & 0 \\
 0 & * & 0 & 0 & 0 & 0 & 0 & 0 \\
 0 & 0 & * & 0 & 0 & 0 & 0 & 0 \\
 0 & 0 & 0 & * & * & 0 & 0 & 0 \\\hline
 0 & 0 & 0 & * & * & 0 & 0 & 0 \\
 0 & 0 & 0 & 0 & 0 & * & 0 & 0 \\
 0 & 0 & 0 & 0 & 0 & 0 & * & 0 \\
 0 & 0 & 0 & 0 & 0 & 0 & 0 & * \\
\end{array}
\right).
\end{equation}}
The 2\(\times\)2 block in the middle provides us with the only
non-abelian residual symmetry. This block is actually rather a
\((1|1)\times(1|1)\) block
supersymmetry, while the rest of (super)symmetries of the overall
\(\text{psu}(2,2|4) \) are broken up to one-dimensional subgroups:
\(\text{psu}(2,2|4)\to ps[ u(1|1) \oplus\text{u(1)} \oplus\text{u(1)} \oplus\text{u(1)} \oplus\text{u(1)}\oplus\text{u}(1)\oplus
\mathbb{R]}\). 
 
We will give below the residual   non-abelian subgroups for other cases of partial twisting which can be easily determined by similar method.

\subsection{The case \texorpdfstring{$\x_1=\x_2$}{x\_1=x\_2}}  Now we consider a slightly more complicated case $\x_1=\x_2=\z$.
  Again using the general formulae (\ref{eq:218}) give the asymptotics \eqref{eq:143} with the coefficients
{\setlength{\belowdisplayskip}{0pt} \setlength{\belowdisplayshortskip}{0pt}
  \begin{align}
  \label{eq:192c}
    \lL&=\lambda_a-\delta_{a,2}\,,&\lM&=\nu_i\,,&
    \uL&=\lambda_a+\delta_{a,1}\,,&\uM&=\nu_i\,,
  \end{align}}{\setlength{\abovedisplayskip}{0pt} \setlength{\abovedisplayshortskip}{0pt}\begin{align}
  A_aA^a&=
  \begin{cases}
    (-1)^a\frac{\bi \prod_{i}(\z-\y_i)}{(\lL[1]-\lL[2])\z^2\prod_{b=3}^4(\z-\x_b)}&\textrm{ if }a\le 2\\
    (-1)^a\frac{\prod_{i}(\x_a-\y_i)}{\x_a(\x_4-\x_3)(\x_a-\z)^2}&\textrm{ if }a\ge 3
  \end{cases}&
  \mathsf{B}_i\mathsf{B}^i&=\frac{\displaystyle \prod_{a}(\x_a-\y_i)}{\displaystyle \y_i\prod_{j\ne i}(\y_j-\y_i)}\,.
 \end{align}}

 The  supersymmetry is here completely broken and the residual non-abelian symmetry is \(SU(2)\).%

   \subsection{Twisted \texorpdfstring{$sl_2$}{sl\_2}-sector}

In this case only the values of Cartan charges \(\Delta\), \(J_1\equiv L\) and \(S_1\equiv S\) are nonzero. We choose accordingly the configuration of twists, using  (\ref{eq:1001}): \(\x_1\x_2=t^2\), \(\x_1\x_3=1\), \(\x_2\x_3=1\), \(\y_1\y_2=1\), \(\y_2\y_4=y^2\), \(\y_2\y_3=1\), or
 \begin{align}
\x_1&=\x_2=\frac{1}{\x_3}=\frac{1}{\x_4}=t,& \y_2&=\frac{1}{\y_1}=\y_4=\frac{1}{\y_3}=y
\end{align}
where \(t^2=e^{i{\cal P}}\) %
and \(y=e^{i\phi}\) is the twist variable related to the conformal spin direction.

Using the general formulae (\ref{eq:218}) give the asymptotics \eqref{eq:143} with the coefficients
{{\setlength{\belowdisplayskip}{0pt} \setlength{\belowdisplayshortskip}{0pt}
\begin{align}
\lL-\lambda_a&=(0,-1,0,-1)\,&\lM-\nu_i&=(0,0,-1,-1)\,,&
\uL-\lambda_a&=(0,1,0,1)\,&\uM-\nu_i&=(1,1,0,0)\,,
\end{align}%
\setlength{\abovedisplayskip}{0pt} \setlength{\abovedisplayshortskip}{0pt}%
\begin{align}
\lambda_1&=\lambda_2=-\lambda_3=-\lambda_4=\frac{J_1}2\,,&
\nu_4=-\nu_1&=\frac{\Delta-S_1}2\,,&
\nu_3=-\nu_2&=\frac{\Delta+S_1}2\,,
\end{align}}\setlength{\abovedisplayskip}{0pt} \setlength{\abovedisplayshortskip}{0pt}%
\begin{align}
  \mathsf{A}_a\mathsf{A}^a&=(-1)^a\bi\frac{(t-y)^2 (t y-1)^2}{\left(t^2-1\right)^2 y^2}\,,&
  \mathsf{B}_1\mathsf{B}^1=\mathsf{B}_2\mathsf{B}^2=-\mathsf{B}_3\mathsf{B}^3=-\mathsf{B}_4\mathsf{B}^4&=\bi\frac{(t-y)^2(t y-1)^2}{(\Delta -1) t^2 \left(y^2-1\right)^2}\,.&
\end{align}
}

The superconformal symmetry is completely broken: \(\text{psu}(2,2|4)\to ps[\text{u}(1,1)\oplus\text{u(2)}\oplus \text{u}(1)\oplus \text{u}(1)]\)    unless we put the total momentum to zero: \({\cal P}=0\). Then the \(\mathcal{N}=2\) supersymmetry gets restored.

\paragraph{\text{sl(2)} with only \(AdS_5\) twists:} If we twist  only the \(AdS_5\) in the previous example, we should  put \(t=1\) and we get instead
{\setlength{\belowdisplayskip}{0pt} \setlength{\belowdisplayshortskip}{0pt}
\begin{align}
\lL-\lambda_a&=(0,-1,-2,-3)&
\lM-\nu_i&=(0,0,-1,-1)&
\uL-\lambda_a&=(3,2,1,0)&
\lM-\nu_i&(1,1,0,0)&
\end{align}%
{\setlength{\abovedisplayskip}{0pt} \setlength{\abovedisplayshortskip}{0pt}%
\begin{align}
  \mathsf{A}_1\mathsf{A}^1&=-\mathsf{A}_4\mathsf{A}^4=\bi\frac{(y-1)^4}{(J_1+3)(J_1+2)y^2}\,,&
  \mathsf{A}_2\mathsf{A}^2&=-\mathsf{A}_3\mathsf{A}^3=\bi\frac{(y-1)^4}{(J_1+2)(J_1+1)y^2}\,,
\end{align}
}}{\setlength{\abovedisplayskip}{0pt} \setlength{\abovedisplayshortskip}{0pt}%
\begin{align}
\mathsf{B}_1\mathsf{B}^1&=\mathsf{B}_2\mathsf{B}^2=-\mathsf{B}_3\mathsf{B}^3=-\mathsf{B}_4\mathsf{B}^4=\bi\frac{(y-1)^2}{(\Delta-1)(y+1)^2}
\end{align}
}

The left-over symmetry is \(\text{psu}(2,2|4)\to ps[\text{u}(1,1)\oplus\text{u(2)}\oplus \text{u}(4)]\)  so that   the supersymmetry is completely broken.

\paragraph{\text{sl(2)} with only \(S^5\) twists:} If we twist here only the \(S^5\), putting \(y=1\) we get instead
{\setlength{\belowdisplayskip}{0pt} \setlength{\belowdisplayshortskip}{0pt}
\begin{align}
\lL&=\lambda_a-(a+1 \mathrm{mod} 2)&
\lM&=\nu_i+1-a&
\uL&=\lambda_a-(a \mathrm{mod} 2)&
\lM&=\nu_i+4-i&
\end{align}%
{\setlength{\abovedisplayskip}{0pt} \setlength{\abovedisplayshortskip}{0pt}%
\begin{align}
\mathsf{A}_a\mathsf{A}^a&=(-1)^a\bi\frac{(t-1)^2}{\left(t+1\right)^2}\,,
&  \mathsf{B}_1\mathsf{B}^1&=-\mathsf{B}_4\mathsf{B}^4=\bi\frac{(t-1)^4}{(\Delta-S_1-3)(\Delta-2)(S_1+1)t^2}\,,\\&&
  \mathsf{B}_2\mathsf{B}^2&=-\mathsf{B}_3\mathsf{B}^3=-\bi\frac{(t-1)^4}{(\Delta-S_1-1)(\Delta-2)(S_1+1)t^2}\,.
\end{align}
}

The left-over symmetry is \(\text{psu}(2,2|4)\to ps[\text{u}(2,2)\oplus\text{u(2)}\oplus \text{u}(2)]\)  so that the theory is conformal.

\section{Links to TBA: mirror Q- T- and Y-functions and the energy}
\label{sec:TBAfromQSC}
In this appendix, we will re-derive the TBA formula for the energy from   the point of view of QSC construction, and relate the mirror Y-functions entering there through the Q-functions  in the gauge in which they enter the  QSC equations. This makes for example our derivation of the single wrapping energy of BMN vacuum of the sec.~\ref{sec:BMNvac} completely consistent within the QSC formalism. 

\subsection{Formulae for T-functions of mirror T-hook.}
\label{sec:TfunctionsfromThook}
As was stressed a number of times in this paper, construction of T-functions on a {\THook}  depends non-trivially on a choice of a basis in a Q-system. A particular ``mirror'' basis \(\bQ_{A|I}\) of Q-functions that reproduces the ``black'' gauge \(\bT_{a,s}\) (introduced in \cite{Gromov2011a,Gromov:2014caa}) and hence the Y-functions entering the TBA equations were linked to the Q-functions of QSC in appendix B of \cite{Gromov:2014caa}. 
The non-trivial part of the construction is the relations valid in the mirror kinematics:
\begin{align}\label{eq:sHrot}
\bQ_{1|\es}&=\bP_1\,,&  \bQ_{2|\es}&=\bP_2\,,& \bQ_{3|\es}&=\mu^{4a}\bP_a\,,& \bQ_{4|\es}&=-\mu^{3a}\bP_a\,.
\end{align}
This identification is  perceived as an H-rotation which should be applied by covariance to all Q-functions, see \secref{sec:rotat-q-funct}. Note that this rotation has a non-unit determinant and one finds \(\bQ_{\bar\es|\bar\es}=(\mu_{12}^+)^{2}\neq 1\), hence the formulae including Hodge-dual functions should be casted in their full form presented in this paper but not in the simplified form with assumption \(Q_{\bar\es|\bar\es}=1\) which is typically used in QSC.

We construct \(\bT_{a,s}\) using the mirror basis of Q-functions \(\bQ_{A|I}\)  and according to the formulae \eqref{eq:155}, with \(K_1=K_2=M_1=M_2=2\), \(\tilde s=s\), \(\tilde a=a\), and \(\varepsilon\) simplified to
\be
\varepsilon_r(a,s)=\varepsilon_l(a,s)=(-1)^{a-s}\,, \quad \varepsilon_u(a,s)=(-1)^{a-s+a\,s}\,;
\ee
and then apply transformation \eqref{eq:sHrot} to express the answer explicitly in terms of Q-functions used in QSC. 

Since the mirror  \(\bT_{a,s}\) are analytic only in the bands of finite width on the complex plane of \(u\) it is very handy to operate simultaneously with both UHPA functions \(Q_{A|I}\) and LHPA functions \(Q^{A}|_{I}\) defined and explained in \cite{Gromov:2014caa}. For instance, according to \eqref{eq:sHrot}, one has \(\bQ_{3|\es}=Q^4|_{\es}\). 

For the upper band of the {\THook}, \(a\geq |s|\), which we will need for the formula for energy, one has the following explicit formulae
\be\label{bTasder}
\bT_{a,s}&=&(-1)^{a-s+as}\sum_{|I|=2-s}\epsilon^{I\bar I}\bQ_{12|I}^{[+a]}\bQ_{34|\bar I}^{[-a]}=
(-1)^{a-s+as}\sum_{|I|=2-s}\epsilon^{I\bar I}Q_{12|I}^{[+a]}(\mu^{3c}\mu^{4d})^{[a+s+1]}Q_{cd|\bar I}^{[-a]}
\nonumber\\
&=&(-1)^{a-s+as}\sum_{|I|=2-s}\epsilon^{I\bar I}Q_{12|I}^{[+a]}Q^{34}|_{\bar I}^{[-a]}=(-1)^{a+as}\sum_{|I|=2-s}Q_{12|I}^{[+a]}\left(Q_{12}|^{I}\right)^{[-a]}\,.
\ee
We could also use the property \(\hat Q_{12}|^{I}=(\omega^{II'})^{[|I|+1]}\hat Q_{12|I'}\) valid in the physical kinematics \cite{Gromov:2014caa} to write the answer uniquely in terms of UHPA Q-functions of QSC, as it is done in \eqref{eq:bTa1} and \eqref{eq:bTa0}.

\subsection{Derivation of TBA formula for Energy from QSC}
\label{app:E2}
We recall that the ``black'' gauge \(\bT_{a,s}\) satisfies certain properties: \(\bT_{0,s}=\bT_{0,-s}=\bT_{0,0}^{[s]}=\bT_{0,0}^{[-s]}\) and \(\bT_{a,2}=\bT_{2,a}\,,\quad\bT_{a,-2}=\bT_{-2,-a}\,\, (a\ge 2)\), where all functions are on the sheet with long cuts. In addition, we know that \(\bT_{0,0}=\mu_{12}^2\) and  \cite{Gromov:2014caa}
\(\bT_{1,0}=\tilde\mu_{12}\mu_{12}\).

We will use the standard definition of Y-functions \(1+Y_{a,s}=\frac{\bT_{a,s}^+\bT_{a,s}^-}{\bT_{a+1,s}\bT_{a-1,s}}\), relating the T-system to the Y-system, and the  ``telescoping'' formulae -- chain cancelations of T-functions -- to write \begin{eqnarray*}  
&&\left(\frac{\mu_{12}}{\tilde\mu_{12}}\right)^{[2n]} \prod_{a=1}^n(1+ Y_{a,0}^{[2n-a]})=\left(\frac{\bT_{0,0}^{[-1]}}{ \bT_{1,0}^{[0]}}\right)^{[2n]}\prod_{a=1}^{n}
 \left(\frac{ \bT_{a,0}^{[2n-a+1]}\bT_{a,0}^{[2n-a-1]}}{\bT_{a-1,0}^{[2n-a]}\bT_{a+1,0}^{[2n-a]}}\right)
 =\frac{ \bT_{n,0}^{[n-1]}}{\bT_{n+1,0}^{[n]}}\,.
\end{eqnarray*}  
Now use that functions \(\bT_{n,0}\) are analytic in the strip \(-\frac{n+1}{2}<{\rm Im}(u)<\frac{n+1}{2}\,,\) cf. \eqref{bTasder}. 
This analyticity implies that the quantity \(\frac{\mu_{12}}{\tilde\mu_{12}} \prod\limits_{a=1}^n(1+ Y_{a,0}^{[-a]})\) is analytic for any \(n\) in the strip \(0< \Im m(u)\le \bi\,n\). Hence  \(\frac{\mu_{12}}{\tilde\mu_{12}} \prod\limits_{a=1}^\infty(1+ Y_{a,0}^{[-a]})\) is analytic in the whole upper-half plane \(\Im m(u)>0\). 

The large-\(u\) asymptotics of the discussed quantity is known and is determined by the energy. Indeed,
\begin{equation}
\label{muovermu}\frac{\tilde\mu_{12}}{\mu_{12}}\equiv\frac{\tilde{\hat\mu}_{12}}{\hat\mu_{12}}=\frac{\hat\mu_{12}^{++}}{\hat\mu_{12}}= \frac{\bi (\Delta-J_1)}{u}+{\cal O}\left(\frac{1}{u^{-2}}\right)\,,
\end{equation} 
while  the second term \(\log \prod\limits_{a=1}^n(1+ Y_{a,0}^{[-a]})\) diminishes   at \(u\to\infty\) quicker than \(1/u\), as it can be seen by its asymptotic expression \(Y_{a,0}^{[-a]}\sim \left(\frac{x^{[-0]}}{x^{[2a]}}\right)^{J_1}\simeq\frac{1}{[(u+\bi 0)(u+\bi a)]^{J_1}}\) computed e.g. in the main text. 

Hence we can write the following useful Cauchy representation:
\begin{eqnarray*}  \frac{1}{  \sqrt{u^2-4g^2}}\log\left(\frac{\mu_{12}}{\tilde\mu_{12}} \prod_{a=1}^\infty(1+ Y_{a,0}^{[-a]})\right)= \int_{-\infty}^{\infty}\frac{dv}{2\pi\bi}\frac{1}{  (u-v)\sqrt{v^2-4g^2}} \log\left(\frac{\mu_{12}}{\tilde\mu_{12}} \prod_{a=1}^\infty(1+ Y_{a,0}^{[-a]})\right)\,.
\end{eqnarray*}
The contour of integration can be closed around a big semi-circle surrounding the upper half-plane since the integrand diminishes as \(1/u^2\) at \(u\to\infty\).

Let us expand now both sides of this equation  up to the order \(1/u^2\). The r.h.s. renders \(\frac{\Delta-J_1}{u^2}\) due to \eqref{muovermu}. Comparing it to the \(1/u^2\) term of the l.h.s., we obtain the equation   
\begin{eqnarray} \gamma=\Delta -J_1 = \oint\frac{du}{2\pi\bi}\frac{u}{  \sqrt{u^2-4g^2}} \left(-\log\left(\frac{\tilde\mu_{12}}{\mu_{12}}\right) +\log\prod_{a=1}^\infty(1+ Y_{a,0}^{[-a]})\right)\,.
\end{eqnarray}   Since the l.h.s. is real we can replace the r.h.s. by its real part. The  first term in the integrand will not contribute\footnote{This type of cancellation is a useful trick exploited considerably in \cite{Cavaglia2011}.}   since the product \(\frac{u}{  \sqrt{u^2-4g^2}}\log\frac{\tilde\mu_{12}}{\mu_{12}}\)  is purely imaginary on the whole real axis: inside the interval   \((- 2g,2g)\) -- due to the first factor, and outside of it -- due to the second factor (calculated with long cuts). The second term gives       
\begin{align}\gamma=g\sum_{a=1}^\infty \int_{-\infty}^{\infty}\frac{du}{2\pi}\left[\p_u\left(\check x^{[-0]}-\frac{1}{ \check x^{[-0]}}\right)\log\left(1+ Y_{a,0}^{[a-0]}\right)-[\p_u\left( \check x^{[+0]}-\frac{1}{\check x^{[+0]}}\right)\log\left(1+ Y_{a,0}^{[-a+0]}\right)\right]\,,   
\end{align} where we used the identity \(g\p_u\left(x-\frac{1}{x}\right)=\frac{x+\frac{1}{x}}{  x-\frac{1}{x}}=\frac{u}{  \sqrt{4g^2-u^2}}\). After shifting the integration arguments in the first and second term, respectively, by  \(\pm \bi a\), it gives the TBA formula for energy \eqref{EnergyTBA}~\footnote{ We ignored  the possible logarithmic poles related to the Bethe roots of excited states defined as \(1+\hat Y_{1,0}(u_j)=\mu_{12}(u_j+\bi/2)=0\). The corresponding logarithmic poles would produce  the driving terms \(\sum_{j}\hat p_1( u_j)\), but they are absent for the case of twisted BMN vacuum considered in this work. }.

\pagebreak

\bibliographystyle{JHEPb}
\bibliography{mybib}

\end{document}